\documentclass[runningheads,envcountsect]{llncs}

\usepackage[utf8]{inputenc}
\usepackage{lipsum}
\usepackage{hyperref}
\hypersetup{
    colorlinks=true,
    linkcolor=blue,
    filecolor=magenta,      
    urlcolor=cyan
    }
\usepackage{tabularx}
\usepackage{booktabs}
\usepackage{subcaption}
\usepackage{amsmath,amssymb}
\usepackage{mathrsfs}

\usepackage[type={CC}, modifier={by-nc-nd}, version={4.0}]{doclicense}

\usepackage{etoolbox}

\usepackage[dvipsnames]{xcolor}%

\usepackage[nomargin,inline,index,%
  status=draft 
]{fixme} 
\fxusetheme{colorsig}
\FXRegisterAuthor{fxJM}{anfxJM}{JM}
\FXRegisterAuthor{fxAB}{anfxAB}{AB}
\FXRegisterAuthor{fxNY}{anfxNY}{NY}
\FXRegisterAuthor{fxFZ}{anfxFZ}{FZ}
\FXRegisterAuthor{fxPH}{anfxPH}{PH}

\makeatletter
\DeclareFontFamily{OMX}{MnSymbolE}{}
\DeclareSymbolFont{MnLargeSymbols}{OMX}{MnSymbolE}{m}{n}
\SetSymbolFont{MnLargeSymbols}{bold}{OMX}{MnSymbolE}{b}{n}
\DeclareFontShape{OMX}{MnSymbolE}{m}{n}{
    <-6>  MnSymbolE5
   <6-7>  MnSymbolE6
   <7-8>  MnSymbolE7
   <8-9>  MnSymbolE8
   <9-10> MnSymbolE9
  <10-12> MnSymbolE10
  <12->   MnSymbolE12
}{}
\DeclareFontShape{OMX}{MnSymbolE}{b}{n}{
    <-6>  MnSymbolE-Bold5
   <6-7>  MnSymbolE-Bold6
   <7-8>  MnSymbolE-Bold7
   <8-9>  MnSymbolE-Bold8
   <9-10> MnSymbolE-Bold9
  <10-12> MnSymbolE-Bold10
  <12->   MnSymbolE-Bold12
}{}

\let\llangle\@undefined
\let\rrangle\@undefined
\DeclareMathDelimiter{\llangle}{\mathopen}%
                     {MnLargeSymbols}{'164}{MnLargeSymbols}{'164}
\DeclareMathDelimiter{\rrangle}{\mathclose}%
                     {MnLargeSymbols}{'171}{MnLargeSymbols}{'171}
\makeatother

\usepackage[inline]{enumitem}%

\usepackage{nicefrac}

\usepackage{proof}

\usepackage[most]{tcolorbox}
\newtcolorbox{myframe}[1][]{
  enhanced,
  arc=0pt,
  outer arc=0pt,
  colback=white,
  boxrule=0.5pt,
  boxsep=0mm,
  left=1mm,
  right=1mm,
  top=0.5mm,
  bottom=0.5mm,
  #1
}

\lstdefinelanguage{Scribble}{%
  basicstyle=\footnotesize\ttfamily,
  stringstyle=\color{Blue},
  showstringspaces=false,
  keywords={nested,new,calls,and,as,at,by,catches,choice,continue,do,from,global,import,instantiates,interruptible,local,module,or,par,protocol,rec,role,sig,throws,to,type,with,int,aux,reliable,crash},
  morestring=[b]",
  morestring=[b]',
  morecomment=[l][\color{greencomments}]{//},
}

\lstdefinelanguage{nuScr}{%
  basicstyle=\footnotesize\ttfamily,
  stringstyle=\color{Blue},
  showstringspaces=false,
  keywords={
    nested,new,calls,and,as,at,by,catches,choice,continue,do,from,global,import,instantiates,interruptible,local,module,or,par,protocol,rec,role,sig,throws,to,type,with,int,aux,
    safe
  },
  morestring=[b]",
  morestring=[b]',
  morecomment=[l][\color{greencomments}]{//},
  morecomment=[s][\color{magenta}]{(*}{*)},
}

\lstdefinelanguage{effpi}{
  keywords=[1]{
    case,class,sealed,abstract,object,extends,type,def,val,if,else,new,var,match
  },
  keywords=[2]{
    InChan,OutChan,RecVar,Rec,Out,In,InErr,Loop,
  },
  keywords=[3]{
    rec,send,receive,receiveErr,eval,par,Channel
  },
  keywordstyle=[1]{\color{blue}},
  keywordstyle=[2]{\color{ImperialIris}}, 
  keywordstyle=[3]{\color{OliveGreen}},
  otherkeywords={=>,.type,<:,>>:},
  morecomment=[l][\color{darkgray}]{//},
}

\usepackage{stmaryrd}
\usepackage{thmtools}
\usepackage{arydshln}

\usepackage{xifthen}
\usepackage{xspace}

\usepackage{mathtools}

\usepackage{balance}

\usepackage{hyperref}
\hypersetup{hidelinks}
\usepackage[capitalise]{cleveref}

\definecolor{ImperialBlue}{HTML}{003E74}
\definecolor{ImperialDarkGreen}{HTML}{02893B}
\definecolor{ImperialTangerine}{HTML}{EC7300}
\definecolor{ImperialIris}{HTML}{751E66}

\definecolor{RYB1}{RGB}{141, 211, 199}
\definecolor{RYB2}{RGB}{255, 255, 179}
\definecolor{RYB3}{RGB}{190, 186, 218}
\definecolor{RYB4}{RGB}{251, 128, 114}
\definecolor{RYB5}{RGB}{128, 177, 211}
\definecolor{RYB6}{RGB}{253, 180, 98}
\definecolor{RYB7}{RGB}{179, 222, 105}

\usetikzlibrary{shapes, shapes.geometric, arrows, arrows.meta, calc, decorations, decorations.pathreplacing, calligraphy, positioning, patterns, backgrounds,fit,snakes, automata}
\tikzset{
  >=stealth,
  node distance=2cm,
  every state/.style={thick, fill=gray!10},
  initial text=$ $,
}
\tikzstyle{rect} = [rectangle, rounded corners, minimum width=3cm, minimum height=1cm,text centered, draw=black, fill=black!10]
\tikzstyle{medrect} = [rectangle, rounded corners, minimum height=.7cm,text centered, draw=black, fill=black!10]
\tikzstyle{smallrect} = [rectangle, rounded corners,text centered, draw=black, fill=black!10]
\tikzstyle{circ} = [circle, minimum height=0.55cm,text centered, draw=black, fill=black!10]
\tikzstyle{smallcirc} = [circle,text centered, draw=black, fill=black!10]
\tikzstyle{arrow} = [thick,-{Latex[length=2mm, width=1.5mm]},>=stealth]
\tikzstyle{arrow2} = [thick,{Latex[length=2mm, width=1.5mm]}-{Latex[length=2mm, width=1.5mm]},>=stealth]
\tikzstyle{garrow} = [thick,-{Latex[length=2mm, width=1.5mm]},>=stealth,draw=black!40]
\tikzstyle{garrow2} = [thick,{Latex[length=2mm, width=1.5mm]}-{Latex[length=2mm, width=1.5mm]},>=stealth,draw=black!40]
\tikzstyle{venn} = [preaction={fill, #1},opacity=0.6,anchor=south,rounded corners=2pt,draw=black]
\tikzstyle{exnode} = [circle,draw=black,inner sep=1pt,fill=gray!20]
\usepackage{pgfplots}
\pgfplotscreateplotcyclelist{stackexchange}{
  {RYB1!50!black,fill=RYB1},
  {RYB4!50!black,fill=RYB4},
  {RYB3!50!black,fill=RYB3},
  {RYB2!50!black,fill=RYB2},
  {RYB5!50!black,fill=RYB5},
  {RYB6!50!black,fill=RYB6},
  {RYB7!50!black,fill=RYB7},
}
\pgfplotscreateplotcyclelist{stackexchangePlusOne}{
  {RYB4!50!black,fill=RYB4},
  {RYB3!50!black,fill=RYB3},
  {RYB2!50!black,fill=RYB2},
  {RYB5!50!black,fill=RYB5},
  {RYB6!50!black,fill=RYB6},
  {RYB7!50!black,fill=RYB7},
}
\pgfplotsset{
  compat=1.8,
  /pgfplots/bar cycle list/.style={/pgfplots/cycle list={%
    {brown!60!black,fill=brown!30!white,mark=none},
    {red,fill=red!30!white,mark=none},
    {blue,fill=blue!30!white,mark=none},
    {black,fill=gray,mark=none},
    }
  },
}
\newcolumntype{L}{>{$}l<{$}}
\newcolumntype{C}{>{$}c<{$}}
\newcolumntype{P}[1]{>{\centering\arraybackslash$}p{#1}<{$}}
\usepackage{multirow}

\usepackage[export]{adjustbox}

\Crefname{section}{\S\!}{\S\!}%
\Crefname{subsection}{\S\!}{\S\!}%
\Crefname{subsubsection}{\S\!}{\S\!}%
\Crefname{appendix}{Appendix \S\!}{Appendix \S\!}
\Crefname{definition}{Def.\@}{Defs.\@}%
\Crefname{figure}{Fig.\@}{Figs.\@}%
\Crefname{example}{Ex.\@}{Exs.\@}%
\Crefname{corollary}{Cor.\@}{Cors.\@}%
\Crefname{theorem}{Thm.\@}{Thms.\@}%
\Crefname{proposition}{Prop.\@}{Props.\@}%
\Crefname{lemma}{Lem.\@}{Lems.\@}
\Crefname{equation}{Eq.\@}{Eqs.\@}

\crefname{section}{\S\!}{\S\!}%
\crefname{subsection}{\S\!}{\S\!}%
\crefname{subsubsection}{\S\!}{\S\!}%
\crefname{appendix}{Appendix \S\!}{Appendix \S\!}
\crefname{definition}{Def.\@}{Defs.\@}%
\crefname{figure}{Fig.\@}{Figs.\@}%
\crefname{example}{Ex.\@}{Exs.\@}%
\crefname{corollary}{Cor.\@}{Cors.\@}%
\crefname{theorem}{Thm.\@}{Thms.\@}%
\crefname{proposition}{Prop.\@}{Props.\@}%
\crefname{lemma}{Lem.\@}{Lems.\@}
\crefname{equation}{Eq.\@}{Eqs.\@}

\usepackage[toc]{appendix}
\usepackage{minitoc}

\renewcommand \thepart{}
\renewcommand \partname{}

\usepackage{magicvariables}

\usepackage{lineno}

\usepackage{wrapfig}

\usepackage{orcidlink}
\newif\ifdraft%
  \draftfalse%
  \drafttrue%

\newcommand{\ifempty}[3]{%
  \ifthenelse{\isempty{#1}}{#2}{#3}%
}%

\newtoggle{techreport}%

\newtoggle{cruft}%
\renewcommand{\operatorname}[1]{\color{black}{\sf{{#1}}}}

\newcommand{\dom}[1]{{\color{black}\operatorname{dom}\!\left({#1}\right)}}%
\newcommand{\ran}[1]{{\color{black}\operatorname{ran}\!\left({#1}\right)}}%
\newcommand{\suchthat}{\colon}%
\newcommand{\gtFv}[1]{
\ifempty{#1}
{
\stFmt{\operatorname{fv}}
}
{
\stFmt{\operatorname{fv}}\!\left({#1}\right)
}}

\newcommand{\rank}[1]{\operatorname{rk}\ifempty{#1}{}{\!\left({#1}\right)}}

\newcommand{\rankRel}{\sqsubset}

\newcommand{\obsvNew}[5]{\operatorname{top}\!\left({#1},\mpChanRole{\mpChanRole{#2}{#3}}{#4}{#5}\right)}

\newcommand{\chan}[1]{\operatorname{chan}
	\ifempty{#1}{}{\!\left({#1}\right)}
}
\newcommand{\subj}[1]{\mathsf{subj}\!\left({#1}\right)}
\newcommand{\fcv}[1]{\mathsf{fcv}\ifempty{#1}{}{\!\left({#1}\right)}}
\newcommand{\fev}[1]{\mathsf{fev}\ifempty{#1}{}{\!\left({#1}\right)}}
\newcommand{\fc}[1]{\mathsf{fc}\ifempty{#1}{}{\!\left({#1}\right)}}
\newcommand{\fs}[1]{\mathsf{fs}\ifempty{#1}{}{\!\left({#1}\right)}}
\newcommand{\fpv}[1]{\mathsf{fpv}\ifempty{#1}{}{\!\left({#1}\right)}}
\newcommand{\depth}[1]{
	{\color{black}\operatorname{d}
	\ifempty{#1}{}{\!\left({#1}\right)}}
}

\newcommand{\unfoldOne}[1]{%
  {\color{black}\operatorname{unf}\!\left({#1}\right)}}%
\newcommand{\notImpliedBy}{\mathrel{{\kern .5em}{\not{\kern -1.2em}\impliedby}}}%
\newcommand{\coloncolonequals}{\Coloneqq}
\newcommand{\bnfdef}{\coloncolonequals}%
\newcommand{\bnfsep}{\mathbin{\;\Big|\;}}%
\newcommand{\bnfsepsmall}{\mathbin{\;\big|\;}}%

\newcommand{\eval}[2]{#1 \downarrow #2}

\def\eg{e.g.\@\xspace}%
\def\ie{i.e.\@\xspace}%
\definecolor{ruleColor}{rgb}{0.1, 0.3, 0.1}
\newcommand{\inferrule}[1]{{\color{ruleColor}\textsc{\scriptsize [#1]}}}%
\newcommand{\inference}[3][]{\infer[\ifempty{#1}{}{\inferrule{#1}}]{#3}{#2}}%
\newcommand{\cinference}[3][]{\infer=[\ifempty{#1}{}{\inferrule{#1}}]{#3}{#2}}%
\newcommand{\inferenceSingle}[2][]{{#2}\ifempty{#1}{}{\;\;\inferrule{#1}}}%

\newcommand{\predP}[1][]{\ifempty{#1}{\varphi}{\varphi_{#1}}}%
\newcommand{\predPi}[1][]{\ifempty{#1}{\varphi'}{\varphi'_{#1}}}%
\newcommand{\predPApp}[2][]{\ifempty{#1}{\predP}{\predP[{#1}]}\!\left({#2}\right)}%

\newcommand{\predPPApp}[2][]{\predP^{#1}\!\left({#2}\right)}%
\newcommand{\predPPiApp}[2][]{\predPi^{#1}\!\left({#2}\right)}%

\newcommand{\bind}[2]{\nicefrac{#2}{#1}}%
\newcommand{\substenum}[1]{\mathord{\left\{{#1}\right\}}}%
\newcommand{\subst}[2]{\substenum{\bind{#1}{#2}}}%
\newcommand{\substBig}[3]{\substenum{\bind{#1}{#2}\suchthat{#3}}}%
\definecolor{hlColor}{rgb}{0.65, 1.0, 0.65}%

\newcommand{\lbbar}{\{\kern-0.2em|}
\newcommand{\rbbar}{|\kern-0.2em\}}

\definecolor{tyColorCustom}{rgb}{0.0, 0.0, 0.85}%
\newcommand{\tySubst}[3]{{#1}\tyCol{\subst{#2}{#3}}}%
\newcommand{\tySubstBig}[4]{{#1}\tyCol{\substBig{#2}{#3}{#4}}}%
\newcommand{\tyCol}[1]{{\color{tyColorCustom}{#1}}}%
\newcommand{\tyFont}[1]{{#1}}%
\newcommand{\tyFmt}[1]{\tyCol{\tyFont{#1}}}%
\newcommand{\tyFontC}[1]{\operatorname{#1}}%
\newcommand{\tyFmtC}[1]{\tyCol{\tyFontC{#1}}}%

\newcommand{\tyGround}[1][]{\tyFmt{\ifempty{#1}{B}{B_{#1}}}}%
\newcommand{\tyGroundi}[1][]{\tyFmt{\ifempty{#1}{B'}{B'_{#1}}}}%
\newcommand{\tyGroundii}[1][]{\tyFmt{\ifempty{#1}{B''}{B''_{#1}}}}%
\newcommand{\tyBool}{\tyFmtC{bool}}%
\newcommand{\tyInt}{\tyFmtC{int}}%
\newcommand{\tyString}{\tyFmtC{str}}%

\newcommand{\tySub}{\mathrel{\tyCol{\leqslant}}}%
\newcommand{\tyNot}{\mathrel{\tyCol{\ntriangleleft}}}
\newcommand{\tyNotSub}{\mathrel{\tyCol{\not\leqslant}}}%
\newcommand{\tyEnvComp}{\mathpunct{\tyCol{,}}}%
\newcommand{\tyEnvEmpty}{\tyCol{\emptyset}}%
\newcommand{\tyEnv}[1][]{\tyCol{\ifempty{#1}{\Gamma}{\Gamma_{#1}}}}%
\newcommand{\tyEnvi}[1][]{\tyCol{\ifempty{#1}{\Gamma'}{\Gamma'_{#1}}}}%
\newcommand{\tyEnvii}[1][]{\tyCol{\ifempty{#1}{\Gamma''}{\Gamma''_{#1}}}}%
\newcommand{\tyEnvMap}[2]{\tyCol{{#1}{:}{#2}}}%
\newcommand{\tyEnvApp}[2]{\tyCol{{#1}({#2})}}%

\newcommand{\tyJudge}[3]{%
  {#1} \mathrel{\tyCol{\vdash}} {#2} \mathrel{\tyCol{:}} {#3}%
}%

\newcommand{\tyJudgePrefix}[3]{%
  {#1} \mathrel{\tyCol{\Vdash}} {#2} \mathrel{\tyCol{:}} {#3}%
}%

\newcommand{\tyJudgeRes}[3]{%
  {#1} \mathrel{\tyCol{\vdash^\star}} {#2} \mathrel{\tyCol{:}} {#3}%
}%

\newcommand{\muCol}[1]{{\color{red}#1}}%
\newcommand{\muFmt}[1]{\muCol{\mathsf{#1}}}%

\newcommand{\muJudge}[2]{{#1} \mathrel{\muCol{\models}} {#2}}%

\newcommand{\muVar}[1][]{\muCol{\ifempty{#1}{\muFmt{Z}}{\muFmt{Z}_{#1}}}}%
\newcommand{\muVari}[1][]{\muCol{\ifempty{#1}{\muFmt{Z}'}{\muFmt{Z}'_{#1}}}}%
\newcommand{\muVarii}[1][]{\muCol{\ifempty{#1}{\muFmt{Z}''}{\muFmt{Z}''_{#1}}}}%

\newcommand{\muPred}[1][]{\muCol{\ifempty{#1}{\phi}{\phi_{#1}}}}%
\newcommand{\muAnd}{\mathbin{\muCol{\land}}}%
\newcommand{\muTrue}{\muCol{\top}}%
\newcommand{\muGFP}[2]{\muCol{\nu{#1}\mathbin{\!.\!}{#2}}}%

\newcommand{\muLFP}[2]{\muCol{\mu{#1}\mathbin{\!.\!}{#2}}}%
\newcommand{\muOr}{\mathbin{\muCol{\lor}}}%
\newcommand{\muImplies}{\mathbin{\muCol{\Rightarrow}}}%
\newcommand{\muWordEmpty}[1][]{\muCol{\epsilon}}%
{\centerline{\bf --- Begin Copied From Previous Paper ---} \hrule}%
{\hrule \centerline{\bf --- End Copied From Previous Paper ---}}%

{\centerline{\bf --- Begin Discussion\ifempty{#1}{}{: {#1}} ---}
  \hrule\vspace{1mm}}%
{\hrule\vspace{1mm}\centerline{\bf --- End Discussion ---}}%

\newtcolorbox{cross}{blank,breakable,parbox=false,
  overlay={\draw[red,line width=5pt] (interior.south west)--(interior.north east);
    \draw[red,line width=5pt] (interior.north west)--(interior.south east);}}

\definecolor{roleColor}{rgb}{0.58,0,0.82}
\newcommand{\roleCol}[1]{{\color{roleColor}#1}}%
\newcommand{\roleSet}{\roleCol{\mathcal{R}}}%
\newcommand{\roleFmt}[1]{\ensuremath{{\boldsymbol{\roleCol{\mathtt{#1}}}}}\xspace}%

\newcommand{\roleP}[1][]{%
  \ifempty{#1}{{\color{roleColor}\roleFmt{p}}}{{\color{roleColor}\roleFmt{p}_{#1}}}%
}%
\newcommand{\rolePi}[1][]{%
  \ifempty{#1}{{\color{roleColor}\roleFmt{p}'}}{{\color{roleColor}\roleFmt{p}'_{#1}}}%
}%
\newcommand{\rolePii}[1][]{%
  \ifempty{#1}{{\color{roleColor}\roleFmt{p}''}}{{\color{roleColor}\roleFmt{p}''_{#1}}}%
}%
\newcommand{\roleQ}[1][]{%
  \ifempty{#1}{{\color{roleColor}\roleFmt{q}}}{{\color{roleColor}\roleFmt{q}_{#1}}}%
}%
\newcommand{\roleQi}[1][]{%
  \ifempty{#1}{{\color{roleColor}\roleFmt{q}'}}{{\color{roleColor}\roleFmt{q}'_{#1}}}%
}%
\newcommand{\roleQii}[1][]{%
  \ifempty{#1}{{\color{roleColor}\roleFmt{q}''}}{{\color{roleColor}\roleFmt{q}''_{#1}}}%
}%
\newcommand{\roleR}[1][]{%
  \ifempty{#1}{{\color{roleColor}\roleFmt{r}}}{{\color{roleColor}\roleFmt{r}_{\!#1}}}%
}%
\newcommand{\roleRi}[1][]{%
  \ifempty{#1}{{\color{roleColor}\roleFmt{r}'}}{{\color{roleColor}\roleFmt{r}'_{\!#1}}}%
}%
\newcommand{\roleRii}[1][]{%
  \ifempty{#1}{{\color{roleColor}\roleFmt{r}''}}{{\color{roleColor}\roleFmt{r}''_{\!#1}}}%
}%
\newcommand{\roleS}[1][]{%
  \ifempty{#1}{{\color{roleColor}\roleFmt{s}}}{{\color{roleColor}\roleFmt{s}_{\!#1}}}%
}%
\definecolor{gtColor}{rgb}{0.43, 0.21, 0.1}
\newcommand{\gtFmt}[1]{\ensuremath{{\color{gtColor}#1}}\xspace}%
\newcommand{\gtMsgFmt}[1]{\gtFmt{\labFmt{#1}}}%
\newcommand{\gtLab}[1][]{%
  \ifempty{#1}{\gtMsgFmt{m}}{{\color{gtColor}\gtMsgFmt{m}_{#1}}}%
}%

\newcommand{\gtG}[1][]{\gtFmt{\ifempty{#1}{G}{G_{#1}}}}%
\newcommand{\gtGi}[1][]{\gtFmt{\ifempty{#1}{G'}{G'_{#1}}}}%
\newcommand{\gtGii}[1][]{\gtFmt{\ifempty{#1}{G''}{G''_{#1}}}}%

\newcommand{\gtSeq}{\mathbin{\gtFmt{.}}}%

\newcommand{\gtRoles}[1]{{\ifempty{#1}{\color{roleColor}\operatorname{roles}}{{\color{roleColor}\operatorname{roles}}({#1})}}}%

\newcommand{\guarded}[2]{{
{\color{roleColor} \operatorname{guarded}}
\ifempty{#1}{}{
({#1},{#2})}
}}%

\newcommand{\gtMove}[1][\phantom{\stEnvAnnotGenericSym}]{\gtFmt{\xrightarrow{#1}}} 
\newcommand{\gtMoveStar}[1][]{\ifempty{#1}{\gtMove[]{}^{\!\gtFmt{*}}}{\gtMove[]{#1}^{\!\gtFmt{*}}}} 
\newcommand{\labFmt}[2][]{\ensuremath{\ifempty{#1}{\mathtt{#2}}{\mathtt{#2}\textsubscript{#1}}}\xspace}%

\definecolor{mergeColor}{rgb}{0.6, 0.0, 0.0}
\definecolor{exprColor}{rgb}{0.6, 0.0, 0.0}
\definecolor{projColor}{rgb}{0.54, 0.15, 0.9}
\definecolor{stColor}{rgb}{0, 0, 0.9}
\newcommand{\stFmt}[1]{\ensuremath{{\color{stColor}#1}}\xspace}%

\definecolor{compColor}{rgb}{0, 0, 0.9}

\newcommand{\stChoice}[2]{\stLabFmt{#1}\ifempty{#2}{}{({#2})}}%

\newcommand{\stSum}[4]{\stFmt{\Sigma_{#2}\!\left\{\roleFmt{#1}\stFmt{{#4}{#3}}\right\}}}

\newcommand{\stSeq}{\mathbin{\!\stFmt{.}\!}}%
\newcommand{\stIntSum}[3]{\roleFmt{#1}\stFmt{\oplus\!\left\{#3\right\}_{#2}}}%
\newcommand{\stExtSum}[3]{\roleFmt{#1}\stFmt{\&\!\left\{#3\right\}_{#2}}}%
\newcommand{\stRec}[2]{\stFmt{\mu{#1}.{#2}}}%
\newcommand{\stEnd}{\stFmt{\mathbf{end}}}%

\newcommand{\stLabFmt}[1]{\stFmt{\labFmt{#1}}}%
\newcommand{\stLab}[1][]{%
  \ifempty{#1}{\stLabFmt{m}}{\stLabFmt{m}_{{\color{stColor}#1}}}
}%
\newcommand{\stLabi}[1][]{%
  \ifempty{#1}{\stLabFmt{m'}}{\stLabFmt{m'}_{{\color{stColor}#1}}}
}%
\newcommand{\stLabii}[1][]{%
  \ifempty{#1}{\stLabFmt{m''}}{\stLabFmt{m''}_{{\color{stColor}#1}}}
}%
\newcommand{\stLabiii}[1][]{%
  \ifempty{#1}{\stLabFmt{m'''}}{\stLabFmt{m'''}_{{\color{stColor}#1}}}
}%

\newcommand{\stS}[1][]{\stFmt{\ifempty{#1}{S}{S_{#1}}}}%
\newcommand{\stSi}[1][]{\stFmt{\ifempty{#1}{S'}{S'_{#1}}}}%
\newcommand{\stSii}[1][]{\stFmt{\ifempty{#1}{S''}{S''_{#1}}}}%
\newcommand{\stSiii}[1][]{\stFmt{\ifempty{#1}{S'''}{S'''_{#1}}}}%

\newcommand{\stT}[1][]{\stFmt{\ifempty{#1}{T}{T_{#1}}}}%
\newcommand{\stTi}[1][]{\stFmt{\ifempty{#1}{T'}{T'_{#1}}}}%
\newcommand{\stTii}[1][]{\stFmt{\ifempty{#1}{T''}{T''_{#1}}}}%
\newcommand{\stTiii}[1][]{\stFmt{\ifempty{#1}{T'''}{T'''_{#1}}}}%
\newcommand{\stRecVarBase}{\stFmt{\mathbf{t}}}%
\newcommand{\stRecVar}[1][]{\stFmt{\ifempty{#1}{\stRecVarBase}{\stRecVarBase_{#1}}}}%
\newcommand{\stRecVari}[1][]{\stFmt{\ifempty{#1}{\stRecVar'}{\stRecVar'_{#1}}}}%
\newcommand{\stRecVarii}[1][]{\stFmt{\ifempty{#1}{\stRecVar''}{\stRecVar''_{#1}}}}%

\newcommand{\stBinMergeStar}{\mathbin{\setbox0\hbox{\stFmt{\sqcap}}\rlap{\hbox to \wd0{\hss $\star$\hss}}\box0}}%

\newcommand{\ltsSend}[5]{\mpChanRole{#1}{#2}{!}\roleFmt{#3}\mathord{:}\labFmt{#4}\stFmt{({#5})}}

\newcommand{\ltsSendRecv}[3]{{#1}{#2}\mathord{:}#3}
\newcommand{\ltsSendRecvS}[4]{\mpFmt{{#1}[{#2}][{#3}]}\mathord{:}{#4}}

\newcommand{\ltsRecv}[5]{\mpChanRole{#1}{#2}?\roleFmt{#3}\mathord{:}\labFmt{#4}\stFmt{({#5})}}

\newcommand{\ltsSubject}[1]{{\color{roleColor} \operatorname{sbj}({#1})}}%
\definecolor{mpColor}{rgb}{0, 0, 0}
\newcommand{\mpFmt}[1]{{\color{mpColor}#1}}%

\newcommand{\mpLabFmt}[1]{\mpFmt{\labFmt{#1}}}%

\newcommand{\mpNat}{\mpFmt{\text{\texttt{n}}}}%

\newcommand{\mpNum}[1]{\mpFmt{\text{\texttt{#1}}}}

\newcommand{\mpTrue}{\mpFmt{\text{\texttt{true}}}}%
\newcommand{\mpFalse}{\mpFmt{\text{\texttt{false}}}}%
\newcommand{\mpSucc}[1]{\text{\texttt{succ}}({#1})}%
\newcommand{\mpNeg}[1]{\text{\texttt{neg}}({#1})}%

\newcommand{\mpChanRole}[2]{{{#1}\mpFmt{[}{#2}\mpFmt{]}}}%

\newcommand{\mpNil}{\mpFmt{\mathbf{0}}}%
\newcommand{\mpSeq}{\mathbin{\mpFmt{\!.\!}}}%
\newcommand{\mpIf}[3]{%
  \mpFmt{\mathsf{if}\,{#1}\,\mathsf{then}\,{#2}\,\mathsf{else}\,{#3}}%
}%
\newcommand{\mpChoice}[3]{%
  \mpFmt{%
    \mpLabFmt{#1}\ifempty{#2}{}{({#2})}\ifempty{#3}{}{\vphantom{x}\mpSeq {#3}}%
  }%
}%
\newcommand{\mpChoiceNoBind}[3]{%
  \mpFmt{%
    \mpLabFmt{#1}\ifempty{#2}{}{\langle{#2}\rangle}\ifempty{#3}{}{\vphantom{x}\mpSeq {#3}}%
  }%
}%
\newcommand{\mpBranchSingle}[5]{%
  \mpFmt{%
    {#1}[{#2}] \mathbin{\!\ifempty{\sum}{\sum}{?}\!}%
    \mpChoice{#3}{#4}{#5}
  }%
}%
\newcommand{\mpBra}[5]{\mpBranchSingle{#1}{#2}{#3}{#4}{#5}}
\newcommand{\mpSel}[5]{%
  \mpFmt{%
    {#1}[{#2}] \mathbin{\! ! \!}%
    \mpChoiceNoBind{#3}{#4}{#5}%
  }%
}%

\newcommand{\mpPrefix}{\mpFmt{\ensuremath{\pi}}\xspace}

\newcommand{\mpSum}[2]{\mpFmt{\sum_{#2} {#1}}}

\newcommand{\mpPar}{\mathbin{\mpFmt{\mid}}}%
\newcommand{\mpBigPar}[2]{\mathbin{\mpFmt{\Pi_{#1}}{#2}}}%
\newcommand{\mpRes}[2]{\mpFmt{\left(\mathbf{\nu}{#1}\right){#2}}}%
\newcommand{\mpRec}[2]{\mpFmt{\mu{#1}.{#2}}}%

\newcommand{\mpErr}{\mpFmt{\boldsymbol{\mathtt{err}}}}%

\newcommand{\mpCtx}[1][]{\mpFmt{\ifempty{#1}{\mathbb{C}}{\mathbb{C}_{#1}}}}%
\newcommand{\mpCtxi}[1][]{\mpFmt{\ifempty{#1}{\mathbb{C}'}{\mathbb{C}'_{#1}}}}%
\newcommand{\mpCtxHole}{[\,]}%
\newcommand{\mpCtxApp}[2]{{#1}\!\left[{#2}\right]}%

\makeVars{mpE}{mpFmt}{e}

\newcommand{\mpC}[1][]{\mpFmt{\ifempty{#1}{c}{c_{#1}}}}%
\newcommand{\mpCi}[1][]{\mpFmt{\ifempty{#1}{c'}{c'_{#1}}}}%
\newcommand{\mpCii}[1][]{\mpFmt{\ifempty{#1}{c''}{c''_{#1}}}}%
\newcommand{\mpS}[1][]{\mpFmt{\ifempty{#1}{s}{s_{#1}}}}%
\newcommand{\mpSi}[1][]{\mpFmt{\ifempty{#1}{s'}{s'_{#1}}}}%
\newcommand{\mpSii}[1][]{\mpFmt{\ifempty{#1}{s''}{s''_{#1}}}}%
\newcommand{\mpSiii}[1][]{\mpFmt{\ifempty{#1}{s'''}{s'''_{#1}}}}%

\newcommand{\mpx}[1][]{\mpFmt{\ifempty{#1}{x}{x_{#1}}}}%
\newcommand{\mpxi}[1][]{\mpFmt{\ifempty{#1}{x'}{x'_{#1}}}}%
\newcommand{\mpxii}[1][]{\mpFmt{\ifempty{#1}{x''}{x''_{#1}}}}%
\newcommand{\mpxiii}[1][]{\mpFmt{\ifempty{#1}{x'''}{x'''_{#1}}}}%
\newcommand{\mpy}[1][]{\mpFmt{\ifempty{#1}{y}{y_{#1}}}}%
\newcommand{\mpyi}[1][]{\mpFmt{\ifempty{#1}{y'}{y'_{#1}}}}%
\newcommand{\mpyii}[1][]{\mpFmt{\ifempty{#1}{y''}{y''_{#1}}}}%
\newcommand{\mpz}[1][]{\mpFmt{\ifempty{#1}{z}{z_{#1}}}}%
\newcommand{\mpX}[1][]{\mpFmt{\ifempty{#1}{X}{X_{#1}}}}%
\newcommand{\mpXi}[1][]{\mpFmt{\ifempty{#1}{X'}{X'_{#1}}}}%
\newcommand{\mpP}[1][]{\mpFmt{\ifempty{#1}{P}{P_{#1}}}}%
\newcommand{\mpPi}[1][]{\mpFmt{\ifempty{#1}{P'}{P'_{#1}}}}%
\newcommand{\mpPii}[1][]{\mpFmt{\ifempty{#1}{P''}{P''_{#1}}}}%
\newcommand{\mpPiii}[1][]{\mpFmt{\ifempty{#1}{P'''}{P'''_{#1}}}}%
\newcommand{\mpQ}[1][]{\mpFmt{\ifempty{#1}{Q}{Q_{#1}}}}%
\newcommand{\mpQi}[1][]{\mpFmt{\ifempty{#1}{Q'}{Q'_{#1}}}}%
\newcommand{\mpQii}[1][]{\mpFmt{\ifempty{#1}{Q''}{Q''_{#1}}}}%
\newcommand{\mpQiii}[1][]{\mpFmt{\ifempty{#1}{Q'''}{Q'''_{#1}}}}%
\newcommand{\mpR}[1][]{\mpFmt{\ifempty{#1}{R}{R_{#1}}}}%
\newcommand{\mpRi}[1][]{\mpFmt{\ifempty{#1}{R'}{R'_{#1}}}}%
\newcommand{\mpRii}[1][]{\mpFmt{\ifempty{#1}{R''}{R''_{#1}}}}%
\newcommand{\mpRiii}[1][]{\mpFmt{\ifempty{#1}{R'''}{R'''_{#1}}}}%

\newcommand{\mpMove}{\to}%
\newcommand{\mpMoveNot}{\not\to}%

\newcommand{\mpMoveCommS}[5]{\xrightarrow{\ltsSendRecvS{#1}{#2}{#3}{\stChoice{#4}{#5}}}}
\newcommand{\mpMoveTau}{\xrightarrow{\tau}}
\newcommand{\mpMoveTauStar}{\xrightarrow{\tau}^{\!\!\!*}}
\newcommand{\mpMoveErr}{\xrightarrow{\mpErr}}
\newcommand{\mpMoveGen}{\xrightarrow{\stEnvAnnotGenericSym}}
\newcommand{\mpMoveStar}{\mathrel{\mpMove{}^{\!\!\!*}}}%
\iffalse
\newcommand{\stEnv}[1][]{\stFmt{\ifempty{#1}{\Gamma}{\Gamma_{#1}}}}%
\newcommand{\stEnvi}[1][]{\stFmt{\ifempty{#1}{\Gamma'}{\Gamma'_{#1}}}}%
\newcommand{\stEnvii}[1][]{\stFmt{\ifempty{#1}{\Gamma''}{\Gamma''_{#1}}}}%
\newcommand{\stEnviii}[1][]{\stFmt{\ifempty{#1}{\Gamma'''}{\Gamma'''_{#1}}}}%
\fi

\newcommand{\stEnv}[1][]{\stFmt{\ifempty{#1}{\Delta}{\Delta_{#1}}}}%
\newcommand{\stEnvi}[1][]{\stFmt{\ifempty{#1}{\Delta'}{\Delta'_{#1}}}}%
\newcommand{\stEnvii}[1][]{\stFmt{\ifempty{#1}{\Delta''}{\Delta''_{#1}}}}%
\newcommand{\stEnviii}[1][]{\stFmt{\ifempty{#1}{\Delta'''}{\Delta'''_{#1}}}}%
\newcommand{\stEnvEmpty}{\stFmt{\emptyset}}%
\newcommand{\stEnvMap}[2]{\stFmt{\mpFmt{#1}\mathbin{:}{#2}}}%
\newcommand{\stEnvComp}{\mathpunct{\stFmt{,}}}%
\newcommand{\stEnvApp}[2]{\stFmt{#1\!\left(\mpFmt{#2}\right)}}%

\newcommand{\stEnvNew}[1][]{\stFmt{\ifempty{#1}{\Delta}{\Delta_{#1}}}}%
\newcommand{\stEnvNewi}[1][]{\stFmt{\ifempty{#1}{\Delta'}{\Delta'_{#1}}}}%

\newcommand{\assocs}[1]{\stFmt{\sqsubseteq_s}}
\newcommand{\assoca}[1]{\stFmt{\sqsubseteq_a}}

\newcommand{\stEnvAnnotOutSym}{\stFmt{!}}%
\newcommand{\stEnvAnnotQSym}{\stFmt{\dagger}}%
\newcommand{\stEnvAnnotInSym}{\stFmt{?}}%
\newcommand{\stEnvAnnotGenericSym}[1][]{\stFmt{\ifempty{#1}{\alpha}{\alpha_{#1}}}}%
\newcommand{\stEnvAnnotGenericSymi}[1][]{\stFmt{\ifempty{#1}{\alpha'}{\alpha'_{#1}}}}%
\newcommand{\stEnvAnnotGenericSymii}[1][]{\stFmt{\ifempty{#1}{\alpha''}{\alpha''_{#1}}}}%

\newcommand{\stEnvEndPred}{\operatorname{end}}%
\newcommand{\stEnvEndP}[1]{\stEnvEndPred(\stFmt{#1})}%

\newcommand{\stEnvDFPred}{\operatorname{df}}%
\newcommand{\stEnvDFP}[1]{\stEnvDFPred(\stFmt{#1})}%

\newcommand{\stEnvLivePred}{\operatorname{live}}%
\newcommand{\stEnvLiveP}[1]{\stEnvLivePred(\stFmt{#1})}%
\newcommand{\RC}{{rc}\xspace}%
\newcommand{\stEnvSafePred}{\operatorname{safe}}%
\newcommand{\stEnvSafeP}[1]{\stEnvSafePred(\stFmt{#1})}%
\newcommand{\quEmpty}[1][]{\tyCol{\tyFont{\varnothing}}}%
\newcommand{\set}[1]{\left\{#1\right\}}

\newcommand{\act}[1]{\texttt{act}(#1)}

\newcommand{\myparagraph}[1]{\noindent\textbf{#1}}

\newcommand{\pprefix}{\ensuremath{\pi}\xspace}

\newcommand{\choice}[2]{\ensuremath{\sum_{#2} #1}\xspace}

\newcommand{\prestruct}{\Rrightarrow}

\newcommand{\anonSubst}[1]{\stFmt{\operatorname{subst}}\left({#1}\right)}

\newcommand{\interpret}[1]{
	\left\llbracket{#1}\right\rrbracket
}

\newcommand{\charP}[2]{\mathcal{P}\!\left[{#1},{#2}\right]}

\newcommand{\charPM}[1]{\mathcal{P}\!\left[{#1}\right]}

\makeatletter
\newcommand{\rDelta@scaling}{0.775}\newcommand{\rDelta@kern}{0.3}

\newcommand{\rDelta}{\mathord{\mathpalette\rDelta@\relax\Delta}}
\newcommand{\rDelta@}[2]{%
  \rlap{\scalebox{\rDelta@scaling}{$\m@th#1\mkern \rDelta@kern mu\Delta$}}%
}
\makeatother

\newcommand{\compEnv}[2]{\stFmt{\rDelta}\!\left[{#1},{#2}\right]}

\newcommand{\syncSel}[2]{\stFmt{\operatorname{sync!}}\!\left({#1},{#2}\right)}

\newcommand{\syncBra}[2]{\stFmt{\operatorname{sync?}}\!\left({#1},{#2}\right)}

\newcommand{\syncSch}[2]{\stFmt{\operatorname{sync}}\!\left({#1},{#2}\right)}

\newcommand{\charTAct}[4]{\stFmt{\operatorname{pfm}\!\left({#1},{#2},{#3},{#4}\right)}}

\newcommand{\charTActii}[4]{\stFmt{\operatorname{pfm'}\!\left({#1},{#2},{#3},{#4}\right)}}

\newcommand{\charTActi}[2]{\stFmt{\operatorname{det}\!\left({#1},{#2}\right)}}

\newcommand{\charTwait}[1]{\stFmt{\operatorname{wait}\!\left({#1}\right)}}

\newcommand{\charSch}[1]{\stFmt{\operatorname{sch}\!\left({#1}\right)}}

\newcommand{\charSchTest}[3]{\stFmt{\operatorname{test}\!\left({#1},{#2},
\ifempty{#3}{\right.}{{#3}\right)}
}}

\newcommand{\charSchTrigger}[4]{\stFmt{\operatorname{trigger}\!\left({#1},{#2},{#3},
\ifempty{#4}{\right.}{{#4}\right)}
}}

\newcommand{\charSchTriggeri}[4]{\stFmt{\operatorname{trigger'}\!\left({#1},{#2},{#3},
\ifempty{#4}{\right.}{{#4}\right)}
}}

\newcommand{\barb}[1]{\operatorname{barb}\!\left({#1}\right)}
\newcommand{\obs}[1]{\operatorname{obs}\!\left({#1}\right)}

\newcommand{\reach}[3]{\operatorname{R}\!\left({#1},{#2},{#3}\right)}

\newcommand{\reachi}[3]{\operatorname{R'}\!\left({#1},{#2},{#3}\right)}

\newcommand{\reachii}[4]{\operatorname{R''}\!\left({#1},{#2},{#3},{#4}\right)}

\newcommand{\argmin}{\operatorname{argmin}}

\newcommand{\mcmp}{{\small\sf{MCM}}$\pi$\xspace}

\newcommand{\textbnf}[1]{\text{\emph{\footnotesize{#1}}}}

\newcommand{\preCalc}{\prec}
\newcommand{\preType}{\mathrel{\stFmt{\prec}}}

\newcommand{\prefix}[1]{\preCalc{#1}}
\newcommand{\threeparticipantlock}[2]{\mathsf{lock}_{#2}\!\left({#1}\right)}

\newcommand{\MC}{\mathit{MC}}
\newcommand{\BI}{\mathit{B}}
\newcommand{\MS}{\mathit{MS}}
\newcommand{\SC}{\mathit{SC}}
\newcommand{\DM}{\mathit{DM}}
\newcommand{\SE}{\mathit{S}}

\newcommand{\hideapp}[1]{{#1}}

\begin{document}


\title{Mixed Choice Multiparty Session Types, Precisely}
\titlerunning{Mixed Choice Multiparty Session Types, Precisely}

\author{Jake Masters\orcidlink{0000-0002-3783-6149}
\and
Nobuko Yoshida\orcidlink{0000-0002-3925-8557}}
\authorrunning{Jake Masters and Nobuko Yoshida}

\institute{University of Oxford, UK\\\email{\{jake.masters,nobuko.yoshida\}@cs.ox.ac.uk}}

\pagestyle{headings}

\maketitle

\begin{abstract}
A \emph{precise} (sound and complete) subtyping
relation $\tySub$ specifies that $\stTi$ is a subtype 
of $\stT$ if and only if 
a program of type $\stTi$ can
always safely replace a program of type $\stT$ without compromising
the safety of a larger program.
This paper formulates and proves preciseness of subtyping for 
\emph{mixed choice multiparty session types} with
\emph{session delegation}, \emph{creation}, and \emph{interleaving}.
We prove soundness by
developing the first general type system for a
full mixed choice multiparty session $\pi$-calculus. 
To prove completeness,
we introduce the \emph{three-party lock}, 
which is a minimal and general form of liveness
error for 
handling interleaved sessions, and
we establish a new proof technique based on a construction 
of \emph{scheduler processes} which enable exhaustive 
detection for all failures of the subtyping relation.
We then extend the preciseness results
to a family of mixed choice multiparty session types. 
Algorithms for checking (1)  subtyping and 
(2) safety, deadlock-freedom,
and liveness of typing contexts are
fully implemented and optimised to run 
in quadratic time with respect to 
the size of the state space and typing context, and
have been evaluated with (mixed choice) case studies from the literature.
\end{abstract}

\section{Introduction}
\label{sec:intro}

\smallskip 

\myparagraph{Precise Subtyping.\ }
Liskov and Wing's substitution principle \cite{10.1145/197320.197383}
defines the notion of \emph{subtyping} 
by specifying when one type safely substitutes for another, \ie,   
if type $\stTi$ is a subtype of $\stT$ (denoted by $\stTi\tySub\stT$), 
then a program $\mpP$ of type $\stTi$
can always be used in place of $\mpPi$ of type $\stT$.
This principle is integrated into 
a typing system usually with a \emph{subsumption rule}: if
program $\mpP$ is typed by $\stTi$ 
and $\stTi\tySub\stT$, then $\mpP$ is typed by $\stT$. 
We then wish our type system to be \emph{maximally expressive}, 
while remaining \emph{sound}.  
Larger subtyping relations empower the subsumption rule,
enlarging the set of typable programs, 
but adding too many relations to the subtyping relation 
makes the type system unsound. For instance, 
if we let $\tyBool\tySub\tyInt$, then we
mistakenly type the expression $\mpTrue+\mpNum{5}$, 
which fails to evaluate and causes a run-time error.

For a type system ($\stFmt{\vdash}$),
write
$\mpCtx[\stT]$ for a
context
such that
$\tyJudge{\set{\tyEnvMap{\mpX}{\set{\stEnvMap{\mpC}{\stT}}}}}{\mpCtxApp{\mpCtx[\stT]}{X}}{\stEnvEmpty}$
for channel $\mpC$ and variable $X$.  
Subtyping relation ($\tySub$) is \emph{precise} 
if: 
\begin{enumerate}
\item (\emph{sound}) using $\tySub$ for subsumption
ensures that typed processes are correct:
$\stTi\tySub \stT$ implies that
if $\tyJudge{}{\mpP}
{\set{\stEnvMap{\mpC}{\stTi}}}$,  then $\mpCtxApp{\mpCtx[\stT]}{\mpP}$ 
does not behave badly; 
\item (\emph{complete})
it cannot be extended without becoming unsound: 
$\stTi \tyNotSub \stT$ implies that for some $\mpCtx[\stT]$
and $\mpP$, we have
$\tyJudge{}{\mpP}
{\set{\stEnvMap{\mpC}{\stTi}}}$ and $\mpCtxApp{\mpCtx[\stT]}{\mpP}$ behaves badly. 
\end{enumerate}
Preciseness is used to \emph{test} the canonicity 
of subtyping with respect to the operational
semantics of a target language.
The problem of determining precise subtyping relations
has been widely studied for various typed $\lambda$-calculi
\cite{Barendregt1983-qz,van2000minimal,Ishihara2002-vx,Vouillon2004-lu,Dezani-Ciancaglini2014-iz,10.1145/2994596},
and later for \emph{session types} \cite{Honda1998,Honda1993,Takeuchi1994},
where subtyping plays an essential role in their integration into programming languages, beginning with the initial work on Java
\cite{HYH2008}.
Preciseness was first studied 
for synchronous and asynchronous binary session types 
\cite{CDSY2017} for a calculus with session delegations.

\smallskip 

\myparagraph{Mixed Choice Multiparty Session Types. \ } 
\emph{Multiparty Session Types} (MPST) 
\cite{HYC2008,HYC2016,DBLP:conf/icdcit/YoshidaG20,CDPY2015}
are a typing disciple for message passing processes 
that are distributed across multiple participants.
Typing ensures adherence to a specified protocol,  
used to guarantee \emph{communication safety} (type and choice safety), 
\emph{deadlock-freedom} (no process becomes stuck during execution) and   
\emph{liveness} (a communication, which is ready, eventually happens) 
of concurrent and distributed systems. 
MPST has been integrated in over 26 programming languages,
adapted to static 
or dynamic checking, type inference and code generation
\cite{Y2024}. 

An extension of MPST,  
\emph{mixed choice multiparty session types} \cite{PY2024}
provide \emph{non-deterministic choices} between 
sets of inputs and outputs directed to multiple participants.
Mixed choice in MPST 
\emph{strictly} increases the expressivity
of typable protocols \cite{PY2024}.
Although the original $\pi$-calculus \cite{Milner1992-rk}
included non-deterministic choice,
the study of
mixed choice MPST
is a recent development,
and its need is growing.  
Recent work
on mixed choice multiparty session types include
probability \cite{DBLP:conf/ictac/BlechschmidtPN25}, modelling 
federated learning protocols \cite{PPGSY2025},
automata-based analyses \cite{Giusto2025}, 
and integration into a functional language \cite{10.1145/3798224} and  
Erlang \cite{DBLP:journals/corr/abs-2602-23927}. 

\smallskip 

\myparagraph{Precise Mixed Choice Multiparty Session Subtyping.\ }
Preciseness of subtyping has been studied 
for synchronous and asynchronous 
MPST \cite{GJPSY2018,GPPSY2023}, but 
two constructions remain untreated: (1) \cite{GJPSY2018,GPPSY2023} are limited to calculi without session delegation, creation, or interleaving;
and (2) \cite{GJPSY2018,GPPSY2023} do not treat 
mixed choice. 
This paper tackles these two key constructions, determining the precise
subtyping relation $\tySub$ for mixed choice multiparty synchronous
session types with session delegation, creation, and interleaving. 
We consider the subtyping first presented in \cite{PY2024} 
and its restrictions to a family of subcalculi.

While soundness immediately follows
from type safety, 
proving completeness with these two extensions is
non-trivial since the previous methods in 
\cite{CDSY2017,GJPSY2018,GPPSY2023} are not applicable. 
To explain this problem,
we give a general strategy (in a contrapositive form)
for proving completeness. 

\begin{itemize}
	\item\textbf{[Step 1]}
	For each type $\stT$, define
	a complementary typing context $\stEnv[\stT]$
	to schedule the execution of $\stT$.
	\item\textbf{[Step 2]}
	Define
	a $\stT$-typed \emph{characteristic process} $\mpP[\stT]$
	and
	a \emph{characteristic context} $\mpCtx[\stT]$ which
        behaves well when filled with
	$\stT$-typed processes. 
	\item\textbf{[Step 3]}
	Characterise the negation of the subtyping relation.
	\item\textbf{[Step 4]}
	Show that if $\stTi\tyNotSub\stT$, 
	then $\mpCtx[\stT][\mpP]$ behaves badly. 
\end{itemize}

\smallskip 

\myparagraph{The first challenge} is to define what ``behaving badly'' means.
In the binary case \cite{CDSY2017}, simple error rules 
are sufficient 
because each characteristic process 
is composed with a characteristic context which has a dual channel type. 
``Bad behaviours'' in \cite{GJPSY2018,GPPSY2023} are
``deadlock'' and ``livelock'' for a \emph{single} multiparty session,  
neither of which are useful for a full session $\pi$-calculus 
since there is no standard typing system
which can ensure deadlock-freedom
or liveness for interleaved sessions. 
In this paper, we solve this issue by
introducing a minimal but general
liveness error, 
called a \emph{three-party lock}. 
Here ``minimal'' refers to the number of participants engaging with a lock, and
``general''
means that it captures all livelock patterns induced by the
complementary typing context's scheduling.
Thus interactions between characteristic processes and contexts
of incompatible types, induce noticeably ``bad behaviour''.
A three-party lock occurs when each participant is prevented
from communicating with another participant due to
circular dependency. 
Process $\mpP$ below is a simple example
of a three-party lock: 
$\mpP=
	\mpBra{\mpChanRole{\mpS}{\roleP[0]}}{\roleP[1]}{}{\mpx}{\mpNil}
	\mpPar
	\mpBra{\mpChanRole{\mpS}{\roleP[1]}}{\roleP[2]}{}{\mpx}{\mpNil}
	\mpPar
	\mpBra{\mpChanRole{\mpS}{\roleP[2]}}{\roleP[0]}{}{\mpx}{\mpNil}$ 
where 
$\mpChanRole{\mpS}{\roleP}$ denotes
session $\mpS$ with participant $\roleP$ and 
$?$ denotes input.
In $\mpP$,  $\roleP[0]$ is dependent on $\roleP[1]$ 
to communicate, $\roleP[1]$ on $\roleP[2]$, and
$\roleP[2]$ on $\roleP[0]$.
Later we prove that the 2-party lock used in \cite{CDSY2017}
is \emph{insufficient} as an error 
to prove completeness for 
multiparty mixed choice,
highlighting the difference between binary and multiparty. 

\smallskip 

\myparagraph{The second challenge} is to define
characteristic processes and contexts. 
Mixed choice allows subtyping to fail in
many more ways than with separate choice (where choice is either input or
output) in \cite{CDSY2017,GJPSY2018,GPPSY2023}.
Resultantly,
the size of characteristic contexts grows
exponentially in the size of their types. 
To capture these failures and avoid the complexity of reasoning, 
we build \emph{complementary contexts},
$\compEnv{\mpS}{\stT}$ for each type $\stT$,  
with a \emph{scheduler process} to monitor all possible failures of subtyping
non-deterministically.
The scheduler keeps track of the expected shape of $\stT$
during execution,
and non-deterministically controls the other participants
to execute $\stT$ and
test for incorrect behaviours,
forcing errors and three-party locks
if any are found.
We use complementary contexts to construct
characteristic processes and contexts (see \cref{ex:compl-types} and \cref{ex:char-proc}). 

\smallskip 

\myparagraph{Contributions.\ }
In this work, we
study the preciseness of
the mixed choice multiparty subtyping relation $\tySub$
(\cref{def:subtyping}).
First, we introduce \emph{three-party locks}
for both processes (\cref{def:three-lock}) and typing contexts (\cref{def:three-lock-type}).
We show that $\tySub$ in \cite{PY2024} is 
a canonical subtyping for mixed choice multiparty session types,
proving 
\emph{operational} (\cref{thm:operational-precise}) and
\emph{denotational
preciseness} (\cref{thm:precise-denote}) 
with respect to \emph{safety} and \emph{three-party lock-freedom} in
the mixed choice multiparty $\pi$-calculus
(\cref{def:full-calc-def}).
This is the first preciseness result involving, not only mixed choice,
but also session delegation in a multiparty session $\pi$-calculus.

We base our typing system on the bottom-up approach
\cite{POPL19LessIsMore} instead of using the top-down approach 
starting from global protocols \cite{GJPSY2018}. 
A key technique to prove completeness is constructing 
\emph{complementary contexts} (\cref{def:char-ctx}), 
which exhaustively check for the greater variety of failures of
subtyping. 

Our approach is generalised
to a family of calculi defined in 
\cite{PY2024}
(\cref{def:mixed-subsystem,thm:operational-precise-sub}),
including 
the calculus in \cite{GJPSY2018}   
(\cref{thm:operational-precise-live}),  
demonstrating extensibility of our approach
to various MPST calculi with/without mixed choice. 

Further, we provide new and efficient algorithms
with implementations,
to decide safety, deadlock-freedom, and liveness of typing contexts. 
Their complexities are quadratic
in the
state space and context, 
an improvement over the
exponential
algorithm \cite[Theorem 6.20]{UY2025}
and model checking
approach \cite[\S~6]{POPL19LessIsMore}.

We have identified and corrected small but crucial errors in several 
existing work \cite{PY2024,POPL19LessIsMore,GJPSY2018}\hideapp{, which are
reported in \cref{rem:subtyping-trivial}, \cref{prop:empty-ctx-safe}
and \cref{app:1-level-fails}}. 
Additionally,
to our best knowledge,
this paper offers the first typing system which
guarantees communication safety, deadlock-freedom and liveness  
in the presence of mixed choice and session delegations. 

\smallskip 

The appendices contain detailed proofs of all the statements in
this paper and further explanations and examples.
An implementation of the algorithms and benchmarks
in \cref{sec:liveness-algo} 
is provided in the repository \cite{anonymous_2026_20428250}.

\section{Mixed Choice Multiparty Session Calculus}
\label{sec:calc}
This section defines the mixed choice multiparty session
$\pi$-calculus (\mcmp),   
extending
the calculus with delegation and hiding in \cite{POPL19LessIsMore} with mixed choice
\cite{PY2024}.
After defining properties of processes, we introduce the notion of 
a \emph{three-party lock} for processes, 
which plays a key role in the proof of preciseness. 

\smallskip 
\myparagraph{Syntax and Semantics. \ }
We first define the syntax of the \mcmp-calculus. 

\begin{definition}[Mixed Choice Multiparty $\pi$-Calculus]
\label{def:full-calc-def}
\rm 
The syntax of the \emph{mixed choice multiparty $\pi$-calculus} (\mcmp)
is defined by:\\[1mm]
{\small
$
\begin{array}{ll}
v\bnfdef\mpx{} \bnfsep \mpNat
\bnfsep\mpTrue
\bnfsep\mpFalse
& \textbnf{(variable, nat, bools)}\\[1mm]
e\bnfdef v
 	\bnfsep e = e'
 	\bnfsep e < e'
 	\bnfsep \mpSucc{e}
 	\bnfsep \mpNeg{e}
& \textbnf{(value, expressions)}\\[1mm]
\mpC \bnfdef \mpy \bnfsep \mpChanRole{\mpS}{\roleP}
&\textbnf{(chan variable, chan with role $\roleP$)}\\[1mm]
\mpPrefix
\bnfdef \mpSel{\mpC}{\roleP}{\stLab}{\mpCi}{}
	\bnfsep \mpSel{\mpC}{\roleP}{\stLab}{e}{}
	\bnfsep \mpBra{\mpC}{\roleP}{\stLab}{\mpy}{}
	\bnfsep \mpBra{\mpC}{\roleP}{\stLab}{\mpx}{}		
& \textbnf{(output to/input prefix from $\roleP$)}\\[1mm]
\mpP
        \bnfdef \mpNil
        \bnfsep \mpRes{\mpS}{\mpP}
        \bnfsep \mpX
   	\bnfsep \mpRec{\mpX}{\mpP}
       	\bnfsep \mpP \mpPar \mpPi
	&\textbnf{(nil, hiding, proc var, recursion, par)}\\
\qquad \bnfsep
\mpIf{e}{\mpP}{\mpQ} \bnfsep 
\mpSum{\mpPrefix_i \mpSeq \mpP[i]}{i\in I}
       \bnfsep \mpErr &\textbnf{(sum with $I\neq\emptyset$, error)}
\end{array}
$}
\end{definition}
\emph{Values} ($v,v',w,...$) are
\emph{expression variables} ($x,x',\dots$),
integers, or booleans.
\emph{Expressions} ($e,e',...$) are values,
comparisons ($=$ and $<$),
successors, or negations.
\emph{Channels} ($c,c',...$) are \emph{channel variables} ($y,y',\dots$)
or channels with roles $\mpChanRole{\mpS}{\roleP}$,
representing endpoints whose users plays role $\roleP$ in the session $\mpS$.
The \emph{prefix}
$\mpSel{\mpC}{\roleP}{\stLab}{\mpCi}{}$
(resp.~$\mpSel{\mpC}{\roleP}{\stLab}{e}{}$)
denotes \emph{sending} a message through \emph{subject} 
$\mpC$ with label $\stLab$
towards role $\roleP$ 
with the channel $\mpCi$
(resp. the expression $e$)
as the payload.
The prefixes
$\mpBra{\mpC}{\roleP}{\stLab}{\mpy}{}$ and
$\mpBra{\mpC}{\roleP}{\stLab}{\mpx}{}$
denote \emph{receiving} messages,
dual to the above prefixes.
$\subj{\mpPrefix}$ denotes the subject of $\mpPrefix$.
Other constructors are standard.

In \emph{processes} ($\mpP, \mpQ, \mpR, ...$), 
the only non-standard constructor is the \emph{sum} constructor,
which implements mixed choice. 
\emph{Sum} $\mpSum{\mpPrefix_i \mpSeq \mpP[i]}{i\in I}$
denotes the non-deterministic choice of the
actions $\{\mpPrefix_i\}_{i\in I}$
with continuation processes $\{\mpP[i]\}_{i\in I}$,
where received payloads are substituted into the continuation.
Given two sum processes, 
we often write their sum,
$\mpSum{\mpPrefix_i \mpSeq \mpP[i]}{i\in I}
\mpFmt{+}\mpSum{\mpPrefix_i \mpSeq \mpP[i]}{i\in J}
=
\mpSum{\mpPrefix_i \mpSeq \mpP[i]}{i\in I\cup J}$.
We write $\dagger$ for $!$ or $?$
and $\overline{!}={?}$ and $\overline{?}={!}$.
\emph{Session restriction} (hiding), $\mpRes{\mpS}{\mpP}$,
creates a new session $\mpS$ with scope $\mpP$.
\emph{Process variable}, $\mpX$,
and
\emph{recursive process}, $\mpRec{\mpX}{\mpP}$,
represent recursion. We assume $\mpP$ is guarded\hideapp{ (\cref{def:calc-guarded})}, i.e., 
$\mpRec{\mpX}{\mpX}$ is not allowed. Other processes are standard. 
$\mpErr$ denotes the error process.

Restriction, branching, and recursive binders,
act as binders.
$\fpv{}/\chan{}$ is the \emph{free process variable/channel} 
functions\hideapp{ (\cref{def:free-var-calc})}.   
We assume the Barendregt convention
where all bound sessions and variables are assumed pairwise distinct,
and distinct from any free sessions or variables.
We assume
the existence of capture-avoiding substitution
operators on processes and expressions.
We define 
\emph{reduction contexts} by
$\mpCtxApp{\mpCtx}{} \bnfdef
\mpCtxHole
\bnfsep
\mpRes{\mpS[1],\dots,\mpS[n]}{\left(\mpCtxHole\mpPar\mpP\right)}$ 
and the \emph{prefix} notation: 
$\mpSel{\mpC}{\roleP}{\stLab}{}{}\prefix{\mpSel{\mpC}{\roleP}{\stLab}{d}{}}$,
$\mpBra{\mpC}{\roleP}{\stLab}{}{}\prefix{\mpBra{\mpC}{\roleP}{\stLab}{\mpz}{}}$,
$\mpFmt{\mpChanRole{\mpC}{\roleP}{\dagger}\stLab}
\prefix{\mpSum{\mpPrefix_i\mpSeq\mpP[i]}{i\in I}}$
if $\mpFmt{\mpChanRole{\mpC}{\roleP}{\dagger}\stLab}
\prefix{\mpPrefix_k}$ for some $k\in I$, and
similarly for $\mpChanRole{\mpC}{\roleP}{\dagger}$.
We write $\mpPrefix\not\prefix{\mpPrefix'}$ etc. if
$\mpPrefix\prefix{\mpPrefix'}$ is not true. 

We now introduce our running example,
the \emph{leader election} protocol from
\cite[Example 2.2]{PY2024},
extended with \emph{session delegations}.  

\begin{example}[Leader Election with Delegation]
\label{ex:elect}\ \\
\begin{minipage}{1\textwidth}
\begin{wrapfigure}[6]{r}{0.3\textwidth}
	\begingroup
	\small
	\centering
	\vspace{-13mm}
	\begin{tikzpicture}[
	sty/.style={rectangle, draw=black, minimum size = 2mm}
	]
		\node (centre) {};
		\path (centre) ++(90:1) node[sty] (0) {0};
		\path (centre) ++(162:1) node[sty] (4) {4};
		\path (centre) ++(234:1) node[sty] (3) {3};
		\path (centre) ++(306:1) node[sty] (2) {2};
		\path (centre) ++(18:1) node[sty] (1) {1};
		
		\draw[->,color=blue] (0) -- (4);
		\draw[->,color=blue] (1) -- (0);
		\draw[->,color=blue] (2) -- (1);
		\draw[->,color=blue] (3) -- (2);
		\draw[->,color=blue] (4) -- (3);
		
		\draw[->,color=red] (0) -- (2);
		\draw[->,color=red] (1) -- (3);
		\draw[->,color=red] (2) -- (4);
		\draw[->,color=red] (3) -- (0);
		\draw[->,color=red] (4) -- (1);
		
		\draw[->,color={rgb:red,0.1;green,0.3;blue,0.1}] (0) to[out=180, in=144, looseness=1.6] (3);
		\draw[->,color={rgb:red,0.1;green,0.3;blue,0.1}] (1) to[out=108, in=72, looseness=1.6] (4);
		\draw[->,color={rgb:red,0.1;green,0.3;blue,0.1}] (2) to[out=36, in=0, looseness=1.6] (0);
		\draw[->,color={rgb:red,0.1;green,0.3;blue,0.1}] (3) to[out=324, in=288, looseness=1.6] (1);
		\draw[->,color={rgb:red,0.1;green,0.3;blue,0.1}] (4) to[out=252, in=216, looseness=1.6] (2);
	\end{tikzpicture}
	\endgroup
	\label{fig:election}
\end{wrapfigure}
	Five participants
	elect a leader in three stages of voting.
	The first round is indicated by {\color{blue}blue},
	the second by {\color{red}red}, and the third by {\color{ruleColor}green}.
	In \underbar{round one}, the $i$th participant either votes using the token $\mpChanRole{\mpS[i]}{\roleP}$ or
	is voted for.
	In \underbar{round two}, if a participant is voted for, then it either votes or is voted for again.
	In \underbar{round three}, the participant with two votes
	receives a final vote from the remaining participant.
	After taking part in the election,
	the participants redeem their tokens.
	\end{minipage}
	\\[1mm]
	\centerline{\small
	\(
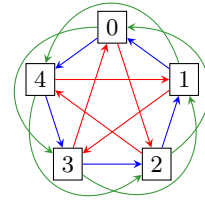
		\begin{array}{rcl}
			\mpP[i]&=&
				\mpSel{\mpChanRole{\mpS}{\roleP[i]}}{\roleP[(i+4)\%5]}{\stLabFmt{elect}}{\mpChanRole{\mpS[i]}{\roleP}}{}+
				\mpBra{\mpChanRole{\mpS}{\roleP[i]}}{\roleP[(i+1)\%5]}{\stLabFmt{elect}}{\mpy}{\mpPi[i]\!\left(\mpy\right)}+
				\\&&
				\mpSel{\mpChanRole{\mpS}{\roleP[i]}}{\roleP[(i+3)\%5]}{\stLabFmt{elect}}{\mpChanRole{\mpS[i]}{\roleP}}{}
			\\[1mm]
			\mpPi[i]\!\left(\mpC\right)&=&
					\mpSel{\mpChanRole{\mpS}{\roleP[i]}}{\roleP[(i+2)\%5]}{\stLabFmt{elect}}{\mpChanRole{\mpS[i]}{\roleP}}{
					\mpSel{\mpC}{\roleS}{\stLabFmt{token}}{\mpNum{i}}{}
					} +
					\\&&
					\mpBra{\mpChanRole{\mpS}{\roleP[i]}}{\roleP[(i+3)\%5]}{\stLabFmt{elect}}{\mpyi}{
						\mpBra{\mpChanRole{\mpS}{\roleP[i]}}{\roleP[(i+2)\%5]}{\stLabFmt{elect}}{\mpyii}{\mpPii[i]\!\left(\mpC,\mpyi,\mpyii\right)}}
						\\[1mm]
			\mpPii[i]\!\left(\mpC,\mpCi,\mpCii\right)&=&\mpSel{\mpChanRole{\mpS[i]}{\roleP}}{\roleS}{\stLabFmt{token}}{\mpNum{i}}{
						\mpSel{\mpC}{\roleS}{\stLabFmt{token}}{\mpNum{i}}{
						\mpSel{\mpCi}{\roleS}{\stLabFmt{token}}{\mpNum{i}}{
						\mpSel{\mpCii}{\roleS}{\stLabFmt{token}}{\mpNum{i}}{}
						}
						}
						}\ (0\leq i \leq 4)
			\\[1mm]
			\mpQ&=&\mpBigPar{0\leq i\leq 4}{
			\mpBra{\mpChanRole{\mpS[i]}{\roleS}}{\roleP}{\stLabFmt{token}}{\mpx}{}
			}\qquad 
			\mpP[\text{\tiny lead}]\ = \ \mpP[0]\mpPar\mpP[1]\mpPar\mpP[2]\mpPar\mpP[3]\mpPar\mpP[4]\mpPar\mpQ
		\end{array}
	\)
	}
\end{example}

\begin{figure}[t!]
	\centering
$
\arraycolsep=7pt
\small
\begin{array}{ll}
		\inferrule{R-Val}&
		{\eval{e}{v}}
		\implies
		{
				(\mpBra{\mpChanRole{\mpS}{\roleQ}}{\roleP}{\stLab}{\mpx}{\mpQ}+\mpQi)
				\mpPar
				(\mpSel{\mpChanRole{\mpS}{\roleP}}{\roleQ}{\stLab}{e}{\mpP}+\mpPi)
				\mpMove
				\mpQ\subst{\mpx}{v}\mpPar\mpP
			}
		\\[1ex]
		\inferrule{R-Chan}&
		{
				(\mpBra{\mpChanRole{\mpS}{\roleQ}}{\roleP}{\stLab}{\mpy}{\mpQ}+\mpQi)
				\mpPar
				(\mpSel{\mpChanRole{\mpS}{\roleP}}{\roleQ}{\stLab}{\mpChanRole{\mpSi}{\rolePi}}{\mpP}+\mpPi)
				\mpMove
				\mpQ\subst{\mpy}{\mpChanRole{\mpSi}{\rolePi}}\mpPar\mpP
			}
		\\[1ex]
		\inferrule{R-Cond}&
		\eval{e}{v}\;\wedge\; v\in\{\mpTrue,\mpFalse\}\implies\mpIf{e}{\mpP[\mpTrue]}{\mpP[\mpFalse]}\mpMove\mpP[v]
		\\[1ex]
		\inferrule{R-Ctx}&
		{\mpP\prestruct\mpPi\;\wedge\; \mpPi\mpMove\mpQi\;\wedge\; \mpQi\prestruct\mpQ}
		\implies
		{\mpCtxApp{\mpCtx}{\mpP}\mpMove\mpCtxApp{\mpCtx}{\mpQ}}
		\\[1ex]
		\inferrule{E-Cond}&
		{\eval{e}{v}\;\wedge\; v\not\in\{\mpTrue,\mpFalse\}}
		\implies
			{\mpIf{e}{\mpP}{\mpQ}\mpMove\mpErr}
			\\[1ex]
			\inferrule{E-Eval}&
			{\eval{e}{\mpErr}}
			\implies
			{\mpSel{\mpC}{\roleP}{\stLab}{e}{\mpP}+\mpPi
			\mpMove
			\mpErr}
			\\[1ex]
			\inferrule{E-M-I}&
			{(\mpBra{\mpChanRole{\mpS}{\roleQ}}{\roleP}{\stLab}{\mpx}{\mpQ}+\mpQi)
			\mpPar
			(\mpSel{\mpChanRole{\mpS}{\roleP}}{\roleQ}{\stLab}{\mpC}{\mpP}+\mpPi)
			\mpMove\mpErr}
			\\[1ex]
			\inferrule{E-M-II}&
			{(\mpBra{\mpChanRole{\mpS}{\roleQ}}{\roleP}{\stLab}{\mpy}{\mpQ}+\mpQi)
			\mpPar
			(\mpSel{\mpChanRole{\mpS}{\roleP}}{\roleQ}{\stLab}{e}{\mpP}+\mpPi)
			\mpMove\mpErr}
			\\[1ex]
			\inferrule{E-M-III}&
(\mpSel{\mpChanRole{\mpS}{\roleP}}{\roleQ}{\stLab}{}{} \prefix{\mpP}
				\,\wedge\,
				\mpBra{\mpChanRole{\mpS}{\roleQ}}{\roleP}{\stLabi}{}{} \prefix{\mpQ}
				\,\wedge\,
				\mpBra{\mpChanRole{\mpS}{\roleQ}}{\roleP}{\stLab}{}{} \not\prefix{\mpQ})
\implies
			{
			\mpP\mpPar
			\mpQ
			\mpMove\mpErr
			}
		\end{array}
$
	\caption{Semantics of the Mixed Choice Multiparty
	$\pi$-Calculus}
	\label{fig:proc-red}
	\label{fig:semantics-calculus}
\end{figure}
\begin{definition}[Operational Semantics]
\label{def:reduction} \rm
We use a reflexive and transitive \emph{structural precongruence}
on processes, 
written $\mpP\prestruct\mpPi$,
defined by the
standard rules, including 
$\mpRec{\mpX}{\mpP}\prestruct\mpP\subst{\mpX}{\mpRec{\mpX}{\mpP}}$\hideapp{ (\cref{def:str})}. 
We write $\mpP\equiv\mpPi$
if $\mpP\prestruct\mpPi$ and $\mpPi\prestruct\mpP$.
\emph{Reduction rules} on processes, $\mpMove$,
are defined by the rules in 
\cref{fig:proc-red} where we assume 
the standard expression evaluation relation, $\eval{e}{v}$ from 
\cite{GJPSY2018}\hideapp{ (\cref{def:expression-eval})}. 
We say that $\mpP$ \emph{has an error}
iff $\mpP\prestruct\mpCtxApp{\mpCtx}{\mpErr}$
for some context $\mpCtxApp{\mpCtx}{}$.
We define reduction of contexts,
written
$\mpCtxApp{\mpCtx}{}\mpMove\mpCtxApp{\mpCtxi}{}$
iff
$\forall\mpP\suchthat\mpCtxApp{\mpCtx}{\mpP}\mpMove\mpCtxApp{\mpCtxi}{\mpP}$.
We define $\mpMoveStar$ to be the reflexive and transitive closure
of $(\mpMove\cup\prestruct)$. $\mpP \mpMove$ denotes
$\exists \mpPi.\ \mpP \mpMove \mpPi$.
\end{definition}
Most rules are standard \cite{YH2024}.
\inferrule{R-Val} and \inferrule{R-Chan}
allow for the communication of values and channels
across parallel components,
with the received payload being substituted into the receiver's continuation.
\inferrule{E-Cond} says that conditioning on a non-boolean value
causes an error.
Similarly, \inferrule{E-Eval} says that expression errors are errors.
\inferrule{E-M-I} (resp. \inferrule{E-M-II})
says that using a channel (resp. value) as the payload when a value (resp. channel) must be received
causes an error.
\inferrule{E-M-III} is the standard label mismatch error;
it says that if two endpoints can communicate but the
sent label is not expected, then an error occurs.
\begin{definition}[Properties]
\label{def:calculus-prop}\rm 
Let $\mpP$ be a process.
We say that $\mpP$ is:
\begin{itemize}
	\item deadlocked iff $\mpP\mpMoveNot$
	and $\mpP\not\prestruct\mpNil$; and
	\item livelocked/has a livelock iff
		$\mpP=\mpCtxApp{\mpCtx}{\mpSum{\mpPrefix_i\mpSeq\mpP[i]}{i\in I}}$
		and
	it is not the case that:\\
	$\mpCtx{\mpCtxHole}\mpMoveStar
	\mpCtxApp{\mpCtxi}{\mpCtxHole\mpPar\mpSum{\mpPrefix'_j\mpSeq\mpQ[j]}{j\in J}}$
	and there exist
	$\mpS$, $\roleP$, $\roleQ$, $\dagger$, $\stLab$, $i\in I$, and $j\in J$
	such that
	$\mpChanRole{\mpChanRole{\mpS}{\roleP}}{\roleQ}{\dagger}\stLab\prefix{\mpPrefix_i}$
	and
	$\mpChanRole{\mpChanRole{\mpS}{\roleQ}}{\roleP}\overline{\dagger}\stLab\prefix{\mpPrefix'_j}$.
\item\emph{safe} iff for all $\mpP\mpMoveStar\mpPi$,
$\mpPi$ has no error;
\item\emph{deadlock-free} iff for all $\mpP\mpMoveStar\mpPi$,
$\mpPi$ is not deadlocked; and
\item\emph{live} iff for all
$\mpP\mpMoveStar\mpPi$,
$\mpPi$ is not livelocked.
\end{itemize}
\end{definition}
A process is \emph{deadlock-free}
if it only stops reducing at $\mpNil$.
Process
$\mpSel{\mpChanRole{\mpS}{\roleP}}{\roleQ}{\stLab}{\mpNum{5}}{\mpNil}$
is deadlocked but safe.
Liveness means that if $\mpP$ is waiting to perform
an action, 
the rest of processes always
move to a point where the action in $\mpP$ can be performed.
All deadlocked processes are livelocked. Process
$
\mpBra{\mpChanRole{\mpS}{\roleR}}{\roleP}{\stLab}{\mpx}{
\mpNil}$
$\mpPar
\mpRec{\mpX}{\left(
	\mpBra{\mpChanRole{\mpS}{\roleQ}}{\roleP}{\stLab}{\mpx}{
\mpX
}
\right)}
\mpPar
\mpRec{\mpX}{\left(
	\mpSel{\mpChanRole{\mpS}{\roleP}}{\roleQ}{\stLab}{\mpNum{0}}{
\mpX
}
\right)}
$
is deadlock-free, but not live. 

\smallskip 

\myparagraph{Three-Party Locks (3-plocks)}
are small and detectable livelocks.
Intuitively, $\mpRes{\mpS}{\mpQ}$ has a 3-plock
if: exactly one component can use $\mpChanRole{\mpS}{\roleP}$;
it is being used in a choice;
and every endpoint that it could communicate with,
must first communicate with another endpoint,
which must first communicate with $\mpChanRole{\mpS}{\roleP}$.
This ensures that the endpoint
$\mpChanRole{\mpS}{\roleP}$ can perform no actions.

\begin{definition}[Three-Party Lock (3-plock)]
\label{def:three-lock}\rm 
	We define the predicate\\
	$\obsvNew{\mpP}{\mpS}{\roleP[1]}{\roleP[2]}{\dagger}$
	by the inductive rules:\\
$        \small
	\begin{array}{c}
		\inference[]{}{\obsvNew{\mpNil}{\mpS}{\roleP[1]}{\roleP[2]}{\dagger}}
		\quad
			\inference[]{}{\obsvNew{\mpX}{\mpS}{\roleP[1]}{\roleP[2]}{\dagger}}
			\quad
			\inference[]{\obsvNew{\mpQ}{\mpS}{\roleP[1]}{\roleP[2]}{\dagger}}{\obsvNew{\mpRes{\mpSi}{\mpQ}}{\mpS}{\roleP[1]}{\roleP[2]}{\dagger}}
			\\[1ex]
			\inference[]{}
			{\obsvNew{\mpBra{\mpChanRole{\mpS}{\roleP[1]}}{\roleP[2]}{\stLab}{\mpz}{\mpQ}}{\mpS}{\roleP[1]}{\roleP[2]}{?}}
\quad 
			\inference[]{}
			{\obsvNew{\mpSel{\mpChanRole{\mpS}{\roleP[1]}}{\roleP[2]}{\stLab}{w}{\mpQ}}{\mpS}{\roleP[1]}{\roleP[2]}{!}}
			\\[1ex]
	\inference[]{\obsvNew{\mpQ}{\mpS}{\roleP[1]}{\roleP[2]}{\dagger}
&
                      \mpC\neq\mpChanRole{\mpS}{\roleP[1]}}
		{\obsvNew{\mpBra{\mpC}{\roleQ}{\stLab}{\mpz}{\mpQ}}{\mpS}{\roleP[1]}{\roleP[2]}{\dagger}}
\quad 
		\inference[]{\obsvNew{\mpQ}{\mpS}{\roleP[1]}{\roleP[2]}{\dagger}
			& \mpC\neq\mpChanRole{\mpS}{\roleP[1]}
			&w\neq\mpChanRole{\mpS}{\roleP[1]}}
			{\obsvNew{\mpSel{\mpC}{\roleP[3]}{\stLab}{w}{\mpQ}}{\mpS}{\roleP[1]}{\roleP[2]}{\dagger}}
			\\[1ex]
			\inference[]{\obsvNew{\mpQ}{\mpS}{\roleP[1]}{\roleP[2]}{\dagger}}{\obsvNew{\mpRec{\mpX}{\mpQ}}{\mpS}{\roleP[1]}{\roleP[2]}{\dagger}}
\quad 
			\inference[]{\obsvNew{\mpQ_i}{\mpS}{\roleP[1]}{\roleP[2]}{\dagger}\ (i=1,2)}
		{\obsvNew{\mpQ_1\mpPar\mpQ_2}{\mpS}{\roleP[1]}{\roleP[2]}{\dagger}}
\quad 
			\inference[]{\forall i\in I&\obsvNew{\mpPrefix_i\mpSeq\mpQ}{\mpS}{\roleP[1]}{\roleP[2]}{\dagger}}{\obsvNew{\mpSum{\mpPrefix_i\mpSeq\mpQ}{i\in I}}{\mpS}{\roleP[1]}{\roleP[2]}{\dagger}}
		\end{array}
$\\[2mm]
	We say that $\mpCtxApp{\mpCtx}{\mpRes{\mpS}{\mpQ}}$ 
	has a \emph{three-party lock (3-plock)} from $\mpChanRole{\mpS}{\roleP}$ if
	$\mpQ=\mpSum{\mpPrefix_i\mpSeq\mpQ[i]}{i\in I}\mpPar\mpP$,
	$\mpChanRole{\mpS}{\roleP}\not\in\chan{\mpP}$, and
	there are
        $\roleQ[i]$, $\roleR[i]$ and $\dagger'_i$ for $i\in I$ such that 
	$\mpChanRole{\mpChanRole{\mpS}{\roleP}}{\roleQ[i]}{\dagger_i}\prefix{\mpPrefix_i}$,
	$\obsvNew{\mpQ}{\mpS}{\roleQ[i]}{\roleR[i]}{\dagger'_i}$, and:
(1)
                if $\roleR[i]=\roleP$ then
                $\dagger'_i\neq\overline{\dagger_i}$;
                and
(2) 
                if $\roleR[i]\neq\roleP$ then
	there is $\dagger''_i$ such that
	$\obsvNew{\mpQ}{\mpS}{\roleR[i]}{\roleP}{\dagger''_i}$.

We say that $\mpP$ is 3-plock free iff
for all $\mpP\mpMoveStar\mpPi$,
$\mpPi$ has no 3-plock.
\end{definition}
If $\mpChanRole{\mpS}{\roleP[1]}\not\in\chan{\mpP}$
then $\obsvNew{\mpP}{\mpS}{\roleP[1]}{\roleP[2]}{\dagger}$
is vacuously true.
All 3-plocks are livelocks, but not all livelocks are 3-plocks.
$\mpP[\text{\tiny lead}]$ (\cref{ex:elect}) is 3-plock free.
This is verified in \cref{ex:leader-typed-prop}
using typing.
Chen et al. \cite{CDSY2017}
introduced similar error conditions
for an input/output session calculus without mixed choice,
which we generalise to \emph{two-participant locks (2-plocks)}\hideapp{
(\cref{def:two-lock})}.
The negative result \cref{prop:two-lock}
shows that
2-plocks are insufficient to define 
`bad behaviour' for multiparty sessions.

\begin{example}[Three-Party Lock]
\label{ex:three-lock-calc}
Consider the following processes:\\[1mm]
\centerline{
$\begin{array}{c}
	\mpR[1] = \mpRes{\mpS}{\mpSel{\mpChanRole{\mpS}{\roleP}}{\roleQ}{\stLab}{\mpNum{1}}{}}
	\quad
	\mpR[2] = \mpRes{\mpS}{\left(\mpSel{\mpChanRole{\mpS}{\roleP}}{\roleQ}{\stLab}{\mpNum{2}}{}\mpPar\mpSel{\mpChanRole{\mpS}{\roleQ}}{\roleP}{\stLab}{\mpNum{2}}{}\right)}
	\\[1ex]
	\mpR[3] = \mpRes{\mpS}{\left(\mpSel{\mpChanRole{\mpS}{\roleP}}{\roleQ}{\stLab}{\mpNum{3}}{}\mpPar\mpBra{\mpChanRole{\mpS}{\roleQ}}{\roleR}{\stLab}{\mpx}{\mpNil}\mpPar\mpSel{\mpChanRole{\mpS}{\roleR}}{\roleP}{\stLab}{\mpNum{3}}{}\right)}
\end{array}$
}\\[1mm]
$\mpR[1]$, $\mpR[2]$, and $\mpR[3]$ all
have 3-plocks:
$\mpR[1]$ because $\mpChanRole{\mpS}{\roleP}$ wants to communicate with
$\mpChanRole{\mpS}{\roleQ}$, which is not present;
$\mpR[2]$ because $\mpChanRole{\mpS}{\roleP}$ wants to send a message to
$\mpChanRole{\mpS}{\roleQ}$, which must first send a message to
$\mpChanRole{\mpS}{\roleP}$; and 
$\mpR[3]$ because $\mpChanRole{\mpS}{\roleP}$ wants to communicate with
$\mpChanRole{\mpS}{\roleQ}$, which must first communicate with
$\mpChanRole{\mpS}{\roleR}$, which must first communicate with
$\mpChanRole{\mpS}{\roleP}$.
$\mpR[1]$, $\mpR[2]$ and $\mpR[3]$ 
are deadlocked, but three participant locks can be deadlock-free: \eg
for $\mpR\in\{\mpR[1],\mpR[2],\mpR[3]\}$,
$\mpR
\mpPar
\mpRec{\mpX}{\mpSel{\mpChanRole{\mpS}{\roleFmt{a}}}{\roleFmt{b}}{\stLab}{\mpTrue}{\mpX}}
\mpPar
\mpRec{\mpX}{\mpBra{\mpChanRole{\mpS}{\roleFmt{b}}}{\roleFmt{a}}{\stLab}{\mpx}{\mpX}}
$
is deadlock-free, but has a three-party lock.
\end{example}
\begin{restatable}[3-Plocks]{proposition}{propThreeLockLiveLock}
	If $\mpP$ has a 3-plock,
	then $\mpP$ is not live.
\end{restatable}

\section{Types, Subtyping and Typing System for \mcmp}
\label{sec:types}
This section first introduces 
types (\cref{def:types}), subtyping (\cref{def:subtyping}) and 
properties of typing contexts (\cref{def:type-prop}), together with  
the \emph{three-party lock} for typing contexts (\cref{def:three-lock-type}).
We then define the typing system (\cref{def:type-rules}) and prove
its type soundness (\cref{thm:type-correct}).

\smallskip 
\myparagraph{Syntax and Subtyping.\ }
A \emph{session type} is an abstraction of the communications
that occur over a channel.
A \emph{typing context} is a partial map from channels to closed session types,
which mirrors the behaviour of processes. 
\begin{definition}[Types and Typing Contexts]
\label{def:types}\rm 
The syntax of \emph{basic types}
($\tyGround,\tyGroundi,...$),
\emph{session types} ($\stT,\stTi,\dots$) 
and \emph{types}  ($\stS,\stSi,\dots$) 
are defined as:\\[1mm]
\centerline{\(
\begin{array}{l}
\tyGround\bnfdef\tyBool\bnfsepsmall\tyInt
		\quad 
		\stS\bnfdef\tyGround\bnfsepsmall\stT
\quad
		\stT\bnfdef
		\stSum{\roleP[i]}{i\in I}{\stChoice{\stLab[i]}{\stS[i]}\stSeq\stT[i]}{\dagger_i}
		\bnfsepsmall
		\stEnd
		\bnfsepsmall
		\stRec{\stRecVar}{\stT}
		\bnfsepsmall
		\stRecVar
		\end{array}
\)}
\\[1mm]
where 
$I\neq\emptyset$,
$\stFmt{\dagger_i}\in\{\stFmt{!},\stFmt{?}\}$ and 
$\{\roleP[i]\stFmt{\dagger_i}\stLab[i]\}_{i\in I}$ are distinct,
and payload types are closed\hideapp{ (\cref{def:type-free-var})}, and
all recursive types are guarded\hideapp{ (\cref{def:type-guarded})}.
$\gtFv{}$ is the free variable function\hideapp{ (\cref{def:type-free-var})},
$\gtRoles{\stT}$ is the set of participants
in $\stT$
and unfolding is defined by
$\unfoldOne{\stRec{\stRecVar}{\stT}}=\tySubst{\stT}{\stRecVar}{\stRec{\stRecVar}{\stT}}$ and $\unfoldOne{\stT}=\stT$ otherwise.
\emph{Typing contexts}
are partial functions from
channels to closed session types, and their syntax is
defined as:
$\stEnv\bnfdef\stEnvEmpty\bnfsepsmall
			\stEnv\stEnvComp\stEnvMap{\mpC}{\stT}$. 
For context $\stEnv$, 
	$\stEnv[\mpS]$ is the restriction to a session $\mpS$,
	$\stEnv\setminus\mpS$ is the context with channels over $\mpS$ removed, and
	$\stEnv\setminus\mpC$ is the context with channel $\mpC$ removed.
	For $\stT=\stSum{\roleP[i]}{i\in I}{\stChoice{\stLab[i]}{\stS[i]}\stSeq\stT[i]}{\dagger_i}$,
	write $\roleP_k\stFmt{\dagger_k}\stChoice{\stLab[k]}{\stS[k]}\stSeq\stT[k]
	\preType\stT$,
	$\roleP_k\stFmt{\dagger_k}\stChoice{\stLab[k]}{\stS[k]}\preType\stT$,
	$\roleP_k\stFmt{\dagger_k}\stLab[k]\preType\stT$,
	$\roleP_k\stFmt{\dagger_k}\preType\stT$
	for $k\in I$, and
	$\roleP\stFmt{\dagger}\stChoice{\stLab}{\stS}\stSeq\stTi
	\not\preType\stT$ if 
        $\neg(\roleP\stFmt{\dagger}\stChoice{\stLab}{\stS}\stSeq\stTi
	\preType\stT)$.
$|\stT|/|\stEnv|$ is the number of constructors in $\stT/\stEnv$\hideapp{
(\cref{def:type-size})}.
\emph{We assume that all session types are well-formed.}
\end{definition}  

\begin{definition}[Mixed Choice Subtyping]\label{def:subtyping}
\rm 
The subtyping relation $\tySub$ between closed types
is coinductively defined as:\\[2mm]
	\centerline{\(
		\begin{array}{c}
		\cinference[B]
		{}{\tyGround\tySub\tyGround}
		\quad
		\cinference[SEnd]{}{\stEnd\tySub\stEnd}
		\quad
		\cinference[S$\Sigma$]
		{|K|>1 & \forall k\in K & \stT[k]\tySub\stTi[k]}
		{\stFmt{\Sigma_{k\in K}\stT[k]\tySub\Sigma_{k\in K}\stTi[k]}}
		\\[1ex]
		\cinference[SSI]
		{\forall i\in I & \stT[i]\tySub\stTi[i] & \stSi[i]\tySub\stS[i]}
		{\stSum{\roleP}{i\in I}{\stChoice{\stLab[i]}{\stS[i]}\stSeq\stT[i]}{\stFmt{ !}}
		\tySub
		\stSum{\roleP}{i\in I\cup J}{\stChoice{\stLab[i]}{\stSi[i]}\stSeq\stTi[i]}{\stFmt{ !}}}
		\quad
		\cinference[S$\mu$L]
		{\tySubst{\stT[1]}{\stRecVar}{\stRec{\stRecVar}{\stT[1]}}\tySub\stT[2]}
		{\stRec{\stRecVar}{\stT[1]}\tySub\stT[2]}
		\\[1ex]
		\cinference[SBr]
		{\forall i\in I & \stT[i]\tySub\stTi[i] & \stS[i]\tySub\stSi[i]}
		{\stSum{\roleP}{i\in I\cup J}{\stChoice{\stLab[i]}{\stS[i]}\stSeq\stT[i]}{\stFmt{ ?}}
		\tySub
		\stSum{\roleP}{i\in I}{\stChoice{\stLab[i]}{\stSi[i]}\stSeq\stTi[i]}{\stFmt{ ?}}}
		\quad
		\cinference[S$\mu$R]
		{\stT[1]\tySub\tySubst{\stT[2]}{\stRecVar}{\stRec{\stRecVar}{\stT[2]}}}
		{\stT[1]\tySub\stRec{\stRecVar}{\stT[2]}}
	\end{array}
	\)}
	\\[2mm]
	We write $\stEnv\tySub\stEnvi$ iff
	$\dom{\stEnv}=\dom{\stEnvi}$
	and $\forall \mpC\in\dom{\stEnv}$, 
	$\stEnvApp{\stEnv}{\mpC}\tySub\stEnvApp{\stEnvi}{\mpC}$; 
and $\stEnvEndP{\stEnv}$
	if $\forall \mpC\in\dom{\stEnv}$,
        $\stEnvApp{\stEnv}{\mpC}\tySub\stEnd$; and
	$\stEnv\setminus\stEnd$ for $\{\stEnvMap{\mpC}{\stT}\in\stEnv\suchthat\stT\tyNotSub\stEnd\}$.
\end{definition}
Subtyping  
$\stT\tySub\stTi$ means that
a process implementing $\stT$ is
``less demanding of its channel''
than one implementing $\stTi$ \cite{GJPSY2018,PY2024,DBLP:conf/concur/DemangeonH11}.
Rule \inferrule{S$\Sigma$} from
\cite{PY2024} states that 
subtyping is componentwise;   
the side condition $|K|>1$ prevents 
degenerate derivations\hideapp{ 
(\cref{rem:subtyping-trivial})}.
Other rules are standard: 
in \inferrule{SSI}, fewer internal choices is less demanding;
in \inferrule{SBr}, more external choices is less demanding; and
in \inferrule{SEnd}, terminated types are subtypes of themselves.
\inferrule{S$\mu$L} and \inferrule{S$\mu$R} are the standard 
recursion rules.
$\tySub$ is a preorder
on types and typing contexts\hideapp{ (\cref{lem:sub-pre})}.

\smallskip

\myparagraph{Semantics and Properties. } 
The LTS semantics of session types given below 
defines behaviours of typed channels.   
Then the LTS for typing contexts
represent behaviours of typed processes, and
will be used for typing processes. 

\begin{definition}[The LTS semantics]\rm
\label{def:type-sem}
Let us define \emph{actions} ($\stEnvAnnotGenericSym$,$\stEnvAnnotGenericSym[i]$,\dots) as:\\
$\stEnvAnnotGenericSym
	\bnfdef \roleP\stEnvAnnotQSym\stChoice{\stLab}{\stS}
	\bnfsepsmall \ltsSendRecvS{\mpS}{\roleP}{\roleQ}{\stChoice{\stLab}{\stS}}$. 
The labelled transition relations on
session types and typing contexts 
are defined as:
\\[1mm]
\centerline{
$
	\begin{array}{c}
	\inference[L$\Sigma$]
        {k\in I}
	{\stSum{\roleP[i]}{i\in I}{\stChoice{\stLab[i]}{\stS[i]}\stSeq\stT[i]}{\dagger_i}\gtMove[{\roleP[k]\stFmt{\dagger_k}\stChoice{\stLab[k]}{\stS[k]}}]\stT[k]}
	\quad
	\inference[L$\mu$]
	{\tySubst{\stT}{\stRecVar}{\stRec{\stRecVar}{\stT}}\gtMove[\stEnvAnnotGenericSym]\stTi}
	{\stRec{\stRecVar}{\stT}\gtMove[\stEnvAnnotGenericSym]\stTi}
\\[3mm]
	\inference[LCnt]
	{\stT[\roleP]\gtMove[\roleQ\stEnvAnnotOutSym\stChoice{\stLab}{\stS}]\stTi[\roleP]
	&
	\stT[\roleQ]\gtMove[\roleP\stEnvAnnotInSym\stChoice{\stLab}{\stSi}]\stTi[\roleQ]
	&
	\stSi\tySub\stS
	}
	{\stEnv\stEnvComp\stEnvMap{\mpChanRole{\mpS}{\roleP}}{\stT[\roleP]}\stEnvComp\stEnvMap{\mpChanRole{\mpS}{\roleQ}}{\stT[\roleQ]}
	\gtMove[\ltsSendRecvS{\mpS}{\roleP}{\roleQ}{\stChoice{\stLab}{\stS}}]
	\stEnv\stEnvComp\stEnvMap{\roleP}{\stTi[\roleP]}\stEnvComp\stEnvMap{\roleQ}{\stTi[\roleQ]}}
\end{array}
$}
\end{definition}
\noindent
$\roleP\stEnvAnnotOutSym\stChoice{\stLab}{\stS}$
denotes sending with label $\stLab$ and
payload of type $\stS$ to $\roleP$; the label
$\roleP\stEnvAnnotInSym\stChoice{\stLab}{\stS}$
is its dual; and
$\ltsSendRecvS{\mpS}{\roleP}{\roleQ}{\stChoice{\stLab}{\stS}}$
a communication from $\roleP$ to $\roleQ$ using session $\mpS$
with label $\stLab$ and payload typed by $\stS$.
A \emph{subject of an action} 
$\subj{\stEnvAnnotGenericSym}$ is defined as:
$\subj{\roleP\stEnvAnnotQSym\stChoice{\stLab}{\stS}}=\{\roleP\}$
and
$\subj{\ltsSendRecvS{\mpS}{\roleP}{\roleQ}{\stChoice{\stLab}{\stS}}}
=\{\mpChanRole{\mpS}{\roleP},\mpChanRole{\mpS}{\roleQ}\}$.
Rule \inferrule{L$\Sigma$} chooses one branch,
rule \inferrule{L$\mu$} is for a recursion;
and rule \inferrule{L$\Sigma$} is for a communication between dual
actions.
$\stEnv\gtMove[\stEnvAnnotGenericSym]$
(resp. $\stT\gtMove[\stEnvAnnotGenericSym]$)
denotes
$\exists \stEnvi.\stEnv\gtMove[\stEnvAnnotGenericSym]\stEnvi$
(resp. $\exists\stTi.\stT\gtMove[\stEnvAnnotGenericSym]\stTi$).
We often write $\gtMove$ by omitting an action.
$\gtMoveStar$ is the reflexive and transitive
closure of $\gtMove$.

The correctness properties in \cref{def:calculus-prop}
correspond to typing context properties,
which are parameters to our typing system.
\begin{definition}[Properties]
\label{def:type-prop}\rm
The following are common
correctness
properties:
for typing contexts:
\begin{itemize}
\item {\bf Safety.} 
We say that $\stEnv$ is \emph{safe}, written $\stEnvSafeP{\stEnv}$,
iff for all $\stEnv\gtMoveStar\stEnvi$,\\
  $\stEnvApp{\stEnvi}{\mpChanRole{\mpS}{\roleP}}\gtMove[{\roleQ\stEnvAnnotOutSym\stChoice{\stLab}{\stS}}]$
  and 
  $\stEnvApp{\stEnvi}{\mpChanRole{\mpS}{\roleQ}}\gtMove[{\roleP\stEnvAnnotInSym\stChoice{\stLabi}{\stSi}}]$, implies
  $\stEnvi\gtMove[\ltsSendRecvS{\mpS}{\roleP}{\roleQ}{\stChoice{\stLab}{\stS}}]$.
\item {\bf Deadlock-freedom.} 
We say that $\stEnv$ is \emph{deadlock-free},
written $\stEnvDFP{\stEnv}$, iff for all $\stEnv\gtMoveStar\stEnvi$,
$\stEnvEndP{\stEnvi}$ or $\stEnvi\,\gtMove$.
\item {\bf Liveness.} We say that a path $\{\stEnv[n]\}_{n\in N}$ is:
\begin{itemize}
\item
\emph{fair}
iff for all $n\in N$,
if
$\stEnv[n]\gtMove[\ltsSendRecvS{\mpS}{\roleP}{\roleQ}{\stChoice{\stLab}{\stS}}]$,
then there exist $k, \stEnvAnnotGenericSym$ such that 
$n\leq k \in N$,
$\{\mpChanRole{\mpS}{\roleP},\mpChanRole{\mpS}{\roleQ}\}\cap\ltsSubject{\stEnvAnnotGenericSym}\neq\emptyset$, and
$\stEnv[k]\gtMove[\stEnvAnnotGenericSym]\stEnv[k+1]$;
\item
\emph{live}
iff for all $n\in N$, if
$\stEnvApp{\stEnv[n]}{\roleP}\gtMove[]$, then there exist $k, \stEnvAnnotGenericSym$ s.t.
$n\leq k \in N$,
$\mpChanRole{\mpS}{\roleP}\in\ltsSubject{\stEnvAnnotGenericSym}$
and $\stEnv[k]\gtMove[\stEnvAnnotGenericSym]$.
\end{itemize}
We say that $\stEnv$ is \emph{live}, written $\stEnvLiveP{\stEnv}$,
iff every fair path starting at $\stEnv$ is live.
\end{itemize}
We say that a property $\predP$ of typing contexts is
\emph{reduction closed} (\RC) if $\predPApp{\stEnvEmpty}$, and
$\predPApp{\stEnv}$ and $\stEnv\gtMove\stEnvi$ implies
$\predPApp{\stEnvi}$.
If $\predP$ is \RC and $\predP\subseteq\stEnvSafePred$,
then we say that $\predP$ is \RC-safe
\eg $\stEnvSafePred$, $\stEnvDFPred$, and $\stEnvLivePred$
are \RC, and
$\stEnvSafePred$, $\stEnvDFPred\cap \stEnvSafePred$, and $\stEnvLivePred\cap \stEnvSafePred$ are \RC-safe.
\end{definition}
\noindent

These properties mirror those for processes (\cref{def:calculus-prop}).
We say $\stEnvSafePred$ typing contexts are \emph{error-free}.
We say a typing context $\stEnv$ has:
a
\emph{label mismatch} if 
$\stEnvApp{\stEnv}{\mpChanRole{\mpS}{\roleP}}\gtMove[{\roleQ\stEnvAnnotOutSym\stChoice{\stLab}{\stS}}]$,
  $\stEnvApp{\stEnvNew}{\mpChanRole{\mpS}{\roleQ}}\gtMove[{\roleP\stEnvAnnotInSym\stChoice{\stLabi}{\stSi}}]$, and
  for all $\stSii$, 
  $\stEnvApp{\stEnvNew}{\mpChanRole{\mpS}{\roleQ}}\quad\not\!\!\!\!\!\!\!\!\!\gtMove[{\roleP\stEnvAnnotInSym\stChoice{\stLab}{\stSii}}]$;
and a \emph{payload mismatch} 
if
$\stEnvApp{\stEnv}{\mpChanRole{\mpS}{\roleP}}\gtMove[{\roleQ\stEnvAnnotOutSym\stChoice{\stLab}{\stS}}]$,
  $\stEnvApp{\stEnvNew}{\mpChanRole{\mpS}{\roleQ}}\gtMove[{\roleP\stEnvAnnotInSym\stChoice{\stLab}{\stSi}}]$, and
  $\stSi\tyNotSub\stS$.

A context is \emph{safe} if no errors ever occur; 
it is \emph{deadlock-free}
if the only reachable contexts with no further transitions
are $\stEnd$-valued;
and 
it is \emph{live} if, assuming fair scheduling,
all participants that want to communicate
will be a subject of some potential communication.
A \emph{fair scheduler} ensures that 
if $\roleP$ and $\roleQ$ can communicate along $\mpS$ then
eventually either (1) $\roleP$ and $\roleQ$ is scheduled to
communicate along $\mpS$, 
or (2) one of them communicates with another participant $\roleR$ along $\mpS$,
disabling a communication.
See \cref{ex:fairness-needed} for why the fair scheduler is needed.

These properties are decidable.
See \cite{POPL19LessIsMore} and \cref{sec:liveness-algo}.

\begin{restatable}[Downwards Closed]{lemma}{lemDownClosed}
\label{lem:prop-down}
For $\predP\in\{\stEnvSafePred,\stEnvSafePred\cap\stEnvDFPred,\stEnvSafePred\cap\stEnvLivePred\}$,
if $\stEnvi\tySub\stEnv$ and $\predPApp{\stEnv}$, then 
	$\predPApp{\stEnvi}$.
\end{restatable}

\begin{example}[Simplified Sum Subtyping Rule is Incorrect]
\label{ex:io-sub-needed}
If we use the following
\inferrule{S$\Sigma$Bad}, 
as discussed in \cite{PY2024}, 
it violates the subject reduction theorem:
\\[1mm]
\centerline{
\(
\inference[S$\Sigma$Bad]{
\forall i\in I \quad  
\stT[i]\tySub\stTi[i]  
\quad
\text{if } \dagger_i = \stFmt{?},
\text{ then } 
\stS[i]\tySub\stSi[i]
\text{ else } \stSi[i]\tySub\stS[i]
	}{
\stSum{\roleP[i]}{i\in I}{\stChoice{\stLab[i]}{\stS[i]}\stSeq\stT[i]}{\dagger_i}\tySub
\stSum{\roleP[i]}{i\in I\cup J}{\stChoice{\stLab[i]}{\stSi[i]}\stSeq\stTi[i]}{\dagger_i}
}
\)}
\\[1mm]
Consider the types:
$\stT[\roleP]=
	(\roleQ\stFmt{?}\stLab+\roleQ\stFmt{?}\stLabi)$, 
$\stTi[\roleP]=\roleQ\stFmt{?}\stLabi$ and 
$\stT[\roleQ]=\roleP\stFmt{!}\stLab$ 
and contexts: 
$\stEnv =
		\set{\stEnvMap{\mpChanRole{\mpS}{\roleP}}{\stT[\roleP]}
		\stEnvComp
		\stEnvMap{\mpChanRole{\mpS}{\roleQ}}{\stT[\roleQ]}}$
                and 
$\stEnvi =
		\set{\stEnvMap{\mpChanRole{\mpS}{\roleP}}{\stTi[\roleP]}
		\stEnvComp
		\stEnvMap{\mpChanRole{\mpS}{\roleQ}}{\stT[\roleQ]}}$. 
Using \inferrule{S$\Sigma$Bad},
$\stTi[\roleP]$ would be a subtype of $\stT[\roleP]$.
But $\stEnv$ is safe while $\stEnvi$ is not.  
Hence, unsafe processes typed by $\stEnvi$
become typable with $\stEnv$ after applying the subsumption rule 
(\inferrule{Sub}).  
\end{example}

\begin{example}[Weak Liveness is not Downwards Closed]
\label{ex:fairness-needed}
Fair paths are necessary in \cref{def:type-prop}.
Say that $\stEnv$ is weak live if
$\forall\stEnv\gtMoveStar\stEnvi$ and
$\forall\mpC\in\dom{\stEnv}$,
if $\stEnvApp{\stEnvi}{\mpC}\gtMove$, then $\stEnvi\gtMoveStar\stEnvii\gtMove[\stEnvAnnotGenericSym]$
with $\mpC\in\ltsSubject{\stEnvAnnotGenericSym}$.
	Weak liveness is not downwards closed.
	Consider:
	$\stTi[\roleP]=
		\stRec{\stRecVar}{\roleQ!\stLab\stSeq\stRecVar}
		\tySub
		\stT[\roleP]=
		\stRec{\stRecVar}{
		\stSum{}{}{
		\roleQ!\stLab\stSeq\stRecVar,\,
		\roleQ!\stLabi\stSeq\roleR!\stLab
		}{}}$,
		$\stT[\roleQ]=
		\stRec{\stRecVar}{
		\stSum{}{}{
			\roleP?\stLab\stSeq\stRecVar,\,
			\roleP?\stLabi
		}{}
		}{}
		$, and
		$\stT[\roleR]=\stFmt{\roleP?\stLab}$.
	The context
	$\stEnv=\set{\stEnvMap{\mpChanRole{\mpS}{\roleP}}{\stT[\roleP]}\stEnvComp
	\stEnvMap{\mpChanRole{\mpS}{\roleQ}}{\stT[\roleQ]}\stEnvComp
	\stEnvMap{\mpChanRole{\mpS}{\roleR}}{\stT[\roleR]}}$
	is weak live, but the context
	$\stEnvi=\set{\stEnvMap{\mpChanRole{\mpS}{\roleP}}{\stTi[\roleP]}\stEnvComp
	\stEnvMap{\mpChanRole{\mpS}{\roleQ}}{\stT[\roleQ]}\stEnvComp
	\stEnvMap{\mpChanRole{\mpS}{\roleR}}{\stT[\roleR]}}\tySub\stEnv$
	is not.
	Downwards closedness is important
        for typing (with \inferrule{Sub})
        to guarantee process liveness (\cref{ex:need-fair-proc}).
\end{example}

\begin{example}[Leader Election with Delegation]
\label{ex:election-types}
We give a typing context for the leader election process
in \cref{ex:elect}:
$
\stEnv[\text{\tiny lead}] =
\cup_{0\leq i\leq 4}
\set{\stEnvMap{\mpChanRole{\mpS[i]}{\roleP}}{\stT}
	\stEnvComp
	\stEnvMap{\mpChanRole{\mpS[i]}{\roleS}}{\stTi}\stEnvComp
        \stEnvMap{\mpChanRole{\mpS}{\roleP[i]}}{\stT[i]}}
$
where for $0\leq i \leq 4$,
$\stT=\roleS\stFmt{!}\stChoice{\stLabFmt{token}}{\tyInt}$, and\\
\centerline{$
\stTi=\stSum{}{}{
			\roleP[(i+4)\%5]!\stChoice{\stLabFmt{elect}}{\stT},\,
			\roleP[(i+3)\%5]!\stChoice{\stLabFmt{elect}}{\stT},\,
			\roleP[(i+1)\%5]?\stChoice{\stLabFmt{elect}}{\stT}\stSeq\stTi[i]
		}{}
$,}
\centerline{\(
	\begin{array}{rcl}
		\stT[i]&=&
		\stSum{}{}{
			\roleP[(i+4)\%5]!\stChoice{\stLabFmt{elect}}{\stT},\,
			\roleP[(i+3)\%5]!\stChoice{\stLabFmt{elect}}{\stT},\,
			\roleP[(i+1)\%5]?\stChoice{\stLabFmt{elect}}{\stT}\stSeq\stTi[i]
		}{}
		\\[1mm]
		\stTi[i]&=&
		\stSum{}{}{
			\roleP[(i+2)\%5]!\stChoice{\stLabFmt{elect}}{\stT},\,
			\roleP[(i+3)\%5]?\stChoice{\stLabFmt{elect}}{\stT}\stSeq
			\roleP[(i+2)\%5]?\stChoice{\stLabFmt{elect}}{\stT}
		}{}
\end{array}        
\)}
Observe that
$\stT\tySub
\stTii=\stSum{}{}{
\roleS!\stChoice{\stLabFmt{token}}{\tyInt},\roleS!\stChoice{\stLabFmt{token}'}{\tyInt}
}{}
$,
thus
\\[1mm]
\centerline{
\(
\begin{array}{rcl}
\stTi[0]&\tyNotSub&
		\stSum{}{}{
			\roleP[2]!\stChoice{\stLabFmt{elect}}{\stT},\,
			\roleP[3]?\stChoice{\stLabFmt{elect}}{\stTii}\stSeq
			\roleP[2]?\stChoice{\stLabFmt{elect}}{\stTii}
		}{}\tySub\stTi[0]
		\\[1mm]
		\stTi[0]&\tySub&
		\stSum{}{}{
		\begin{array}{l}
			\roleP[2]!\stChoice{\stLabFmt{elect}}{\stTii},\,
			\roleP[3]?\stChoice{\stLabFmt{elect}}{\stT}\stSeq
			\roleP[2]?\stChoice{\stLabFmt{elect}}{\stT}
		\end{array}
		}{}
\end{array}
\)}
\\[1mm]
As with the calculus, $\stEnv[\text{\tiny lead}]$
is safe, deadlock-free, and live.
\end{example}
\myparagraph{Three-Party Locks}
in typing contexts are defined below. 

\begin{definition}[Three-Party Lock (3-plock)]
\label{def:three-lock-type}\rm 
 $\stEnv$ has a \emph{three-party lock (3-plock)}
iff
$\unfoldOne{\stEnvApp{\stEnv}{\mpChanRole{\mpS}{\roleP}}}
=\stSum{\roleQ[i]}{i\in
I}{\stChoice{\stLab[i]}{\stS[i]}\stSeq\stT[i]}{\dagger_i}$, 
and for all $i\in I$, if
$\mpChanRole{\mpS}{\roleQ[i]}\in\dom{\stEnv}$, 
either:
\begin{enumerate*}
	\item[(1)] $\unfoldOne{\stEnvApp{\stEnv}{\mpChanRole{\mpS}{\roleQ[i]}}}
=\stEnd
$; or
	\item[(2)] $\unfoldOne{\stEnvApp{\stEnv}{\mpChanRole{\mpS}{\roleQ[i]}}}
=\stSum{\roleR[i]}{j\in J_i}{\stChoice{\stLab[i,j]}{\stS[i,j]}\stSeq\stT[i,j]}{\dagger'_i}$
and if $\mpChanRole{\mpS}{\roleR[i]}\in\dom{\stEnv}$ then either:
\begin{enumerate*}
	\item $\roleR[i]=\roleP$ and $\dagger'_i=\dagger_i$; or
	\item $\unfoldOne{\stEnvApp{\stEnv}{\mpChanRole{\mpS}{\roleR[i]}}}
=\stEnd
$; or 
\item $\unfoldOne{\stEnvApp{\stEnv}{\mpChanRole{\mpS}{\roleR[i]}}}
=\stSum{\roleP}{j\in J'_i}{\stChoice{\stLabi[i,j]}{\stSi[i,j]}\stSeq\stTi[i,j]}{\dagger''_i}$.
	\end{enumerate*}
\end{enumerate*}
The context $\stEnv$ is 3-plock free
iff for all $\stEnv\gtMoveStar\stEnvi$,
$\stEnvi$ has no 3-plock.
\end{definition}
These mirror the conditions in \cref{def:three-lock}.
$\obsvNew{\mpP}{\mpS}{\roleP}{\roleQ}{\dagger}$
says that the only action that $\roleP$ can take on channel $\mpS$,
if there are any,
is $\dagger$ towards $\roleQ$.
The equivalent for types is 
$\mpChanRole{\mpS}{\roleP}\not\in\dom{\stEnv}$, or
$\unfoldOne{\stEnvApp{\stEnv}{\mpChanRole{\mpS}{\roleP}}}=\stEnd$,
or
$\unfoldOne{\stEnvApp{\stEnv}{\mpChanRole{\mpS}{\roleP}}}
=\stSum{\roleQ}{i\in I}{\stChoice{\stLab[i]}{\stS[i]}\stSeq\stT[i]}{\dagger}$.
\cref{def:three-lock-type} corresponds to \cref{def:three-lock}
with $\obsvNew{\mpP}{\mpS}{\roleP}{\roleQ}{\dagger}$
replaced by its equivalent for types.
\begin{example}[Three-Party Lock]
	\label{ex:three-lock-type}
	We provide typing contexts for \cref{ex:three-lock-calc}:\\[1mm]
\centerline{
		$\begin{array}{c}
			\stEnv[1]=\set{
			\stEnvMap{\mpChanRole{\mpS}{\roleP}}
			{\roleQ!\stChoice{\stLab}{\tyInt}}
			\stEnvComp
			\stEnvMap{\mpChanRole{\mpS}{\roleQ}}
			{\stEnd}}
			\quad
			\stEnv[2]=\set{
			\stEnvMap{\mpChanRole{\mpS}{\roleP}}
			{\roleQ!\stChoice{\stLab}{\tyInt}}
			\stEnvComp
			\stEnvMap{\mpChanRole{\mpS}{\roleQ}}
			{\roleP!\stChoice{\stLab}{\tyInt}}}
			\\[1mm]
			\stEnv[3]=\set{
			\stEnvMap{\mpChanRole{\mpS}{\roleP}}
			{\roleQ!\stChoice{\stLab}{\tyInt}}
			\stEnvComp
			\stEnvMap{\mpChanRole{\mpS}{\roleQ}}
			{\roleR?\stChoice{\stLab}{\tyInt}}
			\stEnvComp
			\stEnvMap{\mpChanRole{\mpS}{\roleR}}
			{\roleP!\stChoice{\stLab}{\tyInt}}}
		\end{array}$
}\\[1mm]
	All three have 3-plocks.
$\stEnv[1]$ has a 3-plock
		by \cref{def:three-lock-type}(1) since $\stEnvApp{\stEnv[1]}{\mpChanRole{\mpS}{\roleQ}}=\stEnd$.
		Participant $\roleP$ cannot communicate along $\mpS$
		since it can only communicate with $\roleQ$,
		which has no further behaviour.
                Similarly,
$\stEnv[2]$ and $\stEnv[3]$ have 3-plocks
		by \cref{def:three-lock-type}(2-a) and (2-c),
                respectively.
\end{example}

\begin{figure}[t!]
	\centering
$
		\begin{array}{c}
		\inference[Rec]
		{
		\tyJudge{\tyEnv\tyEnvComp\tyEnvMap{\mpX}{\stEnv}}{\mpP}{\stEnv}
		&
		\dom{\stEnv\setminus\stEnd}\subseteq\chan{\mpP,\tyEnv}
		}
		{\tyJudge{\tyEnv}{\mpRec{\mpX}{\mpP}}{\stEnv}}
		\quad
		\begin{array}[b]{c}
		\inferenceSingle[Var]{\tyJudge{\tyEnv\tyEnvComp\tyEnvMap{\mpX}{\stEnv}}{\mpX}{\stEnv}}
		\\[1ex]
		\inferenceSingle[Nil]{\tyJudge{\tyEnv}{\mpNil}{\stEnvEmpty}}
		\end{array}
		\\[1ex]
		\inference[If]
		{\tyJudge{\tyEnv}{e}{\tyBool}&
		\tyJudge{\tyEnv}{\mpP}{\stEnv}&
		\tyJudge{\tyEnv}{\mpPi}{\stEnv}}
		{\tyJudge{\tyEnv}{\mpIf{e}{\mpP}{\mpPi}}{\stEnv}}
		\qquad
		\inference[Par]
		{
			\tyJudge{\tyEnv}{\mpP}{\stEnv}
			&
			\tyJudge{\tyEnv}{\mpPi}{\stEnvi}
		}
		{\tyJudge{\tyEnv}{\mpP\mpPar\mpPi}{\stEnv
		\stEnvComp\stEnvi
		}}
		\\[1ex]
		\inference[Res]
		{
		\tyJudge{\tyEnv}{\mpP}{\stEnv}
		&
		\predPApp{\stEnv[\mpS]}
		}
		{\tyJudge{\tyEnv}{\mpRes{\mpS}{\mpP}}{\stEnv\setminus\mpS}}
		\qquad
		\inference[Sub]{\tyJudge{\tyEnv}{\mpP}{\stEnv}&\stEnv\tySub\stEnvi&\stEnvEndP{\stEnvii}}
{\tyJudge{\tyEnv}{\mpP}{\stEnvi\stEnvComp\stEnvii}}
		\\[1ex]
		\begin{array}{cc}
		\inference[\colorbox{yellow}{$!$Chan}]
		{
			\begin{array}{c}
				\mpC\neq\mpCi
				\quad
				\stFmt{\roleQ{!}\stChoice{\stLab}{\stEnvApp{\stEnv}{\mpCi}}\stSeq\stTii}\preType\stEnvApp{\stEnv}{\mpC}
				\\
				\tyJudge{\tyEnv}{\mpP}{
				(\stEnv\setminus\mpC\setminus\mpCi)
				\stEnvComp \stEnvMap{\mpC}{\stTii}
				}
			\end{array}
		}
		{\tyJudgePrefix{\tyEnv}{\mpSel{\mpC}{\roleQ}{\stLab}{\mpCi}{\mpP}}{\stEnv}}
		&
		\inference[\colorbox{yellow}{$!$Val}]
		{
			\begin{array}{c}
				\stFmt{\roleQ{!}\stChoice{\stLab}{\tyGround}\stSeq\stTii}
				\preType{\stEnvApp{\stEnv}{\mpC}}
				\\
				\tyJudge{\tyEnv}{e}{\tyGround}
				\quad
				\tyJudge{\tyEnv}{\mpP}{
				(\stEnv\setminus\mpC)
				\stEnvComp \stEnvMap{\mpC}{\stTii}
				}
			\end{array}
		}
		{\tyJudgePrefix{\tyEnv}{\mpSel{\mpC}{\roleQ}{\stLab}{e}{\mpP}}{\stEnv}}
		\\[1ex]
		\inference[\colorbox{yellow}{$?$Chan}]
		{
			\begin{array}{c}
				\stFmt{\roleQ{?}\stChoice{\stLab}{\stTi}\stSeq\stTii}
				\preType{\stEnvApp{\stEnv}{\mpC}}
				\\
				\tyJudge{\tyEnv}{\mpP}{
				(\stEnv\setminus\mpC)
				\stEnvComp \stEnvMap{\mpC}{\stTii}
				\stEnvComp \stEnvMap{\mpy}{\stTi}
				}
			\end{array}
		}
		{\tyJudgePrefix{\tyEnv}{\mpBra{\mpC}{\roleQ}{\stLab}{\mpy}{\mpP}}{\stEnv}}
		&
		\inference[\colorbox{yellow}{$?$Val}]
		{
			\begin{array}{c}
				\stFmt{\roleQ{?}\stChoice{\stLab}{\tyGround}\stSeq\stTii}
				\preType{\stEnvApp{\stEnv}{\mpC}}
				\\
				\tyJudge{\tyEnv
				\tyEnvComp\tyEnvMap{\mpx}{\tyGround}
				}{\mpP}{
				(\stEnv\setminus\mpC)
				\stEnvComp \stEnvMap{\mpC}{\stTii}
				}
			\end{array}
		}
		{\tyJudgePrefix{\tyEnv}{\mpBra{\mpC}{\roleQ}{\stLab}{\mpx}{\mpP}}{\stEnv}}
		\end{array}
                \\
                \\
		\inference[\colorbox{yellow}{Sum}]
		{
		\forall j\in I\left(
			\begin{array}{c}
				\tyJudgePrefix{\tyEnv}{\mpPrefix_j\mpSeq\mpP[j]}{\stEnv}
				\wedge\left(
				\mpC=\subj{\mpPrefix_j}\implies\right.\\
				\left.
				\forall
				\stFmt{\roleQ{\dagger}\stChoice{\stLab}{\stS}\stSeq\stT}
				\preType{\stEnvApp{\stEnv}{\mpC}}.
				\;
				\mpFmt{\mpChanRole{\mpC}{\roleQ}{\dagger}\stLab}
				\preCalc{\mpSum{\mpPrefix_i\mpSeq\mpP[i]}{i\in I}}
				\right)
			\end{array}\right)
		}
		{
        \tyJudge{\tyEnv}{
        \mpSum{\mpPrefix_i\mpSeq\mpP[i]}{i\in I}
		}{\stEnv}}
		\end{array}
$
\caption{Typing Rules. Parameterised by typing context property
	$\predP$.}  
	\label{fig:typing-rules}
\end{figure}
\myparagraph{Typing System and its Type Soundness. \ } We define the
typing system and prove subject reduction and error-freedom. 
\begin{definition}[Type Judgements]
\label{def:type-rules}\rm 
Let $\tyEnv$ denote a \emph{typing environment} ($\tyEnv$,
$\tyEnvi$,\dots)
defined by $\tyEnv \bnfdef \emptyset \bnfsepsmall \tyEnv,
X:\stEnv \bnfsepsmall \tyEnv, x:\tyGround$. 
Judgement $\tyJudge{\tyEnv}{e}{\tyGround}$
is standard for typing expressions \cite[Table 4]{GJPSY2018}\hideapp{ (\cref{def:type-expressions})}.
Let $\predP$ in \inferrule{Res} in \cref{fig:typing-rules}
be \RC-safe.
We define $
\chan{\mpP,\tyEnv}
=
\chan{\mpP}\cup\bigcup_{\mpX\in\fpv{\mpP}}{\dom{\tyEnvApp{\tyEnv}{\mpX}}}$,
the channels appearing in $\mpP$ with reference to an environment.
We define the process typing judgement,
written $\tyJudge{\tyEnv}{\mpP}{\stEnv}$,
inductively by the rules in \cref{fig:typing-rules}.
\end{definition}
\noindent
The new typing rules (highlighted in \colorbox{yellow}{yellow}) are
\inferrule{Rec},
\inferrule{Sum},
\inferrule{!Chan},
\inferrule{!Val},
\inferrule{?Chan}, and
\inferrule{?Val};
\inferrule{Rec} handles recursion, and the others handle mixed choice.
\inferrule{Rec} types recursive processes with a context.
\inferrule{Sum} consists of two parts:
(1) if the channel $\mpC$ occurs in a sum,
then each choice in
$\stEnvApp{\stEnv}{\mpC}$
must
occur in the sum;
(2)
each choice over $\mpC$ in the sum must
occur in $\stEnvApp{\stEnv}{\mpC}$,
which is checked by
the judgement
$\tyJudgePrefix{\tyEnv}{\mpPrefix\mpSeq\mpP}{\stEnv}$.
The judgement
$\tyJudgePrefix{\tyEnv}{\mpPrefix\mpSeq\mpP}{\stEnv}$
is defined by the rules
\inferrule{!Chan},
\inferrule{!Val},
\inferrule{?Chan}, and
\inferrule{?Val}.
\inferrule{!Chan} and \inferrule{!Val} handle sending channels and
values,
respectively; and 
\inferrule{?Chan} and
\inferrule{?Val} dually handle receiving.
Note that in the session delegation (\inferrule{!Chan}),
after $\mpC$
delegates session endpoint $\mpCi$
to its receiver, 
$\mpP$ can no longer hold $\mpCi$
($\stEnv\setminus\mpC\setminus\mpC'$)
since $\mpCi$ must be used linearly. 

\begin{example}[Typing Leader Election]
	\label{ex:typing-rules}
	Recall $\mpP[\text{\tiny lead}]$
	from \cref{ex:elect}
	and
	$\stEnv[\text{\tiny lead}]$
	from \cref{ex:election-types}.
	Derive
	$\tyJudge{\tyEnvEmpty}{\mpP[\text{\tiny lead}]}{\stEnv[\text{\tiny lead}]}$
	by:
	\inferrule{Nil} at $\mpNil$
	then \inferrule{Sub}
	to introduce $\stEnd$ typed channels;
	\inferrule{Sum}
	at each sum subterm; and
	\inferrule{Par} at each composition subterm.\hideapp{
	Full details are available in \cref{ex:typing-rules-full}.}
\end{example}
Our typing system satisfies subject reduction, yielding
error-freedom.

\begin{restatable}[Subject Reduction]{theorem}{subjectReduction}
\label{thm:subj-red}
Let $\predP$ in \inferrule{Res} in \cref{fig:typing-rules}
be \RC-safe.
Suppose that $\tyJudge{\tyEnv}{\mpP}{\stEnv}$
	and $\stEnvSafeP{\stEnv}$.
	\begin{enumerate*}
		\item If $\mpP\prestruct\mpPi$, then $\tyJudge{\tyEnv}{\mpPi}{\stEnv}$.
		\item $\mpP\mpMove\mpPi$, then
	either $\tyJudge{\tyEnv}{\mpPi}{\stEnv}$
	or there exists $\stEnvi$ such that 
	$\stEnv\gtMove\stEnvi$ and $\tyJudge{\tyEnv}{\mpPi}{\stEnvi}$.
	\end{enumerate*}
\end{restatable}
\begin{restatable}[Error-freedom]{corollary}{errorFreedom}
\label{cor:errorfree}
Let $\predP$ in \inferrule{Res} in \cref{fig:typing-rules}
be \RC-safe.
If $\tyJudge{\tyEnv}{\mpP}{\stEnv}$
with $\stEnvSafeP{\stEnv}$ and
$\mpP\mpMoveStar\mpPi$,
then $\mpPi$ contains no errors.
\end{restatable}
\begin{restatable}[Correctness]{theorem}{subjectCorrectness}
\label{thm:type-correct}
	Let $\predP$ in \inferrule{Res} in \cref{fig:typing-rules}
be \RC-safe. 
	If $\tyJudge{\tyEnv}{\mpP}{\stEnv}$
with $\stEnvSafeP{\stEnv}$, then
 	$\mpP$ is safe.
	If $\predP\subseteq\stEnvLivePred$, then
	$\mpP$ is 3-plock free.
\end{restatable}
Recall
that $\stEnvSafePred\cap\stEnvLivePred$
is \RC-safe (\cref{def:type-prop})
and
$\stEnvSafePred\cap\stEnvLivePred\subseteq\stEnvLivePred$.
Using $\stEnvSafePred\cap\stEnvLivePred$
in \inferrule{Res} in \cref{fig:typing-rules},   
typable processes are 3-plock free.
This does \emph{not}
guarantee liveness of
typed process $\mpP$,
since $\mpP$ might contain interleaved sessions.
See \cref{sec:extensions}.

\begin{example}[Interleaving]
\label{ex:interleaved}
Let
$\mpP=
\mpBra{\mpChanRole{\mpS}{\roleQ}}{\roleP}{\stLab}{}{
\mpSel{\mpChanRole{\mpSi}{\roleP}}{\roleQ}{\stLab}{}{}
}
\mpPar
\mpBra{\mpChanRole{\mpSi}{\roleQ}}{\roleP}{\stLab}{}{
\mpSel{\mpChanRole{\mpS}{\roleP}}{\roleQ}{\stLab}{}{}
}
$
and
$
\stEnv=
\stEnvMap{\mpChanRole{\mpS}{\roleP}}{\stFmt{\roleQ{!}\stLab}}
\stEnvComp
\stEnvMap{\mpChanRole{\mpS}{\roleQ}}{\stFmt{\roleP{?}\stLab}}
\stEnvComp
\stEnvMap{\mpChanRole{\mpSi}{\roleP}}{\stFmt{\roleQ{!}\stLab}}
\stEnvComp
\stEnvMap{\mpChanRole{\mpSi}{\roleQ}}{\stFmt{\roleP{?}\stLab}}
$.
We derive the judgement: $\tyJudge{}{\mpP}{\stEnv}$.
Although $\stEnvSafeP{\stEnv}\cap\stEnvDFP{\stEnv}\cap\stEnvLiveP{\stEnv}$,
$\mpP$ is neither deadlock-free nor live.
\end{example}

\begin{example}[Leader Election with Delegation]
\label{ex:leader-typed-prop}
From \cref{ex:typing-rules} and \cref{thm:type-correct},
$\mpP[\text{\tiny lead}]$
is safe and is 3-plock free.
\end{example}

\section{Operational and Denotational Preciseness}
\label{sec:precise}
This section defines and proves 
soundness and completeness (hence preciseness) of subtyping
both
operationally (\cref{thm:operational-precise})
and denotationally (\cref{thm:precise-denote}).

\subsection{Operational Preciseness}
\label{sec:operational}

We first define \emph{operational preciseness}
following \cite{10.1145/197320.197383,CDSY2017,GJPSY2018,GPPSY2023}:    
a subtyping is \emph{sound} if all its instances are sound; 
a subtyping is \emph{complete} if it contains \emph{all} sound instances; 
and a subtyping is \emph{precise} if it is sound and complete.

\begin{definition}[Operational Preciseness]
\label{def:precise}\rm
Let $\predP=\stEnvLivePred\cap\stEnvSafePred$ 
in \inferrule{Res} in \cref{fig:typing-rules}.
We say that a subtyping instance $(\stTi,\stT)$ is \emph{sound} iff for
all roles $\roleP\not\in\gtRoles{\stT}$, sessions $\mpS$, contexts
$\mpCtxApp{\mpCtx}{}$,
processes $\mpP$, and typing contexts $\stEnv$:
\begin{quote}
if \qquad ($\forall Q.\
\tyJudge{\tyEnvEmpty}{\mpQ}{\set{\stEnvMap{\mpChanRole{\mpS}{\roleP}}{\stT}}}
\ \Longrightarrow \ 
\tyJudge{\tyEnvEmpty}{\mpCtxApp{\mpCtx}{\mpQ}}{\stEnvEmpty})$\\
then \ \ 
($\tyJudge{\tyEnvEmpty}{\mpP}{\set{\stEnvMap{\mpChanRole{\mpS}{\roleP}}{\stTi}}}$ 
$\Longrightarrow$
$\mpCtxApp{\mpCtx}{\mpP}$ is safe and
3-plock free.)
\end{quote}
Let $\trianglelefteq$ be a preorder on closed session types.
We say that $\trianglelefteq$ is
\emph{sound} if $\stTi\trianglelefteq\stT$
implies that $(\stTi,\stT)$ is a sound subtyping instance;
$\trianglelefteq$ is 
\emph{complete} if for all sound subtyping instances $(\stTi,\stT)$,
$\stTi\trianglelefteq\stT$; and
$\trianglelefteq$ is \emph{precise} if it is sound and complete.
\end{definition}
\noindent

\begin{theorem}[Operational Soundness]
\label{thm:prec-sound}
The subtyping $\tySub$ is sound.
\end{theorem}
\begin{proof}
Suppose that $\stTi\tySub\stT$ and
for
all roles $\roleP\not\in\gtRoles{\stT}$, sessions $\mpS$, contexts
$\mpCtxApp{\mpCtx}{}$, processes $\mpQ$, and typing contexts $\stEnv$,
if
$\tyJudge{\tyEnvEmpty}{\mpQ}{\set{\stEnvMap{\mpChanRole{\mpS}{\roleP}}{\stT}}}$
then
$\tyJudge{\tyEnvEmpty}{\mpCtxApp{\mpCtx}{\mpQ}}{\stEnvEmpty}$.

By \inferrule{Sub},
if $\tyJudge{\tyEnvEmpty}{\mpP}{\set{\stEnvMap{\mpChanRole{\mpS}{\roleP}}{\stTi}}}$
then
$\tyJudge{\tyEnvEmpty}{\mpP}{\set{\stEnvMap{\mpChanRole{\mpS}{\roleP}}{\stT}}}$.
By assumption,
$\tyJudge{\tyEnvEmpty}{\mpCtxApp{\mpCtx}{\mpP}}{\stEnvEmpty}$.
Thus by \cref{thm:type-correct},
$\mpCtxApp{\mpCtx}{\mpP}$ is safe and 3-plock free.
\qed
\end{proof}

\smallskip

\label{sec:char}
We prove completeness in four main steps.
\begin{description}
\item \textbf{[Step 1]:}
We construct live contexts for every type, $\stT$,
which check for every disallowed behaviour.
The detailed construction is described in \cref{def:char-ctx}.
The contexts without $\stT$ are called \emph{complementary contexts}.
\item \textbf{[Step 2]:}
Using \textbf{Step 1}, we construct \emph{characteristic processes} for every type $\stT$
and channel $\mpC$ (\cref{def:char-proc}).
\item \textbf{[Step 3]:}
We show that the negation of subtyping has an inductive definition
(\cref{thm:subtype-neg}).
\item \textbf{[Step 4]:}
By induction on \textbf{Step 3}, the constructions in
\textbf{Step 2} and \textbf{Step 1}
provide contexts witnessing completeness.
\end{description}

\myparagraph{Complementary Contexts. }
\textbf{Step 1} has a subtle point since we must be able to check for
every disallowed behaviour.
However, types can only encode finitely many choices.
Therefore, we must limit the tested types 
with a known and finite set of labels, payloads and roles.

Let $\mathcal{T}$ be a finite set of session types
such that for all $\stT\in\mathcal{T}$,
if $\stTi$ appears in $\stT$ as a payload then 
$\stTi\in\mathcal{T}$.
Let $L/\roleSet$ be the set of labels/participants
appearing within $\mathcal{T}$.
We say that a type $\stT$ is \emph{$\mathcal{T}$-valid}
iff all payload session types in $\stT$ are in $\mathcal{T}$,
all labels in $\stT$ are in $L$,
and 
all participants in $\stT$ are in $\roleSet$.
We use $\mathcal{T}$-valid types
to handle all possible mixed choices that may appear.

\begin{definition}[Complementary Types]
\label{def:char-ctx} \rm
	Fix a finite set of session types, $\mathcal{T}$.
	We use $\mathcal{T}$
	to enumerate the possible messages, $\stChoice{\stLab}{\stS}$,
	and construct a collection of message sets, $Act$.
	
	We define $\charTwait{\roleQ}$ to be the type
	of $\roleQ$ in a complementary context; and 
	$\charSch{\stT}$ for $\mathcal{T}$-valid types $\stT$
	to be the type of the scheduler in a complementary context for $\stT$.
	We write $\compEnv{\mpS}{\stT}=
	\stEnvMap{\mpChanRole{\mpS}{\roleFmt{sch}}}{\charSch{\stT}}
	\stEnvComp
	\bigcup_{\roleQ\in\roleSet}\stEnvMap{\mpChanRole{\mpS}{\roleQ}}{\charTwait{\roleQ}}$,
	the complementary context for $\stT$ and $\mpS$,
	for $\mathcal{T}$-valid $\stT$.
	The full definition is given below:

	Fix a finite set of session types $\mathcal{T}$
	such that for all $\stT\in\mathcal{T}$,
	if $\stTi$ is a payload type in $\stT$ then $\stTi\in\mathcal{T}$.	
	Let $L$ be the set of labels appearing within $\mathcal{T}$.
	Let $\roleSet$ be the set of participants appearing within $\mathcal{T}$.
	Enumerate:\\[1mm]
	\centerline{\(
	\roleSet=\{\roleR[i]\}_{0\leq i\leq n}
	\qquad
	\{\stChoice{\stLab}{\stS}\suchthat\stLab\in L,\stS\in\mathcal{T}\cup\{\tyInt,\tyBool\}\}
	=\{\stChoice{\stLab[i]}{\stS[i]}\}_{0\leq i \leq N-1}
	\)}
	\\[1mm]
	Let
	$
	Act=\{I\subseteq\{0,\dots,N-1\}\suchthat \forall i\neq j\in I.\; \stLab[i]\neq\stLab[j]\}
	$,
	the set of all message sets, where messages have distinct labels.
	Let $\roleP$ and $\roleFmt{sch}$ be fresh roles,
	and $\stLabii$ be a fresh label
	\ie
	$\roleP,\roleFmt{sch}\not\in\roleSet$ and
	$\stLabii\not\in L$. We first define type $\charTwait{\roleQ}$:\\[1mm] 
	\begingroup
	\small
$			\begin{array}{rcl}
\charTwait{\roleQ}&=&
					\stRec{\stRecVar}{\left(
						\stSum{\roleFmt{sch}}{
							I \in Act,\dagger\in\{!,?\}
						}{
							\stLab[(I,\dagger)]\stSeq\charTAct{\roleQ}{I}{\dagger}{\stRecVar}
						}{?}
						+
						\stSum{\roleFmt{sch}}{
							\dagger\in\{!,?\}
						}{
							\stLabi[\dagger]\stSeq\charTActi{\dagger}{\stRecVar}
						}{?}
						\right.}
						\\
						& &\qquad\qquad\stFmt{+}\stSum{\roleFmt{sch}}{
							\roleR\in\roleSet\setminus\{\roleQ\}
						}{
							\stLab[i]\stSeq
							\syncBra{\roleR[i]}{\stRecVar}
						}{?}
						\stFmt{+}\stFmt{
							\left.
							\roleFmt{sch}?\stLab[\text{\tiny end}]
							\right)
						}
			\end{array}$\\[1mm]
$        \begin{array}{ll@{\quad}l}
\text{with}\ &	\syncBra{\roleQ}{\stT}=
				\roleQ\stFmt{?}\stLab\stSeq\stT
		&
				\syncSel{\roleR[k]}{\stT} = 
				\roleR[0]\stFmt{!}{\stLab}\stSeq
				\dots
				\roleR[k-1]\stFmt{!}\stLab\stSeq
				\roleR[k+1]\stFmt{!}\stLab\stSeq
				\dots
				\roleR[n]\stFmt{!}\stLab\stSeq\stT
				\\
&	\charTActi{\dagger}{\stT} = 
 					\stFmt{
 					\roleFmt{sch}!\stLab\stSeq\stT
 					+
 						\roleP{\dagger}\stLabii
 					}
	&
	\charTAct{\roleQ}{I}{\dagger}{\stT}=
 				\stSum{\roleP}{i\in I}{
 					\stChoice{\stLab[i]}{\stS[i]}\stSeq
 					\syncSel{\roleQ}{
 						\roleFmt{sch}\stFmt{!}\stLab[i]\stSeq\stT
 					}
 				}{\dagger}
	\end{array}
$
		\endgroup
                
\noindent Next we define $\charSch{\stT}$ for $\mathcal{T}$-valid types $\stT$,
	by the following recursive definition:\\
	\centerline{\(
        \begin{array}{l}
				\charSch{\stEnd}=
					\roleR[0]\stFmt{!}{\stLab[\text{\tiny end}]}
					\dots
					\roleR[n]\stFmt{!}\stLab[\text{\tiny end}]
					\qquad
				\charSch{\stRecVar}=\stRecVar
				\qquad
				\charSch{\stRec{\stRecVar}{\stT}}=\stRec{\stRecVar}{\charSch{\stT}}\\[1mm]
				\charSch{\stSum{\roleQ}{(\roleQ,\dagger)\in J,\;i\in I_{(\roleQ,\dagger)}}{
					
						\stChoice{\stLab[i]}{\stS[i]}\stSeq\stT[i,\roleQ,\dagger]
					
				}{\dagger}}=\\
				\quad\stFmt{
					\sum_{(\roleQ,\dagger)\in\roleSet\times\{!,?\}\setminus J}
					\charSchTest{\roleQ}{\overline{\dagger}}{
						\stFmt{\sum_{(\roleQ,\dagger)\in J}
						{\charSchTrigger{\roleQ}{I_{(\roleQ,\dagger)}}{\overline{\dagger}}{\{\charSch{\stT[i,\roleQ,\dagger]}\}_{i\in I_{(\roleQ,\dagger)}}}}
					}}
				}
	\end{array}
	\)}
\centerline{\small\(\begin{array}{ll}
\text{with }	& \charSchTrigger{\roleQ}{I}{\dagger}{\{\stT[i]\}_{i\in I}} = 
				\stFmt{
						\roleQ!
						\stLab[(I,\dagger)]\stSeq
						\syncSch{\roleQ}{
							\stSum{\roleQ}{i\in I}{\stLab[i]\stSeq\stT[i]}{?}
						}
				}
                       \;\;
                       \charSchTest{\roleQ}{\dagger}{\stT}
					= 
				\stFmt{
					\roleQ!\stLabi[\dagger]\stSeq\roleQ?\stLab\stSeq\stT
				}
				\\
&				\syncSch{\roleR[k]}{\stT} = 
				\roleR[0]\stFmt{!}{\stLab[k]}
				\dots
				\roleR[k-1]\stFmt{!}\stLab[k]\stSeq
				\roleR[k+1]\stFmt{!}\stLab[k]
				\dots
				\roleR[n]\stFmt{!}\stLab[k]\stSeq\stT
		\end{array}
\)}
\end{definition}
\begin{itemize}
\item The information in $\mathcal{T}$
is used to restrict the possible payloads,
labels, and
roles, that
the complementary context must account for.
\item
Our live context for $\stT$ is
$\compEnv{\mpS}{\stT}\stEnvComp\stEnvMap{\mpChanRole{\mpS}{\roleP}}{\stT}$.
\item
For all $\roleQ\in\roleSet$,
$\charTwait{\roleQ}$ responds to commands from $\roleFmt{sch}$
to either:
terminate;
check if $\roleP$ erroneously contains the action
$\roleQ\stFmt{\overline{\dagger}}$;
communicate with $\roleP$ with $\dagger$
and the message set $\{\stChoice{\stLab[i]}{\stS[i]}\}_{i\in I}$;
or synchronise with another participant.
\item In $\charTwait{\roleQ}$, 
$\syncSel{\roleQ}{\stT}$ 
sends a message from $\roleQ$ to the other participants; 
$\syncBra{\roleQ}{\stT}$ is a dual receiver;
$\charTAct{\roleQ}{I}{\dagger}{\stT}$ allows $\roleQ$ to communicate with
$\roleP$ with action $\stFmt{\dagger}$, and
$\charTActi{\dagger}{\stT}$ allows $\roleQ$ to test
for the action of $\roleQ\stFmt{\overline{\dagger}}$ in $\roleP$. 
\item $\charSch{\stT}$ is the scheduling type.
It tells other participants how to communicate with $\roleP$
to check erroneous actions and
to make any all acceptable actions.
\item In $\charSch{\stT}$, 
$\syncSch{\roleQ}{\stT}$ lets
$\roleFmt{sch}$ command all $\roleP\in \roleSet\setminus\{\roleQ\}$  
to receive a synchronisation from $\roleQ$; 
$\charSchTest{\roleQ}{\dagger}{\stT}$ lets $\roleFmt{sch}$ command 
$\roleQ$ to test 
$\roleQ\stFmt{\overline{\dagger}}$ at $\roleP$; and 
$\charSchTrigger{\roleQ}{I}{\dagger}{\{\stT[i]\}_{i\in I}}$
lets $\roleFmt{sch}$ command 
$\roleQ$ to communicate with $\roleP$ with $\stFmt{\dagger}$. 
\end{itemize}
\hideapp{Further explanations can be found in \cref{def:char-ctx-app}.}
\begin{example}[Complementary Types]
\label{ex:compl-types}
	Consider the types
	$\stT=\stFmt{\roleQ!\stChoice{\stLab}{\stEnd}}$ and
	$\stTi=\stFmt{\roleR!\stChoice{\stLab}{\stEnd}}$.
	Let $\mathcal{T}=\{\stT,\stTi,\stEnd\}$
	so that $\stT$ and $\stTi$ are $\mathcal{T}$-valid,
	$\roleSet=\{\roleQ,\roleR\}$, and
	$Act=\left\{
	\{\stChoice{\stLab}{\stEnd}\},
	\{\stChoice{\stLab}{\tyInt}\},
	\{\stChoice{\stLab}{\tyBool}\}
	\right\}$.
	We have that:\\[1mm]
	\begingroup
\centerline{
$\begin{array}{rcl}
		\charSch{\stT}&=&
		\stSum{}{}{
			\roleQ!\stLabi[!]\stSeq\roleQ?\stLab\stSeq\stTii,\,
			\roleR!\stLabi[!]\stSeq\roleR?\stLab\stSeq\stTii,\,
			\roleR!\stLabi[?]\stSeq\roleR?\stLab\stSeq\stTii
		}{}
		\\[1mm]
		\stTii&=&
		\roleQ!\stLab[(\{\stChoice{\stLab}{\stEnd}\}, ?)]\stSeq
		\roleR!\stLab[\roleQ]\stSeq
		\roleQ?\stLab\stSeq
		\roleQ!\stLab[\text{\tiny end}]\stSeq
		\roleR!\stLab[\text{\tiny end}]
		\end{array}
$}\\[1mm]
	\endgroup
	The scheduler will first choose an action not present in $\stT$
	\ie not a prefix of $\unfoldOne{\stT}$
	($\roleQ!$, $\roleR!$, or $\roleR?$),
	and tell its participant
	to test for it.
	If $\roleP$ implements the chosen action,
	then a label mismatch error occurs.
	Otherwise, the participant returns back to the scheduler,
	which continues with $\stTii$.
	The scheduler tells $\roleQ$ to receive
	the message set $\{\stChoice{\stLab}{\stEnd}\}$
	from $\roleP$, and
	tells $\roleR$ to synchronise with $\roleQ$.
	Once $\roleQ$ receives a message from $\roleP$
	and synchronises with $\roleR$,
	$\roleQ$ informs the scheduler of the outcome of the communication.
	The scheduler now believes that $\roleP$
	should have the type $\stEnd$,
	so tells $\roleQ$ and $\roleR$ to terminate.
	
	Note that this differentiates $\stTi$ and $\stT$,
	since $\stTi$ contains the action $\roleR!$.
	The scheduler may tell $\roleR$
	to check for $\roleR!$,
	inducing a label mismatch.
\end{example}

\myparagraph{Characteristics Processes. }
For \textbf{Step 2}, we introduce characteristic processes
for types and contexts.
For this purpose, we introduce the following helper values and processes
to check the types of values.
Below $v_{\tyGround}$ is a value of type $\tyGround$ and
$\mpP[\tyGround]\!\left(\mpx\right)$
is a process that is unsafe iff $\mpx$ is not of type $\tyGround$
and terminates otherwise.

\begin{definition}[Basic Type Characteristics]
\label{def:basic-char}
Let $v_{\tyBool}=\mpTrue$, $v_{\tyInt}=\mpNum{0}$,
$\mpP[\tyInt]\!\left(\mpx\right)=\mpIf{\mpSucc{\mpx}}{\mpNil}{\mpNil}$,
$\mpP[\tyBool]\!\left(\mpx\right)=\mpIf{\mpx}{\mpNil}{\mpNil}$.
\end{definition}

The characteristic process of a session type $\stT$,
as defined below in \cref{def:char-proc},
is a process that:
\begin{enumerate*}
\item behaves as specified by $\stT$; and
\item checks that received values and channels
behave as expected in $\stT$.
\end{enumerate*}

\begin{definition}[Characteristic Processes]
\label{def:char-proc}
For channel $\mpC$ and
type $\stT$, with $\mathcal{T}$-valid payload types,
we define the process $\charP{\mpC}{\stT}$
by recursion on $\stT$.

For a context $\stEnv$,
	we write $\charPM{\stEnv}=\mpBigPar{\mpC\in\stEnv}{\charP{\mpC}{\stEnvApp{\stEnv}{\mpC}}}$.\\[1mm]
	\centerline{\(\begin{array}{c}
			\charP{\mpC}{\stEnd}=\mpNil
			\qquad
			\charP{\mpC}{\stRec{\stRecVar}{\stT}}=\mpRec{\mpX[\stRecVar]}{\charP{\mpC}{\stT}}
			\qquad
			\charP{\mpC}{\stRecVar}=\mpX[\stRecVar]\\[1mm]
			\charP{\mpC}{\stSum{\roleQ[i]}
			{i \in I}
			{\stChoice{\stLab[i]}{\stS[i]}\stSeq\stT[i]}{\dagger_i}}=
				\mpRes{\mpS[i]\suchthat i\in I'}{
				\left(
				\mpSum{\mpPrefix_i\mpSeq\mpP[i]}{i\in I}
				\mpPar
				\mpBigPar{i\in I''}{
					\charPM{\compEnv{\mpS[i]}{\stS[i]}}
					}\right)}
	\end{array}\)}
	\\[1mm]
	where
	$I'=\{i\in I\suchthat \dagger_i={!},\stS[i]\not\in\{\tyInt,\tyBool\}\}$,
	$I'' = \{i\in I'\suchthat \stS[i]\neq\stEnd\}$
	\\[1mm]
	\centerline{\(
        \small
			\mpPrefix_i\mpSeq\mpP[i]=\left\{
			\begin{array}{ll}
				\mpSel{\mpC}{\roleQ[i]}{\stLab[i]}{\mpChanRole{\mpS[i]}{\roleP}}{\left(
						\charP{\mpC}{\stT[i]}
						\mpPar
						\charPM{\bigcup_{j\in I'\setminus\{i\}}\stEnvMap{\mpChanRole{\mpS[j]}{\roleP}}{\stS[j]}}
					\right)}
				& \text{\footnotesize $\dagger_i={!}$,$\stS[i]\not\in\{\tyInt,\tyBool\}$}\\
				\mpSel{\mpC}{\roleQ[i]}{\stLab[i]}{v_{\stS[i]}}{\left(
						\charP{\mpC}{\stT[i]}
						\mpPar
						\charPM{\bigcup_{j\in I'}\stEnvMap{\mpChanRole{\mpS[j]}{\roleP}}{\stS[j]}}
					\right)}
					& \text{\footnotesize $\dagger_i={!}$,$\stS[i]\in\{\tyInt,\tyBool\}$}
					\\
				\mpBra{\mpC}{\roleQ[i]}{\stLab[i]}{\mpy}{\left(
						\charP{\mpC}{\stT[i]}
						\mpPar
						\charP{\mpy}{\stS[i]}
						\mpPar
						\charPM{\bigcup_{j\in I'}\stEnvMap{\mpChanRole{\mpS[j]}{\roleP}}{\stS[j]}}
					\right)} & \text{\footnotesize$\dagger_i={?}$,$\stS[i]\not\in\{\tyInt,\tyBool\}$}\\
				\mpBra{\mpC}{\roleQ[i]}{\stLab[i]}{\mpx}{\left(
						\charP{\mpC}{\stT[i]}
						\mpPar
						\mpP[{\stS[i]}]\!\left(\mpx\right)
						\mpPar
						\charPM{\bigcup_{j\in I'}\stEnvMap{\mpChanRole{\mpS[j]}{\roleP}}{\stS[j]}}
					\right)}\;
					& \text{\footnotesize$\dagger_i={?}$,$\stS[i]\in\{\tyInt,\tyBool\}$}
			\end{array}
			\right.
	\)}
\end{definition}
Characteristic processes are designed to mimic semantics
of a type and
a complementary context whenever a session endpoint
is created for using a payload.

Characteristic process $\charP{\mpC}{\stT}$
follows the structure of $\stT$ along channel $\mpC$.
To output values of type $\tyGround$,
the characteristic value $v_{\tyGround}$ is used.
To output channels of type $\stTi$,
a new session, $\mpSi$, is created
with channel $\mpChanRole{\mpSi}{\roleP}$
having type $\stTi$ as enforced by
the introduction of characteristic process of
a complementary context of $\stTi$.
Received values are verified at runtime by
the characteristic processes of basic types.
Received channels are verified at runtime by
characteristic process of their expected types.
Leftover channel endpoints, created for sends that did not occur,
are used by characteristic processes of their expected types.

\begin{restatable}[Characteristic Processes]{proposition}{propCharProcWellDef}
\label{prop:char-proc}
For channel $\mpC$ and
type $\stT$, with $\mathcal{T}$-valid payload types,
the recursive definition for $\charP{\mpC}{\stT}$
is terminates.
\end{restatable}

\begin{restatable}[Complementary Context]{proposition}{propComplCtx}
\label{prop:compl-live}
	If $\stT$ is $\mathcal{T}$-valid and closed, then
	$
	\left(\stEnvSafePred\cap\stEnvLivePred\right)
	\!\left(
	\compEnv{\mpS}{\stT}
	\stEnvComp
	\stEnvMap{\mpChanRole{\mpS}{\roleP}}{\stT}
	\right)
	$.
\end{restatable}

A session type can be used to type its
characteristic process.

\begin{restatable}[Characteristic Processes]{theorem}{thmCharProc}
\label{thm:char-proc-type}
	Let $\stT$ be $\mathcal{T}$-valid.\\
	Write $\gtFv{\stT}=\{\stRecVar[1],\dots\stRecVar[n]\}$
	and let $\stT[1],\dots,\stT[n]$ be closed and $\mathcal{T}$-valid.
	\\[1mm]
	\centerline{\(
	\tyJudge{
		\{\tyEnvMap{\mpX_{\stRecVar[i]}}{
			\set{\stEnvMap{\mpC}{\stT[i]}}
		}\}_{i\leq n}
	}{
		\charP{\mpC}{\stT}
	}{\set{
		\stEnvMap{\mpC}
		{
			\stT\stFmt{\subst{\stRecVar[1]}{\stT[1]}}\dots\stFmt{\subst{\stRecVar[n]}{\stT[n]}}
		}
	}}
	\)}
\end{restatable}
\noindent Specifically, 
	$\tyJudge{\tyEnvEmpty}{\charP{\mpC}{\stT}}{\set{\stEnvMap{\mpC}{\stT}}}$
for closed and $\mathcal{T}$-valid $\stT$.
By subject reduction, this says that
$\stT$ captures the behaviour
of its characteristic process $\charP{\mpC}{\stT}$.
The reverse is also true:
\cref{thm:char-fid}
states that $\charP{\mpC}{\stT}$
captures behaviours of $\stT$.

\begin{example}[Characteristic Process]
\label{ex:char-proc}
	Consider the type:\\[1mm]
	\centerline{$
	\stT=\stSum{}{}{
		\roleQ!\stChoice{\stLabi}{\stEnd}
		,\,
		\roleQ!\stChoice{\stLab}{\tyInt}
		\stSeq
		\roleQ?\stChoice{\stLab}{
			\roleR!\stChoice{\stLab}{\tyBool}
		}
	}{}$}
	We construct its characteristic process by performing the specified behaviour and
	handling the received channel:\\[1mm]
        \centerline{
	$\charP{\mpChanRole{\mpS}{\roleP}}{\stT}=
		\mpRes{\mpSi}{
			{\left(
					\mpSel{\mpChanRole{\mpS}{\roleP}}{\roleQ}{\stLabi}{\mpChanRole{\mpSi}{\roleP}}{}
					+
					\mpSel{\mpChanRole{\mpS}{\roleP}}{\roleQ}{\stLabi}{\mpNum{0}}{
						\mpBra{\mpChanRole{\mpS}{\roleP}}{\roleQ}{\stLab}{\mpy}{
							\mpSel{\mpy}{\roleR}{\stLab}{\mpTrue}{}
						}
					}
			\right)}{}
		}
$}\\[1mm]
	Observe that:
$
		\tyJudge{\tyEnvEmpty}{\mpSel{\mpy}{\roleR}{\stLab}{\mpTrue}{}}{\set{\stEnvMap{\mpy}{\roleR!\stChoice{\stLab}{\tyBool}}}}
		\quad
		\tyJudge{\tyEnvEmpty}{\charP{\mpChanRole{\mpS}{\roleP}}{\stT}}{\set{\stEnvMap{\mpChanRole{\mpS}{\roleP}}{\stT}}}$.\\
	We use \inferrule{Nil} at $\mpNil$,
	$\inferrule{Sub}$ to introduce $\stEnd$-valued channels,
	$\inferrule{Sum}$ at all the sum processes,
	and \inferrule{Res} at $\mpRes{\mpSi}{}$
	since
	$\stEnvLiveP{\set{\stEnvMap{\mpChanRole{\mpSi}{\roleP}}{\stEnd}}}$.
\end{example}

\begin{theorem}[Two-Party Locks are Insufficient]
\label{prop:two-lock}
\label{thm:two-lock}
	There are types $\stTi\tyNotSub\stT$ such that
	for all $\stEnv$, if
	$
	\stEnv
	\stEnvComp
	\stEnvMap{\mpChanRole{\mpS}{\roleP}}{\stT}
	$
	is live,
	then
	$
	\stEnv
	\stEnvComp
	\stEnvMap{\mpChanRole{\mpS}{\roleP}}{\stTi}
	$
	is 2-plock free.
\end{theorem}

\begin{proof}
	$\stEnv$ has a two-participant lock (2-plock)
iff
$\unfoldOne{\stEnvApp{\stEnv}{\mpChanRole{\mpS}{\roleP}}}
=\stSum{\roleQ[i]}{i\in I}{\stChoice{\stLab[i]}{\stS[i]}\stSeq\stT[i]}{\dagger_i}$
and,
for all $i\in I$ if $\mpChanRole{\mpS}{\roleQ[i]}\in\dom{\stEnv}$ then,
$\unfoldOne{\stEnvApp{\stEnv}{\mpChanRole{\mpS}{\roleQ[i]}}}
=\stEnd
$ or
$\unfoldOne{\stEnvApp{\stEnv}{\mpChanRole{\mpS}{\roleQ[i]}}}
=\stSum{\roleP}{j\in J_i}{\stChoice{\stLab[i,j]}{\stS[i,j]}\stSeq\stT[i,j]}{\dagger}$.

	Let $\stT=(\roleQ\stFmt{?}\stLab\stSeq\roleR\stFmt{?}\stLabi
	+\roleR\stFmt{?}\stLab\stSeq\roleQ\stFmt{?}\stLabi)$
	and
	$\stTi=\roleQ\stFmt{?}\stLab\stSeq\roleR\stFmt{?}\stLabi$.
	Suppose that $\stEnv\stEnvComp\stEnvMap{\mpChanRole{\mpS}{\roleP}}{\stT}$
	is live.
	If $\stEnv\gtMoveStar\stEnvi$ and
	$\stEnvi\stEnvComp\stEnvMap{\mpChanRole{\mpS}{\roleP}}{\stTi}\gtMove[\ltsSendRecvS{\mpS}{\roleQ}{\roleP}{\stLab}]\stEnvii$,
	then $\stEnv\stEnvComp\stEnvMap{\mpChanRole{\mpS}{\roleP}}{\stT}\gtMoveStar\stEnvii$, so $\stEnvii$ is live and is 2-plock free.
	$\stEnvi\stEnvComp\stEnvMap{\mpChanRole{\mpS}{\roleP}}{\stTi}$
	has no two-participant locks:
	$\stEnvApp{\stEnvi}{\mpChanRole{\mpS}{\roleQ}}\gtMoveStar\stTii$
	without $\roleP$ s.t.
	$\stTii\gtMove[\roleP\stFmt{!}\stLabii]$,
	by following a fair transition sequence of
	$\stEnvi\stEnvComp\stEnvMap{\mpChanRole{\mpS}{\roleP}}{\stT}$
	until $\mpChanRole{\mpS}{\roleP}$ terminates;
	so $\unfoldOne{\stEnvApp{\stEnvi}{\mpChanRole{\mpS}{\roleQ}}}\neq\stEnd$
	and
	$\unfoldOne{\stEnvApp{\stEnvi}{\mpChanRole{\mpS}{\roleQ}}}\neq\stSum{\roleP}{i\in I}{\stChoice{\stLab[i]}{\stS[i]}}{?}$.
	\qed
\end{proof}
Note that the proof of
\cref{thm:type-correct}
shows that if a context has no 2-plocks,
then neither does any process that it types.
Thus, \cref{prop:two-lock}
also proves that 2-plocks for processes
are insufficient for preciseness.

\smallskip

\label{sec:complete}

\begin{figure}[t]
  \[
\small  
		\begin{array}{c}
			\inference[$\mu$L]{\tySubst{\stTi}{\stRecVar}{\stRec{\stRecVar}{\stTi}}\tyNot\stT}
			{\stRec{\stRecVar}{\stTi}\tyNot\stT}
			\quad
			\inference[$!$L]
			{
			\roleQ\stFmt{!}\stLab\preType\stTi&
			\roleQ\stFmt{!}\stLab\not\preType\stT
			}
			{
			\stTi
			\tyNot
			\stT
			}
			\quad
			\inference[$?$L]
			{
			\roleQ\stFmt{?} \preType \stTi
			&
			\roleQ\stFmt{?}\not\preType\stT
			}
			{
			\stTi
			\tyNot
			\stT
			}
			\\[1ex]
			\inference[$\mu$R]{\stTi\tyNot\tySubst{\stT}{\stRecVar}{\stRec{\stRecVar}{\stT}}}
			{\stTi\tyNot\stRec{\stRecVar}{\stT}}
			\quad
			\inference[$?$R]
			{
			\roleQ\stFmt{?}\stLab\preType \stT&
			\roleQ\stFmt{?}\stLab\not\preType\stTi
			}
			{
			\stTi
			\tyNot
			\stT
			}
			\quad
			\inference[$!$R]
			{
			\roleQ\stFmt{!}\preType\stT&
			\roleQ\stFmt{!}\not\preType\stTi
			}
			{
			\stTi
			\tyNot
			\stT
			}
			\\[1ex]
			\inference[LR$?$]
			{
			\roleQ\stFmt{?}\stChoice{\stLab}{\stSi}\preType\stTi
			&
			\roleQ\stFmt{?}\stChoice{\stLab}{\stS}\preType\stT
			&
			\stSi\tyNot\stS
			}
			{
			\stTi
			\tyNot
			\stT
			}
			\quad
			\inference[$\stEnd$-L]{}{\stEnd\tyNot\stFmt{\Sigma_{i\in I}\stT[i]}}
			\quad
			\inference[$\tyGround$-L]{\stS\neq\tyGround}{\tyGround\tyNot\stS}
			\\[1ex]
			\inference[LR$!$]
			{
			\roleQ\stFmt{!}\stChoice{\stLab}{\stSi}\preType\stTi
			&
			\roleQ\stFmt{!}\stChoice{\stLab}{\stS}\preType\stT
			&
			\stS\tyNot\stSi
			}
			{
			\stTi
			\tyNot
			\stT
			}
			\quad
			\inference[$\stEnd$-R]{}{\stFmt{\Sigma_{i\in I}\stT[i]}\tyNot\stEnd}
			\quad
			\inference[$\tyGround$-R]{\stS\neq\tyGround}{\stS\tyNot\tyGround}
			\\[1ex]
			\inference[LR]
			{
			\roleQ\stFmt{\dagger}\stChoice{\stLab}{\stSi}\stSeq\stTiii\preType\stTi
			&
			\roleQ\stFmt{\dagger}\stChoice{\stLab}{\stS}\stSeq\stTii\preType\stT
			&
			\stTiii\tyNot\stTii
			}
			{
			\stTi
			\tyNot
			\stT
			}
		\end{array}
\]
\caption{Negation of Subtyping Rules}
\label{fig:neg-sub}
\end{figure}

\myparagraph{Negation of Subtyping. }
As for \textbf{Step 3},
recall that subtyping is defined coinductively (\cref{def:types}).
The negation of coinductive relations are inductive,
suggesting that we can find inductive rules for $\tyNotSub$.
Indeed, \cref{thm:subtype-neg} proves this.

\begin{restatable}[Negation of Subtyping]{theorem}{thmNegationSubtyping}
\label{thm:subtype-neg}
	Define the binary relation $\stTi\tyNot\stT$
	on closed session types inductively by the rules in \cref{fig:neg-sub},
	then $\stTi\tyNot\stT$ iff $\stTi\tyNotSub\stT$.
\end{restatable}
\begin{example}[Negation of Subtyping]
\label{ex:neg-sub}
Consider the types
	$\stT=\stFmt{\roleQ!\stChoice{\stLab}{\stEnd}}$ and
	$\stTi=\stFmt{\roleR!\stChoice{\stLab}{\stEnd}}$.
$\stTi\tyNot\stT$ by:
\inferrule{$!$L} as $\roleR!\stLab\preType\stTi$ and
$\roleR!\stLab\not\preType\stT$; or by
\inferrule{$!$R} as $\roleQ!\preType\stT$ and
$\roleQ!\not\preType\stTi$.
Thus,
$\stFmt{\roleQ?\stChoice{\stLabi}{\stTi}}\tyNot\stFmt{\roleQ?\stChoice{\stLabi}{\stT}}$
by \inferrule{LR$?$}.

\end{example}
\noindent
A similar theorem is crucial in \cite{GJPSY2018,CDSY2017}
where
preciseness was proved by induction
on `failing derivations' of subtyping.

\smallskip 
\myparagraph{Operational Completeness. }
To proceed with \textbf{Step 4},
we observe that
the semantics of characteristic processes
derive from the semantics of their types (\cref{thm:char-fid}).
Also observe that
complementary contexts
fully explore their types\hideapp{ (\cref{thm:comp-fid})},
in the sense that
every transition of $\stT$ is used within the
some path from $\compEnv{\mpS}{\stT}\stEnvComp\stEnvMap{\mpChanRole{\mpS}{\roleP}}{\stT}$.
From this, we deduce operational completeness.

\begin{restatable}[Characteristic Fidelity]{lemma}{thmCharFidelity}
\label{thm:char-fid} 
	(1) If $\stEnv\gtMove\stEnvi$, then
		$\charPM{\stEnv}\mpMoveStar\mpQ\mpPar\charPM{\stEnvi}$;
	(2) If $\roleQ\stFmt{!}\stChoice{\stLab}{\stT}\preType\stEnvApp{\stEnv}{\mpChanRole{\mpS}{\roleR}}$ and
		$\roleR\stFmt{?}\stChoice{\stLab}{\stTi}\preType\stEnvApp{\stEnv}{\mpChanRole{\mpS}{\roleQ}}$, then
		$\charPM{\stEnv}\mpMoveStar\mpQ\mpPar\mpRes{\mpSi}{
		\charPM{\compEnv{\mpSi}{\stT}\stEnvComp\stEnvMap{\mpChanRole{\mpSi}{\roleP}}{\stTi}}}$
		if $\stT\neq\stEnd$ and
		$\charPM{\stEnv}\mpMoveStar\mpQ\mpPar\mpRes{\mpSi}{\charP{\mpChanRole{\mpSi}{\roleP}}{\stTi}}$
		if $\stT=\stEnd$;
	(3) If $\stEnv$ has a label 
		or payload mismatch, with one being a basic type,
		then
		$\charPM{\stEnv}\mpMoveStar\mpCtxApp{\mpCtx}{\mpErr}$;
		and 
	(4) If $\stEnv$ has a 3-plock on $\mpS$, then so does
		$\mpRes{\mpS}{\charPM{\stEnv}}$.
\end{restatable}

The behaviours of a type $\stT$ (up to subtyping) are
precisely those making $\compEnv{\mpS}{\stT}$
safe and live\hideapp{ (\cref{thm:comp-fid})}
\ie if $\stTi$ does not behave according to $\stT$ (is not a subtype),
then $\compEnv{\mpS}{\stT}$ can detect this,
causing an error in the characteristic process.

\begin{restatable}[Characteristic Strictness]{theorem}{thmCharStrict}
\label{thm:char-strict}
	If $\stTi\tyNotSub\stT$ are $\mathcal{T}$-valid and closed, then
	there is $\mpP$ such that
	$\mpRes{\mpS}{
	\left(
		\charPM{\compEnv{\mpS}{\stT}}
		\mpPar
		\charP{\mpChanRole{\mpS}{\roleP}}{\stTi}
	\right)
	}\mpMoveStar\mpP$
	and
	$\mpP$ has an error or a 3-plock.
\end{restatable}

\begin{proof}
	Induction on the derivation of $\stTi\tyNot\stT$,
	using\hideapp{ \cref{thm:comp-fid} and}
	\cref{thm:char-fid}.
	\qed
\end{proof}

\begin{theorem}[Preciseness]
\label{thm:operational-precise}
	The subtyping $\tySub$
	is operationally precise.
\end{theorem}
\begin{proof}
Let $\stTi\tyNotSub\stT$
and fix $\mathcal{T}$ making
$\stTi$ and $\stT$ $\mathcal{T}$-valid.
If $\tyJudge{\tyEnvEmpty}{\mpQ}{\set{\stEnvMap{\mpChanRole{\mpS}{\roleP}}{\stT}}}$,
then
$\tyJudge{\tyEnvEmpty}{
	\charPM{\compEnv{\mpS}{\stT}}
	\mpPar
	\mpQ
	}{(\compEnv{\mpS}{\stT}\stEnvComp\stEnvMap{\mpChanRole{\mpS}{\roleP}}{\stT})}$, 
hence 
	$\tyJudge{\tyEnvEmpty}{
		\mpRes{\mpS}{
	\left(
		\charPM{\compEnv{\mpS}{\stT}}
		\mpPar
		\mpQ
	\right)
	}
	}{\stEnvEmpty}$.
	$\tyJudge{\tyEnvEmpty}{\charP{\mpChanRole{\mpS}{\roleP}}{\stTi}}{\set{\stEnvMap{\mpChanRole{\mpS}{\roleP}}{\stTi}}}$
	and
	$\mpRes{\mpS}{
	\left(
		\charPM{\compEnv{\mpS}{\stT}}
		\mpPar
		\charP{\mpChanRole{\mpS}{\roleP}}{\stTi}
	\right)
	}$
	is not safe or is not 3-plock free by \cref{thm:char-strict}.
	Therefore, $(\stTi,\stT)$
	is not a sound subtyping instance.
	Therefore, $\tySub$ is complete.
	By \cref{thm:prec-sound}, $\tySub$ is sound,
	so $\tySub$ is precise.
	\qed
\end{proof}

\subsection{Denotational Preciseness}
\label{sec:denotation}

We define the interpretation of closed type $\stT$ to be
the set of processes
which $\stT$ types:
$\interpret{\stT}=\{\mpP\suchthat\tyJudge{\tyEnvEmpty}{\mpP}{\set{\stEnvMap{\mpy}{\stT}}}\}$.
We say that a preorder $\trianglelefteq$
is {\bf denotationally precise}
if for all $\stTi$ and $\stT$,
$\stTi\trianglelefteq\stT$ iff $\interpret{\stTi}\subseteq\interpret{\stT}$.
\begin{theorem}[Denotational Preciseness]
\label{thm:precise-denote}
The subtyping
	$\tySub$ is denotationally precise.
\end{theorem}
\begin{proof}
	If $\stTi\tySub\stT$, then
	$\interpret{\stTi}\subseteq\interpret{\stT}$
	by \inferrule{Sub}.
	If $\interpret{\stTi}\subseteq\interpret{\stT}$,
	then fix $\mathcal{T}$
	with $\stTi$ and $\stT$
	$\mathcal{T}$-valid.
	By \inferrule{Sub},
	$\charP{\mpy}{\stTi}\in\interpret{\stT}$,
	so
	$\tyJudge{\tyEnvEmpty}{
		\mpRes{\mpS}{
	\left(
		\charPM{\compEnv{\mpS}{\stT}}
		\mpPar
		\charP{\mpChanRole{\mpS}{\roleP}}{\stTi}
	\right)
	}
	}{\stEnvEmpty}$,
	thus 
	$\mpRes{\mpS}{
	\left(
		\charPM{\compEnv{\mpS}{\stT}}
		\mpPar
		\charP{\mpChanRole{\mpS}{\roleP}}{\stTi}
	\right)}$
	is safe and 3-plock free.
	By \cref{thm:char-strict},
	$\stTi\tySub\stT$.
	\qed
\end{proof}

\section{Extensions to Mixed Choice Multiparty Subcalculi}
\label{sec:extensions}
We define a typing system (\cref{def:restricted-type-rule})
for processes without session interleaving and delegation
(\cref{prop:restricted-no-delegation}) and
show that type checking
guarantees liveness and deadlock-freedom
(\cref{thm:typing-prop-df-live}).
Using this typing system, 
we define preciseness with respect to liveness (\cref{def:precise-live})
and show that the subtyping, $\tySub$,
is precise with respect to liveness (\cref{thm:operational-precise-live}).
We conclude this section with further extensions of our preciseness result
(\cref{def:precise-sub})
to four multiparty session subcalculi of mixed choice
given in \cite{PY2024} (\cref{def:mixed-subsystem}).

\smallskip

\myparagraph{Session Fidelity, Deadlock-Freedom and Liveness}
As in the previous work, 
deadlock-freedom
and liveness of typed processes are not generally guaranteed. 
We limit the calculus 
to one without session delegation and interleaving by using 
the following restricted type judgement. 

\begin{definition}[Restricted Type Judgement]
\label{def:restricted-type-rule}
	We define the typing judgement,
	$\tyJudgeRes{\tyEnv}{\mpP}{\stEnv}$,
	inductively by the rules in \cref{fig:typing-rules},
	replacing \inferrule{Sum} by:
	\\[1mm]
	\centerline{\small
	\(
		\inference[$\text{Sum}^*$]
		{
			\begin{array}{c}
				\forall j\in I\;
				\forall
				\stFmt{\roleQ{\dagger}\stChoice{\stLab}{\stS}\stSeq\stTi}
				\preType \stT
				\left(
				\stEnvEndP{\stEnv}\wedge
				\tyJudgePrefix{\tyEnv}{\mpPrefix_j\mpSeq\mpP[j]}{\left(\stEnv\stEnvComp\stEnvMap{\mpC}{\stT}\right)}
				\wedge
				\mpFmt{\mpChanRole{\mpC}{\roleQ}{\dagger}\stLab}
				\preCalc{\mpSum{\mpPrefix_i\mpSeq\mpP[i]}{i\in I}}
				\right)
			\end{array}
		}
		{
        \tyJudgeRes{\tyEnv}{
        \mpSum{\mpPrefix_i\mpSeq\mpP[i]}{i\in I}
		}{\left(\stEnv\stEnvComp\stEnvMap{\mpC}{\stT}\right)}}
	\)}
\end{definition}
\noindent
\inferrule{$\text{Sum}^*$}
differs from \inferrule{Sum}
since it disallows non-deterministic choice
over distinct channels and
the only channel whose reducts are hidden by a
communication involving $\mpC$
is $\mpC$ itself.
Therefore, we minimally remove
session interleaving and
non-trivial session delegation.
This is less restrictive than the
`plays role' restriction in \cite{POPL19LessIsMore},
and retains all the same guarantees.

\begin{restatable}[Session Delegation]{proposition}{propNoSessionDelegation}
\label{prop:restricted-no-delegation}
	If $\tyJudgeRes{\tyEnv}{\mpP}{\stEnv}$,
	then all instances of session delegation in
	the derivation are trivial
	\ie
	if $\tyJudgeRes{\tyEnvi}{\mpPi}{\stEnvi}$
	is a subderivation
	and $\mpSel{\mpCi}{\roleP}{\stLab}{\mpC}{}\preCalc{\mpPi}$,
	then $\stEnd\tySub\stEnvApp{\stEnvi}{\mpC}$.
\end{restatable}

\begin{example}[Unbounded Leader Election wihout Delegation]
\label{ex:leader-election-res}
Let $\predP=\stEnvSafePred\cap\stEnvLivePred$ in \inferrule{Res} in \cref{fig:typing-rules}.
Recall $\mpP[\text{\tiny lead}]$
from \cref{ex:elect}.
This process contains non-trivial session delegation
(sent channels are used once received),
thus it is not typable with
the restricted type judgement (\cref{prop:restricted-no-delegation}).
We remove delegation to reattain
typability.
	This restricted typed calculus is still interesting,
        as it can type 
	a revised leader election with unbounded 
	session creations. 
	Consider an extension,
	where the elected participant
	spawns a new session
	running the protocol.\\[1mm]
\centerline{
	$\mpQ[\text{\tiny lead}]=\mpRec{\mpX}{
	\mpRes{\mpS}{\left(
	\mpQ[0]\mpPar\mpQ[1]\mpPar\mpQ[2]\mpPar\mpQ[3]\mpPar\mpQ[4]\right)}}\quad\text{where, for}\,0\leq i\leq 4
$}\\[1mm]
\begingroup
\centerline{
	\small
		$\begin{array}{rcl}
			\mpQ[i]&=&
				\mpSel{\mpChanRole{\mpS}{\roleP[i]}}{\roleP[(i+4)\%5]}{\stLabFmt{elect}}{\mpNum{i}}{}+
				\mpBra{\mpChanRole{\mpS}{\roleP[i]}}{\roleP[(i+1)\%5]}{\stLabFmt{elect}}{\mpx}{\mpQi[i]}+
				\mpSel{\mpChanRole{\mpS}{\roleP[i]}}{\roleP[(i+3)\%5]}{\stLabFmt{elect}}{\mpNum{i}}{}
			\\
			\mpQi[i]&=&
					\mpSel{\mpChanRole{\mpS}{\roleP[i]}}{\roleP[(i+2)\%5]}{\stLabFmt{elect}}{\mpNum{i}}{} +
					\mpBra{\mpChanRole{\mpS}{\roleP[i]}}{\roleP[(i+3)\%5]}{\stLabFmt{elect}}{\mpxi}{
						\mpBra{\mpChanRole{\mpS}{\roleP[i]}}{\roleP[(i+2)\%5]}{\stLabFmt{elect}}{\mpxii}{
						\mpX
						}}
		\end{array}$
}\\[1mm]
	\endgroup
	We derive $\tyJudgeRes{\tyEnvEmpty}{\mpQ[\text{\tiny lead}]}{\stEnvEmpty}$.\hideapp{
	See \cref{ex:leader-election-res-detail}
	for the full detail.}
\end{example}

\begin{restatable}[Subject Reduction]{theorem}{subjectReductionRes}
\label{thm:restricted-sr}
Let $\predP$ in \inferrule{Res} in \cref{fig:typing-rules}
be \RC-safe.
Suppose that $\tyJudgeRes{\tyEnv}{\mpP}{\stEnv}$
	and $\stEnvSafeP{\stEnv}$.
	\begin{enumerate*}
		\item If $\mpP\prestruct\mpPi$, then $\tyJudgeRes{\tyEnv}{\mpPi}{\stEnv}$.
		\item $\mpP\mpMove\mpPi$, then
	either $\tyJudgeRes{\tyEnv}{\mpPi}{\stEnv}$
	or there exists $\stEnvi$ s.t.
	$\stEnv\gtMove\stEnvi$ and $\tyJudgeRes{\tyEnv}{\mpPi}{\stEnvi}$.
	\end{enumerate*}
\end{restatable}

\begin{restatable}[Session Fidelity]{theorem}{sessionFidelity}
\label{thm:sess-fid}
Let $\predP$ in \inferrule{Res} in \cref{fig:typing-rules}
be \RC-safe.
	Suppose that $\mpP$ is guarded, $\stEnvSafeP{\stEnv}$, and
	$\tyJudgeRes{\tyEnvEmpty}{\mpP}{\stEnv}$.
	If $\stEnv\gtMove$, then
	there exist $\mpPi$ and $\stEnvi$ such that $\stEnv\gtMove\stEnvi$,
	$\mpP\mpMoveStar\mpPi$, and 
	$\tyJudgeRes{\tyEnvEmpty}{\mpPi}{\stEnvi}$.
\end{restatable}

\begin{restatable}[Properties]{theorem}{thmTypingProperties}
\label{thm:typing-prop-df-live}
Let $\predP$ in \inferrule{Res} in \cref{fig:typing-rules}
be \RC-safe.
	Suppose that $\mpP$ is guarded,
	$\tyJudgeRes{\tyEnvEmpty}{\mpP}{\stEnv}$, and $\predPApp{\stEnv}$:
	(1) If $\predP\subseteq\stEnvDFPred$, then $\mpP$ is deadlock-free;
	(2) If $\predP\subseteq\stEnvLivePred$, then $\mpP$ is live.
\end{restatable}

\begin{example}[Leader Election without Delegation]
	From \cref{ex:leader-election-res}
	and \cref{thm:typing-prop-df-live},
	we deduce that the process
	$\mpQ[\text{\tiny lead}]$ is
	live.
\end{example}

\begin{example}[Fairness is Necessary for Liveness]
	\label{ex:need-fair-proc}
	Due to \inferrule{Sub}
	in \cref{fig:typing-rules},
	the fact that $\stEnvSafePred\cap\stEnvLivePred$
	is downwards closed (\cref{lem:prop-down})
	is crucial.
	Safe and
	weak live
	(liveness without fairness)
	is not downwards closed
	(\cref{ex:fairness-needed}),
	so it can type processes that are not live.
	Recall $\stEnvi\tySub\stEnv$ from \cref{ex:fairness-needed},
	where $\stEnv$ is weak live, $\stEnvi$ is not,
	$\stEnvSafeP{\stEnv}$, and $\stEnvSafeP{\stEnvi}$.
	Consider the process\\[1mm]
\centerline{
		$\mpP=
		\mpRec{\mpX}{\mpRes{\mpSi}{\mpSel{\mpChanRole{\mpS}{\roleP}}{\roleQ}{\stLab}{\mpChanRole{\mpSi}{\roleP}}{\mpX}}}
		\mpPar
		\mpRec{\mpX}{
		\mpSum{\left(
			\begin{array}{l}
				\mpBra{\mpChanRole{\mpS}{\roleQ}}{\roleP}{\stLab}{\mpy}{\mpX}
				\\
				\mpBra{\mpChanRole{\mpS}{\roleQ}}{\roleP}{\stLabi}{\mpy}{\mpNil}
			\end{array}
		\right)}{}}
		\mpPar
		\mpBra{\mpChanRole{\mpS}{\roleR}}{\roleP}{\stLabi}{\mpy}{\mpNil}
$}\\[1mm]
	$\mpP$ is not live;
	the reduct $\mpBra{\mpChanRole{\mpS}{\roleR}}{\roleP}{\stLabi}{\mpy}{\mpNil}$
	can never be used.
	However,
	$\tyJudgeRes{\tyEnvEmpty}{\mpP}{\stEnvi}$
	so by \inferrule{Sub},
	$\tyJudgeRes{\tyEnvEmpty}{\mpP}{\stEnv}$
	(independent of $\predP$ in \inferrule{Res} in \cref{fig:typing-rules}).
\end{example}

\smallskip
\myparagraph{Preciseness of a Familiy of Mixed Choice
  $\pi$-Calculi. \ }
\label{sec:family}
Without session delegation,
the subtyping $\tySub$
is precise w.r.t. liveness.

\begin{definition}[Operational Preciseness w.r.t. Liveness]
\label{def:precise-live}\rm 
	We say that a subtyping instance $(\stTi,\stT)$
	is \emph{sound w.r.t. liveness}
	similarly to \cref{def:precise} but with:
$\predP=\stEnvLivePred\cap\stEnvSafePred\cap\{\stEnv\suchthat\stEnv\text{ contains no session payloads}\}$;
	$\tyJudgeRes{\tyEnv}{\mpP}{\stTii}$;
	$\stT$ and $\stTi$ only having basic typed payloads; and
	requiring process liveness.
	
	Let $\trianglelefteq$ be a preorder on closed session types.
	We say that $\trianglelefteq$ is:
	\emph{sound w.r.t. liveness} if ($\stTi\trianglelefteq\stT$ and
	$\stTi$ and $\stT$ only have basic typed payloads)
	implies that $(\stTi,\stT)$ is a sound w.r.t. liveness subtyping instance;
	\emph{complete w.r.t. liveness} if for all sound w.r.t. liveness subtyping instances $(\stTi,\stT)$,
	$\stTi\trianglelefteq\stT$; and
	\emph{precise w.r.t. liveness} if it is both sound and complete w.r.t. liveness.
\end{definition}

\begin{remark}[Precise w.r.t. Liveness]
\label{rem:prec-live}
The terminology `precise w.r.t. liveness'
is taken from \cite{GPPSY2023}.
It is used as in \cref{def:precise-live},
where the typing system is sound only if
typed processes are live.
Although safety/error-freedom is not mentioned,
it is still required for soundness.
\end{remark}

\begin{theorem}[Preciseness w.r.t. Liveness]
\label{thm:operational-precise-live}
	The subtyping $\tySub$
	is operationally and denotationally precise w.r.t. liveness.
\end{theorem}

\begin{proof}
	\cref{thm:typing-prop-df-live} implies soundness.
	Replace the $\stEnd$ valued payloads in \cref{def:char-ctx}
	by $\tyInt$ and remove all $I\in Act$ s.t. $\exists i\in I$
        s.t. $\stS[i]$ is a session type.
	Following the proof in \cref{sec:complete}, 
	we derive completeness
	since 3-plocks
	are livelocks.
\qed	
\end{proof}

We now extend our result to the family of multiparty subcalculi in \cite{PY2024}:
\begin{definition}[Subcalculi of Mixed Choice]\rm
\label{def:mixed-subsystem}
\begin{enumerate}
	\item Mixed Separate Choice per Participant ($\MS$).
	Each participant in a sum type cannot be both send to and received from \ie
	in $\stSum{\roleQ[i]}{i\in I}{\stChoice{\stLab[i]}{\stS[i]}\stSeq\stT[i]}{\dagger_i}$,
	$\forall i,j\in I.\roleQ[i]=\roleQ[j]\implies\stFmt{\dagger_i}=\stFmt{\dagger_j}$.
	\item Separate Choice ($\SC$).
	In a sum type, either every choice is a send, or every choice is a receive \ie
	the sum syntax is given by $\stSum{\roleQ[i]}{i\in I}{\stChoice{\stLab[i]}{\stS[i]}\stSeq\stT[i]}{\dagger}$.
	\item Directed Mixed Choice ($\DM$).
	All choices in a sum type are with the same participant \ie
	the sum syntax is given by $\stSum{\roleQ}{i\in I}{\stChoice{\stLab[i]}{\stS[i]}\stSeq\stT[i]}{\dagger_i}$.
	\item Separated Choice ($\SE$).
	Sum types are either a sum of outputs to one participant, or  
	a sum of inputs from one participant \ie
	$\stSum{\roleQ}{i\in I}{\stChoice{\stLab[i]}{\stS[i]}\stSeq\stT[i]}{\dagger}$.
\end{enumerate}
\end{definition}
Note that these subcalculi allow session delegations  
unless specified.
	
For each $\mathbb{F}\in\{\MS,\SC,\DM,\SE\}$,
the subtyping $\tySub$,
restricted to types in $\mathbb{F}$,
is precise w.r.t. $\mathbb{F}$.
If we further restrict $\tySub$
to types without session-typed payloads,
the subtyping $\tySub$
is precise w.r.t. $\mathbb{F}$ and liveness.

\begin{definition}[Operational Preciseness with respect to Subcalculi]
\label{def:precise-sub}\rm
	Let $\mathbb{F}\in\{\MS,\SC,\DM,\SE\}$
	be a subcalculus of mixed choice.
	We say that a subtyping instance $(\stTi,\stT)$
	is \emph{sound} w.r.t. $\mathbb{F}$ 
	similarly to \cref{def:precise} but with:
	$\predP=\stEnvLivePred\cap\stEnvSafePred\cap\{\stEnv\suchthat\stEnv\text{ is in }\mathbb{F}\}$; and
	$\stT$ and $\stTi$ are the type syntax defined in $\mathbb{F}$.
	Let $\trianglelefteq$ be a preorder on session types.
	We say that $\trianglelefteq$ is:
	\emph{sound w.r.t. $\mathbb{F}$} if ($\stTi\trianglelefteq\stT$ and
	$\stT$, $\stTi$ are in $\mathbb{F}$) 
	implies that $(\stTi,\stT)$ is a sound w.r.t. $\mathbb{F}$ subtyping instance;
	\emph{complete w.r.t. $\mathbb{F}$} if for all sound w.r.t. $\mathbb{F}$ subtyping instances $(\stTi,\stT)$,
	$\stTi\trianglelefteq\stT$; and
	\emph{precise w.r.t. $\mathbb{F}$} if it is both sound and complete w.r.t. $\mathbb{F}$.
\end{definition}

\begin{theorem}[Preciseness w.r.t. Subsystems]
\label{thm:operational-precise-sub}
	For $\mathbb{F}\in\{\MS,\SC,\DM,\SE\}$,
	the subtyping, $\tySub$,
	is operationally precise and denotationally w.r.t. $\mathbb{F}$
	and
	operationally and denotationally precise w.r.t. $\mathbb{F}$ and liveness.
\end{theorem}
For $\MS$, it is immediate since complementary contexts (\cref{def:char-ctx})
are $\MS$.
For $\mathbb{F}\in\{\MS,\SC,\DM,\SE\}$, remove the
disallowed syntax from 
\cref{def:char-ctx}
and proceed as in \cref{sec:complete}.
\hideapp{
See Appendix \ref{app:extension} for 
details of these changes. }

\section{Implementations of Subtyping and Properties}
\label{sec:implementation}
This section first gives the algorithms for checking 
properties and their complexity results, and then  
explains our tool implementation and benchmark results. 

\subsection{Deciding Context Properties}

\label{sec:liveness-algo}
For a context $\stEnv$,
let: 
$\barb{\stEnv}=\dom{\stEnv\setminus\stEnd}$, the set of barbs;
$\obs{\stEnv}=\{\ltsSubject{\stEnvAnnotGenericSym}\suchthat\stEnv\gtMove[\stEnvAnnotGenericSym]\}$, the set of observations; and
$\reach{\stEnv}{\stEnvi}{\mpC}$ be true
iff there is a path from $\stEnv$ to $\stEnvi$,
where $\mpC$ is never observed \ie
$\reach{\stEnv}{\stEnvi}{\mpC}$
iff there exists a path $\{\stEnv[i]\}_{0\leq i\leq n}$
such that $\stEnv=\stEnv[0]$,
$\stEnvi=\stEnv[n]$, and
for all $i\leq n$,
$\mpC\not\in\bigcup\obs{\stEnv[i]}$.
Similarly to \cite[Lem. 6.17]{UY2025},
we can rewrite fair and not live paths
as a finite cycle;
we can divide this cycle into
smaller cycles.
From this we derive
\cref{prop:live}.
\begin{restatable}[Liveness]{proposition}{propLiveness}
\label{prop:live}
	$\stEnv$ is not live iff
	there exists $\stEnvi$ s.t. $\mpC\in\barb{\stEnvi}$
	and $\stEnv\gtMoveStar\stEnvi$ and
	for all $X \in \obs{\stEnvi}$ there exist
	$\stEnvii\,\gtMove[\stEnvAnnotGenericSym]\,\stEnviii$
	s.t.
	$\reach{\stEnvi}{\stEnvii}{\mpC}$,
	$\reach{\stEnviii}{\stEnvi}{\mpC}$, and
	$X\cap\ltsSubject{\stEnvAnnotGenericSym}\neq \emptyset$.
\end{restatable}

Deciding subtyping is necessary
to construct the LTS
(\inferrule{LCnt} in \cref{def:type-sem}).
We apply the subtyping algorithm in \cite[Thm.~3.13]{UY2025},
extended with mixed choice.

\begin{theorem}[Deciding Subtyping]
\label{thm:sub-decide}
Let $\stT$ and $\stTi$ be type graphs (\cite[Def.~3.6]{UY2025}).
Checking $\stTi\tySub\stT$
has worst-case complexity $\mathcal{O}\left(|\stTi|\cdot|\stT|\right)$.
\end{theorem}

\begin{restatable}[Deciding Properties]{theorem}{thmDecide}
\label{thm:property-decide}
	Let $\stEnv$ be a typing context
	with $|\stEnv|=n$, $m=|\{\stEnvi\suchthat\stEnv\gtMoveStar\stEnvi\}|$,
	and $|\dom{\stEnv}|=r$.
	Safety, deadlock-freedom, and liveness
	can all be decided in either:
	(1) time $\mathcal{O}\left(rn^2m^2\right)$
	and space $\mathcal{O}\left(n^2+rm^2\right)$; or
	(2) time $\mathcal{O}\left(rn^2m^2\log{n}\right)$
	and space $\mathcal{O}\left(n^2+rnm\right)$.
\end{restatable}

\noindent The time complexities are achieved
by enumerating the state space as it is generated,
recording $\obs{\stEnvi}$, $\barb{\stEnvi}$, and
the states adjacent to $\stEnvi$, as $\stEnvi$ is explored.
(1) is achieved by using adjacency arrays; and 
(2) is by using adjacency sets.
Safety and deadlock-freedom
are decided by checking each state for
label/payload mismatches and
termination while having barbs,
respectively.
Liveness is decided by checking the LTS against
the predicate in \cref{prop:live} directly.

\subsection{Implementation and Benchmarks}

\label{sec:impl}

We implemented the
session type and typing context
syntax and semantics, from this paper,
in Scala~3.82.
For typing context semantics,
we implemented the subtyping algorithm in
\cref{thm:sub-decide}.
Using these, we implemented the liveness predicate from \cref{prop:live}
as well as safety and deadlock-freedom.
First, we construct an LTS from a typing context;
during the construction, we
infer whether the context is safe/deadlock-free.
Then, we compute the reachability relation.
Finally, we loop over all roles, $\roleP$; for each one
we compute the `reachability without observing $\roleP$'
relation, and loop over all contexts to check if the inner
predicate is satisfied.
This decides if the predicate is satisfied in time
$\mathcal{O}(rn^2m^2)$,
where $r$ is the size of the domain,
$n$ is the size of the context,
and $m$ is the size of the state space.
The optimisation of using type graphs,
instead of syntactic types that are manually unfolded,
is provided,
and is used during benchmarking.

Our machine configurations are:
Windows 11 Home Version 25H2;
AMD Ryzen 7 4700U @ 2.00 GHz;
16 GB RAM.
Each benchmark is run multiple times before
recording 20 benchmarks,
from which we calculate a mean.

\hideapp{Typing contexts for each example are available in
Appendix \ref{app:impl} (some of which are parameterised).} 
Benchmarks evaluating our tool 
are given in the \cref{table:benchmarks}. 
For each example we report:
which subcalculi the example is within;
the parameter used if applicable;
the number of states in the LTS;
the mean time to determine safety in ms; 
the mean time to determine deadlock-freedom in ms; and
the mean time to determine liveness in ms.

\begingroup
\setlength{\tabcolsep}{2pt}
\renewcommand{\arraystretch}{1.1}

\begin{table}[htbp]
\centerline{
\fontsize{8}{10}\selectfont
\begin{tabular}{|l||c|c|c|c||r|r|r||r|r|r|}
\hline
	Example&$\MC$&$\MS$&$\SC$&$\BI$&\#$N$&\#states&\stEnvSafePred&$\stEnvDFPred$&$\stEnvLivePred$\\
	\hline\hline
OAuth2 \cite{POPL19LessIsMore} & & & & &N/A & 5 &
$0.9$&$0.9$&$0.9$
\\\hline
TwoBuyer \cite{POPL19LessIsMore}& & & & &N/A & 7 & 
$0.1$&$0.1$&$0.1$
\\\hline
MapReduce \cite{POPL19LessIsMore}& & & & & 3  & 20 & 
$1.1$&$1.1$&$1.25$
\\
 & & & & & 5  & 42 & 
$0.75$&$0.75$&$1.05$
\\
 & & & && 10 & 132 & 
$5.35$&$5.35$&$9.85$
\\
 & & & && 50 & 2652 &
$5065.15$&$5065.25$&$13786.5$
\\\hline
MPWorker \cite{POPL19LessIsMore}& & & && 3 & 156 & 
$5.65$&$5.65$&$11.05$
\\
 & & & & &5 & 3906 & 
$3219.75$&$3219.85$&$8005.5$
\\\hline
BinaryCounter & & & && 3 & 125 & 
$1.85$&$1.85$&$3.3$
\\
& & & & & 5 & 3125 & 
$1000.0$&$1000.05$&$1991.55$
\\\hline
BadBinaryCounter& \checkmark& & & & 3 & 1375 & 
$157.65$&$157.65$&$186.25$
\\\hline
LeaderElection \cite{PY2024} & \checkmark & & & &N/A & 22 & 
$0.15$&$0.15$&$0.3$
\\\hline
ServerRequests \cite{JY2020} & & &\checkmark &&  2 & 5 & 
$0.05$&$0.05$&$0.05$
\\
 & & & &&10 & 21 & 
$0.2$&$0.2$&$0.55$
\\
 & & & &&100 & 201 & 
$181.25$&$181.25$&$237.75$
\\\hline
LoadBalancing \cite{majumdar_et_al:LIPIcs.CONCUR.2021.35} & & &\checkmark & & 2 & 4 & 
$0.1$&$0.1$&$0.1$
\\
& & & & & 10 & 12 &
$0.1$&$0.1$&$0.25$
\\
 & & & & & 100 & 102 & 
$57.8$&$57.8$&$74.45$
\\\hline
Calculator \cite{FASE16EndpointAPI} & & & & \checkmark&N/A & 5 & 
$0.05$&$0.05$&$0.05$
\\\hline
SMTP \cite{FASE16EndpointAPI} & & & & \checkmark&N/A & 8 & 
$0.05$&$0.05$&$0.05$
\\\hline
CircuitBreaker \cite{DBLP:conf/ecoop/LagaillardieNY22}& & & \checkmark& &N/A & 14 & 
$0.15$&$0.15$&$0.2$
\\\hline
DistributedLogging \cite{DBLP:conf/ecoop/LagaillardieNY22}& & & & \checkmark&N/A & 4 & 
$0.0$&$0.0$&$0.05$
\\\hline
TravelAgent \cite{HYH2008} & & & & \checkmark&N/A & 6 & 
$0.0$&$0.0$&$0.0$
\\\hline
OnlineWallet \cite{NYH2013}& & & & & N/A & 10 & 
$0.05$&$0.05$&$0.05$
\\\hline
LeaderElection \cite{10.1145/3798224} & &\checkmark & & & 3 & 40 & $1.25$&$1.25$&$1.35$\\
 & & & && 5 & 467 & $16.45$&$16.55$&$27.45$\\
 & & & & & 8 & 18350 & $14428.35$&$14428.6$&$30586.9$\\\hline
Running \cite{DBLP:conf/ictac/BlechschmidtPN25} & & &\checkmark & &N/A & 6 & 
$0.0$&$0.0$&$0.0$
\\\hline
ClientServerWorkers \cite{DBLP:journals/corr/abs-2604-06872}& & \checkmark& & &N/A& 11 & 
$0.1$&$0.1$&$0.1$
\\\hline
TimeOut \cite{DBLP:journals/corr/abs-2604-06872}& & \checkmark& & &N/A & 6 &
$0.05$&$0.05$&$0.05$
\\\hline
Fire \cite{DBLP:journals/corr/abs-2604-06872}& \checkmark& & & &N/A & 13 & 
$0.05$&$0.05$&$0.05$
\\\hline
LeaderElection
(\cref{ex:election-types})
&&\checkmark&&&N/A&17&
$0.1$&$0.1$&$0.2$
\\\hline
Cycle
(\cref{ex:cycle-sem-type})
& \checkmark &&&&2&3&
$0.05$&$0.05$&$0.05$\\
&&&&&3&33&
$0.2$&$0.2$&$0.25$\\
&&&&&4&125&
$2.0$&$2.0$&$2.85$\\
&&&&&5&359&
$12.9$&$12.9$&$14.4$\\
&&&&&10&10743&
$15640.3$&$15640.45$&$17455.35$
\\\hline
\end{tabular}
}
\caption{\small
Benchmarks for liveness algorithm implementation.
$\MC$ denotes that full mixed choice is necessary;
$\MS$, mixed separate choice per participant;
$\SC$, separate choice;
$\BI$, binary (2-party) sessions; and 
\#$N$ denotes the parameter for the example, if applicable. 
Mean times are given in ms.
}
\label{table:benchmarks}
\end{table}

\endgroup

Our implementation is nearly instant on
all handwritten examples.
While property checking is not instant on our large parameterised
examples,
it is still practical with
the largest mean being checking liveness of
the leader election protocol from \cite{10.1145/3798224}
with $8$ participants.
This context's LTS has $18350$ states and
the property checker terminates in
$31$ seconds.
Larger examples, taking longer,
exceed the memory capacity of
personal computers for both
this work and \texttt{mpstk} (a tool used in \cite{POPL19LessIsMore})
before time becomes a practical issue.
Benchmarks for \texttt{mpstk} are omitted
as the non mixed choice benchmarks are available
in \cite{POPL19LessIsMore,BSYZ2022,PY2026}
and are around $100$ times slower than our implementation.

Notice that the average times for checking
$\stEnvSafePred$ and $\stEnvDFPred$
are similar, but lower than
the times for $\stEnvLivePred$.
This is because
$\stEnvSafePred$ and $\stEnvDFPred$
can be immediately inferred
during construction of the LTS, 
whereas checking $\stEnvLivePred$
requires constructing the LTS and
performing further processing.

\section{Related Work and Conclusion}
\label{sec:conclude}

\myparagraph{Preciseness of Subtyping. }
The canonicity of subtyping relations in functional calculi
has been assessed by denotational soundness and completeness,
defined by interpretations of types. 
Denotational completeness of the $\lambda$-calculus
was first considered in \cite{Barendregt1983-qz} for arrow and
intersection types,  
followed by \cite{van2000minimal,Ishihara2002-vx,Vouillon2004-lu} 
extending to union and pair types. 
Operational preciseness was first defined by
Ligatti et al. \cite{10.1145/2994596}
for the call-by-value $\lambda$-calculus
with iso-recursive types. Following this work, 
Dezani and Ghilezan \cite{Dezani-Ciancaglini2014-iz}
show operational preciseness of subtyping
for a concurrent $\lambda$-calculus with arrow, intersection, and union types.

The first work on preciseness for session types 
\cite{CDSY2017}  
has shown that the
synchronous \cite{DBLP:journals/toplas/CarboneHY12} and asynchronous
subtyping
relations
for binary session types with session delegation
are precise with respect to safety and
`co-channel duality'
in their $\pi$-calculus.
Their paper \cite{CDY2024}
summarises the current state of the art on precise subtyping. 
The technique developed for binary session types is insufficient
for multiparty communication
(\cref{prop:two-lock}):
we require 3-plocks
to take the place of `co-channel duality'
(2-plocks).

Ghilezan et al.~\cite{GJPSY2018,GPPSY2023}
proved preciseness with respect to safety and liveness,
for synchronous and
asynchronous subtypings
in multiparty session calculi.
The top-down approach for precise asynchronous subtyping
is proposed in \cite{pischke2026asynchronousglobalprotocolsprecisely}.
The proof strategy in 
\cite{GJPSY2018,GPPSY2023}
follows the steps given in
\cref{sec:intro}.
There are, however, significant technical differences from the proofs 
in this paper.
In Step 1, 
we use a scheduler that has a collection of possible messages
(\cref{def:char-ctx}) 
in contrast to \cite{GJPSY2018,GPPSY2023} which use cyclic
communications to give participants control over their communications.
Cyclic communications do not work in mixed choice
as multiple participants may act simultaneously.
Related to Step 2,
\cite{GJPSY2018,GPPSY2023} leave 
session delegation and creation and interleaving
in characteristic processes as an open problem, 
which has been closed in this paper (\cref{def:char-proc}).
In Step 3, we directly characterise the negation of subtyping
instead of using an intermediate algorithm as in \cite{GJPSY2018}.
In Step 4, the use of a full $\pi$-calculus with interleaving
required us to introduce 3-plocks (\cref{def:three-lock})
instead of simply using errors or referring to processes getting `stuck'.
This allows us to have a more general type system and
requires us to handle delegation and manipulation of
3-plocks in our lemmas (\cref{thm:char-fid}).  
Additionally, mixed choice causes the negation of subtyping
to have more cases than in \cite{GJPSY2018},
adding comprehensive proof cases which do not occur in separated choice.
Furthermore, we have discovered
that \cite{GJPSY2018}'s operational soundness is flawed because 
their subject reduction theorem fails for their projection\hideapp{;
see Appendix~\ref{app:1-level-fails} for a counterexample}.
The full merge coinductive projection from \cite{UY2025}
might fix this issue.

Li et al.~\cite{Li2024-mu} propose a
decidable subtyping relation
for communicating automata \cite{Brand1983-wu}
which is sound and complete with respect to
\emph{subprotocol fidelity} and
deadlock-freedom.
This work considers
correctness w.r.t. implementing global types, 
which is not related to either process correctness or typing systems. 
Padovani and Zavattaro~\cite{DBLP:conf/ecoop/PadovaniZ25}
propose a subtyping for binary session types with asynchronous semantics
that is complete with respect to \emph{fair termination}, 
and Padovani et al. \cite{DBLP:conf/concur/BravettiPZ25}
propose one that is complete with respect to
a \emph{convergence} relation.
Both completeness results are w.r.t. type semantics
not their typing system; hence it differs from our preciseness.

\smallskip 

\myparagraph{Bottom-Up Mixed Choice and Tools. }
The non-deterministic sum syntax originated
with the $\pi$-calculus \cite{Milner1992-rk},
and has since incited work analysing its expressive power.
Palamidessi \cite{Palamidessi03}
showed that there is no `good'
encoding of the $\pi$-calculus with guarded mixed choice
into the $\pi$-calculus with guarded separate choice.
This was followed by \cite{PY2024}, 
giving the expressiveness hierarchy
of session calculi with various restrictions of choice.
The subtyping in \cite{PY2024} is the first extension of
the standard subtyping to mixed choice and,
by fixing rule \inferrule{S$\Sigma$}\hideapp{ (\cref{rem:subtyping-trivial})},
is equal to our subtyping.
Prokić et al.~\cite{PPGSY2025} have formalised and typed 
federated learning protocols with asynchronous separated choice.  
Their subtyping is the restriction of our subtyping
to separated choice types,
which is precise under synchronous semantics
but not asynchronous semantics.
Hinrichsen et al.~\cite{10.1145/3798224}
introduced `Mixtris',
a separation logic for multiparty message-passing programs. 
Mixtris is used to reason about
\emph{functional} correctness using dependent binders, 
and neither
formulates nor guarantees
either deadlock-freedom or liveness. 
While their functional correctness is intrinsically undecidable, 
we have: 
proved decidability 
for deadlock-freedom and liveness;
implemented an algorithm to decide 
these properties; and
provided its complexity.
Blechschmidt et al. 
\cite{DBLP:conf/ictac/BlechschmidtPN25}
use a probabilistic mixed choice multiparty session $\pi$-calculus;
this additionally has probabilistic selections
but excludes session delegation and
only allows choice over a single channel.
They introduced the `refinement subtyping',
which permits replacement of channels by
several channels implementing the same behaviour.
This differs from the standard approach of
defining a subtyping on session types
and extending to typing contexts pointwise.

The tool \texttt{mpstk} \cite{POPL19LessIsMore}
and its extension \cite{lmcs2025}
decide typing context properties for bottom-up MPST systems. 
They encode properties
as $\mu$-calculus formulae \cite[Figure 5]{POPL19LessIsMore}
to use mCRL2 for model checking.
This has two main limitations:
\texttt{mpstk} does not support mixed choice;
the formula for liveness in \cite[Figure 5]{POPL19LessIsMore}
is not applicable for mixed choice liveness;
and the correctness of the formulae has been left unproven.
We instead propose algorithms to check
typing context properties, which 
are more efficient than the approach in
\cite{POPL19LessIsMore}.
Our algorithm has no external dependencies, 
thus it is more extendable for a future adaptation.

Typing systems for $\pi$-calculi often fail to guarantee
deadlock-freedom and liveness
due to interleaving and delegation (\cref{ex:interleaved}).
Work to overcome this is inapplicable to this paper:
\cite{CDYP2015} relies on global types,
\cite{Padovani2014-zb,Van_den_Heuvel2022-ex}
use binary session types and
none of them use mixed choice.

\smallskip 
\myparagraph{Top-Down Mixed Choice. \ }
The first work providing a top-down approach with mixed choice 
is \cite{DY2012},
where they explore a relationship between 
communicating automata \cite{Brand1983-wu} and multiparty session types. 
Later,
Lange et al.~\cite{LTY2015} 
studied a synthesis algorithm 
which builds a global choreography  
with mixed choice from communicating automata.  
Castagna et al. \cite{lmcs:773}
provide a semantic description of well-formedness
for asynchronous mixed choice global types
and a semantic notion of projection
with a decidable approximation.
Jongmans and Yoshida \cite{JY2020}
extended global types with mixed choice,
existential quantification,
and unrestricted parallel composition.
They verify well-formedness by
checking bisimilarity between projected contexts and global types.
Majumdar et al. \cite{majumdar_et_al:LIPIcs.CONCUR.2021.35}
and 
Li et al.~\cite{Li2023-sx}
provide projection algorithms
for sender-driven global types under asynchronous semantics.
This is a restriction of mixed choice
where each global choice has the same sender.
The algorithm in \cite{Li2023-sx} is complete
for deadlock-free communicating automata \cite{Brand1983-wu}.
Hamers and Jongmans \cite{Hamers2022-ho}
implemented mixed choice session types
in `Discourje',
for run-time verification of Clojure programs.
Cruz-Filipe et al. \cite{10.1145/3167132.3167267}
consider flexible choice for choreographies,
but as `multicoms/multisels' instead of mixed choices.

Realisability of mixed choice global types in asynchronous 
semantics with respect to 
deadlock-free communicating automata \cite{Brand1983-wu}
has been proven 
undecidable in \cite{10.1007/978-3-031-91121-7_13}.  
Bocchi et al. \cite{DBLP:journals/corr/abs-2602-23927}
propose new global syntax and semantics,
which allow for the transient inconsistencies
that may occur during non-deterministic communications.
Di Giusto et al. \cite{10.1145/3756907.3756918}
approach realisability without
projecting.
They provide algorithms, for both
synchronous and asynchronous (p2p)
semantics, that decide
whether a global type is realisable,
given a complementary global type.
These cannot be found in general for mixed choice types.
Barbanera and Dezani 
have considered mixed choice global types,
synchronously \cite{DBLP:journals/corr/abs-2508-13616}
and asynchronously \cite{DBLP:journals/corr/abs-2604-06872}.
They introduced notions of coherence
between sets of global choices and local contexts
to deconstruct sessions into modules.
This differs from standard projection algorithms
as it relates a global type to an entire context at once,
and the algorithm is parameterised by a partition
of the participant set.
This causes
issues with scaling,
since components cannot be considered individually
and useful partitions
are non-trivial to provide.
They use a session calculus (similar to local types),
which lacks session interleaving and delegation.

None of the above works prove the preciseness of subtyping relations.

\smallskip 
\myparagraph{Conclusion and Future Work. \ } 
The preciseness result of this paper
shows that the mixed choice multiparty subtyping $\tySub$
(\cref{def:subtyping})
is the canonical notion of subtyping for the full synchronous multiparty session calculus in
\cref{sec:calc};
it is both operationally precise (\cref{thm:operational-precise})
and denotationally precise (\cref{thm:precise-denote}).

Establishing preciseness requires the introduction of new
machinery and proof methodologies, together with detailed analysis and
technically involved proofs.    
Two of our main technical contributions are 
3-plocks (\cref{def:three-lock}) and
complementary contexts (\cref{def:char-ctx}).
3-plocks are minimal patterns of livelocks and
is the part of context liveness
that is maintained by typing processes with
delegation and session interleaving (\cref{thm:type-correct}).
Complementary context
$\compEnv{\mpS}{\stT}$ for each type $\stT$,
monitors all possible failures of subtyping
non-deterministically and   
forces errors or 3-plocks 
if any incorrect behaviour is detected. 
These proof techniques are not required 
for two-party or separated choice, 
shedding light on the expressiveness of multiparty mixed choice. 

We proposed a new typing system for a mixed choice
multiparty session calculus which ensures safety (\cref{thm:type-correct}), and
deadlock-freedom and liveness (\cref{thm:typing-prop-df-live}).
Our third main technical contribution is the
representation of $\stEnvLivePred$ in \cref{prop:live}.
This substantially improves
the complexity of deciding $\stEnvLivePred$
(\cref{thm:property-decide}),
making it tractable in
the size of the state space and context.
The benchmarking of our implementation (\cref{sec:impl})
has shown that this approach is efficient in practice.

Our future work includes:
preciseness for asynchronous mixed choice
multiparty session types;
and the top-down approach with mixed choice,
building a limited but practical projection algorithm along the line of 
\cite{DBLP:journals/corr/abs-2602-23927}. 

\subsubsection*{Acknowledgements.}
We thank the reviewers for their detailed comments.
This work was supported by
EPSRC
EP/T006544/2,
EP/T014709/2,
EP/Z533749/1,
ARIA and
Horizon EU TaRDIS 101093006 (UKRI number 10066667).

\bibliographystyle{plainurl}

\bibliography{main}

\newpage

\appendix

\section{Appendix for Calculus (\S~2) and Types (\S~3)}
\label{app:calc}
\subsection{Calculus}

\begin{definition}[Expression Evaluation {\cite{GJPSY2018}}]\rm
\label{def:expression-eval}
	We define the evaluation of expressions
	to values, denoted $\eval{e}{v}$, recursively:
	\begin{equation*}
	\begin{array}{c}
		\inference[Val]{}{\eval{v}{v}}
		\quad
		\inference[EqT]{\eval{e}{v}&\eval{e'}{v}}{\eval{e=e'}{\mpTrue}}
		\quad
		\inference[EqF]{\eval{e}{v}&\eval{e'}{v'}&v\neq v'}{\eval{e=e'}{\mpFalse}}
		\\[1ex]
		\inference[IneqT]{\eval{e}{\mpNum{n}} & \eval{e'}{\mpNum{m}} & \mpNum{n} < \mpNum{m}}{\eval{e<e'}{\mpTrue}}
		\quad
		\inference[IneqF]{\eval{e}{\mpNum{n}} & \eval{e'}{\mpNum{m}} & \mpNum{n} \geq \mpNum{m}}{\eval{e<e'}{\mpFalse}}
		\\[1ex]
		\inference[Succ]{\eval{e}{\mpNum{n}}}{\eval{\mpSucc{e}}{\mpNum{n+1}}}
		\quad
		\inference[NegNum]{\eval{e}{\mpNum{n}}}{\eval{\mpNeg{e}}{\mpNum{-n}}}
		\\[1ex]
		\inference[NegT]{\eval{e}{\mpTrue}}{\eval{\mpNeg{e}}{\mpFalse}}
		\quad
		\inference[NegF]{\eval{e}{\mpFalse}}{\eval{\mpNeg{e}}{\mpTrue}}
		\end{array}
	\end{equation*}
	We define the evaluation of expressions to errors, denoted $\eval{e}{\mpErr}$,
	recursively:
	\begin{equation*}
	\small
		\begin{array}{c}
			\inference[]{\eval{e}{v}
			&\text{or}&
			\eval{e'}{v}&
			v\not\in\tyInt}{\eval{e<e'}{\mpErr}}
			\qquad
			\inference[]
			{\eval{e}{v}&v\not\in\tyInt}
			{\eval{\mpSucc{e}}{\mpErr}}
			\\[1ex]
			\inference[]
			{\eval{e}{\mpErr}}
			{
			\begin{array}{c}
			\eval{e=e'}{\mpErr}
			\quad
			\eval{e'=e}{\mpErr}
			\quad
			\eval{e<e'}{\mpErr}
			\quad
			\eval{e'<e}{\mpErr}
			\quad
			\eval{\mpSucc{e}}{\mpErr}
			\quad
			\eval{\mpNeg{e}}{\mpErr}
			\end{array}
			}
		\end{array}
	\end{equation*}
\end{definition}

\begin{definition}[Parallel Composition]\rm
\label{def:canon-par}
	Let $F$ be a choice function on indexing sets,
	we define:
	\[
		\mpBigPar{i\in \{k\}}{\mpP[i]} = \mpP[k]\quad\text{;}\quad
		\mpBigPar{i\in I}{\mpP[i]} = \mpP[F(I)] \mpPar \mpBigPar{i\in I\setminus\{F(I)\}}{\mpP[i]}\text{ otherwise}
	\]
	We say that $``\mpP=\mpBigPar{i\in I}{\mpP[i]}''$ iff:
	$I=I'\cup I''$ with $I',I''\neq\emptyset$, $I'\cap I''=\emptyset$, and
	$\mpP=\mpP[L]\mpPar\mpP[R]$ with
	$``\mpP[L]=\mpBigPar{i\in I'}{\mpP[i]}''$ and
	$``\mpP[R]=\mpBigPar{i\in I''}{\mpP[i]}''$;
	or $I=\{i\}$ and $\mpP=\mpP[i]$.
\end{definition}
\noindent
Although this is omitted in existing work,
this is important to formally define.
The syntax only explicitly permits binary parallel composition
and reordering parallel compositions
does not give equal processes,
they will only be congruent.

\begin{definition}[Subject of Prefix]\rm
\label{def:prefix-subj}
	We define the subject of a prefix by:
	$\ltsSubject{\mpSel{\mpC}{\roleP}{\stLab}{\mpCi}{}}=
	\ltsSubject{\mpSel{\mpC}{\roleP}{\stLab}{e}{}}=
	\ltsSubject{\mpBra{\mpC}{\roleP}{\stLab}{\mpy}{}}=
	\ltsSubject{\mpBra{\mpC}{\roleP}{\stLab}{\mpx}{}}=\mpC$.
\end{definition}

\begin{definition}[Free Variables]
\label{def:free-var-calc}
	We define the {\bf free expression variable} function on processes by:
	\begin{equation*}
		\begin{array}{lcll}
			\fev{\mpNil}&=&\emptyset\\
			\fev{\mpX}&=&\emptyset\\
			\fev{\mpRec{\mpX}{\mpP}}&=&\fev{\mpP}\\
			\fev{\mpIf{e}{\mpP}{\mpPi}}&=&\fev{e}\cup\fev{\mpP}\cup\fev{\mpPi}\\
			\fev{\mpSel{\mpC}{\roleP}{\stLab}{e}{\mpP}}&=&\fev{e}\cup\fev{\mpP}\\
			\fev{\mpSel{\mpC}{\roleP}{\stLab}{\mpCi}{\mpP}}&=&\fev{\mpP}\\
			\fev{\mpBra{\mpC}{\roleP}{\stLab}{\mpx}{\mpP}}&=&\fev{\mpP}\setminus\{\mpx\}\\
			\fev{\mpBra{\mpC}{\roleP}{\stLab}{\mpy}{\mpP}}&=&\fev{\mpP}\\
			\fev{\mpSum{\mpPrefix_i\mpSeq\mpP[i]}{i\in I}}&=&\bigcup_{i\in I}\fev{\pprefix_i\mpSeq\mpP[i]}\\
			\fev{\mpP\mpPar\mpPi}&=&\fev{\mpP}\cup\fev{\mpPi}\\
			\fev{\mpRes{\mpS}{\mpP}}&=&\fev{\mpP}\\
			\fev{\mpErr}&=&\emptyset
		\end{array}
	\end{equation*}
	We define the {\bf free channel variable} function on channels and processes by:
	\begin{equation*}
		\begin{array}{lcll}
			\fcv{\mpChanRole{\mpS}{\roleP}}&=&\emptyset\\
			\fcv{\mpy}&=&\{\mpy\}\\
			\fcv{\mpNil}&=&\emptyset\\
			\fcv{\mpX}&=&\emptyset\\
			\fcv{\mpRec{\mpX}{\mpP}}&=&\fcv{\mpP}\\
			\fcv{\mpIf{e}{\mpP}{\mpPi}}&=&\fcv{\mpP}\cup\fcv{\mpPi}\\
			\fcv{\mpSel{\mpC}{\roleP}{\stLab}{e}{\mpP}}&=&\fcv{\mpC}\cup\fcv{\mpP}\\
			\fcv{\mpSel{\mpC}{\roleP}{\stLab}{\mpCi}{\mpP}}&=&\fcv{\mpC}\cup\fcv{\mpCi}\cup\fcv{\mpP}\\
			\fcv{\mpBra{\mpC}{\roleP}{\stLab}{\mpx}{\mpP}}&=&\fcv{\mpC}\cup\fcv{\mpP}\\
			\fcv{\mpBra{\mpC}{\roleP}{\stLab}{\mpy}{\mpP}}&=&\fcv{\mpC}\cup\fcv{\mpP}\setminus\{\mpy\}\\
			\fcv{\mpSum{\mpPrefix_i\mpSeq\mpP[i]}{i\in I}}&=&\bigcup_{i\in I}\fcv{\pprefix_i\mpSeq\mpP[i]}\\
			\fcv{\mpP\mpPar\mpPi}&=&\fcv{\mpP}\cup\fcv{\mpPi}\\
			\fcv{\mpRes{\mpS}{\mpP}}&=&\fcv{\mpP}\\
			\fcv{\mpErr}&=&\emptyset
		\end{array}
	\end{equation*}
	We define the {\bf free process variable} function on processes by:
	\begin{equation*}
		\begin{array}{lcl}
		\fpv{\mpNil}&=&\emptyset\\
		\fpv{\mpX}&=&\{\mpX\}\\
		\fpv{\mpRec{\mpX}{\mpP}}&=&\fpv{\mpP}\setminus\{\mpX\}\\
		\fpv{\mpIf{e}{\mpP}{\mpPi}}&=&\fpv{\mpP}\cup\fpv{\mpPi}\\
		\fpv{\mpSum{\mpPrefix_i\mpSeq\mpP[i]}{i\in I}}&=&\bigcup_{i\in I}\fpv{\mpP[i]}\\
		\fpv{\mpP\mpPar\mpPi}&=&\fpv{\mpP}\cup\fpv{\mpPi}\\
		\fpv{\mpRes{\mpS}{\mpP}}&=&\fpv{\mpP}\\
		\fpv{\mpErr}&=&\emptyset
		\end{array}
	\end{equation*}
	We define the {\bf free channels with roles} function on channels and processes by:
	\begin{equation*}
		\begin{array}{lcll}
			\fc{\mpChanRole{\mpS}{\roleP}}&=&\{\mpChanRole{\mpS}{\roleP}\}\\
			\fc{\mpy}&=&\emptyset\\
			\fc{\mpNil}&=&\emptyset\\
			\fc{\mpX}&=&\emptyset\\
			\fc{\mpRec{\mpX}{\mpP}}&=&\fc{\mpP}\\
			\fc{\mpIf{e}{\mpP}{\mpPi}}&=&\fc{\mpP}\cup\fc{\mpPi}\\
			\fc{\mpSel{\mpC}{\roleP}{\stLab}{e}{\mpP}}&=&\fc{\mpC}\cup\fc{\mpP}\\
			\fc{\mpSel{\mpC}{\roleP}{\stLab}{\mpCi}{\mpP}}&=&\fc{\mpC}\cup\fc{\mpCi}\cup\fc{\mpP}\\
			\fc{\mpBra{\mpC}{\roleP}{\stLab}{\mpx}{\mpP}}&=&\fc{\mpC}\cup\fc{\mpP}\\
			\fc{\mpBra{\mpC}{\roleP}{\stLab}{\mpy}{\mpP}}&=&\fc{\mpC}\cup\fc{\mpP}\\
			\fc{\mpSum{\mpPrefix_i\mpSeq\mpP[i]}{i\in I}}&=&\bigcup_{i\in I}\fc{\pprefix_i\mpSeq\mpP[i]}\\
			\fc{\mpP\mpPar\mpPi}&=&\fc{\mpP}\cup\fc{\mpPi}\\
			\fc{\mpRes{\mpS}{\mpP}}&=&\fc{\mpP}\setminus\{\mpChanRole{\mpS}{\roleP}\suchthat\roleP\text{ is a participant}\}\\
			\fc{\mpErr}&=&\emptyset
		\end{array}
	\end{equation*}
	We define the {\bf channel} function on processes by
	$\chan{\mpP}=\fcv{\mpP}\cup\fc{\mpP}$.
	We define the {\bf free session} function on processes by
	$\fs{\mpP}=\{\mpS\suchthat\exists\mpChanRole{\mpS}{\roleP}\in\fc{\mpS}\}$.
\end{definition}

\begin{definition}[Precongruence]\rm
\label{def:str}
\label{def:struct-precong}
We define the structural precongruence relation on processes inductively
by the rules below,
where $\mpP\equiv\mpQ$ means
$\mpP\prestruct\mpQ$ and $\mpQ\prestruct\mpP$:
	\[
	\begin{array}{c}
		\inference[]
		{}
		{\mpP\mpPar\mpQ\equiv\mpQ\mpPar\mpP}
		\quad
		\inference[]
		{}
		{\mpP\mpPar(\mpQ\mpPar\mpR)\equiv(\mpP\mpPar\mpQ)\mpPar\mpR}
		\quad
		\inference[]{}{\mpP\mpPar\mpNil\equiv\mpP}
		\quad
		\inference{}{\mpRes{\mpS}{\mpNil}\equiv\mpNil}
		\\[1ex]
		\inference[]{}{\mpRes{\mpS}{\mpRes{\mpSi}{\mpP}}\equiv\mpRes{\mpSi}{\mpRes{\mpS}{\mpP}}}
		\quad
		\inference[]{\mpS\not\in\fs{\mpP}}{\mpRes{\mpS}{\left(\mpP\mpPar\mpQ\right)}\equiv\mpP\mpPar\mpRes{\mpS}{\mpQ}}
		\quad
		\inference[]{}{\mpP\equiv\mpP}
		\\[1ex]
		\inference[]{\mpP\prestruct\mpPi&\mpPi\prestruct\mpPii}{\mpP\prestruct\mpPii}
		\quad
		\inference[]{\mpP\prestruct\mpQ}{\mpCtxApp{\mpCtx}{\mpP}\prestruct\mpCtxApp{\mpCtx}{\mpQ}}
		\quad
		\inference[]{}{\mpRec{\mpX}{\mpP}\prestruct\mpP\subst{\mpX}{\mpRec{\mpX}{\mpP}}}
	\end{array}
	\]
	We define the structural precongruence relation on contexts
	by the rule:
	\[
		\inference
		{
		\mpP\prestruct\mpRes{\mpSi[1],\dots,\mpSi[m]}{\mpPi}
		&
		\{\mpSii[i]\}_{1\leq i\leq k}=\{\mpS[i]\}_{1\leq i\leq n}\cup\{\mpSi[i]\}_{1\leq i\leq m}
		}
		{
			\mpRes{\mpS[1],\dots,\mpS[n]}{\left(\mpCtxHole\mpPar\mpP\right)}
			\prestruct
			\mpRes{\mpSii[1],\dots,\mpSii[k]}{\left(\mpCtxHole\mpPar\mpPi\right)}
		}
	\]
\end{definition}

\begin{lemma}[Context Precongruence]
\label{lem:context-precong}
	(1) Precongruence on contexts is reflexive and transitive.\\
	(2) If $\mpCtxApp{\mpCtx}{}\prestruct\mpCtxApp{\mpCtxi}{}$, then
	$\mpCtxApp{\mpCtx}{\mpQ}\prestruct\mpCtxApp{\mpCtxi}{\mpQ}$.
\end{lemma}

\begin{proof}[1]
	Let
	\[\mpCtxApp{\mpCtx}{}=\mpRes{\mpS[1],\dots,\mpS[n]}{\left(\mpCtxHole\mpPar\mpP\right)}\],
	$\mpP\prestruct\mpP$ so
	\[\mpRes{\mpS[1],\dots,\mpS[n]}{\left(\mpCtxHole\mpPar\mpP\right)}\prestruct
	\mpRes{\mpS[1],\dots,\mpS[n]}{\left(\mpCtxHole\mpPar\mpP\right)}\]
	Suppose
	\[\begin{array}{rcl}\mpRes{\mpS[a,1],\dots,\mpS[a,n_a]}{\left(\mpCtxHole\mpPar\mpP[0]\right)}
	&\prestruct&
	\mpRes{\mpS[b,1],\dots,\mpS[b,n_b]}{\left(\mpCtxHole\mpPar\mpP[1]\right)}
	\\&\prestruct&
	\mpRes{\mpS[c,1],\dots,\mpS[c,n_n]}{\left(\mpCtxHole\mpPar\mpP[2]\right)}
	\end{array}\]
	So
	\[
	\begin{array}{c}
	\mpP[0]\prestruct\mpRes{\mpSi[a,1],\dots,\mpSi[a,n'_a]}{\mpP[1]}
	\\
	\mpP[1]\prestruct\mpRes{\mpSi[b,1],\dots,\mpSi[b,n'_b]}{\mpP[2]}
	\\
	\{\mpS[b,i]\}_{1\leq i\leq n_b}=\{\mpS[a,i]\}_{1\leq i\leq n_a}\cup\{\mpSi[a,i]\}_{1\leq i\leq n'_a}
	\\
	\{\mpS[c,i]\}_{1\leq i\leq n_c}=\{\mpS[b,i]\}_{1\leq i\leq n_b}\cup\{\mpSi[b,i]\}_{1\leq i\leq n'_b}
	\end{array}\]
	So,
	\[
	\begin{array}{c}
	\mpP[0]\prestruct\mpRes{\mpSi[b,1],\dots,\mpSi[b,n'_b]}{\mpRes{\mpSi[a,1],\dots,\mpSi[a,n'_a]}{\mpP[1]}}
	\\
	\{\mpS[c,i]\}_{1\leq i\leq n_c}=\{\mpS[a,i]\}_{1\leq i\leq n_a}\cup\{\mpSi[a,i]\}_{1\leq i\leq n'_a}\cup\{\mpSi[b,i]\}_{1\leq i\leq n'_b}
	\end{array}\]
	So,
	\[\mpRes{\mpS[a,1],\dots,\mpS[a,n_a]}{\left(\mpCtxHole\mpPar\mpP[0]\right)}\prestruct\mpRes{\mpS[c,1],\dots,\mpS[c,n_c]}{\left(\mpCtxHole\mpPar\mpP[2]\right)}\]
\qed\end{proof}

\begin{proof}[2]
	This is immediate from
	reordering restrictions,
	the context rule,
	transitivity, and
	moving restrictions over composition
	(by the Barendregt convention).
\qed\end{proof}

\begin{lemma}
\label{lem:ctx-sem}
	$\mpCtxApp{\mpCtx}{}\prestruct\mpCtxApp{\mpCtxi}{}\implies\forall\mpP.\,\mpCtxApp{\mpCtx}{\mpP}\prestruct\mpCtxApp{\mpCtxi}{\mpP}$
\end{lemma}

\begin{proof}
	Context precongruence is an application of the context rule
	followed by the free session parallel rule and reordering session restrictions.
\qed\end{proof}

\begin{lemma}[Parallel Composition]
\label{lem:parallel-cong}
	If $``\mpP=\mpBigPar{i\in I}{\mpP[i]}''$ and
	$``\mpQ=\mpBigPar{i\in I}{\mpP[i]}''$, then
	$\mpP\equiv\mpQ$.
\end{lemma}

\begin{proof}
	We induct on the $|I|$:
	if $``\mpP=\mpBigPar{i\in I}{\mpP[i]}''$, then,
	for $j\in I$,
	if $|I|>1$ then
	$\mpP\equiv\mpP[j]\mpPar\mpBigPar{i\in I\setminus\{j\}}{\mpP[i]}$
	and if $I=\{j\}$ then $\mpP=\mpP[j]$.
	If $I=\{j\}$ then $\mpP=\mpP[j]$.
	If $|I|>1$, then
	$\mpP=\mpP[l]\mpPar\mpP[r]$
	with $``\mpP[l]=\mpBigPar{i\in I_l}{\mpP[i]}''$,
	$``\mpP[r]=\mpBigPar{i\in I_r}{\mpP[i]}''$,
	$I_l\cup I_r=I$,
	$I_l\cap I_r=\emptyset$, and
	$I_l,I_r\neq\emptyset$.
	If $j\in I_l$, then
	$\mpP[r]\equiv\mpBigPar{i\in I_r}{\mpP[i]}$,
	by I.H. and definition of $\mpBigPar{i\in I_r}{\mpP[i]}$, and
	$\mpP[l]\equiv\mpP[j]\mpBigPar{i\in I_l\setminus\{j\}}{\mpP[i]}$
	if $|I_l|>1$ or
	$\mpP[l]=\mpP[j]$
	by I.H.
	So $\mpP\equiv\mpP[j]\mpPar\mpBigPar{i\in I_l\setminus\{j\}}{\mpP[i]}\mpPar\mpBigPar{i\in I_r}{\mpP[i]}$.
	$\mpBigPar{i\in I_l\setminus\{j\}}{\mpP[i]}\mpPar\mpBigPar{i\in I_r}{\mpP[i]}\equiv
	\mpBigPar{i\in I\setminus\{j\}}{\mpP[i]}$ by I.H., so
	$\mpP\equiv\mpP[j]\mpPar\mpBigPar{i\in I\setminus\{j\}}{\mpP[i]}$.
	
	Now, if $``\mpP=\mpBigPar{i\in I}{\mpP[i]}''$ and
	$``\mpQ=\mpBigPar{i\in I}{\mpP[i]}''$, then,
	by the above and the recursive rule in the canonical parallel composition,
	$\mpP\equiv\mpBigPar{i\in I}{\mpP[i]}\equiv\mpQ$.
\qed\end{proof}

\begin{definition}[Labels]\rm
\label{def:calc-transition-label}
We define the labels of transitions to be:
\begin{equation*}
	\stEnvAnnotGenericSym \bnfdef \ltsSendRecvS{\mpS}{\roleP}{\roleQ}{\stLab}{}
	\bnfsep \tau
	\bnfsep \mpErr
\end{equation*}
$\ltsSendRecvS{\mpS}{\roleP}{\roleQ}{\stLab}$ denotes that
$\roleP$ has sent the message $\stLab$ to $\roleQ$
in session $\mpS$;
$\tau$ denotes an anonymous transition; and
$\mpErr$ denotes an error.
\end{definition}

\begin{definition}[Process Transitions]\rm
\label{def:calc-transitions}
	The transition relation, $\mpMoveGen$,
	on processes is inductively defined by the rules:
	{\small
		\[
		\begin{array}{ll}
		\inferrule{R-Cond}&
		\eval{e}{v}\;\wedge\; v\in\{\mpTrue,\mpFalse\}\implies\mpIf{e}{\mpP[\mpTrue]}{\mpP[\mpFalse]}\mpMoveTau\mpP[v]
		\\[1ex]
		\inferrule{R-Ctx-I}&
		{\mpP\mpMoveTau\mpQ}
		\implies
		{\mpCtxApp{\mpCtx}{\mpP}\mpMoveTau\mpCtxApp{\mpCtx}{\mpQ}}
		\\[1ex]
		\inferrule{R-Ctx-II}&
		{\mpP\mpMoveCommS{\mpS}{\roleP}{\roleQ}{\stLab}{}\mpQ
		\wedge
		\mpS\not\in\{\mpSi[i]\}_{1\leq i\leq n}
		}		
		\\&\quad
		\implies
		{
		\mpRes{\mpSi[1],\dots,\mpSi[n]}{\left(\mpP\mpPar\mpR\right)}
		\mpMoveCommS{\mpS}{\roleP}{\roleQ}{\stLab}{}
		\mpRes{\mpSi[1],\dots,\mpSi[n]}{\left(\mpQ\mpPar\mpR\right)}
		}
		\\[1ex]
		\inferrule{R-Ctx-III}&
		{\mpP\mpMoveCommS{\mpS}{\roleP}{\roleQ}{\stLab}{}\mpQ
		\wedge
		\mpS\in\{\mpSi[i]\}_{1\leq i\leq n}
		}		
		\\&\quad
		\implies
		{
		\mpRes{\mpSi[1],\dots,\mpSi[n]}{\left(\mpP\mpPar\mpR\right)}
		\mpMoveTau
		\mpRes{\mpSi[1],\dots,\mpSi[n]}{\left(\mpQ\mpPar\mpR\right)}
		}
		\\[1ex]
		\inferrule{R-Cong}&
		\mpP\prestruct\mpPi\mpMoveGen\mpQi\prestruct\mpQ
		\implies
		\mpP\mpMoveGen\mpQ
		\\[1ex]
		\inferrule{R-Val}&
		{\eval{e}{v}}
		\implies
		{
				(\mpBra{\mpChanRole{\mpS}{\roleQ}}{\roleP}{\stLab}{\mpx}{\mpQ}+\mpQi)
				\mpPar
				(\mpSel{\mpChanRole{\mpS}{\roleP}}{\roleQ}{\stLab}{e}{\mpP}+\mpPi)
				\mpMoveCommS{\mpS}{\roleP}{\roleQ}{\stLab}{}
				\mpQ\subst{\mpx}{v}\mpPar\mpP
			}
		\\[1ex]
		\inferrule{R-Chan}&
		{
				(\mpBra{\mpChanRole{\mpS}{\roleQ}}{\roleP}{\stLab}{\mpy}{\mpQ}+\mpQi)
				\mpPar
				(\mpSel{\mpChanRole{\mpS}{\roleP}}{\roleQ}{\stLab}{\mpChanRole{\mpSi}{\rolePi}}{\mpP}+\mpPi)
				\mpMoveCommS{\mpS}{\roleP}{\roleQ}{\stLab}{}
				\mpQ\subst{\mpy}{\mpChanRole{\mpSi}{\rolePi}}\mpPar\mpP
			}
		\\[1ex]
		\inferrule{E-Cond}&
		{\eval{e}{v}\;\wedge\; v\not\in\{\mpTrue,\mpFalse\}}
		\implies
			{\mpIf{e}{\mpP}{\mpQ}\mpMoveErr\mpErr}
			\\[1ex]
			\inferrule{E-Eval}&
			{\eval{e}{\mpErr}}
			\implies
			{\mpSel{\mpC}{\roleP}{\stLab}{e}{\mpP}+\mpPi
			\mpMoveErr
			\mpErr}
			\\[1ex]
			\inferrule{E-M-I}&
			{(\mpBra{\mpChanRole{\mpS}{\roleQ}}{\roleP}{\stLab}{\mpx}{\mpQ}+\mpQi)
			\mpPar
			(\mpSel{\mpChanRole{\mpS}{\roleP}}{\roleQ}{\stLab}{\mpC}{\mpP}+\mpPi)
			\mpMoveErr\mpErr}
			\\[1ex]
			\inferrule{E-M-II}&
			{(\mpBra{\mpChanRole{\mpS}{\roleQ}}{\roleP}{\stLab}{\mpy}{\mpQ}+\mpQi)
			\mpPar
			(\mpSel{\mpChanRole{\mpS}{\roleP}}{\roleQ}{\stLab}{e}{\mpP}+\mpPi)
			\mpMoveErr\mpErr}
			\\[1ex]
			\inferrule{E-M-III}&
			{
				\mpSel{\mpChanRole{\mpS}{\roleP}}{\roleQ}{\stLab}{}{} \preCalc{\mpP}
				\;\wedge\;
				\mpBra{\mpChanRole{\mpS}{\roleQ}}{\roleP}{\stLabi}{}{} \preCalc{\mpQ}
				\;\wedge\;
				\mpBra{\mpChanRole{\mpS}{\roleQ}}{\roleP}{\stLab}{}{} \not\preCalc{\mpQ}
			}
			\implies
			{
				\mpQ\mpPar
				\mpP
				\mpMoveErr\mpErr
			}\\[1ex]
			\inferrule{E-Ctx}&
			\mpP\mpMoveErr\mpQ
			\implies
			\mpCtxApp{\mpCtx}{\mpP}\mpMoveErr\mpCtxApp{\mpCtx}{\mpQ}
		\end{array}
		\]}
\end{definition}
\noindent
The reduction relation is the transition relation
with labels removed.
We will use this to provide stronger versions of our theorems.

\begin{definition}[Context Reduction]\rm
\label{def:context-reduction-rules}
	We say that $\mpCtxApp{\mpCtx}{}\mpMoveGen\mpCtxApp{\mpCtxi}{}$
	by the following rules:
	\[
		\begin{array}{c}
			\inference
			{\mpP\mpMoveTau\mpQ}
			{
				\mpRes{\mpSi[1],\dots,\mpSi[n]}{\left(\mpCtxHole\mpPar\mpP\right)}
				\mpMoveTau
				\mpRes{\mpSi[1],\dots,\mpSi[n]}{\left(\mpCtxHole\mpPar\mpQ\right)}
			}
			\\[1ex]
			\inference
			{\mpP\mpMoveCommS{\mpS}{\roleP}{\roleQ}{\stLab}{}\mpQ
			& \mpS\not\in\{\mpSi[i]\}_{1\leq i\leq n}}
			{
				\mpRes{\mpSi[1],\dots,\mpSi[n]}{\left(\mpCtxHole\mpPar\mpP\right)}
				\mpMoveCommS{\mpS}{\roleP}{\roleQ}{\stLab}{}
				\mpRes{\mpSi[1],\dots,\mpSi[n]}{\left(\mpCtxHole\mpPar\mpQ\right)}
			}
			\\[1ex]
			\inference
			{\mpP\mpMoveCommS{\mpS}{\roleP}{\roleQ}{\stLab}{}\mpQ
			& \mpS\in\{\mpSi[i]\}_{1\leq i\leq n}}
			{
				\mpRes{\mpSi[1],\dots,\mpSi[n]}{\left(\mpCtxHole\mpPar\mpP\right)}
				\mpMoveTau
				\mpRes{\mpSi[1],\dots,\mpSi[n]}{\left(\mpCtxHole\mpPar\mpQ\right)}
			}
			\quad
			\inference{\mpCtxApp{\mpCtx}{}\prestruct\mpCtxApp{\mpCtx'}{}\mpMoveGen\mpCtxApp{\mpCtx''}{}\prestruct\mpCtxApp{\mpCtx'''}{}}{
				\mpCtxApp{\mpCtx}{}\mpMoveGen\mpCtxApp{\mpCtx'''}{}
			}
		\end{array}
	\]
\end{definition}

\begin{lemma}[Context Transitions]
	\label{lem:ctx-trans}
	$\mpCtxApp{\mpCtx}{}\mpMoveGen\mpCtxApp{\mpCtxi}{}$
	iff $\forall\mpP.\mpCtxApp{\mpCtx}{\mpP}\mpMoveGen\mpCtxApp{\mpCtxi}{\mpP}$.
\end{lemma}

\begin{proof}
	The forward direction is just an application
	of the rules \inferrule{R-Ctx-I},\inferrule{R-Ctx-II}, and \inferrule{R-Ctx-II}
	followed by \inferrule{R-Cong}.
	
	For the backwards direction,
	consider
	\[\mpRes{\mpS[1],\dots,\mpS[n]}{\left(\mpCtxHole\mpPar\mpQ\right)}\]
	Let $\roleP$ and $\rolePi$ be participants not appearing in $\mpQ$ and let
	\[\mpP=\mpBra{\mpChanRole{\mpS[1]}{\roleP}}{\rolePi}{\stLab}{\mpx}{
	\mpBra{\mpChanRole{\mpS[2]}{\roleP}}{\rolePi}{\stLab}{\mpx}{
	\dots
	\mpBra{\mpChanRole{\mpS[n]}{\roleP}}{\rolePi}{\stLab}{\mpx}{}
	}
	}\]
	Now,
	\[
	\begin{array}{rl}&\mpRes{\mpS[1],\dots,\mpS[n]}{\left(\mpP\mpPar\mpQ\right)}
	\mpMoveGen\mpRes{\mpSi[1],\dots,\mpSi[m]}{\left(\mpP\mpPar\mpQi\right)}
	\\\text{iff}&\mpQ\prestruct\mpRes{\mpSii[1],\dots,\mpSii[k]}{\mpQi}
	\,\text{and}\,\{\mpS[1],\dots,\mpS[n],\mpSii[1],\dots,\mpSii[k]\}=\{\mpSi[1],\dots,\mpSi[m]\}\end{array}\]
\qed\end{proof}

\begin{definition}[Guarded]\rm
\label{def:calc-guarded}
	We define the predicate $\guarded{}{}$
	on processes and sets of process variables recursively by:
	\[
		\begin{array}{c}
			\inference[]
			{}
			{\guarded{\mpNil}{S}}
			\qquad
			\inference[]
			{\guarded{\mpP}{S}}
			{\guarded{\mpRes{\mpS}{\mpP}}{S}}
			\qquad
			\inference[]
			{\mpX\not\in S}
			{\guarded{\mpX}{S}}
			\\[1ex]
			\inference[]
			{\guarded{\mpP}{S\cup\{\mpX\}}}
			{\guarded{\mpRec{\mpX}{\mpP}}{S}}
			\qquad
			\inference[]
			{\guarded{\mpP}{S}&\guarded{\mpPi}{S}}
			{\guarded{\mpIf{e}{\mpP}{\mpPi}}{S}}
			\\[1ex]
			\inference[]
			{\guarded{\mpP}{S}&\guarded{\mpPi}{S}}
			{\guarded{\mpP\mpPar\mpPi}{S}}
			\quad
			\inference[]
			{\forall i\in I\;\guarded{\mpP[i]}{\emptyset}}
			{\guarded{\mpSum{\mpPrefix_i\mpSeq\mpP[i]}{i\in I}}{S}}
		\end{array}
	\]
\end{definition}
\noindent
We say that $\mpP$ is guarded if $\guarded{\mpP}{\emptyset}$.

\begin{lemma}[Guarded]\label{lem:calc-guarded}\ \\
	(1) If $\guarded{\mpP}{S}$ and $\guarded{\mpPi}{\emptyset}$, then
	$\guarded{\mpP\subst{\mpX}{\mpPi}}{S}$.\\
	(2) If $\guarded{\mpP}{\emptyset}$ and $\mpP\prestruct\mpPi$,
	then $\guarded{\mpPi}{\emptyset}$.\\
	(3) If $\guarded{\mpP}{\emptyset}$ and $\mpP\mpMove\mpPi$,
	then $\guarded{\mpPi}{\emptyset}$.
\end{lemma}

\begin{proof}[1]
	Induction on the judgement $\guarded{\mpP}{S}$:\\
	if $\mpP=\mpNil$ then
	$\guarded{\mpNil}{S}$;\\
	if $\mpP=\mpRes{\mpS}{\mpQ}$ then
	$\guarded{\mpQ}{S}$ and by I.H.
	$\guarded{\mpQ\subst{\mpX}{\mpPi}}{S}$
	so\\
	$\guarded{\mpRes{\mpS}{\mpP\subst{\mpX}{\mpPi}}}{S}$;\\
	if $\mpP=\mpX$ then
	$\guarded{\mpPi}{\emptyset}$ so $\guarded{\mpPi}{S}$;\\
	if $\mpP=\mpX'\neq\mpX$ then $\guarded{\mpX'}{S}$;\\
	if $\mpP=\mpRec{\mpX'}{\mpQ}$ then
	$\guarded{\mpQ}{S\cup\{\mpX'\}}$ so by I.H.
	$\guarded{\mpQ\subst{\mpX}{\mpPi}}{S}$ so
	$\guarded{\mpRec{\mpX'}{\mpQ}\subst{\mpX}{\mpPi}}{S}$;\\
	if $\mpP=\mpIf{e}{\mpQ}{\mpQi}$ then
	$\guarded{\mpQ}{S}$ and $\guarded{\mpQi}{S}$, so by I.H.,
	$\guarded{\mpQ\subst{\mpX}{\mpPi}}{S}$ and
	$\guarded{\mpQi\subst{\mpX}{\mpPi}}{S}$, so\\
	$\guarded{(\mpIf{e}{\mpQ}{\mpQi})\subst{\mpX}{\mpPi}}{S}$;\\
	if $\mpP=\mpQ\mpPar\mpQi$ then
	$\guarded{\mpQ}{S}$ and $\guarded{\mpQi}{S}$, so by I.H.,\\
	$\guarded{\mpQ\subst{\mpX}{\mpPi}}{S}$ and
	$\guarded{\mpQi\subst{\mpX}{\mpPi}}{S}$, so
	$\guarded{(\mpQ\mpPar\mpQi)\subst{\mpX}{\mpPi}}{S}$;\\
	if $\mpP=\mpSum{\mpPrefix_i\mpSeq\mpP[i]}{i\in I}$
	then $\guarded{\mpP[i]}{\emptyset}$ for $i\in I$, so
	by I.H.,
	$\guarded{\mpP[i]\subst{\mpX}{\mpPi}}{\emptyset}$ for $i\in I$, so
	so $\guarded{\mpP\subst{\mpX}{\mpPi}}{S}$.
\qed\end{proof}

\begin{proof}[2]
	Induction on $\mpP\prestruct\mpPi$:\\
	if $\guarded{\mpP\mpPar\mpQ}{\emptyset}$,
	then $\guarded{\mpP}{\emptyset}$ and
	$\guarded{\mpQ}{\emptyset}$,\\so
	$\guarded{\mpQ\mpPar\mpP}{\emptyset}$;\\
	if $\guarded{\mpP\mpPar(\mpQ\mpPar\mpR)}{\emptyset}$, then
	$\guarded{\mpP}{\emptyset}$ and
	$\guarded{\mpQ\mpPar\mpR}{\emptyset}$, so\\
	$\guarded{\mpQ}{\emptyset}$ and $\guarded{\mpR}{\emptyset}$, so
	$\guarded{\mpP\mpPar\mpQ}{\emptyset}$,\\so
	$\guarded{(\mpP\mpPar\mpQ)\mpPar\mpR}{\emptyset}$;\\
	if $\guarded{(\mpQ\mpPar\mpR)\mpPar\mpP)}{\emptyset}$, then
	$\guarded{\mpP}{\emptyset}$ and
	$\guarded{\mpQ\mpPar\mpR}{\emptyset}$, so\\
	$\guarded{\mpQ}{\emptyset}$ and $\guarded{\mpR}{\emptyset}$, so
	$\guarded{\mpR\mpPar\mpP}{\emptyset}$,\\so
	$\guarded{\mpQ\mpPar(\mpR\mpPar\mpP)}{\emptyset}$;\\
	$\guarded{\mpNil}{\emptyset}$ so if
	$\guarded{\mpP}{\emptyset}$ then $\guarded{\mpP\mpPar\mpNil}{\emptyset}$;\\
	$\guarded{\mpP\mpPar\mpNil}{\emptyset}$ then $\guarded{\mpP}{\emptyset}$;\\
	if $\guarded{\mpRes{\mpS}{\mpP}}{\emptyset}$ then
	$\guarded{\mpP}{\emptyset}$;\\
	if $\guarded{\mpP}{\emptyset}$ then $\guarded{\mpRes{\mpS}{\mpP}}{\emptyset}$;\\
	if $\guarded{\mpRes{\mpS}{\mpRes{\mpSi}{\mpP}}}{\emptyset}$ then
	$\guarded{\mpRes{\mpSi}{\mpP}}{\emptyset}$ so\\
	$\guarded{\mpP}{\emptyset}$ so
	$\guarded{\mpRes{\mpSi}{\mpRes{\mpS}{\mpP}}}{\emptyset}$;\\
	if $\guarded{\mpRes{\mpS}{\mpQ\mpPar\mpP}}{\emptyset}$
	then
	$\guarded{\mpQ\mpPar\mpP}{\emptyset}$ so
	$\guarded{\mpQ}{\emptyset}$ and\\
	$\guarded{\mpP}{\emptyset}$, so
	$\guarded{\mpRes{\mpS}{\mpP}}{\emptyset}$, so
	$\guarded{\mpQ\mpPar\mpRes{\mpS}{\mpP}}{\emptyset}$;\\
	if $\guarded{\mpQ\mpPar\mpRes{\mpS}{\mpP}}{\emptyset}$ then
	$\guarded{\mpRes{\mpS}{\mpP}}{\emptyset}$ and
	$\guarded{\mpQ}{\emptyset}$, so\\
	$\guarded{\mpP}{\emptyset}$, so
	$\guarded{\mpQ\mpPar\mpP}{\emptyset}$, so
	$\guarded{\mpRes{\mpS}{\mpQ\mpPar\mpP}}{\emptyset}$;\\
	if $\guarded{\mpP}{\emptyset}$ then $\guarded{\mpP}{\emptyset}$;\\
	if $\guarded{\mpP}{\emptyset}$ and $\mpP\prestruct\mpPi\prestruct\mpPii$ then,
	by I.H., $\guarded{\mpPi}{\emptyset}$, so, by I.H.,
	$\guarded{\mpPii}{\emptyset}$;\\
	if $\guarded{\mpRec{\mpX}{\mpP}}{\emptyset}$ then
	$\guarded{\mpP}{\emptyset\cup\{\mpX\}}$,\\so by (1),
	$\guarded{\mpP\subst{\mpX}{\mpRec{\mpX}{\mpP}}}{\emptyset}$;\\
	if $\guarded{\mpRes{\vec{\mpS}}{\left(\mpP\mpPar\mpPi\right)}}{\emptyset}$
	and $\mpP\prestruct\mpQ$, then
	$\guarded{\mpP}{\emptyset}$ and $\guarded{\mpPi}{\emptyset}$ so by I.H.
	$\guarded{\mpQ}{\emptyset}$, so
	$\guarded{\mpRes{\vec{\mpS}}{\left(\mpQ\mpPar\mpPi\right)}}{\emptyset}$.
\qed\end{proof}

\begin{proof}[3]
	Induction on $\mpP\mpMove\mpQ$:\\
	if $\mpP\prestruct\mpPi\mpMove\mpQi\prestruct\mpQ$ then
	by (2) $\guarded{\mpPi}{\emptyset}$ so by I.H.
	$\guarded{\mpQi}{\emptyset}$ so by (2)
	$\guarded{\mpQ}{\emptyset}$;\\
	if $\mpP=\mpCtxApp{\mpCtx}{\mpPi}$,
	$\mpQ=\mpCtxApp{\mpCtx}{\mpQi}$, and
	$\mpPi\mpMove\mpQi$, then
	$\guarded{\mpPi}{\emptyset}$ so by I.H.\\
	$\guarded{\mpQi}{\emptyset}$
	so $\guarded{\mpQ}{\emptyset}$;\\
	if $\mpP=\mpIf{e}{\mpQ[l]}{\mpQ[r]}$
	and $\mpQ\in\{\mpQ[l],\mpQ[r]\}$, then
	$\guarded{\mpQ[l]}{\emptyset}$ and\\$\guarded{\mpQ[r]}{\emptyset}$,
	so $\guarded{\mpQ}{\emptyset}$;\\
	if $\mpP=\mpQ[l]\mpPar\mpQ[r]$
	with $\mpSel{\mpChanRole{\mpS}{\roleP}}{\roleQ}{\stLab}{d}{\mpQi[r]}\preCalc\mpQ[r]$
	and
	$\mpBra{\mpChanRole{\mpS}{\roleQ}}{\roleP}{\stLab}{\mpz}{\mpQi[l]}\preCalc\mpQ[l]$
	and
	$\mpQ=\mpQi[l]\subst{\mpz}{d}\mpPar\mpQi[r]$, then
	$\guarded{\mpQi[l]}{\emptyset}$ and
	$\guarded{\mpQi[r]}{\emptyset}$, so
	$\guarded{\mpQi[l]\subst{\mpz}{d}}{\emptyset}$
	as $\guarded{}{}$ is ignorant of channels and expressions, so
	$\guarded{\mpQ}{\emptyset}$.
\qed\end{proof}

\begin{definition}[Depth]\rm
\label{def:calc-depth}
	We define the depth partial function, $\depth{}$,
	on processes to be the number of constructors in
	between the head and communications or $\mpNil$:
	\[
		\begin{array}{lcl}
			\depth{\mpNil}&=&0\\
			\depth{\choice{\pprefix_i\mpSeq\mpP[i]}{i\in I}}&=&0\\
			\depth{\mpX}&=&\text{undefined}\\
			\depth{\mpRec{\mpX}{\mpP}}&=&1+\depth{\mpP}\\
			\depth{\mpIf{e}{\mpP}{\mpPi}}&=&1+\depth{\mpP}+\depth{\mpPi}\\
			\depth{\mpRes{\mpS}{\mpP}}&=&1+\depth{\mpP}\\
			\depth{\mpP\mpPar\mpPi}&=&1+\depth{\mpP}+\depth{\mpPi}
		\end{array}
	\]
\end{definition}

\begin{lemma}[Depth - I]
\label{def:calc-depth-guarded}
	If $\guarded{\mpP}{\fpv{\mpP}}$,
	then $\depth{\mpP}$ exists.
\end{lemma}

\begin{proof}
	Induction on the derivation of
	$\guarded{\mpP}{\fpv{\mpP}}$:\\
	if $\mpP=\mpNil$, then $\depth{\mpNil}=0$;\\
	if $\mpP\neq\mpX$, as $\mpX\in\fpv{\mpX}$;\\
	if $\mpP=\mpRec{\mpX}{\mpPi}$, then
	$\guarded{\mpPi}{\fpv{\mpP}\cup\{\mpX\}}$
	so $\guarded{\mpPi}{\fpv{\mpPi}}$ so
	$\depth{\mpPi}$ exists so $\depth{\mpP}=\depth{\mpPi}+1$;\\
	if $\mpP=\mpIf{e}{\mpPi}{\mpPii}$, then
	$\guarded{\mpPi}{\fpv{\mpP}}$ and $\guarded{\mpPii}{\fpv{\mpP}}$,
	so $\guarded{\mpPi}{\fpv{\mpPi}}$ and $\guarded{\mpPii}{\fpv{\mpPii}}$,
	so $\depth{\mpPi}$ and $\depth{\mpPii}$ exist, so
	$\depth{\mpP}=1+\depth{\mpPi}+\depth{\mpPii}$;\\
	if $\mpP=\mpRes{\mpS}{\mpPi}$, then
	$\guarded{\mpPi}{\fpv{\mpP}}$ and $\fpv{\mpP}=\fpv{\mpPi}$, so
	$\depth{\mpPi}$ exists, so
	$\depth{\mpP}=1+\depth{\mpPi}$;\\
	if $\mpP=\mpPi\mpPar\mpPii$, then
	$\guarded{\mpPi}{\fpv{\mpP}}$ and $\guarded{\mpPii}{\fpv{\mpP}}$,\\
	so $\guarded{\mpPi}{\fpv{\mpPi}}$ and $\guarded{\mpPii}{\fpv{\mpPii}}$,
	so $\depth{\mpPi}$ and $\depth{\mpPii}$ exist, so
	$\depth{\mpP}=1+\depth{\mpPi}+\depth{\mpPii}$;\\
	if $\mpP$ is a sum process, then $\depth{\mpP}=0$.
\qed\end{proof}

\begin{lemma}[Depth - II]
\label{lem:calc-depth-subst}
	If $\guarded{\mpP}{\fpv{\mpP}}$, then
	$\depth{\mpP\subst{\mpX}{\mpPi}}=\depth{\mpP}$.
\end{lemma}

\begin{proof}
	Induction on $\guarded{\mpP}{\fpv{\mpP}}$:\\
	if $\mpP=\mpNil$, then $\depth{\mpP\subst{\mpX}{\mpPi}}=\depth{\mpP}=\depth{\mpNil}=0$;\\
	if $\mpP=\mpSum{\mpPrefix_i\mpSeq\mpP[i]}{i\in I}$,
	then $\depth{\mpP\subst{\mpX}{\mpPi}}=\depth{\mpP}=0$;\\
	if $\mpP=\mpRec{\mpX'}{\mpPii}$, then
	$\guarded{\mpPii}{\fpv{\mpP}\cup\{\mpX'\}}$
	so
	$\guarded{\mpPii}{\fpv{\mpPii}}$ so, by I.H.,
	$\depth{\mpPii\subst{\mpX}{\mpPi}}=\depth{\mpPii}$, so
	$\depth{\mpP\subst{\mpX}{\mpPi}}=1+\depth{\mpPii\subst{\mpX}{\mpPi}}=1+\depth{\mpPii}=\depth{\mpP}$;\\
	if $\mpP=\mpIf{e}{\mpQ}{\mpQi}$, then
	$\guarded{\mpQ}{\fpv{\mpP}}$ and $\guarded{\mpQi}{\fpv{\mpP}}$,
	so $\guarded{\mpQ}{\fpv{\mpQ}}$ and $\guarded{\mpQi}{\fpv{\mpQii}}$,
	so by I.H.,
	$\depth{\mpQ\subst{\mpX}{\mpPi}}=\depth{\mpQ}$ and
	$\depth{\mpQi\subst{\mpX}{\mpPi}}=\depth{\mpQi}$, so
	$\depth{\mpP\subst{\mpX}{\mpPi}}=
	1+\depth{\mpQ\subst{\mpX}{\mpPi}}+\depth{\mpQi\subst{\mpX}{\mpPi}}
	=
	1+\depth{\mpQ}+\depth{\mpQi}
	=
	\depth{\mpP}$;\\
	if $\mpP=\mpQ\mpPar\mpQi$, then
	$\guarded{\mpQ}{\fpv{\mpP}}$ and $\guarded{\mpQi}{\fpv{\mpP}}$,\\
	so $\guarded{\mpQ}{\fpv{\mpQ}}$ and $\guarded{\mpQi}{\fpv{\mpQii}}$,
	so by I.H.,
	$\depth{\mpQ\subst{\mpX}{\mpPi}}=\depth{\mpQ}$ and
	$\depth{\mpQi\subst{\mpX}{\mpPi}}=\depth{\mpQi}$, so
	$\depth{\mpP\subst{\mpX}{\mpPi}}=
	1+\depth{\mpQ\subst{\mpX}{\mpPi}}+\depth{\mpQi\subst{\mpX}{\mpPi}}
	=
	1+\depth{\mpQ}+\depth{\mpQi}
	=
	\depth{\mpP}$;\\
	if $\mpP=\mpRes{\mpS}{\mpPii}$, then
	$\guarded{\mpPii}{\fpv{\mpP}}$ and $\fpv{\mpP}=\fpv{\mpPii}$, by I.H.,\\
	$\depth{\mpPii\subst{\mpX}{\mpPi}}=\depth{\mpPii}$, so\\
	$\depth{\mpP\subst{\mpX}{\mpPi}}=1+\depth{\mpPii\subst{\mpX}{\mpPi}}=1+\depth{\mpPii}=\depth{\mpP}$.
\qed\end{proof}

\begin{example}[Cycle]
\label{ex:cycle}
	We have five participants connected\\
	\begin{minipage}{1\textwidth}
\begin{wrapfigure}[5]{r}{0.25\textwidth}
	\begingroup
	\centering
	\vspace{-13mm}
	\begin{tikzpicture}[
	act/.style={circle, draw=black, minimum size = 2mm}
	]
		\node (centre) {};
		\path (centre) ++(90:1) node[act] (0) {$\star$};
		\path (centre) ++(162:1) node (4) {$\star$};
		\path (centre) ++(234:1) node[act] (3) {$\star$};
		\path (centre) ++(306:1) node (2) {$\star$};
		\path (centre) ++(18:1) node (1) {$\star$};
		
		\draw[->,color=blue] (0) -- (4);
		\draw[<-,color=blue] (1) -- (0);
		\draw[dashed,color=blue] (2) -- (1);
		\draw[->,color=blue] (3) -- (2);
		\draw[<-,color=blue] (4) -- (3);
	\end{tikzpicture}
	\endgroup
	\label{fig:cycle}
\end{wrapfigure}
	in a cycle.
	Active participants may receive and send to either adjacent participant.
	Inactive participants may only receive.
	A participant becomes active upon receiving, and
	inactive upon sending.
	For $0\leq i\leq4$,
	we define $\mpP[i]$ to be the active process
	and $\mpQ[i]$ to be the inactive process
	for participant $\roleP[i]$.
	\end{minipage}
	\[
		\begin{array}{lcl}
			\mpP[i]&=&\mpRec{\mpX}{\left(
			\mpSum{\mpBra{\mpChanRole{\mpS}{\roleP[i]}}{\roleP[(i+j)\%5]}{\stLab}{\mpx}{\mpX}}{j\in\{-1,1\}}
			+\right.}\\&&\;
			\left.\mpSum{\mpSel{\mpChanRole{\mpS}{\roleP[i]}}{\roleP[(i+j)\%5]}{\stLab}{\mpNum{0}}{
				\mpSum{\mpBra{\mpChanRole{\mpS}{\roleP[i]}}{\roleP[(i+j)\%5]}{\stLab}{\mpx}{\mpX}}{j\in\{-1,1\}}
			}}{j\in\{-1,1\}}
			\right)
			\\
			\mpQ[i]&=&\mpSum{\mpBra{\mpChanRole{\mpS}{\roleP[i]}}{\roleP[(i+j)\%5]}{\stLab}{\mpx}{\mpP[i]}}{j\in\{-1,1\}}
			\\
			\mpP[\text{\tiny cycle}]&=&\mpP[0]\mpPar\mpQ[1]\mpPar\mpP[2]\mpPar\mpQ[3]\mpPar\mpQ[4]
		\end{array}
	\]
\end{example}

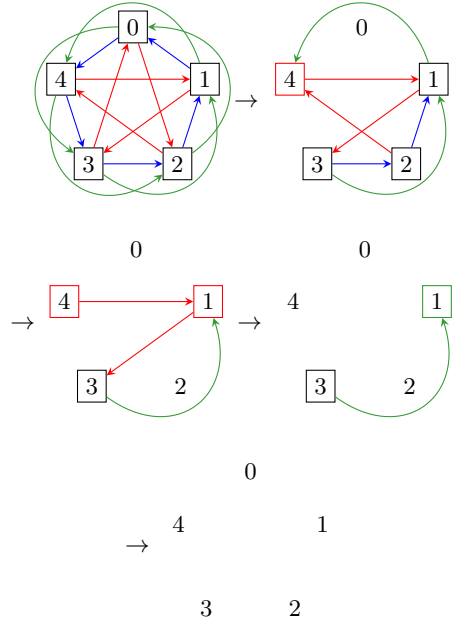
\begin{figure}[t!]
	\centering
	\begin{tabular}{c}
	\begin{tikzpicture}[
	sty/.style={rectangle, draw=black, minimum size = 2mm},
	styi/.style={rectangle, draw=red, minimum size = 2mm},
	styii/.style={rectangle, draw={rgb:red,0.1;green,0.3;blue,0.1}, minimum size = 2mm},
	styiii/.style={rectangle, draw=white, minimum size = 2mm}
	]
		\node (centre) {};
		\path (centre) ++(90:1) node[sty] (0) {0};
		\path (centre) ++(162:1) node[sty] (4) {4};
		\path (centre) ++(234:1) node[sty] (3) {3};
		\path (centre) ++(306:1) node[sty] (2) {2};
		\path (centre) ++(18:1) node[sty] (1) {1};
		
		\draw[->,color=blue] (0) -- (4);
		\draw[->,color=blue] (1) -- (0);
		\draw[->,color=blue] (2) -- (1);
		\draw[->,color=blue] (3) -- (2);
		\draw[->,color=blue] (4) -- (3);
		
		\draw[->,color=red] (0) -- (2);
		\draw[->,color=red] (1) -- (3);
		\draw[->,color=red] (2) -- (4);
		\draw[->,color=red] (3) -- (0);
		\draw[->,color=red] (4) -- (1);
		
		\draw[->,color={rgb:red,0.1;green,0.3;blue,0.1}] (0) to[out=180, in=144, looseness=1.6] (3);
		\draw[->,color={rgb:red,0.1;green,0.3;blue,0.1}] (1) to[out=108, in=72, looseness=1.6] (4);
		\draw[->,color={rgb:red,0.1;green,0.3;blue,0.1}] (2) to[out=36, in=0, looseness=1.6] (0);
		\draw[->,color={rgb:red,0.1;green,0.3;blue,0.1}] (3) to[out=324, in=288, looseness=1.6] (1);
		\draw[->,color={rgb:red,0.1;green,0.3;blue,0.1}] (4) to[out=252, in=216, looseness=1.6] (2);
		
		\path (centre) ++(0:1.5) node {$\mpMove$};
	\end{tikzpicture}
	
	\begin{tikzpicture}[
	sty/.style={rectangle, draw=black, minimum size = 2mm},
	styi/.style={rectangle, draw=red, minimum size = 2mm},
	styii/.style={rectangle, draw={rgb:red,0.1;green,0.3;blue,0.1}, minimum size = 2mm},
	styiii/.style={rectangle, draw=white, minimum size = 2mm}
	]
		\node (centre) {};
		\path (centre) ++(90:1) node[styiii] (0) {0};
		\path (centre) ++(162:1) node[styi] (4) {4};
		\path (centre) ++(234:1) node[sty] (3) {3};
		\path (centre) ++(306:1) node[sty] (2) {2};
		\path (centre) ++(18:1) node[sty] (1) {1};
		
		\draw[->,color=blue] (2) -- (1);
		\draw[->,color=blue] (3) -- (2);
		
		\draw[->,color=red] (1) -- (3);
		\draw[->,color=red] (2) -- (4);
		\draw[->,color=red] (4) -- (1);
		
		\draw[->,color={rgb:red,0.1;green,0.3;blue,0.1}] (1) to[out=108, in=72, looseness=1.6] (4);
		\draw[->,color={rgb:red,0.1;green,0.3;blue,0.1}] (3) to[out=324, in=288, looseness=1.6] (1);
	\end{tikzpicture}
	\\
	\begin{tikzpicture}[
	sty/.style={rectangle, draw=black, minimum size = 2mm},
	styi/.style={rectangle, draw=red, minimum size = 2mm},
	styii/.style={rectangle, draw={rgb:red,0.1;green,0.3;blue,0.1}, minimum size = 2mm},
	styiii/.style={rectangle, draw=white, minimum size = 2mm}
	]
		\node (centre) {};
		\path (centre) ++(180:1.5) node {$\mpMove$};
		\path (centre) ++(90:1) node[styiii] (0) {0};
		\path (centre) ++(162:1) node[styi] (4) {4};
		\path (centre) ++(234:1) node[sty] (3) {3};
		\path (centre) ++(306:1) node[styiii] (2) {2};
		\path (centre) ++(18:1) node[styi] (1) {1};
		
		\draw[->,color=red] (1) -- (3);
		\draw[->,color=red] (4) -- (1);
		
		\draw[->,color={rgb:red,0.1;green,0.3;blue,0.1}] (3) to[out=324, in=288, looseness=1.6] (1);
		
		\path (centre) ++(0:1.5) node {$\mpMove$};
	\end{tikzpicture}
	\begin{tikzpicture}[
	sty/.style={rectangle, draw=black, minimum size = 2mm},
	styi/.style={rectangle, draw=red, minimum size = 2mm},
	styii/.style={rectangle, draw={rgb:red,0.1;green,0.3;blue,0.1}, minimum size = 2mm},
	styiii/.style={rectangle, draw=white, minimum size = 2mm}
	]
		\node (centre) {};
		
		\path (centre) ++(90:1) node[styiii] (0) {0};
		\path (centre) ++(162:1) node[styiii] (4) {4};
		\path (centre) ++(234:1) node[sty] (3) {3};
		\path (centre) ++(306:1) node[styiii] (2) {2};
		\path (centre) ++(18:1) node[styii] (1) {1};
		
		\draw[->,color={rgb:red,0.1;green,0.3;blue,0.1}] (3) to[out=324, in=288, looseness=1.6] (1);
	\end{tikzpicture}
	\\
	\begin{tikzpicture}[
	sty/.style={rectangle, draw=black, minimum size = 2mm},
	styi/.style={rectangle, draw=red, minimum size = 2mm},
	styii/.style={rectangle, draw={rgb:red,0.1;green,0.3;blue,0.1}, minimum size = 2mm},
	styiii/.style={rectangle, draw=white, minimum size = 2mm}
	]
		\node (centre) {};
		\path (centre) ++(180:1.5) node {$\mpMove$};
		\path (centre) ++(90:1) node[styiii] (0) {0};
		\path (centre) ++(162:1) node[styiii] (4) {4};
		\path (centre) ++(234:1) node[styiii] (3) {3};
		\path (centre) ++(306:1) node[styiii] (2) {2};
		\path (centre) ++(18:1) node[styiii] (1) {1};
		
		\draw[->,color=white] (3) to[out=324, in=288, looseness=1.6] (1);
	\end{tikzpicture}
	\end{tabular}
	\caption{Leader Election Semantics}
	\label{fig:leader-sem}
	\end{figure}

\begin{example}[Semantics]
\label{ex:semantics}
	Recall $\mpP[\text{\tiny lead}]$ from \cref{ex:elect},
	this process is safe, deadlock-free, and live.
	We present one of its reduction paths below and in \cref{fig:leader-sem}.
	In this election, $\roleP[0]$ elects $\roleP[4]$,
	$\roleP[2]$ elects $\roleP[1]$,
	$\roleP[4]$ elects $\roleP[1]$, and
	$\roleP[3]$ elects $\roleP[1]$.
	\[\small\begin{array}{cl}
		&\mpP[0]\mpPar\mpP[1]\mpPar\mpP[2]\mpPar\mpP[3]\mpPar\mpP[4]\mpPar\mpQ\\
		\mpMove&
		\mpP[1]\mpPar\mpP[2]\mpPar\mpP[3]\mpPar\mpPi[4]\!\left(\mpChanRole{\mpS[0]}{\roleP}\right)\mpPar\mpQ\\
		\mpMove&
		\mpPi[1]\!\left(\mpChanRole{\mpS[2]}{\roleP}\right)\mpPar\mpP[3]\mpPar\mpPi[4]\!\left(\mpChanRole{\mpS[0]}{\roleP}\right)\mpPar\mpQ\\
		\mpMove&
		\mpBra{\mpChanRole{\mpS}{\roleP[1]}}{\roleP[3]}{\stLabFmt{elect}}{\mpyii}{\mpPii[i]\!\left(\mpChanRole{\mpS[2]}{\roleP},\mpChanRole{\mpS[4]}{\roleP},\mpyii\right)}
		\mpPar
		\mpP[3]
		\mpPar
		\mpSel{\mpChanRole{\mpS[0]}{\roleP}}{\roleS}{\stLabFmt{token}}{\mpNum{4}}{}
		\mpPar
		\mpQ\\
		\mpMove&
		\mpPi[1]\!\left(\mpChanRole{\mpS[2]}{\roleP},\mpChanRole{\mpS[4]}{\roleP},\mpChanRole{\mpS[3]}{\roleP}\right)
		\mpPar
		\mpSel{\mpChanRole{\mpS[0]}{\roleP}}{\roleS}{\stLabFmt{token}}{\mpNum{4}}{}
		\mpPar
		\mpQ\\
		\mpMove&
		\mpPi[1]\!\left(\mpChanRole{\mpS[2]}{\roleP},\mpChanRole{\mpS[4]}{\roleP},\mpChanRole{\mpS[3]}{\roleP}\right)
		\mpPar
		\mpBigPar{1\leq i< 5}{
			\mpBra{\mpChanRole{\mpS[i]}{\roleS}}{\roleP}{\stLabFmt{token}}{\mpx}{}
			}\\
		\mpMove&
		\mpSel{\mpChanRole{\mpS[2]}{\roleP}}{\roleS}{\stLabFmt{token}}{\mpNum{i}}{
						\mpSel{\mpChanRole{\mpS[4]}{\roleP}}{\roleS}{\stLabFmt{token}}{\mpNum{i}}{
						\mpSel{\mpChanRole{\mpS[3]}{\roleP}}{\roleS}{\stLabFmt{token}}{\mpNum{i}}{}
						}
						}
		\mpPar
		\mpBigPar{2\leq i< 5}{
			\mpBra{\mpChanRole{\mpS[i]}{\roleS}}{\roleP}{\stLabFmt{token}}{\mpx}{}
			}\\
		\mpMove&
						\mpSel{\mpChanRole{\mpS[4]}{\roleP}}{\roleS}{\stLabFmt{token}}{\mpNum{i}}{
						\mpSel{\mpChanRole{\mpS[3]}{\roleP}}{\roleS}{\stLabFmt{token}}{\mpNum{i}}{}
						}
		\mpPar
		\mpBigPar{3\leq i< 5}{
			\mpBra{\mpChanRole{\mpS[i]}{\roleS}}{\roleP}{\stLabFmt{token}}{\mpx}{}
			}\\
		\mpMove&
						\mpSel{\mpChanRole{\mpS[3]}{\roleP}}{\roleS}{\stLabFmt{token}}{\mpNum{i}}{}
		\mpPar
			\mpBra{\mpChanRole{\mpS[3]}{\roleS}}{\roleP}{\stLabFmt{token}}{\mpx}{}
		\mpMove\mpNil
	\end{array}
	\]
	Recall $\mpP[\text{\tiny cycle}]$ from \cref{ex:cycle},
	which is similarly safe, deadlock-free, and live.
	We characterise the reductions of $\mpP[\text{\tiny cycle}]$.
	See \cref{fig:cycle-sem} and
	for all choices of $a,b,c\in\{0,1,2,3,4\}$,
	\[
	\mpP[\text{\tiny cycle}]\mpMoveStar
	\mpBigPar{i\in\{a,b\}}{\mpP[i]}\mpPar
	\mpBigPar{i\in\{0,1,2,3,4\}\setminus\{a,b\}}{\mpQ[i]}
	\mpMove
	\mpP[c]\mpPar\mpBigPar{i\in\{0,1,2,3,4\}\setminus\{c\}}{\mpQ[i]}
	\]
	\begin{figure}[t!]
	\tiny
	\centering
	\begin{tabular}{c}
	\begin{tikzpicture}[
	act/.style={circle, draw=black, minimum size = 2mm},
	acti/.style={circle, draw=white, minimum size = 2mm}
	]
		\node (centre) {};
		\path (centre) ++(90:1) node[act] (0) {$\star$};
		\path (centre) ++(162:1) node[acti] (4) {$\star$};
		\path (centre) ++(234:1) node[act] (3) {$\star$};
		\path (centre) ++(306:1) node[acti] (2) {$\star$};
		\path (centre) ++(18:1) node[acti] (1) {$\star$};
		
		\draw[->,color=blue] (0) -- (4);
		\draw[<-,color=blue] (1) -- (0);
		\draw[dashed,color=blue] (2) -- (1);
		\draw[->,color=blue] (3) -- (2);
		\draw[<-,color=blue] (4) -- (3);
		
		\path (centre) ++(0:1.5) node {$\mpMoveStar$};
	\end{tikzpicture}
	
	\begin{tikzpicture}[
	act/.style={circle, draw=black, minimum size = 2mm},
	acti/.style={circle, draw=white, minimum size = 2mm}
	]
		\node (centre) {};
		\path (centre) ++(90:1) node[acti] (0) {$\star$};
		\path (centre) ++(162:1) node[acti] (4) {$\star$};
		\path (centre) ++(234:1) node[acti] (3) {$\star$};
		\path (centre) ++(306:1) node[act] (2) {$\star$};
		\path (centre) ++(18:1) node[act] (1) {$\star$};
		
		\draw[dashed,color=blue] (0) -- (4);
		\draw[->,color=blue] (1) -- (0);
		\draw[<->,color=blue] (2) -- (1);
		\draw[<-,color=blue] (3) -- (2);
		\draw[dashed,color=blue] (4) -- (3);
		
		\path (centre) ++(0:1.5) node {$\mpMoveStar$};
	\end{tikzpicture}
	
	\begin{tikzpicture}[
	act/.style={circle, draw=black, minimum size = 2mm},
	acti/.style={circle, draw=white, minimum size = 2mm}
	]
		\node (centre) {};
		\path (centre) ++(90:1) node[acti] (0) {$\star$};
		\path (centre) ++(162:1) node[act] (4) {$\star$};
		\path (centre) ++(234:1) node[acti] (3) {$\star$};
		\path (centre) ++(306:1) node[acti] (2) {$\star$};
		\path (centre) ++(18:1) node[acti] (1) {$\star$};
		
		\draw[<-,color=blue] (0) -- (4);
		\draw[dashed,color=blue] (1) -- (0);
		\draw[dashed,color=blue] (2) -- (1);
		\draw[dashed,color=blue] (3) -- (2);
		\draw[->,color=blue] (4) -- (3);
	\end{tikzpicture}
	\end{tabular}
	\caption{Cycle Semantics}
	\label{fig:cycle-sem}
	\end{figure}
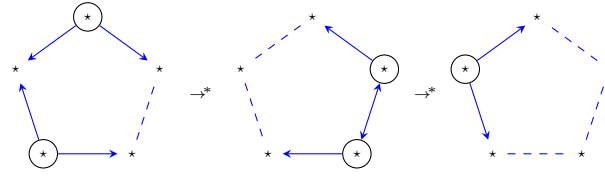
\end{example}

\begin{lemma}[Top Under Transitions]
\label{lem:top-transitions}
Suppose that $\obsvNew{\mpP}{\mpS}{\roleP}{\roleQ}{\dagger}$:
\begin{enumerate}
\item If $\mpP\prestruct\mpPi$, then $\obsvNew{\mpPi}{\mpS}{\roleP}{\roleQ}{\dagger}$.
\item If $\mpP\mpMoveTau\mpPi$, then $\obsvNew{\mpPi}{\mpS}{\roleP}{\roleQ}{\dagger}$.
\item If $\mpP\mpMoveCommS{\mpS}{\rolePi}{\roleQi}{\stLab}{}\mpPi$, then either:
$\roleP\not\in\{\rolePi,\roleQi\}$ and $\obsvNew{\mpPi}{\mpS}{\roleP}{\roleQ}{\dagger}$;
$\roleP=\rolePi$, $\roleQ=\roleQi$, and $\dagger={!}$; or
$\roleP=\roleQi$, $\roleQ=\rolePi$, and $\dagger={?}$.
\end{enumerate}
\end{lemma}

\begin{proof}
	(1) is by induction on the derivation of $\mpP\prestruct\mpPi$;
	(2) is by induction on the derivation of $\mpP\mpMoveTau\mpPi$; and
	(3) is by induction on the derivation of $\mpP\mpMoveCommS{\mpS}{\rolePi}{\roleQi}{\stLab}{}\mpPi$.
\qed\end{proof}

\propThreeLockLiveLock*

\begin{proof}
	Suppose that $\mpP$ has a three-party lock
	\ie
	$\mpP=\mpRes{\mpSi[1],\dots,\mpSi[n]}{\left(\mpRes{\mpS}{\mpQ}\mpPar\mpR\right)}$,
	$\mpSum{\mpPrefix_i\mpSeq\mpQ[i]}{i\in I}\mpPar\mpPi=\mpQ$
	$\mpChanRole{\mpS}{\roleP}\not\in\chan{\mpPi}$ and
	for all $i\in I$:\\
	$\mpChanRole{\mpChanRole{\mpS}{\roleP}}{\roleQ[i]}{\dagger_i}\preCalc\mpPrefix_i$;
	$\obsvNew{\mpQ}{\mpS}{\roleQ[i]}{\roleR[i]}{\dagger'_i}$;
	if $\roleR[i]=\roleP$ then $\dagger_i'\neq\overline{\dagger_i}$; and
	if $\roleR[i]\neq\roleP$ then $\obsvNew{\mpQ}{\mpS}{\roleR[i]}{\roleP}{\dagger''_i}$.
	
	\[
	\mpP\prestruct\mpRes{\mpSi[1],\dots,\mpSi[n],\mpS}{\left(\mpSum{\mpPrefix_i\mpSeq\mpQ[i]}{i\in I}\mpPar\mpPi\mpPar\mpR\right)}
	\]
	$\mpChanRole{\mpS}{\roleP}\not\in\chan{\mpPi\mpPar\mpR}$, so
	for $i\in I$
	$\obsvNew{\mpQ\mpPar\mpR}{\mpS}{\roleQ[i]}{\roleR[i]}{\dagger'_i}$ and\\
	if $\roleR[i]\neq\roleP$ then $\obsvNew{\mpQ\mpPar\mpR}{\mpS}{\roleR[i]}{\roleP}{\dagger''_i}$.
	
	If $\mpP$ is live, then
	there is $i\in I$ such that $\mpPi\mpPar\mpR\mpMoveStar\mpRi$,
	if $\dagger_i={!}$ then
	$\mpSum{\mpPrefix_i\mpSeq\mpQ[i]}{i\in I}\mpPar\mpRi\mpMoveCommS{\mpS}{\roleP}{\roleQ[i]}{}{}$; and
	if $\dagger_i={?}$ then
	$\mpSum{\mpPrefix_i\mpSeq\mpQ[i]}{i\in I}\mpPar\mpRi\mpMoveCommS{\mpS}{\roleQ[i]}{\roleP}{}{}$.
	By \cref{lem:top-transitions},
	$\obsvNew{\mpRi}{\mpS}{\roleQ[i]}{\roleR[i]}{\dagger'_i}$,
	so $\obsvNew{\mpSum{\mpPrefix_i\mpSeq\mpQ[i]}{i\in I}\mpPar\mpRi}{\mpS}{\roleQ[i]}{\roleR[i]}{\dagger'_i}$.
	So \cref{lem:top-transitions} provides a contradiction.
\qed\end{proof}

We define two-participant locks,
which generalise the error conditions
\inferrule{Err-New-Sync},
\inferrule{Err-Out-Out-Sync},
and \inferrule{Err-In-In-Sync}
in
\cite[Table 4]{CDSY2017}
to a multiparty calculus with mixed choice.

\begin{definition}[Two-Party Lock]\rm
\label{def:two-lock}
We say that $\mpCtxApp{\mpCtx}{\mpRes{\mpS}{\mpQ}}$
	has a \emph{two-party lock} from $\mpChanRole{\mpS}{\roleP}$ if
	$\mpQ=\mpSum{\mpPrefix_i\mpSeq\mpQ[i]}{i\in I}\mpPar\mpP$,
	$\mpChanRole{\mpS}{\roleP}\not\in\chan{\mpP}$, and
	there are $\roleQ[i]$ and $\dagger_i$ for $i\in
	I$ such that 
	$\mpChanRole{\mpChanRole{\mpS}{\roleP}}{\roleQ[i]}{\dagger_i}\prefix{\mpPrefix_i}$
	and
	$\obsvNew{\mpQ}{\mpS}{\roleQ[i]}{\roleP}{\dagger_i}$.
\end{definition}

Note that these are three-party locks
where case (1) in \cref{def:three-lock}
always holds.

\subsection{Types}

\begin{definition}[Free Type Variables]\rm
\label{def:type-free-var}
	We define the function $\gtFv{}$ to return the free recursive variables of session types by:
	\begin{equation*}
		\begin{array}{lcl}
			\gtFv{\stEnd}&=&\emptyset\\
			\gtFv{\stRecVar}&=&\{\stRecVar\}\\
			\gtFv{\stRec{\stRecVar}{\stT}}&=&\gtFv{\stT}\setminus\{\stRecVar\}\\
			\gtFv{\stSum{\roleP[i]}{i\in I}{\stChoice{\stLab[i]}{\stS[i]}\stSeq\stT[i]}{\dagger_i}}&=&\bigcup_{i\in I}\gtFv{\stT[i]}
		\end{array}
	\end{equation*}
	We say $\stT$ is closed if $\gtFv{\stT}=\emptyset$.
\end{definition}

\begin{definition}[Guarded]\rm
\label{def:type-guarded}
	We define the predicate $\guarded{}{}$ on session types and sets of recursive type variables by:
\begin{equation*}
	\begin{array}{c}
		\inference[]{}
		{\guarded{\stEnd}{S}}
		\quad
		\inference[]
		{\stRecVar\not\in S}
		{\guarded{\stRecVar}{S}}
		\quad
		\inference[]
		{\guarded{\stT}{S\cup\{\stRecVar\}}}
		{\guarded{\stRec{\stRecVar}{\stT}}{S}}
		\\[2ex]
		\inference[]
		{\forall i\in I & \guarded{\stT[i]}{S}}
		{\guarded{\stSum{\roleP[i]}{i\in I}{\stChoice{\stLab[i]}{\stS[i]}\stSeq\stT[i]}{\dagger_i}}{S}}
	\end{array}
	\end{equation*}
	The session type $\stT$ is guarded iff
	$\guarded{\stT}{\emptyset}$.
\end{definition}

\begin{lemma}[Guarded]\label{lem:type-guarded}\ \\
	(1) If $\guarded{\stT}{S}$ and
	$\guarded{\stTi}{S'}$, then
	$\guarded{\tySubst{\stT}{\stRecVar}{\stTi}}{(S\setminus\{\stRecVar\})\cup S'}$.
	(2) If $\guarded{\stRec{\stRecVar}{\stT}}{S}$, then
	$\guarded{\tySubst{\stT}{\stRecVar}{\stRec{\stRecVar}{\stT}}}{S}$.
\end{lemma}

\begin{proof}
	(1) is immediate from induction on the derivation of $\guarded{\stT}{S}$.\\
	(2) is immediate from (1):

	if $\guarded{\stRec{\stRecVar}{\stT}}{S}$ then
	$\guarded{\stT}{S\cup\{\stRecVar\}}$, so
	$\guarded{\tySubst{\stT}{\stRecVar}{\stRec{\stRecVar}{\stT}}}{S}$
	as $S=((S\cup\{\stRecVar\})\setminus\{\stRecVar\})\cup S$.
\qed\end{proof}

\begin{definition}[Unfolding]\rm
\label{def:unfold}
	We define the unfolding of a session type recursively by:
	$\unfoldOne{\stRec{\stRecVar}{\stT}}=\unfoldOne{\tySubst{\stT}{\stRecVar}{\stRec{\stRecVar}{\stT}}}$; $\unfoldOne{\stT}=\stT$ otherwise.
\end{definition}

\begin{definition}[Depth]\rm
\label{def:type-depth}
	We define the depth partial function on session types:
	\[
		\begin{array}{lcl}
			\depth{\stEnd}&=&0\\
			\depth{\stSum{\roleP[i]}{i\in I}{\stChoice{\stLab[i]}{\stS[i]}\stSeq\stT[i]}{\dagger_i}}&=&0\\
			\depth{\stRecVar}&=&\text{undefined}\\
			\depth{\stRec{\stRecVar}{\stT}}&=&1+\depth{\stT}
		\end{array}
	\]
\end{definition}

\begin{lemma}[Depth]
	\label{lem:depth-type}
	\begin{enumerate}
	\item If $\guarded{\stT}{\gtFv{\stT}}$, then $\depth{\stT}$ exists.
	\item If $\depth{\stT}$ exists, then $\depth{\tySubst{\stT}{\stRecVar}{\stTi}}=\depth{\stT}$.
	\end{enumerate}
	So if $\stT$ is a closed session type, then $\depth{\stT}$ exists.
\end{lemma}

\begin{proof}
\begin{enumerate}
\item
	We induct on the structure of $\stT$.
	\begin{itemize}
		\item If $\stT=\stEnd$ or $\stSum{\roleP[i]}{i\in I}{\stChoice{\stLab[i]}{\stS[i]}\stSeq\stT[i]}{\dagger_i}$, then $\depth{\stT}=0$.
		\item If $\stT=\stRecVar$, then $\neg\guarded{\stT}{\{\stRecVar\}}$.
		\item If $\stT=\stRec{\stRecVar}{\stTi}$ and $\guarded{\stT}{\gtFv{\stT}}$,
		then $\guarded{\stTi}{\{\stRecVar\}\cup\gtFv{\stT}}$.
		So, by I.H.,
		$\depth{\stTi}$ exists so $\depth{\stT}=1+\depth{\stTi}$ exists.
	\end{itemize}
\item
	Suppose that $\depth{\stT}$ exists. We induct on the structure of $\stT$.
	\begin{itemize}
		\item If $\stT=\stEnd$ or $\stSum{\roleP[i]}{i\in I}{\stChoice{\stLab[i]}{\stS[i]}\stSeq\stT[i]}{\dagger_i}$, then $\depth{\stT}=0$ and
		$\tySubst{\stT}{\stRecVar}{\stTi}=\stEnd$ or\\
		$\stSum{\roleP[i]}{i\in I}{\stChoice{\stLab[i]}{\stS[i]}\stSeq\tySubst{\stT[i]}{\stRecVar}{\stTi}}{\dagger_i}$,
		so $\depth{\tySubst{\stT}{\stRecVar}{\stTi}}=0$.
		\item If $\stT=\stRecVar$, then $\depth{\stT}$ does not exist.
		\item If $\stT=\stRec{\stRecVari}{\stTii}$ ($\stRecVari\not\in\{\stRecVar\}\cup\gtFv{\stTii}$), then\\
		$\depth{\stT}=1+\depth{\stTii}$
		and $\tySubst{\stT}{\stRecVar}{\stTii}=\stRec{\stRecVari}{\tySubst{\stTii}{\stRecVar}{\stTi}}$.
		By I.H., $\depth{\tySubst{\stTii}{\stRecVar}{\stTi}}=\depth{\stTii}$, so
		$\depth{\tySubst{\stT}{\stRecVar}{\stTii}}=
		1+\depth{\tySubst{\stTii}{\stRecVar}{\stTi}}=
		1+\depth{\stTii}=
		\depth{\stT}$.
	\end{itemize}
\end{enumerate}
\qed\end{proof}

\begin{definition}[Size]\rm
\label{def:type-size}
For a type $\stS$,
we define $|\stS|$ recursively on $\stS$:
$|\tyBool|=0$;
$|\tyInt|=0$;
$|\stEnd|=1$;
$|\stRecVar|=1$;
$|\stRec{\stRecVar}{\stT}|=1+|\stT|$; and
$|\stSum{\roleQ[i]}{i\in I}{\stChoice{\stLab[i]}{\stS[i]}\stSeq\stT[i]}{\dagger_i}|=
1+\sum_{i\in I}\left(|\stS[i]|+|\stT[i]|\right)$.

For a typing context $\stEnv$,
we define $|\stEnv|=\sum_{(\stEnvMap{\mpC}{\stT})\in\stEnv}|\stT|$.
\end{definition}

\subsection{Subtyping}

\begin{remark}[Subtyping Rules]
\label{rem:subtyping-trivial}
The formulation of \inferrule{S$\Sigma$} in \cite{PY2024}
contains the premise that $\{\stT[i]\}_{i\in K}$
share no prefixes.
This is enforced by the definition of $\Sigma_{k\in K}\stT[k]$
as we assume that types are
well-formed. 
The added condition $K>1$ makes the subtyping non-trivial. 
If $K=1$, it allows instances of:
\[
	\cinference[]{\stT\tySub\stTi}{\stT\tySub\stTi}
\]
so that $\stT\tySub\stTi$ is always true.
\end{remark}

\begin{lemma}[Combined Rules]
	\label{lem:sub-rule}
	We can replace the rules \inferrule{SSI}, \inferrule{SBr}, and \inferrule{S$\Sigma$}
	by the single rule:
	\[
		\cinference[S$\Sigma^\star$]
		{
		\begin{array}{c}
		\forall k\in K
		\quad
		\forall i\in I_k\cap I'_k
		\quad
		\stT[k,i]\tySub\stTi[k,i]
		\\
		(\stFmt{\dagger_k}=\stFmt{{!}}\implies \emptyset\neq I_k\subseteq I'_k
		\text{ and }
		\stSi[k,i]\tySub\stS[k,i])
		\\
		(\stFmt{\dagger_k}=\stFmt{{?}}\implies \emptyset\neq I'_k\subseteq I_k
		\text{ and }
		\stS[k,i]\tySub\stSi[k,i])
		\end{array}
		}
		{\stFmt{\Sigma_{k\in K}\stSum{\roleP[k]}{i\in I_k}{\stChoice{\stLab[k,i]}{\stS[k,i]}\stSeq\stT[k,i]}{\dagger_k}
		\tySub
		\Sigma_{k\in K}\stSum{\roleP[k]}{i\in I'_k}{\stChoice{\stLab[k,i]}{\stSi[k,i]}\stSeq\stTi[k,i]}{\dagger_k}}}
	\]	
\end{lemma}

\begin{proof}
	By fixing $|K|=1$, \inferrule{S$\Sigma$}
	subsumes both \inferrule{SSI} and \inferrule{SBr}.
	
	Suppose that $\stT\tySub\stTi$ is derived using \inferrule{S$\Sigma$},
	so
	$\stT=\stFmt{\Sigma_{k\in K}\stT[k]}$ and
	$\stTi=\stFmt{\Sigma_{k\in K}\stTi[k]}$
	where $|K|>1$ and, for all $k\in K$,
	$\stT[k]\tySub\stTi[k]$.
	We induct on the number of components in the top-level sum of $\stT$ and $\stTi$
	that $\stT\tySub\stTi$ by \inferrule{S$\Sigma^\star$}.
	
	By $\Sigma$, $\{\stT[k]\}_{k\in K}$ and $\{\stTi[k]\}_{k\in K}$
	are all sum types.
	
	Let $k\in K$.
	
	Suppose that $\stT[k]\tySub\stTi[k]$ is derived from \inferrule{S$\Sigma$}.
	As $|K|>1$, the number of components in the top-level sum of $\stT[k]$ and $\stTi[k]$
	is lower than the number of components in the top-level sum of $\stT$ and $\stTi$.
	Therefore, by the I.H.
	this instance of \inferrule{S$\Sigma$} is subsumed by
	\inferrule{S$\Sigma^\star$},
	$\stT[k]=\stFmt{\Sigma_{k'\in K'_k}}\stSum{\roleP[k,k']}{i\in I_{k,k'}}{\stChoice{\stLab[k,k',i]}{\stS[k,k',i]}\stSeq\stT[k,k',i]}{\dagger_{k,k'}}$,\\
	$\stTi[k]=\stFmt{\Sigma_{k'\in K'_k}}\stSum{\roleP[k,k']}{i\in I'_{k,k'}}{\stChoice{\stLab[k,k',i]}{\stSi[k,k',i]}\stSeq\stTi[k,k',i]}{\dagger_{k,k'}}$,
	and for all $k'\in K'_k$ and $i\in I_{k,k'}\cap I'_{k,k'}$,
		$\stT[k,k',i]\tySub\stT[k,k',i]$,
		$\stFmt{\dagger_{k,k'}}=\stFmt{{!}}\implies \emptyset\neq I_{k,k'}\subseteq I'_{k,k'}
		\wedge
		\stSi[k,k',i]\tySub\stS[k,k',i]
		$,
		and
		$\stFmt{\dagger_{k,k'}}=\stFmt{{?}}\implies \emptyset\neq I'_{k,k'}\subseteq I_{k,k'}
		\wedge
		\stS[k,k',i]\tySub\stSi[k,k',i]
		$.
		
	If $\stT[k]\tySub\stTi[k]$ is not derived from \inferrule{S$\Sigma$}, then,
	as they are both sum types, it is derived from \inferrule{SSI} or \inferrule{SBr}.
	Therefore,
	$\stT[k]$ is of the form $\stSum{\roleP}{i\in I}{\stChoice{\stLab[i]}{\stS[i]}\stSeq\stTii[i]}{\dagger}$ and
	$\stTiii[k]$ is of the form $\stSum{\roleP}{i\in I'}{\stChoice{\stLab[i]}{\stSi[i]}\stSeq\stTi[i]}{\dagger}$
	with $\stT[i]\tySub\stTi[i]$ for $i\in I\cap I'$
	and
	($I\subseteq I'$ and $\stS[i]\tySub\stSi[i]$ for $i\in I$)
	if $\stFmt{\dagger}=\stFmt{{!}}$
	and
	($I'\subseteq I$ and $\stSi[i]\tySub\stS[i]$ for $i\in I'$)
	if $\stFmt{\dagger}=\stFmt{{?}}$.
	Let $K'_k=\{0\}$,
	$\roleP[k,0]=\roleP$,
	$I_{k,0}=I$,
	$I'_{k,0}=I'$,
	$\stLab[k,0,i]=\stLab[i]$,
	$\stS[k,0,i]=\stS[i]$,
	$\stSi[k,0,i]=\stSi[i]$,
	$\stT[k,0,i]=\stTii[i]$,
	$\stTi[k,0,i]=\stTiii[i]$, and
	$\dagger_{k,0}=\dagger$.
	
	Now,
	$\stT=\stFmt{\Sigma_{(k,k')\in\bigcup_{k\in K}\{k\}\times K'_k}}
	\stSum{\roleP[k,i]}{i\in I_{k,k'}}{\stChoice{\stLab[k,k',i]}{\stS[k,k',i]}\stSeq\stT[k,k',i]}{\dagger_{k,k'}}$ and\\
	$\stT=\stFmt{\Sigma_{(k,k')\in\bigcup_{k\in K}\{k\}\times K'_k}}
	\stSum{\roleP[k,i]}{i\in I'_{k,k'}}{\stChoice{\stLab[k,k',i]}{\stSi[k,k',i]}\stSeq\stTi[k,k',i]}{\dagger_{k,k'}}$
	so that we can apply \inferrule{S$\Sigma^\star$}
	to subsume this instance of \inferrule{S$\Sigma$}.
	
	In the other direction,
	every instance of \inferrule{S$\Sigma^\star$}
	is an instance of either \inferrule{SSI} or \inferrule{SBr}
	at every $k\in K$, and then
	an instance of \inferrule{S$\Sigma$} (when $|K|>1$).
	\qed
\end{proof}

\begin{lemma}[Inversion of Subtyping - I]
	\label{lem:sub-inv}
	\begin{enumerate}
		\item If $\stRec{\stRecVar}{\stT}\tySub\stTi$, then
		$\tySubst{\stT}{\stRecVar}{\stRec{\stRecVar}{\stT}}\tySub\stTi$.
		\item If $\stT\tySub\stRec{\stRecVar}{\stTi}$, then
		$\stT\tySub\tySubst{\stTi}{\stRecVar}{\stRec{\stRecVar}{\stTi}}$.
	\end{enumerate}
\end{lemma}

\begin{proof}
	\begin{enumerate}
		\item Suppose that $\stRec{\stRecVar}{\stT}\tySub\stTi$,
		we induct on $\depth{\stTi}$ that
		$\tySubst{\stT}{\stRecVar}{\stRec{\stRecVar}{\stT}}\tySub\stTi$.
		
		$\stRec{\stRecVar}{\stT}\tySub\stTi$ can only be derived
		by \inferrule{S$\mu$L} or \inferrule{S$\mu$R}.
		If it is derived from \inferrule{S$\mu$L}, then
		$\tySubst{\stT}{\stRecVar}{\stRec{\stRecVar}{\stT}}\tySub\stTi$.
		Otherwise it is derived from \inferrule{S$\mu$R},
		so $\stTi=\stRec{\stRecVari}{\stTii}$
		and $\stRec{\stRecVar}{\stT}\tySub\tySubst{\stTii}{\stRecVari}{\stTi}$.
		By Lem.~\ref{lem:depth-type},
		$\depth{\tySubst{\stTii}{\stRecVari}{\stTi}}=\depth{\stTi}-1$.
		Therefore, by I.H.,
		$\tySubst{\stT}{\stRecVar}{\stRec{\stRecVar}{\stT}}
		\tySub
		\tySubst{\stTii}{\stRecVari}{\stTi}$.
		By \inferrule{S$\mu$R},
		$\tySubst{\stT}{\stRecVar}{\stRec{\stRecVar}{\stT}}
		\tySub
		\stTi$.
		\item Suppose that $\stT\tySub\stRec{\stRecVar}{\stTi}$,
		we induct on $\depth{\stT}$ that
		$\stTi\tySub\tySubst{\stTi}{\stRecVar}{\stRec{\stRecVar}{\stTi}}$.
		
		$\stT\tySub\stRec{\stRecVar}{\stTi}$ can only be derived
		by \inferrule{S$\mu$L} or \inferrule{S$\mu$R}.
		If it is derived from \inferrule{S$\mu$R}, then
		$\stT\tySub\tySubst{\stTi}{\stRecVar}{\stRec{\stRecVar}{\stTi}}$.
		Otherwise it is derived from \inferrule{S$\mu$L},
		so $\stT=\stRec{\stRecVari}{\stTii}$
		and $\tySubst{\stTii}{\stRecVari}{\stT}\tySub\stRec{\stRecVar}{\stTi}$.
		By Lem.~\ref{lem:depth-type},
		$\depth{\tySubst{\stTii}{\stRecVari}{\stT}}=\depth{\stT}-1$.
		Therefore, by I.H.,
		$\tySubst{\stTii}{\stRecVari}{\stT}
		\tySub
		\tySubst{\stTi}{\stRecVar}{\stRec{\stRecVar}{\stTi}}$.
		By \inferrule{S$\mu$L},
		$\stTi\tySub\tySubst{\stTi}{\stRecVar}{\stRec{\stRecVar}{\stTi}}$.
	\end{enumerate}
\qed\end{proof}

\begin{restatable}[$\tySub$ is a Preorder]{proposition}{subtypingPreorder}
\label{lem:sub-pre}
	The subtyping relation on closed session types
	and on typing contexts
	are reflexive and transitive.
\end{restatable}

\begin{proof}
	Consider the relation $\stT\Re\stTi$ iff $\unfoldOne{\stT}=\unfoldOne{\stTi}$.
	$\Re\subseteq\tySub$ as $\Re$ is backwards closed under the rules of subtyping.
	Therefore, for all $\stT$, $\stT\tySub\stT$.
	
	Consider the relation $\stT\Re\stTii$ iff there exists $\stTi$ such that
	$\stT\tySub\stTi\tySub\stTii$.
	This is backwards closed under the rules of subtyping.
	If $\stT$ or $\stTii$ are recursive, this is by Lem.~\ref{lem:sub-inv}.
	If $\stTi$ is recursive, then apply Lem.~\ref{lem:sub-inv} until it is not
	and then,
	$\stT\tySub\stTi\tySub\stTii$ are derived by \inferrule{S$\Sigma^\star$}
	or \inferrule{S$\stEnd$};
	in either case, we can combine these to see that $\Re$ is backwards closed under them.
	Therefore, $\Re\subseteq\tySub$.
\qed\end{proof}

\begin{lemma}[Semantics of Subtyping]
\label{lem:sub-sem}
	Suppose that $\stT\tySub\stTi$.
	\begin{enumerate}
		\item If $\stT\,\gtMove[{\roleP\stEnvAnnotOutSym\stChoice{\stLab}{\stS}}]\,\stT[1]$, then there exist $\stTi[1]$ and $\stSi$ such that
		$\stTi\,\gtMove[{\roleP\stEnvAnnotOutSym\stChoice{\stLab}{\stSi}}]\,\stTi[1]$,
		$\stT[1]\tySub\stTi[1]$, and
		$\stSi\tySub\stS$.
		\item If $\stT\,\gtMove[{\roleP\stEnvAnnotInSym\stChoice{\stLab}{\stS}}]$, then there exist $\stLabi$, $\stSi$, $\stSii$, $\stT[1]$, and $\stTi[1]$ such that
		$\stT\,\gtMove[{\roleP\stEnvAnnotInSym\stChoice{\stLabi}{\stSi}}]\,\stT[1]$,
		$\stTi\,\gtMove[{\roleP\stEnvAnnotInSym\stChoice{\stLabi}{\stSii}}]\,\stTi[1]$,
		$\stT[1]\tySub\stTi[1]$, and
		$\stSi\tySub\stSii$.
		\item If $\stTi\,\gtMove[{\roleP\stEnvAnnotInSym\stChoice{\stLab}{\stS}}]\,\stTi[1]$, then there exists $\stT[1]$ such that
		$\stT\,\gtMove[{\roleP\stEnvAnnotInSym\stChoice{\stLab}{\stSi}}]\,\stT[1]$,
		$\stSi\tySub\stS$, and
		$\stT[1]\tySub\stTi[1]$.
		\item If $\stTi\,\gtMove[{\roleP\stEnvAnnotOutSym\stChoice{\stLab}{\stS}}]$, then there exist $\stLabi$, $\stSi$, $\stSii$, $\stT[1]$, and $\stTi[1]$ such that
		$\stT\,\gtMove[{\roleP\stEnvAnnotOutSym\stChoice{\stLabi}{\stSii}}]\,\stT[1]$,
		$\stTi\,\gtMove[{\roleP\stEnvAnnotOutSym\stChoice{\stLabi}{\stSi}}]\,\stTi[1]$,
		$\stT[1]\tySub\stTi[1]$, and
		$\stSi\tySub\stSii$.
	\end{enumerate}
\end{lemma}

\begin{proof}
	We use the subtyping rules from Lemma~\ref{lem:sub-rule}.
	Suppose that $\stT\tySub\stTi$.
	
	\begin{enumerate}
		\item We induct on $\depth{\stTi}$.
		\begin{itemize}
			\item If $\depth{\stTi}=0$, then $\stTi=\stEnd$
			or $\stTi=\stSum{\rolePi[i]}{i\in I'}{\stChoice{\stLabi[i]}{\stSi[i]}\stSeq\stTi[i]}{\dagger'_i}$.
			 We induct on the derivation of
		$\stT\,\gtMove[{\roleP\stEnvAnnotOutSym\stChoice{\stLab}{\stS}}]\,\stT[1]$.
			\begin{itemize}
				\item If $\stT=\stSum{\roleP[i]}{i\in I}{\stChoice{\stLab[i]}{\stS[i]}\stSeq\stT[i]}{\dagger_i}$ and there is $k\in I$ s.t.
				$\roleP=\roleP[k]$,
				$\stLab=\stLab[k]$,
				$\stS=\stS[k]$,
				$\stEnvAnnotOutSym=\stFmt{\dagger_k}$, and
				$\stT[1]=\stT[k]$, then, by
				the subtyping rules,
				$\stTi\neq\stEnd$
				so $\stT\tySub\stTi$ is derived by \inferrule{S$\Sigma^\star$}.
				Therefore, there is $k'\in I'$ s.t.
				$\roleP=\rolePi[k']$,
				$\stLab=\stLabi[k']$,
				$\stSi[k']\tySub\stS$,
				$\stEnvAnnotOutSym=\stFmt{\dagger'_{k'}}$, and
				$\stT[k]\tySub\stTi[k']$.
				Let $\stTi[0]=\stTi[k']$, so that
				$\stTi\,\gtMove[{\roleP\stEnvAnnotOutSym\stChoice{\stLab}{\stSi[k']}}]\,\stTi[1]$ with $\stT[1]\tySub\stTi[1]$.
				\item If $\stT=\stRec{\stRecVar}{\stTii}$
				and $\tySubst{\stTii}{\stRecVar}{\stT}\,\gtMove[{\roleP\stEnvAnnotOutSym\stChoice{\stLab}{\stS}}]\,\stT[1]$,
				then, by Lem.~\ref{lem:sub-inv},
				$\tySubst{\stTii}{\stRecVar}{\stT}\tySub\stTi$ and, by I.H.,
				there exists $\stTi[1]$ such that
				$\stTi\,\gtMove[{\roleP\stEnvAnnotOutSym\stChoice{\stLab}{\stSi}}]\,\stTi[1]$ and $\stT[1]\tySub\stTi[1]$ and $\stSi\tySub\stS$.
			\end{itemize}
			\item If $\depth{\stTi}>0$, then
			$\stTi=\stRec{\stRecVar}{\stTii}$.
			By Lem.~\ref{lem:depth-type},
			$\depth{\tySubst{\stTii}{\stRecVar}{\stTi}}=\depth{\stTi}-1$.
			By Lem.~\ref{lem:sub-inv},
			$\stT\tySub\tySubst{\stTii}{\stRecVar}{\stTi}$.
			By the I.H.,
			there exists $\stTi[1]$ such that
				$\tySubst{\stTii}{\stRecVar}{\stTi}\,\gtMove[{\roleP\stEnvAnnotOutSym\stChoice{\stLab}{\stSi}}]\,\stTi[1]$ and $\stT[1]\tySub\stTi[1]$ and $\stSi\tySub\stS$.
		So, $\stTi\,\gtMove[{\roleP\stEnvAnnotOutSym\stChoice{\stLab}{\stSi}}]\,\stTi[1]$.
		\end{itemize}
		\item We induct on $\depth{\stTi}$.
		\begin{itemize}
			\item If $\depth{\stTi}=0$, then $\stTi=\stEnd$
			or $\stTi=\stSum{\rolePi[i]}{i\in I'}{\stChoice{\stLabi[i]}{\stSi[i]}\stSeq\stTi[i]}{\dagger'_i}$.
			 We induct on the derivation of
		$\stT\,\gtMove[{\roleP\stEnvAnnotInSym\stChoice{\stLab}{\stS}}]$.
			\begin{itemize}
				\item If $\stT=\stSum{\roleP[i]}{i\in I}{\stChoice{\stLab[i]}{\stS[i]}\stSeq\stT[i]}{\dagger_i}$ and there is $k\in I$ s.t.
				$\roleP=\roleP[k]$ and
				$\stEnvAnnotInSym=\dagger_k$, then, by
				the subtyping rules,
				$\stTi\neq\stEnd$
				so $\stT\tySub\stTi$ is derived by \inferrule{S$\Sigma^\star$}.
				Therefore, there are $k'\in I'$ and $k''\in I$ s.t.
				$\roleP=\roleP[k'']=\rolePi[k']$,
				$\stLab[k'']=\stLabi[k']$,
				$\stS[k'']\tySub\stSi[k']$,
				$\stEnvAnnotInSym=\stFmt{\dagger'_{k'}}=\stFmt{\dagger_{k''}}$, and
				$\stT[k'']\tySub\stTi[k']$.
				So,
				$\stTi\,\gtMove[{\roleP\stEnvAnnotInSym\stChoice{\stLab[k'']}{\stSi[k'']}}]\,\stTi[k']$,
				$\stT\,\gtMove[{\roleP\stEnvAnnotInSym\stChoice{\stLab[k'']}{\stS[k'']}}]\,\stT[k'']$,
				 and $\stT[k'']\tySub\stTi[k']$.
				\item If $\stT=\stRec{\stRecVar}{\stTii}$
				and $\tySubst{\stTii}{\stRecVar}{\stT}\,\gtMove[{\roleP\stEnvAnnotInSym\stChoice{\stLab}{\stSi}}]$,
				then, by Lem.~\ref{lem:sub-inv},
				$\tySubst{\stTii}{\stRecVar}{\stT}\tySub\stTi$ and, by I.H.,
				there exist $\stTi[1]$, $\stT[1]$, $\stLabi$, $\stSi$, and $\stSii$ such that
				$\stTi\,\gtMove[{\roleP\stEnvAnnotInSym\stChoice{\stLabi}{\stSi}}]\,\stTi[1]$,
				$\tySubst{\stTii}{\stRecVar}{\stT}\,\gtMove[{\roleP\stEnvAnnotInSym\stChoice{\stLabi}{\stSii}}]\,\stT[1]$, and $\stT[1]\tySub\stTi[1]$ and $\stSii\tySub\stSi$.
				So, $\stT\,\gtMove[{\roleP\stEnvAnnotInSym\stChoice{\stLabi}{\stSii}}]\,\stT[1]$.
			\end{itemize}
			\item If $\depth{\stTi}>0$, then
			$\stTi=\stRec{\stRecVar}{\stTii}$.
			By Lem.~\ref{lem:depth-type},
			$\depth{\tySubst{\stTii}{\stRecVar}{\stTi}}=\depth{\stTi}-1$.
			By Lem.~\ref{lem:sub-inv},
			$\stT\tySub\tySubst{\stTii}{\stRecVar}{\stTi}$.
			By the I.H.,
			there exist $\stT[1]$, $\stTi[1]$, $\stLabi$, $\stSi$, and $\stSii$ such that
				$\tySubst{\stTii}{\stRecVar}{\stTi}\,\gtMove[{\roleP\stEnvAnnotInSym\stChoice{\stLabi}{\stSi}}]\,\stTi[1]$,
				$\stT\,\gtMove[{\roleP\stEnvAnnotInSym\stChoice{\stLabi}{\stSii}}]\,\stT[1]$,
		and $\stT[1]\tySub\stTi[1]$ and $\stSii\tySub\stSi$.
		So, $\stTi\,\gtMove[{\roleP\stEnvAnnotInSym\stChoice{\stLabi}{\stSi}}]\,\stTi[1]$.
		\end{itemize}

		\item We induct on $\depth{\stT}$.
		\begin{itemize}
			\item If $\depth{\stT}=0$, then $\stT=\stEnd$
			or $\stT=\stSum{\roleP[i]}{i\in I}{\stChoice{\stLab[i]}{\stS[i]}\stSeq\stTi[i]}{\dagger'_i}$.
			 We induct on the derivation of
		$\stTi\,\gtMove[{\roleP\stEnvAnnotInSym\stChoice{\stLab}{\stSi}}]\,\stTi[1]$.
			\begin{itemize}
				\item If $\stTi=\stSum{\rolePi[i]}{i\in I'}{\stChoice{\stLabi[i]}{\stSi[i]}\stSeq\stTi[i]}{\dagger'_i}$ and there is $k'\in I$ s.t.
				$\roleP=\rolePi[k']$,
				$\stLab=\stLabi[k']$,
				$\stSi=\stSi[k']$,
				$\stEnvAnnotInSym=\stFmt{\dagger'_{k'}}$, and
				$\stTi[1]=\stTi[k']$, then, by
				the subtyping rules,
				$\stT\neq\stEnd$
				so $\stT\tySub\stTi$ is derived by \inferrule{S$\Sigma^\star$}.
				Therefore, there is $k\in I$ s.t.
				$\roleP=\roleP[k]$,
				$\stLab=\stLab[k]$,
				$\stS[k]\tySub\stSi$,
				$\stEnvAnnotInSym=\dagger_{k}$, and
				$\stT[k]\tySub\stTi[k']$.
				Let $\stT[0]=\stT[k]$, so that
				$\stT\,\gtMove[{\roleP\stEnvAnnotInSym\stChoice{\stLab}{\stS[k]}}]\,\stT[1]$ with $\stT[1]\tySub\stTi[1]$.
				\item If $\stTi=\stRec{\stRecVar}{\stTii}$
				and $\tySubst{\stTii}{\stRecVar}{\stTi}\,\gtMove[{\roleP\stEnvAnnotInSym\stChoice{\stLab}{\stSi}}]\,\stTi[1]$,
				by Lem.~\ref{lem:sub-inv},
				$\stTi\tySub\tySubst{\stTii}{\stRecVar}{\stTi}$ and, by I.H.,
				there exists $\stT[1]$ and $\stS$ such that
				$\stT\,\gtMove[{\roleP\stEnvAnnotInSym\stChoice{\stLab}{\stS}}]\,\stT[1]$ and $\stT[1]\tySub\stTi[1]$ and $\stS\tySub\stSi$.
			\end{itemize}
			\item If $\depth{\stT}>0$, then
			$\stT=\stRec{\stRecVar}{\stTii}$.
			By Lem.~\ref{lem:depth-type},
			$\depth{\tySubst{\stTii}{\stRecVar}{\stT}}=\depth{\stT}-1$.
			By Lem.~\ref{lem:sub-inv},
			$\tySubst{\stTii}{\stRecVar}{\stT}\tySub\stTi$.
			By the I.H.,
			there exists $\stT[1]$ and $\stS$ such that
				$\tySubst{\stTii}{\stRecVar}{\stT}\,\gtMove[{\roleP\stEnvAnnotInSym\stChoice{\stLab}{\stS}}]\,\stT[1]$ and $\stT[1]\tySub\stTi[1]$ and $\stS\tySub\stSi$.
		So, $\stT\,\gtMove[{\roleP\stEnvAnnotInSym\stChoice{\stLab}{\stS}}]\,\stT[1]$.
		\end{itemize}
		\item We induct on $\depth{\stT}$.
		\begin{itemize}
			\item If $\depth{\stT}=0$, then $\stT=\stEnd$
			or $\stT=\stSum{\roleP[i]}{i\in I}{\stChoice{\stLab[i]}{\stS[i]}\stSeq\stT[i]}{\dagger_i}$.
			 We induct on the derivation of
		$\stTi\,\gtMove[{\roleP\stEnvAnnotOutSym\stChoice{\stLab}{\stS}}]$.
			\begin{itemize}
				\item If $\stTi=\stSum{\rolePi[i]}{i\in I'}{\stChoice{\stLabi[i]}{\stSi[i]}\stSeq\stTi[i]}{\dagger'_i}$ and there is $k\in I'$ s.t.
				$\roleP=\rolePi[k]$ and
				$\stEnvAnnotOutSym=\dagger'_k$, then, by
				the subtyping rules,
				$\stT\neq\stEnd$
				so $\stT\tySub\stTi$ is derived by \inferrule{S$\Sigma^\star$}.
				Therefore, there are $k'\in I'$ and $k''\in I$ s.t.
				$\roleP=\roleP[k'']=\rolePi[k']$,
				$\stLab[k'']=\stLabi[k']$,
				$\stSi[k']\tySub\stS[k'']$,
				$\stEnvAnnotOutSym=\stFmt{\dagger'_{k'}}=\stFmt{\dagger_{k''}}$, and
				$\stT[k'']\tySub\stTi[k']$.
				So,
				$\stTi\,\gtMove[{\roleP\stEnvAnnotOutSym\stChoice{\stLab[k'']}{\stS[k'']}}]\,\stTi[k']$,
				$\stT\,\gtMove[{\roleP\stEnvAnnotOutSym\stChoice{\stLab[k'']}{\stSi[k']}}]\,\stT[k'']$,
				 and $\stT[k'']\tySub\stTi[k']$.
				\item If $\stTi=\stRec{\stRecVar}{\stTii}$
				and $\tySubst{\stTii}{\stRecVar}{\stTi}\,\gtMove[{\roleP\stEnvAnnotOutSym\stChoice{\stLab}{\stS}}]$,
				then, by Lem.~\ref{lem:sub-inv},
				$\stT\tySub\tySubst{\stTii}{\stRecVar}{\stTi}$ and, by I.H.,
				there exist $\stTi[1]$, $\stT[1]$, $\stLabi$, $\stSi$, and $\stSii$ such that
				$\stT\,\gtMove[{\roleP\stEnvAnnotOutSym\stChoice{\stLabi}{\stSii}}]\,\stT[1]$,
				$\tySubst{\stTii}{\stRecVar}{\stTi}\,\gtMove[{\roleP\stEnvAnnotOutSym\stChoice{\stLabi}{\stSi}}]\,\stTi[1]$, $\stT[1]\tySub\stTi[1]$, and $\stSi\tySub\stSii$.
				So, $\stTi\,\gtMove[{\roleP\stEnvAnnotOutSym\stChoice{\stLabi}{\stSi}}]\,\stTi[1]$.
			\end{itemize}
			\item If $\depth{\stT}>0$, then
			$\stT=\stRec{\stRecVar}{\stTii}$.
			By Lem.~\ref{lem:depth-type},
			$\depth{\tySubst{\stTii}{\stRecVar}{\stT}}=\depth{\stT}-1$.
			By Lem.~\ref{lem:sub-inv},
			$\tySubst{\stTii}{\stRecVar}{\stT}\tySub\stTi$.
			By the I.H.,
			there exist $\stT[1]$, $\stTi[1]$, $\stLabi$, $\stSi$, and $\stSii$ such that
				$\tySubst{\stTii}{\stRecVar}{\stT}\,\gtMove[{\roleP\stEnvAnnotOutSym\stChoice{\stLabi}{\stSii}}]\,\stT[1]$,
				$\stTi\,\gtMove[{\roleP\stEnvAnnotOutSym\stChoice{\stLabi}{\stSi}}]\,\stTi[1]$, $\stT[1]\tySub\stTi[1]$, and $\stSi\tySub\stSii$.
		So, $\stT\,\gtMove[{\roleP\stEnvAnnotOutSym\stChoice{\stLabi}{\stSii}}]\,\stT[1]$.
		\end{itemize}
	\end{enumerate}
\qed\end{proof}

\begin{lemma}[Subtyping End]
\label{lem:sub-end}
	If $\stT\tySub\stEnd$ or $\stEnd\tySub\stT$, then $\unfoldOne{\stT}=\stEnd$
\end{lemma}

\begin{proof}
	Suppose that $\stT\tySub\stEnd$ or $\stEnd\tySub\stT$.
	By Lem.~\ref{lem:sub-inv},
	\begin{equation*}
	\unfoldOne{\stT}\tySub\stEnd\text{ or }\stEnd\tySub\unfoldOne{\stT}
	\end{equation*}
	$\unfoldOne{\stT}$ is not a sum type, as then Lem.~\ref{lem:sub-sem}
	would conclude that $\stEnd\,\gtMove$.
	$\unfoldOne{\stT}$ is not a recursive type, by unfolding.
	Therefore, $\unfoldOne{\stT}=\stEnd$.
\qed\end{proof}

\begin{lemma}[Inversion of Subtyping - II]
\label{lem:sub-inv-2}
	Suppose that $\stSi\tySub\stS$.
	\begin{enumerate}
		\item $\unfoldOne{\stSi}=\stEnd$
		iff
		$\unfoldOne{\stS}=\stEnd$.
		\item If $\unfoldOne{\stSi}=\stSum{\roleQ[i]}{i\in I'}{\stChoice{\stLab[i]}{\stS[i]}\stSeq\stT[i]}{\dagger_i}$
		and
		$\unfoldOne{\stS}=\stSum{\roleQ[i]}{i\in I}{\stChoice{\stLab[i]}{\stS[i]}\stSeq\stT[i]}{\dagger_i}$, then
		\begin{itemize}
			\item For all $i\in I$ such that
			$\stFmt{\dagger_i}=\stFmt{?}$,
			there exists $i'\in I'$ such that
			$\roleQ[i]=\roleQ[i']$,
			$\stLab[i]=\stLab[i']$,
			$\stFmt{\dagger_{i'}}=\stFmt{?}$,
			$\stS[i']\tySub\stS[i]$, and
			$\stT[i']\tySub\stT[i]$.
			\item For all $i\in I$ such that
			$\stFmt{\dagger_i}=\stFmt{!}$,
			there exist $i'\in I'$ and $i''\in I$ such that
			$\roleQ[i]=\roleQ[i']=\roleQ[i'']$,
			$\stLab[i'']=\stLab[i']$,
			$\stFmt{\dagger_{i'}}=\stFmt{\dagger_{i''}}=\stFmt{!}$,
			$\stS[i'']\tySub\stS[i']$, and
			$\stT[i']\tySub\stT[i'']$.
			\item For all $i'\in I'$ such that
			$\stFmt{\dagger_{i'}}=\stFmt{!}$,
			there exists $i\in I$ such that
			$\roleQ[i]=\roleQ[i']$,
			$\stLab[i]=\stLab[i']$,
			$\stFmt{\dagger_{i}}=\stFmt{!}$,
			$\stS[i]\tySub\stS[i']$, and
			$\stT[i']\tySub\stT[i]$.
			\item For all $i''\in I'$ such that
			$\stFmt{\dagger_i}=\stFmt{?}$,
			there exist $i\in I'$ and $i'\in I$ such that
			$\roleQ[i]=\roleQ[i']=\roleQ[i'']$,
			$\stLab[i]=\stLab[i']$,
			$\stFmt{\dagger_{i'}}=\stFmt{\dagger_{i}}=\stFmt{?}$,
			$\stS[i']\tySub\stS[i]$, and
			$\stT[i']\tySub\stT[i]$.
		\end{itemize}
		\item $\stS$ is a basic type
		iff
		$\stSi$ is a basic type,
		in which case
		$\stS=\stSi$.
	\end{enumerate}
\end{lemma}

\begin{proof}
	Since the unfolding of a type is specified by its transitions,
	(1) and (2) are rephrasings of Lem.~\ref{lem:sub-sem}.
	
	(3) holds because the only rule from which
	$\stS\tySub\stSi$ can be derived, 
	if one of $\stS$ and $\stSi$ are basic types,
	is \inferrule{$\tyGround$}.
\qed\end{proof}

\subsection{Type Semantics}

\begin{proposition}[Fair Paths]
\label{prop:fair-path}
	For all contexts $\stEnv$,
	there is a fair path from $\stEnv$.
\end{proposition}

\begin{proof}
	Assume that we have some way of choosing from
	choices of labels,
	possibly depending on previous choices
	(this parameterisation will be useful in \cref{thm:typing-prop-df-live},
	but such a choice exists as there will always be finitely many choices).
	
	We construct a sequence of contexts $\{\stEnv[i]\}_{i\in N}$
	and a sequence of transition labels $\{\stEnvAnnotGenericSym[i]\}_{i,i+1\in N}$
	s.t. $\stEnv[i]\gtMove[{\stEnvAnnotGenericSym[i]}]\stEnv[i+1]$
	if $i,i+1\in N$
	recursively.
	For the sake of notation, we say that
	$\min{\emptyset}=-\inf$:
	
	$\stEnv[0]=\stEnv$.
	
	Given $\stEnv[j]$ for $0\leq j \leq i$ and $\stEnvAnnotGenericSym[j]$ for $0\leq j<i$,
	let $C_i=\bigcup\{\ltsSubject{\stEnvAnnotGenericSym}\suchthat \stEnv[i]\gtMove[\stEnvAnnotGenericSym]\}$.
	If $C_i=\emptyset$, then $N=\{0,\dots,i\}$ and we are done.
	Otherwise, pick $\mpChanRole{\mpS}{\roleP}\in C$
	such that $\max{\{j\suchthat j<i\wedge\mpChanRole{\mpS}{\roleP}\in\ltsSubject{\stEnvAnnotGenericSym[j]}\}}$
	is minimal;
	choose $\mpChanRole{\mpS}{\roleQ}\in C_i$ s.t.
	there exists $\stEnvAnnotGenericSym$ with
	$\stEnv[i]\gtMove[\stEnvAnnotGenericSym]$
	and $\ltsSubject{\stEnvAnnotGenericSym}=\{\mpChanRole{\mpS}{\roleP},\mpChanRole{\mpS}{\roleQ}\}$,
	choose whether $\roleP$ is sending or receiving, if both are possible,
	and choose which label is used, using the parameterisation mentioned at the start.
	We now get $\stEnv[i]\gtMove[\stEnvAnnotGenericSym]\stEnvi$
	with $\stEnvAnnotGenericSym$
	being a transition over $\mpS$ with $\roleP$ communicating with $\roleQ$,
	as specified, with specified label.
	Let $\stEnvAnnotGenericSym[i]=\stEnvAnnotGenericSym$
	and $\stEnv[i+1]=\stEnvi$.
	
	For all $i\in N$,
	if
	\[
	|\argmin_{\mpC\in C_i}{\max{\{j\suchthat j<i\wedge\mpC\in\ltsSubject{\stEnvAnnotGenericSym[j]}\}}}|=1
	\]
	then
	\[
	\min_{\mpC\in C_{i+1}}{\max{\{j\suchthat j<i\wedge\mpC\in\ltsSubject{\stEnvAnnotGenericSym[j]}\}}}<\min_{\mpC\in C_i}{\max{\{j\suchthat j<i+1\wedge\mpC\in\ltsSubject{\stEnvAnnotGenericSym[j]}\}}}
	\]
	and if
	\[
	|\argmin_{\mpC\in C_i}{\max{\{j\suchthat j<i\wedge\mpC\in\ltsSubject{\stEnvAnnotGenericSym[j]}\}}}|=n+2
	\]
	then
	\[
	|\argmin_{\mpC\in C_{i+1}}{\max{\{j\suchthat j<i+1\wedge\mpC\in\ltsSubject{\stEnvAnnotGenericSym[j]}\}}}|=n+1
	\]
	Therefore, the path is fair:
	if $\stEnv[i]\gtMove[\stEnvAnnotGenericSym]$
	with $\ltsSubject{\stEnvAnnotGenericSym}=\{\mpCi,\mpCii\}$
	then there exists $j\geq i$ such that
	$\ltsSubject{\stEnvAnnotGenericSym[j]}\cap\{\mpCi,\mpCii\}\neq\emptyset$
	by induction on
	\[
	\begin{array}{rcl}
	(|\dom{\stEnv}|+1)&\cdot&\left(
	\begin{array}{cl}
	&\min_{\mpC\in C_i}{\max{\{j\suchthat j<i\wedge\mpC\in\ltsSubject{\stEnvAnnotGenericSym[j]}\}}}
	\\
	-&
	\max{\{j\suchthat j<i\wedge\{\mpCi,\mpCii\}\cap\ltsSubject{\stEnvAnnotGenericSym[j]}\neq\emptyset\}}
	\end{array}
	\right)
	\\&+&
	|\argmin_{\mpC\in C_i}{\max{\{j\suchthat j<i\wedge\mpC\in\ltsSubject{\stEnvAnnotGenericSym[j]}\}}}|
	\end{array}
	\]
\qed\end{proof}

\begin{proposition}[Live Contexts]
\label{prop:live-ctx}
	\begin{enumerate}
	\item If $\stEnv$ is a context without mixed choice
	(each sum constructor contains exactly one of $\stFmt{!}$ and $\stFmt{?}$,
	and exactly one participant),
	the $\stEnv$ is live
	iff it is live in the sense of \cite{YHK2026}.
	\item
	If $\stEnv$ is live and $\unfoldOne{\stEnvApp{\stEnv}{\mpChanRole{\mpS}{\roleP}}}=
	\stSum{\roleP[i]}{i\in I}{\stChoice{\stLab[i]}{\stS[i]}\stSeq\stT[i]}{\dagger_i}
	$,
	then
	there exists $i\in I$ such that
	$\unfoldOne{\stEnvApp{\stEnv}{\mpChanRole{\mpS}{\roleP[i]}}}\neq \stEnd$.
	\end{enumerate}
\end{proposition}

\begin{proof}[1]
	Suppose that $\stEnv$ is a context without mixed choice.
	
	Suppose that $\stEnv$ is live in the sense of \cite{YHK2026}.
	Let $\{\stEnv[i]\}_{i\in N}$ is a fair path;
	this is fair in the sense of \cite{YHK2026}
	as if $\stEnv[i]\gtMove[\ltsSendRecvS{\mpS}{\roleP}{\roleQ}{\stChoice{\stLab}{\stS}}]$
	then there is $j\geq i$ such that
	$\stEnv[j]\gtMove[\stEnvAnnotGenericSymi]\stEnv[j+1]$
	and $\ltsSubject{\stEnvAnnotGenericSym}\cap\ltsSubject{\stEnvAnnotGenericSymi}\neq\emptyset$.
	Assume that $j\geq i$ is minimal,
	so $\stEnvApp{\stEnv[i]}{\roleP}=\stEnvApp{\stEnv[j]}{\roleP}$
	and
	$\stEnvApp{\stEnv[i]}{\roleQ}=\stEnvApp{\stEnv[j]}{\roleQ}$.
	Since $\stEnvApp{\stEnv[j]}{\roleP}$ and $\stEnvApp{\stEnv[i]}{\roleQ}=\stEnvApp{\stEnv[j]}{\roleQ}$
	have no mixed choice, the only action that
	$\stEnvApp{\stEnv[j]}{\roleP}$ can do is $\roleQ\stFmt{!}$
	and the only action that
	$\stEnvApp{\stEnv[j]}{\roleQ}$ can do is $\roleP\stFmt{?}$,
	so $\stEnvAnnotGenericSymi=\ltsSendRecvS{\mpS}{\roleP}{\roleQ}{\stChoice{\stLabi}{\stSi}}$
	for some $\stChoice{\stLabi}{\stSi}$,
	so $\{\stEnv[i]\}_{i\in N}$ is fair in the sense of \cite{YHK2026}
	so it is live.
	So, if $\stEnvApp{\stEnv[i]}{\roleP}\gtMove[\roleQ\stFmt{!}\stChoice{\stLab}{\stS}]$, then there is $j\geq i$ and $\stChoice{\stLabi}{\stSi}$ such that
	$\stEnv[j]\gtMove[\ltsSendRecvS{\mpS}{\roleP}{\roleQ}{\stChoice{\stLabi}{\stSi}}]\stEnv[j+1]$
	and if $\stEnvApp{\stEnv[i]}{\roleP}\gtMove[\roleQ\stFmt{?}\stChoice{\stLab}{\stS}]$, then there is $j\geq i$ and $\stChoice{\stLabi}{\stSi}$ such that
	$\stEnv[j]\gtMove[\ltsSendRecvS{\mpS}{\roleQ}{\roleP}{\stChoice{\stLabi}{\stSi}}]\stEnv[j+1]$.
	Therefore, $\{\stEnv[i]\}_{i\in N}$ is live.
	
	Suppose that $\stEnv$ is live
	and $\{\stEnv[i]\}_{i\in N}$ is fair in the sense of \cite{YHK2026}.
	$\{\stEnv[i]\}_{i\in N}$ is fair, since
	if $\stEnv[i]\gtMove[\ltsSendRecvS{\mpS}{\roleP}{\roleQ}{\stChoice{\stLab}{\stS}}]$,
	there is $j\geq i$ s.t.
	$\stEnv[j]\gtMove[\ltsSendRecvS{\mpS}{\roleP}{\roleQ}{\stChoice{\stLabi}{\stSi}}]$
	and
	$\ltsSubject{\ltsSendRecvS{\mpS}{\roleP}{\roleQ}{\stChoice{\stLab}{\stS}}}
	\cap
	\ltsSubject{\ltsSendRecvS{\mpS}{\roleP}{\roleQ}{\stChoice{\stLabi}{\stSi}}}=
	\{\mpChanRole{\mpS}{\roleP},\mpChanRole{\mpS}{\roleQ}\}\neq\emptyset$.
	So $\{\stEnv[i]\}_{i\in N}$ is live.
	If
	$\stEnvApp{\stEnv[i]}{\roleP}\gtMove[\roleQ\stFmt{\dagger}\stChoice{\stLab}{\stS}]$,
	then there is $j\geq i$ s.t. $\stEnv[j]\gtMove[\stEnvAnnotGenericSym]$
	and $\mpChanRole{\mpS}{\roleP}\in\ltsSubject{\stEnvAnnotGenericSym}$.
	$\{\stEnv[i]\}_{i\in N}$ is fair in the sense of \cite{YHK2026}, so
	there is $j'\geq j$ s.t.
	$\stEnv[j']\gtMove[\stEnvAnnotGenericSymi]\stEnv[j'+1]$
	and $\mpChanRole{\mpS}{\roleP}\in\ltsSubject{\stEnvAnnotGenericSymi}$.
	Let $j''\geq i$ be minimal s.t. there exists
	$\stEnvAnnotGenericSymii$
	s.t.
	$\stEnv[j'']\gtMove[\stEnvAnnotGenericSymii]\stEnv[j''+1]$
	and $\mpChanRole{\mpS}{\roleP}\in\ltsSubject{\stEnvAnnotGenericSymii}$.
	As $\stEnv$ contains no mixed choice,
	if $\dagger=\stFmt{!}$ then
	$\stEnvAnnotGenericSymii=\ltsSendRecvS{\mpS}{\roleP}{\roleQ}{\stChoice{\stLabii}{\stSii}}$,
	and if $\dagger=\stFmt{?}$ then
	$\stEnvAnnotGenericSymii=\ltsSendRecvS{\mpS}{\roleQ}{\roleP}{\stChoice{\stLabii}{\stSii}}$.
	Therefore,
	$\{\stEnv[i]\}_{i\in N}$ is live in the sense of \cite{YHK2026}.
\qed\end{proof}

\begin{proof}[2]
	Let $\stEnv$ be live and $\unfoldOne{\stEnvApp{\stEnv}{\mpChanRole{\mpS}{\roleP}}}=
	\stSum{\roleP[i]}{i\in I}{\stChoice{\stLab[i]}{\stS[i]}\stSeq\stT[i]}{\dagger_i}
	$.
	By \cref{prop:fair-path},
	let $\{\stEnv[i]\}_{i\in N}$
	be a fair path from $\stEnv$,
	so it is also live.
	So there is $i\in N$ s.t.
	$\stEnv[i]\gtMove[\stEnvAnnotGenericSym]$
	with $\mpChanRole{\mpS}{\roleP}\in\ltsSubject{\stEnvAnnotGenericSym}$.
	Take $i$ to be minimal, so that
	$\ltsSubject{\stEnvAnnotGenericSym}=\{\mpChanRole{\mpS}{\roleP},
	\mpChanRole{\mpS}{\roleP[j]}\}$ for some $j\in I$,
	since $\stEnvApp{\stEnv}{\mpChanRole{\mpS}{\roleP}}=\stEnvApp{\stEnv[i]}{\mpChanRole{\mpS}{\roleP}}$.
	So $\stEnvApp{\stEnv}{\mpChanRole{\mpS}{\roleP[j]}}\gtMoveStar
	\stEnvApp{\stEnv[i]}{\mpChanRole{\mpS}{\roleP[j]}}\gtMove$.
	So, $\unfoldOne{\stEnvApp{\stEnv}{\mpChanRole{\mpS}{\roleP[j]}}}\neq\stEnd$.
\qed\end{proof}

\lemDownClosed*

\begin{lemma}[Subtyping Contexts]
\label{lem:suptying-ctx}
\begin{enumerate}
\item Suppose that $\stEnv\tySub\stEnvi$ and $\stEnvi$ is safe.
	If $\stEnv\gtMove[\ltsSendRecvS{\mpS}{\roleP}{\roleQ}{\stChoice{\stLab}{\stS}}]\stEnvii$, then there exist $\stEnviii$ and $\stSi$ such that
	$\stEnvi\gtMove[\ltsSendRecvS{\mpS}{\roleP}{\roleQ}{\stChoice{\stLab}{\stSi}}]\stEnviii$,
	$\stSi\tySub\stS$, and
	$\stEnvii\tySub\stEnviii$.
	If $\stEnvi\gtMove[\ltsSendRecvS{\mpS}{\roleP}{\roleQ}{\stChoice{\stLabi}{\stS}}]$, then there exist $\stLab$, $\stSi$, $\stSii$, $\stEnvii$, $\stEnviii$ such that
	$\stEnv\gtMove[\ltsSendRecvS{\mpS}{\roleP}{\roleQ}{\stChoice{\stLab}{\stSii}}]\stEnvii$,
	$\stEnvi\gtMove[\ltsSendRecvS{\mpS}{\roleP}{\roleQ}{\stChoice{\stLab}{\stSi}}]\stEnviii$,
	$\stSi\tySub\stSii$, and
	$\stEnvii\tySub\stEnviii$.
\item
Suppose that $\stEnv\tySub\stEnvi$.
For $\predP\in\{\stEnvSafePred,\stEnvSafePred\cap\stEnvDFPred,\stEnvSafePred\cap\stEnvLivePred\}$,
if $\predPApp[]{\stEnvi}$ then $\predPApp[]{\stEnv}$.
\end{enumerate}
\end{lemma}

\begin{proof}[1]
	Suppose that $\stEnv\tySub\stEnvi$ and $\stEnvi$ is safe.
	\begin{itemize}
		\item Suppose that
		$\stEnv\,\gtMove[\ltsSendRecvS{\mpS}{\roleP}{\roleQ}{\stChoice{\stLab}{\stS}}]\,\stEnvii$.\\
		So,
		$\stEnvApp{\stEnv}{\mpChanRole{\mpS}{\roleP}}\,\gtMove[\roleQ\stEnvAnnotOutSym\stChoice{\stLab}{\stS}]\,\stEnvApp{\stEnvii}{\mpChanRole{\mpS}{\roleP}}$,
		$\stEnvApp{\stEnv}{\mpChanRole{\mpS}{\roleQ}}\,\gtMove[\roleP\stEnvAnnotInSym\stChoice{\stLab}{\stSi}]\,\stEnvApp{\stEnvii}{\mpChanRole{\mpS}{\roleQ}}$,
		$\stSi\tySub\stS$
		and, for $\mpC\in\dom{\stEnv}\setminus\{\mpChanRole{\mpS}{\roleP},\mpChanRole{\mpS}{\roleQ}\}$,
		$\stEnvApp{\stEnv}{\mpC}=\stEnvApp{\stEnvii}{\mpC}$.
		By Lem.~\ref{lem:sub-sem}, there exist $\stT[\roleP]$, $\stLabi$, $\stSii$, and $\stSiii$
		such that
		$\stEnvApp{\stEnvi}{\mpChanRole{\mpS}{\roleP}}\,\gtMove[\roleQ\stEnvAnnotOutSym\stChoice{\stLab}{\stSii}]\,\stT[\roleP]$,
		$\stEnvApp{\stEnvi}{\mpChanRole{\mpS}{\roleQ}}\,\gtMove[\roleP\stEnvAnnotInSym\stChoice{\stLabi}{\stSiii}]$,
		$\stSii\tySub\stS$, and
		$\stEnvApp{\stEnvii}{\mpChanRole{\mpS}{\roleP}}\tySub\stT[\roleP]$.
		By safety of $\stEnvi$, there exists $\stEnviii$ such that
		$\stEnvi\,\gtMove[\ltsSendRecv{\roleP}{\roleQ}{\stChoice{\stLab}{\stSii}}]\,\stEnviii$.
		The transition relation is deterministic, so $\stT[\roleP]=\stEnvApp{\stEnviii}{\mpChanRole{\mpS}{\roleP}}$ and, by Lem.~\ref{lem:sub-sem},
		$\stEnvApp{\stEnvii}{\mpChanRole{\mpS}{\roleQ}}\tySub\stEnvApp{\stEnviii}{\mpChanRole{\mpS}{\roleQ}}$.
		Also, for $\mpC\in\dom{\stEnv}\setminus\{\mpChanRole{\mpS}{\roleP},\mpChanRole{\mpS}{\roleQ}\}$,
		$\stEnvApp{\stEnvii}{\mpC}=\stEnvApp{\stEnv}{\mpC}
		\tySub\stEnvApp{\stEnvi}{\mpC}=\stEnvApp{\stEnviii}{\mpC}$.
		Therefore, $\stEnvii\tySub\stEnviii$.
		\item Suppose that
		$\stEnvi\,\gtMove[\ltsSendRecvS{\mpS}{\roleP}{\roleQ}{\stChoice{\stLabi}{\stSi}}]$.
		So, $\stEnvApp{\stEnvi}{\mpChanRole{\mpS}{\roleP}}\,\gtMove[\roleQ\stEnvAnnotOutSym\stChoice{\stLabi}{\stSi}]$
		and
		$\stEnvApp{\stEnvi}{\mpChanRole{\mpS}{\roleQ}}\,\gtMove[\roleP\stEnvAnnotInSym\stChoice{\stLabi}{\stSii}]$.\\
		By Lem.~\ref{lem:sub-sem},
		there exists $\stT[\roleP]$, $\stTi[\roleP]$, $\stLab$, $\stS[0]$, $\stSi[0]$
		such that
		$\stEnvApp{\stEnv}{\mpChanRole{\mpS}{\roleP}}\,\gtMove[\roleQ\stEnvAnnotOutSym\stChoice{\stLab}{\stS[0]}]\,\stT[\roleP]$,\\
		$\stEnvApp{\stEnvi}{\mpChanRole{\mpS}{\roleP}}\,\gtMove[\roleQ\stEnvAnnotOutSym\stChoice{\stLab}{\stSi[0]}]\,\stTi[\roleP]$,
		$\stSi[0]\tySub\stS[0]$, and
		$\stT[\roleP]\tySub\stTi[\roleP]$.
		By safety of $\stEnvi$,
		there exists $\stEnviii$ such that
		$\stEnvi\,\gtMove[\ltsSendRecvS{\mpS}{\roleP}{\roleQ}{\stChoice{\stLab}{\stSi[0]}}]\,\stEnviii$.
		The transition relation is deterministic, so $\stEnvApp{\stEnviii}{\mpChanRole{\mpS}{\roleP}}=\stTi[\roleP]$.
		$\stEnvApp{\stEnvi}{\mpChanRole{\mpS}{\roleQ}}\,\gtMove[\roleP\stEnvAnnotInSym\stChoice{\stLab}{\stSi[1]}]\,\stEnvApp{\stEnviii}{\mpChanRole{\mpS}{\roleQ}}$ for some $\stSi[0]\tySub\stSi[1]$, so,
		by Lem.~\ref{lem:sub-sem},
		there exist $\stT[\roleQ]$ and $\stS[1]$ such that
		$\stEnvApp{\stEnv}{\mpChanRole{\mpS}{\roleQ}}\,\gtMove[\roleP\stEnvAnnotInSym\stChoice{\stLab}{\stS[1]}]\,\stT[\roleQ]$,
		$\stS[1]\tySub\stSi[1]$, and
		$\stT[\roleQ]\tySub\stEnvApp{\stEnviii}{\mpChanRole{\mpS}{\roleQ}}$.\\
		Let $\stEnvii=\{\stEnvMap{\roleR}{\stEnvMap{\stEnv}{\mpC}}\}_{\mpC\in\dom{\stEnv}\setminus\{\mpChanRole{\mpS}{\roleP},\mpChanRole{\mpS}{\roleQ}\}}
		\stEnvComp
		\stEnvMap{\mpChanRole{\mpS}{\roleP}}{\stT[\roleP]}
		\stEnvComp
		\stEnvMap{\mpChanRole{\mpS}{\roleQ}}{\stT[\roleQ]}$, so that\\
		$\stEnv\,\gtMove[\ltsSendRecvS{\mpS}{\roleP}{\roleQ}{\stChoice{\stLab}{\stS[0]}}]\,\stEnvii$ and $\stEnvii\tySub\stEnviii$.
	\end{itemize}
\qed\end{proof}

\begin{proof}[2]
	\begin{enumerate}
		\item Let $\stEnv\tySub\stEnvi$.
		\begin{itemize}
				\item Suppose that
				$\stEnvApp{\stEnvNew}{\mpChanRole{\mpS}{\roleP}}\,\gtMove[{\roleQ\stEnvAnnotOutSym\stChoice{\stLab}{\stS[0]}}]$
				and
				$\stEnvApp{\stEnvNew}{\mpChanRole{\mpS}{\roleQ}}\,\gtMove[{\roleP\stEnvAnnotInSym\stChoice{\stLabi}{\stSi}}]$.\\
				By Lem.~\ref{lem:sub-sem},
				$\stEnvApp{\stEnvNewi}{\mpChanRole{\mpS}{\roleP}}\,\gtMove[{\roleQ\stEnvAnnotOutSym\stChoice{\stLab}{\stSi[0]}}]$,
				$\stSi[0]\tySub\stS[0]$, and
				$\stEnvApp{\stEnvNewi}{\mpChanRole{\mpS}{\roleQ}}\,\gtMove[{\roleP\stEnvAnnotInSym\stChoice{\stLabii}{\stSii}}]$.
				By safety,
				$\stEnvNewi\gtMove[\ltsSendRecvS{\mpS}{\roleP}{\roleQ}{\stChoice{\stLab}{\stSi[0]}}]$,
				so
				$\stEnvApp{\stEnvNewi}{\mpChanRole{\mpS}{\roleQ}}\,\gtMove[{\roleP\stEnvAnnotInSym\stChoice{\stLab}{\stSi[1]}}]$
				for some $\stSi[1]\tySub\stSi[0]$.\\
				By Lem.~\ref{lem:sub-sem},
				$\stEnvApp{\stEnvNew}{\mpChanRole{\mpS}{\roleQ}}\,\gtMove[{\roleP\stEnvAnnotInSym\stChoice{\stLab}{\stS[1]}}]$
				for some $\stS[1]\tySub\stSi[1]$.
				Therefore,
				$\stEnvNew\gtMove[\ltsSendRecvS{\mpS}{\roleP}{\roleQ}{\stChoice{\stLab}{\stS[0]}}]$.
			\item By (1), if $\stEnv\gtMove\stEnvii$
			then there is $\stEnviii$ such that
			$\stEnvSafeP{\stEnviii}$ and $\stEnvii\tySub\stEnviii$.
		\end{itemize}
		So, if $\stEnv\tySub\stEnvi$ and $\stEnvi$ is safe, then $\stEnv$ is safe.
		
		\item Suppose that $\stEnv\tySub\stEnvi$ and $\stEnvDFP{\stEnvi}$ and $\stEnvSafeP{\stEnvi}$.
		By above, $\stEnvSafeP{\stEnv}$.
		Suppose that $\stEnv\,\gtMoveStar\,\stEnvii\!\not\!\!\!\!\gtMove[]$.
		By (1), there exists $\stEnviii$ such that
		$\stEnvi\,\gtMoveStar\,\stEnviii\!\not\!\!\!\!\gtMove[]$
		and $\stEnvii\tySub\stEnviii$.
		$\stEnvi$ is deadlock-free, so
		for all $\mpC\in\dom{\stEnviii}=\dom{\stEnvii}$,
		$\unfoldOne{\stEnvApp{\stEnviii}{\mpC}}=\stEnd$.
		By Lem.~\ref{lem:sub-inv},
		$\stEnvApp{\stEnvii}{\mpC}\tySub\unfoldOne{\stEnvApp{\stEnviii}{\mpC}}=\stEnd$.
		By Lem.~\ref{lem:sub-end},
		$\unfoldOne{\stEnvApp{\stEnvii}{\mpC}}=\stEnd$.
		
		Therefore, $\stEnvDFP{\stEnv}$.
		
		\item Suppose that $\stEnv\tySub\stEnvi$ and $\stEnvLiveP{\stEnvi}$ and $\stEnvSafeP{\stEnvi}$.
		By above, $\stEnvSafeP{\stEnv}$.
		Suppose that $\{\stEnv[i]\}_{i\in N}$
		is a fair path starting at $\stEnv$.
		Say that $\{\stEnvAnnotGenericSym[i]\}_{i,i+1\in N}$
		is such that $\stEnv[i]\,\gtMove[{\stEnvAnnotGenericSym[i]}]\,\stEnv[i+1]$
		for $i,i+1\in N$
		and that if $\stEnv[i]\,\gtMove[\stEnvAnnotGenericSym]$
		then there is $j\geq i$
		such that $\ltsSubject{\stEnvAnnotGenericSym}\cap\ltsSubject{\stEnvAnnotGenericSym[j]}\neq\emptyset$.
		If there is no choice in the label, then this is obvious.
		If we have a choice of label, pick one containing
		a role with the earliest last occurrence,
		so that these labels can be used in the definition of fairness.
		By (1), there exists a path $\{\stEnvi[i]\}_{i\in N}$ starting at $\stEnvi$
		such that $\stEnvi[i]\,\gtMove[{\stEnvAnnotGenericSymi[i]}]\,\stEnvi[i+1]$
		for $i,i+1\in N$ and
		$\stEnv[i]\tySub\stEnvi[i]$,
		where $\stEnvAnnotGenericSymi[i]$
		and $\stEnvAnnotGenericSym[i]$ differ only in payload type.
		If $\stEnvi[i]\,\gtMove[\stEnvAnnotGenericSym]$, then, by (1),
		$\stEnv[i]\,\gtMove[\stEnvAnnotGenericSymi]$
		with
		$\ltsSubject{\stEnvAnnotGenericSym}=\ltsSubject{\stEnvAnnotGenericSymi}$.
		There is $j\geq i$
		such that $\ltsSubject{\stEnvAnnotGenericSym}\cap\ltsSubject{\stEnvAnnotGenericSym[j]}\neq\emptyset$.
		Therefore, $\{\stEnvi[i]\}_{i\in N}$ is fair.
		$\stEnvi$ is live, so $\{\stEnvi[i]\}_{i\in N}$ is live.
		If $\stEnvApp{\stEnv[i]}{\mpC}\,\gtMove[]$, then,
		by Lem.~\ref{lem:sub-sem},
		$\stEnvApp{\stEnvi[i]}{\mpC}\,\gtMove[]$.
		So, there exists $j\geq i$ and $\stEnvAnnotGenericSymii[0]$ such that
		$\mpC\in\ltsSubject{\stEnvAnnotGenericSymii[0]}$
		and $\stEnvi[j]\,\gtMove[{\stEnvAnnotGenericSymii[0]}]$,
		so, by (1), there exists $\stEnvAnnotGenericSymii[1]$
		such that
		$\ltsSubject{\stEnvAnnotGenericSymii[0]}=\ltsSubject{\stEnvAnnotGenericSymii[1]}$
		and $\stEnv[j]\,\gtMove[{\stEnvAnnotGenericSymii[1]}]$.
	\end{enumerate}
\qed\end{proof}

\begin{example}[Leader Election with Delegation]
\label{ex:election-type-sem-app}
We give a typing context to mirror the semantics of \cref{ex:elect}.
\begingroup
\small
\[
	\begin{array}{lclclcl}
		\stT&=&
		\roleS\stFmt{!}\stChoice{\stLabFmt{token}}{\tyInt}
		&\quad&
		\stT[i]&=&
		\stSum{}{}{
		\begin{array}{l}
			\roleP[(i+4)\%5]!\stChoice{\stLabFmt{elect}}{\stT}\\
			\roleP[(i+3)\%5]!\stChoice{\stLabFmt{elect}}{\stT}\\
			\roleP[(i+1)\%5]?\stChoice{\stLabFmt{elect}}{\stT}\stSeq\stTi[i]
		\end{array}
		}{}\\
		\stTi&=&
		\roleP\stFmt{?}\stChoice{\stLabFmt{token}}{\tyInt}
		&\quad&
		\stTi[i]&=&
		\stSum{}{}{
		\begin{array}{l}
			\roleP[(i+2)\%5]!\stChoice{\stLabFmt{elect}}{\stT}\\
			\roleP[(i+3)\%5]?\stChoice{\stLabFmt{elect}}{\stT}\stSeq
			\roleP[(i+2)\%5]?\stChoice{\stLabFmt{elect}}{\stT}
		\end{array}
		}{}
	\end{array}
\]
\[
	\stEnv[\text{\tiny lead}] =
	\bigcup_{0\leq i \leq 4}\set{
	\stEnvMap{\mpChanRole{\mpS[i]}{\roleP}}{\stT}
	\stEnvComp
	\stEnvMap{\mpChanRole{\mpS[i]}{\roleS}}{\stTi}
	\stEnvComp
	\stEnvMap{\mpChanRole{\mpS}{\roleP[i]}}{\stT[i]}
	}
\]
\endgroup
Observe that
$\stT\tySub
\stTii=\stSum{}{}{
\roleS!\stChoice{\stLabFmt{token}}{\tyInt},\roleS!\stChoice{\stLabFmt{to'}}{\tyInt}
}{}
$,
so
\begingroup
\[\begin{array}{lcl}
	\stTi[0]&\tySub&
		\stSum{}{}{
		\begin{array}{l}
			\roleP[2]!\stChoice{\stLabFmt{elect}}{\stTii}\\
			\roleP[3]?\stChoice{\stLabFmt{elect}}{\stT}\stSeq
			\roleP[2]?\stChoice{\stLabFmt{elect}}{\stT}
		\end{array}
		}{}
	\\
	\stTi[0]&\tyNotSub&
		\stSum{}{}{
		\begin{array}{l}
			\roleP[2]!\stChoice{\stLabFmt{elect}}{\stT}\\
			\roleP[3]?\stChoice{\stLabFmt{elect}}{\stTii}\stSeq
			\roleP[2]?\stChoice{\stLabFmt{elect}}{\stTii}
		\end{array}
		}{}\tySub\stTi[0]
\end{array}
\]
\endgroup
We trace the same election as in \cref{ex:semantics}
from $\stEnv[\text{\tiny lead},\mpS]$.
\begingroup
\small
\[
	\begin{array}{cl}
	&\stEnv[\text{\tiny lead},\mpS]\\
	\gtMove[\ltsSendRecvS{\mpS}{\roleP[0]}{\roleP[4]}{\stChoice{\stLabFmt{elect}}{\stT}}]&
	\stEnvMap{\mpChanRole{\mpS}{\roleP[0]}}{\stEnd}
	\stEnvComp
	\stEnvMap{\mpChanRole{\mpS}{\roleP[4]}}{
	\stTi[4]
	}
	\stEnvComp
	\bigcup_{1\leq i < 4}
	\stEnvMap{\mpChanRole{\mpS}{\roleP[i]}}{\stT[i]}\\
	\gtMove[\ltsSendRecvS{\mpS}{\roleP[2]}{\roleP[1]}{\stChoice{\stLabFmt{elect}}{\stT}}]&
	\stEnvMap{\mpChanRole{\mpS}{\roleP[0]}}{\stEnd}
	\stEnvComp
	\stEnvMap{\mpChanRole{\mpS}{\roleP[2]}}{\stEnd}
	\stEnvComp
	\stEnvMap{\mpChanRole{\mpS}{\roleP[4]}}{
	\stTi[4]
	}
	\stEnvComp
	\stEnvMap{\mpChanRole{\mpS}{\roleP[1]}}{
	\stTi[1]
	}
	\stEnvComp
	\stEnvMap{\mpChanRole{\mpS}{\roleP[3]}}{\stT[3]}\\
	\gtMove[\ltsSendRecvS{\mpS}{\roleP[4]}{\roleP[1]}{\stChoice{\stLabFmt{elect}}{\stT}}]&
	\stEnvMap{\mpChanRole{\mpS}{\roleP[0]}}{\stEnd}
	\stEnvComp
	\stEnvMap{\mpChanRole{\mpS}{\roleP[2]}}{\stEnd}
	\stEnvComp
	\stEnvMap{\mpChanRole{\mpS}{\roleP[4]}}{\stEnd}
	\stEnvComp
	\stEnvMap{\mpChanRole{\mpS}{\roleP[1]}}{
	\roleP[3]?\stChoice{\stLabFmt{elect}}{\stT}
	}
	\stEnvComp
	\stEnvMap{\mpChanRole{\mpS}{\roleP[3]}}{\stT[3]}\\
	\gtMove[\ltsSendRecvS{\mpS}{\roleP[3]}{\roleP[1]}{\stChoice{\stLabFmt{elect}}{\stT}}]&
	\bigcup_{0\leq i < 5}
	\stEnvMap{\mpChanRole{\mpS}{\roleP[i]}}{\stEnd}
	\end{array}
\]
\endgroup
As with the calculus, $\stEnv[\text{\tiny lead}]$
is safe, deadlock-free, and live.
\end{example}

\begin{example}[Cycle]
\label{ex:cycle-sem-type}
We give a typing context to mirror the semantics of \cref{ex:cycle}.
\begingroup
\small
\[\begin{array}{lcl}
	\stT[i]&=&
	\stRec{\stRecVar}{\left(
		\stSum{\roleP[(i+j)\%5]}{j\in\{-1,1\}}{\stChoice{\stLab}{\tyInt}\stSeq\stRecVar}{?}\right.}\\&\stFmt{+}&\left.
		\stSum{\roleP[(i+j)\%5]}{j\in\{-1,1\}}{\stChoice{\stLab}{\tyInt}\stSeq
		\stSum{\roleP[(i+j)\%5]}{j\in\{-1,1\}}{\stChoice{\stLab}{\tyInt}\stSeq\stRecVar}{?}
		}{!}\right)
	\\
	\stTi[i]&=&\stSum{\roleP[(i+j)\%5]}{j\in\{-1,1\}}{\stChoice{\stLab}{\tyInt}\stSeq\stT[i]}{?}
	\end{array}
\]
\[
	\stEnv[\text{\tiny cycle}] =\set{
	\stEnvMap{\mpChanRole{\mpS}{\roleP[0]}}{\stT[0]}
	\stEnvComp
	\stEnvMap{\mpChanRole{\mpS}{\roleP[1]}}{\stTi[1]}
	\stEnvComp
	\stEnvMap{\mpChanRole{\mpS}{\roleP[2]}}{\stT[2]}
	\stEnvComp
	\stEnvMap{\mpChanRole{\mpS}{\roleP[3]}}{\stTi[3]}
	\stEnvComp
	\stEnvMap{\mpChanRole{\mpS}{\roleP[4]}}{\stTi[4]}}
\]
\endgroup
While this is safe and deadlock-free,
it is not live since
there is a fair execution path
where $\roleP[3]$ never has an action enabled.
\begingroup
\small
\[
	\begin{array}{lc@{\{}l@{\}}cl}
	\stEnv[\text{\tiny cycle}]&\gtMove&
	\stEnvMap{\mpChanRole{\mpS}{\roleP[0]}}{\stT[0]}
	\stEnvComp
	\stEnvMap{\mpChanRole{\mpS}{\roleP[1]}}{\stT[1]}
	\stEnvComp
	\stEnvMap{\mpChanRole{\mpS}{\roleP[2]}}{\stTi[2]}
	\stEnvComp
	\stEnvMap{\mpChanRole{\mpS}{\roleP[3]}}{\stTi[3]}
	\stEnvComp
	\stEnvMap{\mpChanRole{\mpS}{\roleP[4]}}{\stTi[4]}
	&=&\stEnvi[\text{\tiny cycle}]
	\\
	&\gtMove&
	\stEnvMap{\mpChanRole{\mpS}{\roleP[0]}}{\stT[0]}
	\stEnvComp
	\stEnvMap{\mpChanRole{\mpS}{\roleP[1]}}{\stTi[1]}
	\stEnvComp
	\stEnvMap{\mpChanRole{\mpS}{\roleP[2]}}{\stTi[2]}
	\stEnvComp
	\stEnvMap{\mpChanRole{\mpS}{\roleP[3]}}{\stTi[3]}
	\stEnvComp
	\stEnvMap{\mpChanRole{\mpS}{\roleP[4]}}{\stTi[4]}
	&=&
	\stEnvii[\text{\tiny cycle}]\\
	&\gtMove&
	\stEnvMap{\mpChanRole{\mpS}{\roleP[0]}}{\stTi[0]}
	\stEnvComp
	\stEnvMap{\mpChanRole{\mpS}{\roleP[1]}}{\stT[1]}
	\stEnvComp
	\stEnvMap{\mpChanRole{\mpS}{\roleP[2]}}{\stTi[2]}
	\stEnvComp
	\stEnvMap{\mpChanRole{\mpS}{\roleP[3]}}{\stTi[3]}
	\stEnvComp
	\stEnvMap{\mpChanRole{\mpS}{\roleP[4]}}{\stTi[4]}
	&=&
	\stEnviii[\text{\tiny cycle}]\gtMove\stEnvii[\text{\tiny cycle}]
	\end{array}
\]
\endgroup
The path $\stEnv[\text{\tiny cycle}],\stEnvi[\text{\tiny cycle}],\left(\stEnvii[\text{\tiny cycle}],\stEnviii[\text{\tiny cycle}]\right)^*$
is fair because, once the loop is entered,
each enabled action is disabled by the transition,
and it is not live because $\roleP[3]$
is never a subject of any enabled actions.
For more details on why we need to consider fair paths,
see \cite{POPL19LessIsMore}.

Let us generalise this construction to
cycles with $n$ participants for $n\geq 2$,
starting with $2$ active participants.
\begingroup
\small
\[\begin{array}{lcl}
	\stT[i,n]&=&
	\stRec{\stRecVar}{\left(
		\stSum{\roleP[(i+j)\%n]}{j\in\{-1,1\}}{\stChoice{\stLab}{\tyInt}\stSeq\stRecVar}{?}\right.}\\&\stFmt{+}&\left.
		\stSum{\roleP[(i+j)\%n]}{j\in\{-1,1\}}{\stChoice{\stLab}{\tyInt}\stSeq
		\stSum{\roleP[(i+j)\%n]}{j\in\{-1,1\}}{\stChoice{\stLab}{\tyInt}\stSeq\stRecVar}{?}
		}{!}\right)
	\\
	\stTi[i,n]&=&\stSum{\roleP[(i+j)\%n]}{j\in\{-1,1\}}{\stChoice{\stLab}{\tyInt}\stSeq\stT[i,n]}{?}
	\end{array}
\]
\[
	\stEnv[\text{\tiny cycle},n] =\set{
	\stEnvMap{\mpChanRole{\mpS}{\roleP[0]}}{\stT[0,n]}
	\stEnvComp
	\stEnvMap{\mpChanRole{\mpS}{\roleP[1]}}{\stT[1,n]}}
	\cup
	\bigcup_{2\leq i \leq n-1}
	\set{
	\stEnvMap{\mpChanRole{\mpS}{\roleP[i]}}{\stTi[i,n]}}
\]
\endgroup
For all $n\geq 2$, $\stEnv[\text{\tiny cycle},n]$ is safe and deadlock-free.

For $n\in\{2,3,4\}$,
$\stEnv[\text{\tiny cycle},n]$ is live
since every participant either has a communication
enabled or will have a communication enabled
after one step.

For $n\geq 5$,
$\stEnv[\text{\tiny cycle},n]$ is not live
for the same reason (and fair path)
as $\stEnv[\text{\tiny cycle}]$.
\end{example}

\subsection{Typing System}

\begin{definition}[Typing Rules for Expressions]\rm
\label{def:type-expressions}
	We define the typing judgement on expressions,
	written $\tyJudge{\tyEnv}{e}{\tyGround}$,
	inductively by the rules:
	\begin{equation*}
	\begin{array}{c}
		\inference[Var]{}{\tyJudge{\tyEnv\tyEnvComp\tyEnvMap{\mpx}{\tyGround}}{\mpx}{\tyGround}}
		\qquad
		\inference[Val]{v\in\tyGround}{\tyJudge{\tyEnv}{v}{\tyGround}}
		\\[2ex]
		\inference[Eq]{\tyJudge{\tyEnv}{e}{\tyGround} & \tyJudge{\tyEnv}{e'}{\tyGround}}{\tyJudge{\tyEnv}{e=e'}{\tyBool}}
		\qquad
		\inference[Ineq]{\tyJudge{\tyEnv}{e}{\tyInt}&\tyJudge{\tyEnv}{e'}{\tyInt}}{\tyJudge{\tyEnv}{e<e'}{\tyBool}}
		\\[2ex]
		\inference[Succ]{\tyJudge{\tyEnv}{e}{\tyInt}}{\tyJudge{\tyEnv}{\mpSucc{e}}{\tyInt}}
		\qquad
		\inference[Neg]{\tyJudge{\tyEnv}{e}{\tyGround}&\tyGround\in\{\tyInt,\tyBool\}}{\tyJudge{\tyEnv}{\mpNeg{e}}{\tyGround}}
		\end{array}
	\end{equation*}
\end{definition}

\inferrule{Var} says that variables are typed according to contexts;
\inferrule{Val} that values of a type are typed by that type;
\inferrule{Eq} and \inferrule{Ineq} that equality and inequality checking
should return booleans if the inputs make sense;
\inferrule{Non-Det} that a non-deterministic choice between
two expressions of the same type should return a value of the shared type;
\inferrule{Succ} that the successor of an integer is an integer; and
\inferrule{Neg} that the negation of an integer/boolean is an integer/boolean.

\begin{lemma}[Strengthening Environments - Expressions]
\label{lem:str-expr}
$\tyJudge{\tyEnv}{e}{\tyGround}$
iff $\tyJudge{\{\tyEnvMap{\mpx}{\tyEnvApp{\tyEnv}{\mpx}}\}_{\mpx\in\fev{e}}}{e}{\tyGround}$.
\end{lemma}

\begin{proof}
	By induction on $e$.
	\begin{itemize}[leftmargin=1.5cm]
		\item[\inferrule{Var}]$\tyJudge{\tyEnv\tyEnvComp\tyEnvMap{\mpx}{\tyGround}}{\mpx}{\tyGround}$.
		We have that $\tyJudge{\tyEnvMap{\mpx}{\tyGround}}{\mpx}{\tyGround}$.
		
		\item[\inferrule{Val}]Suppose that $v\in\tyGround$ and $\tyJudge{\tyEnv}{v}{\tyGround}$.
		We have that $\tyJudge{}{v}{\tyGround}$.
		
		\item[\inferrule{Eq}] Suppose that $\tyJudge{\tyEnv}{e}{\tyGround}$ and
		$\tyJudge{\tyEnv}{e'}{\tyGround}$.
		By I.H.,
		$\tyJudge{\{\tyEnvMap{\mpx}{\tyEnvApp{\tyEnv}{\mpx}}\}_{\mpx\in\fev{e}}}{e}{\tyGround}$ and
		$\tyJudge{\{\tyEnvMap{\mpx}{\tyEnvApp{\tyEnv}{\mpx}}\}_{\mpx\in\fev{e'}}}{e}{\tyGround}$.
		By I.H.,
		$\tyJudge{\{\tyEnvMap{\mpx}{\tyEnvApp{\tyEnv}{\mpx}}\}_{\mpx\in\fev{e=e'}}}{e}{\tyGround}$ and
		$\tyJudge{\{\tyEnvMap{\mpx}{\tyEnvApp{\tyEnv}{\mpx}}\}_{\mpx\in\fev{e=e'}}}{e}{\tyGround}$.
		Therefore,
		$\tyJudge{\{\tyEnvMap{\mpx}{\tyEnvApp{\tyEnv}{\mpx}}\}_{\mpx\in\fev{e=e'}}}{e=e'}{\tyBool}$.
		
		\item[\inferrule{Ineq}] Suppose that $\tyJudge{\tyEnv}{e}{\tyInt}$ and
		$\tyJudge{\tyEnv}{e'}{\tyInt}$.
		By I.H.,
		$\tyJudge{\{\tyEnvMap{\mpx}{\tyEnvApp{\tyEnv}{\mpx}}\}_{\mpx\in\fev{e}}}{e}{\tyInt}$ and
		$\tyJudge{\{\tyEnvMap{\mpx}{\tyEnvApp{\tyEnv}{\mpx}}\}_{\mpx\in\fev{e'}}}{e}{\tyInt}$.
		By I.H.,
		$\tyJudge{\{\tyEnvMap{\mpx}{\tyEnvApp{\tyEnv}{\mpx}}\}_{\mpx\in\fev{e<e'}}}{e}{\tyInt}$ and
		$\tyJudge{\{\tyEnvMap{\mpx}{\tyEnvApp{\tyEnv}{\mpx}}\}_{\mpx\in\fev{e<e'}}}{e}{\tyInt}$.
		Therefore,
		$\tyJudge{\{\tyEnvMap{\mpx}{\tyEnvApp{\tyEnv}{\mpx}}\}_{\mpx\in\fev{e<e'}}}{e<e'}{\tyBool}$.
		
		\item[\inferrule{Succ}] Suppose that $\tyJudge{\tyEnv}{e}{\tyInt}$.
		By I.H.,
		$\tyJudge{\{\tyEnvMap{\mpx}{\tyEnvApp{\tyEnv}{\mpx}}\}_{\mpx\in\fev{e}}}{e}{\tyInt}$.
		Therefore,
		$\tyJudge{\{\tyEnvMap{\mpx}{\tyEnvApp{\tyEnv}{\mpx}}\}_{\mpx\in\fev{e}}}{\mpSucc{e}}{\tyInt}$.
		
		\item[\inferrule{Neg}] Suppose that $\tyJudge{\tyEnv}{e}{\tyGround}$ and
		$\tyGround\in\{\tyInt,\tyBool\}$.
		By I.H.,
		$\tyJudge{\{\tyEnvMap{\mpx}{\tyEnvApp{\tyEnv}{\mpx}}\}_{\mpx\in\fev{e}}}{e}{\tyGround}$.
		Therefore,
		$\tyJudge{\{\tyEnvMap{\mpx}{\tyEnvApp{\tyEnv}{\mpx}}\}_{\mpx\in\fev{e}}}{\mpNeg{e}}{\tyGround}$.
	\end{itemize}
\qed\end{proof}

\begin{lemma}[Typing Substitution]
\label{lem:sub-expr}
	If $\tyJudge{\tyEnv}{e}{\tyGround}$ and
	$\tyJudge{\tyEnvi}{e'}{\tyEnvApp{\tyEnv}{\mpx}}$, then
	let $\tyEnvii=\tyEnv\setminus\{\tyEnvMap{\mpx}{\tyEnvApp{\tyEnv}{\mpx}}\}\cup\tyEnvi$.
	If $\tyEnvii$ is an environment, then $\tyJudge{\tyEnvii}{e\subst{\mpx}{e'}}{\tyGround}$.
\end{lemma}

\begin{proof}
	Induction on $e$.
	\begin{itemize}[leftmargin=1.5cm]
		\item[$\mpx$]
			$\tyGround=\tyEnvApp{\tyEnv}{\mpx}$.
			$\tyJudge{\tyEnvi}{e'}{\tyGround}$
			so
			$\tyJudge{\{\tyEnvMap{\mpx}{\tyEnvApp{\tyEnv}{\mpx}}\}_{\mpx\in\fev{e'}}}{e'}{\tyGround}$
			so
			$\tyJudge{\tyEnvii}{e'}{\tyGround}$
			and $e'=e\subst{\mpx}{e'}$.
			
		\item[$v$]
			$v\in \tyGround$.
			$\tyJudge{}{v}{\tyGround}$, so
			$\tyJudge{\tyEnvii}{v}{\tyGround}$
			and $v=e\subst{\mpx}{e'}$.
			
		\item[$e_l=e_r$]
			$\tyGround=\tyBool$
			and there is $\tyGroundii$
			such that
			$\tyJudge{\tyEnv}{e_l}{\tyGroundii}$
			and
			$\tyJudge{\tyEnv}{e_r}{\tyGroundii}$.
			By I.H.,
			$\tyJudge{\tyEnvii}{e_l\subst{\mpx}{e'}}{\tyGroundii}$
			and
			$\tyJudge{\tyEnvii}{e_r\subst{\mpx}{e'}}{\tyGroundii}$.
			So,
			$\tyJudge{\tyEnvii}{e\subst{\mpx}{e'}}{\tyBool}$.
			
		\item[$e_l<e_r$]
			$\tyGround=\tyBool$
			and
			$\tyJudge{\tyEnv}{e_l}{\tyInt}$
			and
			$\tyJudge{\tyEnv}{e_r}{\tyInt}$.
			By I.H.,
			$\tyJudge{\tyEnvii}{e_l\subst{\mpx}{e'}}{\tyInt}$
			and
			$\tyJudge{\tyEnvii}{e_r\subst{\mpx}{e'}}{\tyInt}$.
			So,
			$\tyJudge{\tyEnvii}{e\subst{\mpx}{e'}}{\tyBool}$.
		\item[$\mpSucc{e''}$]
			$\tyGround=\tyInt$
			and
			$\tyJudge{\tyEnv}{e''}{\tyInt}$.
			By I.H.,
			$\tyJudge{\tyEnvii}{e''\subst{\mpx}{e'}}{\tyInt}$.
			So,
			$\tyJudge{\tyEnvii}{e\subst{\mpx}{e'}}{\tyInt}$.
			
		\item[$\mpNeg{e''}$]
			$\tyGround=\tyInt$ or $\tyBool.$
			and
			$\tyJudge{\tyEnv}{e''}{\tyGround}$.
			By I.H.,
			$\tyJudge{\tyEnvii}{e''\subst{\mpx}{e'}}{\tyGround}$.
			So,
			$\tyJudge{\tyEnvii}{e\subst{\mpx}{e'}}{\tyGround}$.
	\end{itemize}
\qed\end{proof}

\begin{lemma}[Evaluation]
\label{lem:eval-type}
	If $\tyJudge{\tyEnv}{e}{\tyGround}$ and $\fev{e}=\emptyset$,
	then:
	\begin{enumerate}
		\item there exists $v$ such that $\eval{e}{v}$;
		\item if $\eval{e}{v}$, then $v\in\tyGround$;
		\item $\neg\eval{e}{\mpErr}$.
	\end{enumerate}
\end{lemma}

\begin{proof}
	Induction on $e$
	\begin{itemize}[leftmargin=1.5cm]
		\item[$v$]
			$v\in \tyGround$.
			$\eval{v}{v}$
			and this is the only rule concluding with expression $v$.
			No rule concluding $\eval{v}{\mpErr}$.
			
		\item[$e_l=e_r$]
			$\tyGround=\tyBool$
			and there is $\tyGroundi$
			such that
			$\tyJudge{\tyEnv}{e_l}{\tyGroundi}$
			and
			$\tyJudge{\tyEnv}{e_r}{\tyGroundi}$.
			By I.H.
			there are $v_l,v_r\in \tyGroundi$ such that
			$\eval{e_l}{v_l}$ and
			$\eval{e_r}{v_r}$.
			If $v_l=v_r$, then $\eval{e}{\mpTrue}$
			otherwise,
			$\eval{e}{\mpFalse}$.
			
			If $\eval{e}{v'}$, then $v'\in\{\mpTrue,\mpFalse\}=\tyBool$.
			
			By I.H., $\neg\eval{e_l}{\mpErr}$
			and $\neg\eval{e_r}{\mpErr}$,
			so no rule concludes $\eval{e}{\mpErr}$.
			
		\item[$e_l<e_r$]
			$\tyGround=\tyBool$
			and
			$\tyJudge{\tyEnv}{e_l}{\tyInt}$
			and
			$\tyJudge{\tyEnv}{e_r}{\tyInt}$.
			By I.H.
			there are $v_l,v_r\in \tyInt$ such that
			$\eval{e_l}{v_l}$ and
			$\eval{e_r}{v_r}$.
			If $v_l<v_r$, then $\eval{e}{\mpTrue}$
			otherwise,
			$\eval{e}{\mpFalse}$.
			
			If $\eval{e}{v'}$, then $v'\in\{\mpTrue,\mpFalse\}=\tyBool$.
			
			By I.H., $\neg\eval{e_l}{\mpErr}$,
			$\neg\eval{e_r}{\mpErr}$, and if
			$\eval{e_l}{v_l}$ and
			$\eval{e_r}{v_r}$, then
			$v_l,v_r\in \tyInt$,
			so no rule concludes $\eval{e}{\mpErr}$.
		\item[$\mpSucc{e'}$]
			$\tyGround=\tyInt$.
			By I.H.,
			there exists $n\in\tyInt$ such that
			$\eval{e'}{n}$, so $\eval{e'}{n+1}$.
			
			If $\eval{e}{v}$, then there exists $v'$ s.t.
			$\eval{e'}{v'}$ and $v=v'+1$ so by I.H.
			$v'\in\tyInt$
			so $v\in\tyInt$.
			
			By I.H., $\neg\eval{e'}{\mpErr}$
			and if $\eval{e'}{v}$, then $v\in\tyInt$.
			So, no rule concludes $\eval{e}{\mpErr}$.
			
		\item[$\mpNeg{e'}$]
			Suppose that $\tyGround=\tyInt$.
			By I.H.,
			there exists $n\in\tyInt$ such that
			$\eval{e'}{n}$, so $\eval{e'}{-n}$.
			
			If $\eval{e}{v}$, then there exists $v'$ s.t.
			$\eval{e'}{v'}$.
			By I.H.
			$v'\in\tyInt$
			so $v\in\tyInt$ and $v=-v'$.
			
			Suppose that $\tyGround=\tyBool$.
			By I.H.,
			$\eval{e'}{\mpTrue}$ or $\eval{e'}{\mpFalse}$,
			so $\eval{e}{\mpFalse}$ or $\eval{e}{\mpTrue}$.
			
			If $\eval{e}{v}$, then there exists $v'$ s.t.
			$\eval{e'}{v'}$
			By I.H.,
			$v'\in\tyBool$.
			If $v'=\mpFalse/\mpTrue$, then
			so $v=\mpTrue/\mpFalse\in\tyBool$.
			
			By I.H., $\neg\eval{e'}{\mpErr}$
			so no rule concludes $\eval{e}{\mpErr}$.
	\end{itemize}
\qed\end{proof}

\begin{lemma}[Error-Free]
\label{lem:pi-no-err}
	If $\tyJudge{\tyEnv}{\mpP}{\stEnv}$,
	then $\mpP$ contains no $\mpErr$.
\end{lemma}

\begin{proof}
	Induction on the typing judgement.
	Every subterm must be typable,
	and no rule types $\mpErr$.
\qed\end{proof}

\begin{lemma}[Context Domains]
\label{lem:type-ctx-domains}
	If $\tyJudge{\tyEnv}{\mpP}{\stEnv}$,
	then
	$\dom{\tyEnv}\supseteq\fev{\mpP}\cup\fpv{\mpP}$, and
	$\dom{\stEnv}\supseteq\chan{\mpP}$.
\end{lemma}

\begin{proof}
	We induct on the derivation of $\tyJudge{\tyEnv}{e}{\tyGround}$
	to show that $\dom{\tyEnv}\supseteq\fev{e}$:
	if $\mpx\in\fev{e}$ then
	the judgement
	$\tyJudge{\tyEnv}{\mpx}{\tyGroundi}$
	will occur in the derivation,
	so $\mpx\in\dom{\tyEnv}$.
	
	We induct on the derivation of $\tyJudge{\tyEnv}{\mpP}{\stEnv}$:\\
	If $\mpx\in\fev{\mpP}$, then
	a judgement of the form
	$\tyJudge{\tyEnvi}{e}{\tyGround}$ occurs with $\mpx\in\fev{e}$,
	so $\mpx\in\dom{\tyEnvi}$.
	By the Barendregt convention, this is not a bound occurrence
	of $\mpx$ and so is never removed from the context by any inductive rule,
	so $\mpx\in\dom{\tyEnv}$.\\
	If $\mpX\in\fpv{\mpP}$, then
	a judgement of the form
	$\tyJudge{\tyEnvi}{\mpX}{\stEnvi}$ occurs,
	so $\mpX\in\dom{\tyEnvi}$.
	By the Barendregt convention, this is not a bound occurrence
	of $\mpX$ and so is never removed from the context by any inductive rule,
	so $\mpX\in\dom{\tyEnv}$.\\
	If $\mpC\in\chan{\mpP}$, then
	it appears in the judgement of \inferrule{Sum},
	so there is $\stEnvi$ appearing in the derivation
	with $\mpC\in\dom{\stEnvi}$.
	By the Barendregt convention, this is not a bound occurrence
	of $\mpC$ and so is never removed from the domain of the
	context by any inductive rule,
	so $\mpC\in\dom{\stEnv}$.
\qed\end{proof}

\begin{lemma}[Context Domain - II]
\label{lem:type-used-channels}
	If $\tyJudge{\tyEnv}{\mpP}{\stEnv}$, then
	$\dom{\stEnv\setminus\stEnd}\subseteq\chan{\mpP,\tyEnv}$
\end{lemma}

\begin{proof}
	Let $\mpC\in\dom{\stEnv\setminus\stEnd}$,
	we show that $\mpC\in\chan{\mpP,\tyEnv}$
	by induction on the derivation of
	$\tyJudge{\tyEnv}{\mpP}{\stEnv}$:\\
	\inferrule{Rec}. Hypothesis says $\dom{\stEnv\setminus\stEnd}\subseteq\chan{\mpP,\tyEnv}$.\\
	\inferrule{Nil}. $\dom{\stEnv\setminus\stEnd}=\emptyset$.\\
	\inferrule{Var}. $\mpC\in\dom{\stEnv}=\dom{\tyEnvApp{\tyEnv}{\mpC}}\subseteq\chan{\mpP,\tyEnv}$.\\
	\inferrule{If}. $\mpP=\mpIf{e}{\mpPi}{\mpPii}$ and
	$\tyJudge{\tyEnv}{\mpPi}{\stEnv}$ so by I.H.,
	$\mpC\in\chan{\mpPi,\tyEnv}\subseteq\chan{\mpP,\tyEnv}$.\\
	\inferrule{Par}. $\mpP=\mpP[0]\mpPar\mpP[1]$,
$\tyJudge{\tyEnv}{\mpP[0]}{\stEnv[0]}$,
$\tyJudge{\tyEnv}{\mpP[1]}{\stEnv[1]}$, and
$\stEnv=\stEnv[0]\stEnvComp\stEnv[1]$.
$\mpC\in\dom{\stEnv[i]\setminus\stEnd}$
for $i=0$ or $i=1$, so by I.H.,
$\mpC\in\chan{\mpP[i],\tyEnv}\subseteq\chan{\mpP,\tyEnv}$.\\
\inferrule{Sub}. $\tyJudge{\tyEnv}{\mpP}{\stEnvi}$,
$\stEnvEndP{\stEnvii}$, and $\stEnv=\stEnvi\stEnvComp\stEnvii$.
$\mpC\in\dom{\stEnvi\setminus\stEnd}$, so by I.H.,
$\mpC\in\chan{\mpP,\tyEnv}$.\\
\inferrule{Res}. $\tyJudge{\tyEnv}{\mpPi}{\stEnvi}$
with $\stEnv=\stEnvi\setminus\mpS$, so by I.H.,
$\mpC\in\chan{\mpPi,\tyEnv}$.
$\mpC$ is not over $\mpS$, so
$\mpC\in\chan{\mpP,\tyEnv}$.\\
\inferrule{Sum}. $\mpP=\mpSum{\mpPrefix_i\mpSeq\mpP[i]}{i\in I}$.
Fix $j\in I$, 
$\tyJudge{\tyEnvi}{\mpP[j]}{\stEnvi}$
with $\tyEnv$ and $\tyEnvi$ agreeing on process variables,
and $\stEnvi$ and $\stEnv$ agreeing on $\mpCi\not\in\chan{\mpPrefix_j}$.
If $\mpC\in\chan{\mpPrefix_j}$, then $\mpC\in\chan{\mpP,\tyEnv}$.
Otherwise, $\mpC\in\chan{\mpP[j],\tyEnvi}\subseteq\chan{\mpP,\tyEnv}$ by I.H.
\qed\end{proof}

\begin{lemma}[Unused Channels]
\label{lem:typing-remove-chan}
	If $\tyJudge{\tyEnv}{\mpP}{\stEnv}$,
	$\mpC\not\in\chan{\mpP}$, and
	$\forall \stEnvi\in\ran{\tyEnv}\cup\{\stEnv\}:
	\mpC\not\in \stEnvi$ or $\stEnvApp{\stEnvi}{\mpC}\tySub\stEnd$,
	then
	$\tyJudge{\tyEnvi}{\mpP}{\stEnv\setminus\mpC}$,
	where $\tyEnvi$ is $\tyEnv$ with all contexts $\stEnvi$
	replaced by $\stEnvi\setminus\mpC$.
\end{lemma}

\begin{proof}
	Induction on the typing judgement
	$\tyJudge{\tyEnv}{\mpP}{\stEnv}$.
	Each rule is invariant upon removing
	$\mpC$ from all contexts and environments,
	assuming that it has an end type and does not appear in the processes.

	In particular, if $\mpC\not\in \chan{\mpP,\tyEnv}$ and
	$\stEnvApp{\stEnv}{\mpC}\tySub\stEnd$, then
	$\tyJudge{\tyEnv}{\mpP}{\stEnv\setminus\mpC}$.
\qed\end{proof}

\begin{lemma}[Weakening]
\label{lem:pi-weak}
	If $\tyEnvApp{\tyEnv[1]}{\mpx}=\tyEnvApp{\tyEnv[2]}{\mpx}$
	for $\mpx\in\fev{\mpP}$, and
	$\tyEnvApp{\tyEnv[1]}{\mpX}=\tyEnvApp{\tyEnv[2]}{\mpX}$
	for $\mpX\in\fpv{\mpP}$,
	then
	$\tyJudge{\tyEnv[1]}{\mpP}{\stEnv}$
	implies
	$\tyJudge{\tyEnv[2]}{\mpP}{\stEnv}$.
\end{lemma}

\begin{proof}
	We proceed by induction on the typing judgement,
	$\tyJudge{\tyEnv[1]}{\mpP}{\stEnv}$:\\
	\inferrule{Rec}. $\mpP=\mpRec{\mpX}{\mpPi}$,
	$\tyJudge{\tyEnv[1]\tyEnvComp\tyEnvMap{\mpX}{\stEnv}}{\mpPi}{\stEnv}$,
	and $\dom{\stEnv\setminus\stEnd}\subseteq\chan{\mpP,\tyEnv[1]}$,
	so by I.H.,
	$\tyJudge{\tyEnv[2]\tyEnvComp\tyEnvMap{\mpX}{\stEnv}}{\mpPi}{\stEnv}$.
	$\chan{\mpPi,\tyEnv[1]}=\chan{\mpPi,\tyEnv[2]}$
	as it depends on the free process variables of $\mpPi$,
	so $\tyJudge{\tyEnv[2]}{\mpP}{\stEnv}$.\\
	\inferrule{Var}. $\mpP=\mpX$ and $\tyEnvApp{\tyEnv[1]}{\mpX}=\stEnv$,
	so $\tyEnvApp{\tyEnv[2]}{\mpX}=\stEnv$,
	so $\tyJudge{\tyEnv[2]}{\mpP}{\stEnv}$.\\
	\inferrule{Nil}. $\mpP=\mpNil$ and $\stEnv=\stEnvEmpty$, so
	$\tyJudge{\tyEnv[2]}{\mpNil}{\stEnvEmpty}$.\\
	\inferrule{If}. $\mpP=\mpIf{e}{\mpPi}{\mpPii}$,
	$\tyJudge{\tyEnv[1]}{e}{\tyBool}$,
	$\tyJudge{\tyEnv[1]}{\mpPi}{\stEnv}$, and
	$\tyJudge{\tyEnv[1]}{\mpPii}{\stEnv}$.
	By \cref{lem:str-expr},
	$\tyJudge{\tyEnv[2]}{e}{\tyBool}$.
	$\fpv{\mpPi}\cup\fev{\mpPi},\fpv{\mpPii}\cup\fev{\mpPii}\subseteq\fpv{\mpP}\cup\fev{\mpP}$.
	By I.H.,
	$\tyJudge{\tyEnv[2]}{\mpPi}{\stEnv}$ and
	$\tyJudge{\tyEnv[2]}{\mpPii}{\stEnv}$.
	So,
	$\tyJudge{\tyEnv[2]}{\mpP}{\stEnv}$.\\
	\inferrule{Par}.
	$\mpP=\mpPi\mpPar\mpPii$,
	$\tyJudge{\tyEnv[1]}{\mpPi}{\stEnvi}$,
	$\tyJudge{\tyEnv[1]}{\mpPii}{\stEnvii}$, and
	$\stEnv=\stEnvi\stEnvComp\stEnvii$.
	$\fpv{\mpPi}\cup\fev{\mpPi},\fpv{\mpPii}\cup\fev{\mpPii}\subseteq\fpv{\mpP}\cup\fev{\mpP}$.
	By I.H.,
	$\tyJudge{\tyEnv[2]}{\mpPi}{\stEnvi}$ and
	$\tyJudge{\tyEnv[2]}{\mpPii}{\stEnvii}$.
	So,
	$\tyJudge{\tyEnv[2]}{\mpP}{\stEnv}$.\\
	\inferrule{Res}.
	$\mpP=\mpRes{\mpS}{\mpPi}$,
	$\tyJudge{\tyEnv[1]}{\mpPi}{\stEnvi}$,
	$\predPApp{\stEnvi[\mpS]}$, and
	$\stEnvi\setminus\mpS=\stEnv$.
	By I.H.,
	$\tyJudge{\tyEnv[2]}{\mpPi}{\stEnvi}$.
	So, $\tyJudge{\tyEnv[2]}{\mpRes{\mpS}{\mpPi}}{\stEnv}$.\\
	\inferrule{Sum}.
	$\mpP=\mpSum{\mpPrefix_i\mpSeq\mpP[i]}{i\in I}$,
	for all $j\in I$ $\tyJudgePrefix{\tyEnv[1]}{\mpPrefix_j\mpSeq\mpP[j]}{\stEnv}$;
	the other hypotheses are independent of $\tyEnv[1]$
	so it suffices to show that
	$\tyJudgePrefix{\tyEnv[2]}{\mpPrefix_j\mpSeq\mpP[j]}{\stEnv}$
	for $j\in I$.
	
	If $\mpPrefix_j=\mpSel{\mpC}{\roleQ}{\stLab}{\mpCi}{}$, then
	$\mpC\neq\mpCi$,
	$\stFmt{\roleQ{!}\stChoice{\stLab}{\stEnvApp{\stEnv}{\mpCi}}\stSeq\stTii}\preType\stEnvApp{\stEnv}{\mpC}$,
	and
	$\tyJudge{\tyEnv[1]}{\mpP[j]}{
				(\stEnv\setminus\mpC\setminus\mpCi)
				\stEnvComp \stEnvMap{\mpC}{\stTii}
				}$.
	By I.H.,
	$\tyJudge{\tyEnv[2]}{\mpP[j]}{
				(\stEnv\setminus\mpC\setminus\mpCi)
				\stEnvComp \stEnvMap{\mpC}{\stTii}
				}$.
	So, $\tyJudgePrefix{\tyEnv[2]}{\mpPrefix_j\mpSeq\mpP[j]}{\stEnv}$.
	
	If $\mpPrefix_j=\mpSel{\mpC}{\roleQ}{\stLab}{e}{}$, then
	$\stFmt{\roleQ{!}\stChoice{\stLab}{\tyGround}\stSeq\stTii}
				\preType \stEnvApp{\stEnv}{\mpC}$,
	$\tyJudge{\tyEnv[1]}{e}{\tyGround}$, and
				$
				\tyJudge{\tyEnv[1]}{\mpP[j]}{
				(\stEnv\setminus\mpC)
				\stEnvComp \stEnvMap{\mpC}{\stTii}
				}$.
	By I.H.,
	$\tyJudge{\tyEnv[2]}{\mpP[j]}{
				(\stEnv\setminus\mpC)
				\stEnvComp \stEnvMap{\mpC}{\stTii}
				}$.
	By \cref{lem:str-expr},
	$\tyJudge{\tyEnv[2]}{e}{\tyGround}$.
	So, $\tyJudgePrefix{\tyEnv[2]}{\mpPrefix_j\mpSeq\mpP[j]}{\stEnv}$.
	
	If $\mpPrefix_j=\mpBra{\mpC}{\roleQ}{\stLab}{\mpy}{}$, then
	$\stFmt{\roleQ{?}\stChoice{\stLab}{\stTi}\stSeq\stTii}
				\preType \stEnvApp{\stEnv}{\mpC}$ and
				$\tyJudge{\tyEnv[1]}{\mpP[j]}{
				(\stEnv\setminus\mpC)
				\stEnvComp \stEnvMap{\mpC}{\stTii}
				\stEnvComp \stEnvMap{\mpy}{\stTi}
				}$.
	By I.H., $\tyJudge{\tyEnv[2]}{\mpP[j]}{
				(\stEnv\setminus\mpC)
				\stEnvComp \stEnvMap{\mpC}{\stTii}
				\stEnvComp \stEnvMap{\mpy}{\stTi}
				}$.
	So, $\tyJudgePrefix{\tyEnv[2]}{\mpPrefix_j\mpSeq\mpP[j]}{\stEnv}$.
	
	If $\mpPrefix_j=\mpBra{\mpC}{\roleQ}{\stLab}{\mpx}{}$, then
	$\stFmt{\roleQ{?}\stChoice{\stLab}{\tyGround}\stSeq\stTii}
				\preType \stEnvApp{\stEnv}{\mpC}$ and
				$\tyJudge{\tyEnv[1]
				\tyEnvComp\tyEnvMap{\mpx}{\tyGround}
				}{\mpP[j]}{
				(\stEnv\setminus\mpC)
				\stEnvComp \stEnvMap{\mpC}{\stTii}
				}$.
	By I.H., $\tyJudge{\tyEnv[2]
				\tyEnvComp\tyEnvMap{\mpx}{\tyGround}
				}{\mpP[j]}{
				(\stEnv\setminus\mpC)
				\stEnvComp \stEnvMap{\mpC}{\stTii}
				}$.
	So, $\tyJudgePrefix{\tyEnv[2]}{\mpPrefix_j\mpSeq\mpP[j]}{\stEnv}$.\\
	\inferrule{Sub}.
	$\stEnv=\stEnvi\stEnvComp\stEnvii$,
	$\stEnviii\tySub\stEnvi$,
	$\stEnvEndP{\stEnvii}$, and
	$\tyJudge{\tyEnv[1]}{\mpP}{\stEnviii}$.
	By I.H.,
	$\tyJudge{\tyEnv[2]}{\mpP}{\stEnviii}$
	so
	$\tyJudge{\tyEnv[2]}{\mpP}{\stEnvi\stEnvComp\stEnvii}$.
\qed\end{proof}

\begin{lemma}[Substitution]
\label{lem:typing-subst}
	\begin{enumerate}
		\item If $\tyJudge
		{\tyEnv\tyEnvComp\tyEnvMap{\mpx}{\tyGround}}
		{\mpP}
		{\stEnv}$
		and
		$\tyJudge{\tyEnv}{e}{\tyGround}$,
		then
		$\tyJudge{\tyEnv}{\mpP\subst{\mpx}{e}}{\stEnv}$.
		\item If $\tyJudge
		{\tyEnv\tyEnvComp\tyEnvMap{\mpX}{\stEnvi}}
		{\mpP}
		{\stEnv}$
		and
		$\tyJudge{\tyEnv}{\mpQ}{\stEnvi}$,
		then
		$\tyJudge{\tyEnv}{\mpP\subst{\mpX}{\mpQ}}{\stEnv}$.
		\item If $\tyJudge
		{\tyEnv}
		{\mpP}
		{\stEnv}$ and $\sigma$ is a permutation of free channels,
		then
		$\tyJudge{\tyEnv\sigma}{\mpP\sigma}{\stEnv\sigma}$.
	\end{enumerate}
\end{lemma}

\begin{proof}[1]
	We proceed by induction on the typing judgement,
	$\tyJudge{\tyEnv\tyEnvComp\tyEnvMap{\mpx}{\tyGround}}{\mpP}{\stEnv}$:\\
	\inferrule{Rec}. $\mpP=\mpRec{\mpX}{\mpPi}$,
	$\tyJudge{\tyEnv\tyEnvComp\tyEnvMap{\mpx}{\tyGround}\tyEnvComp\tyEnvMap{\mpX}{\stEnv}}{\mpPi}{\stEnv}$,
	and $\dom{\stEnv\setminus\stEnd}\subseteq\chan{\mpPi,\tyEnv\tyEnvComp\tyEnvMap{\mpx}{\tyGround}}$,
	so by I.H.,
	$\tyJudge{\tyEnv\tyEnvComp\tyEnvMap{\mpX}{\stEnv}}{\mpPi\subst{\mpx}{e}}{\stEnv}$.
	$\chan{\mpPi,\tyEnv\tyEnvComp\tyEnvMap{\mpx}{\tyGround}}=\chan{\mpPi\subst{\mpx}{e},\tyEnv}$,
	so $\tyJudge{\tyEnv}{\mpP\subst{\mpx}{e}}{\stEnv}$.\\
	\inferrule{Var}. $\mpP=\mpX$ so $\mpP\subst{\mpx}{e}=\mpX$.
	and $\tyEnvApp{\tyEnv\tyEnvComp\tyEnvMap{\mpx}{\tyGround}}{\mpX}=\stEnv$,
	so $\tyEnvApp{\tyEnv}{\mpX}=\stEnv$,
	so $\tyJudge{\tyEnv}{\mpP}{\stEnv}$.\\
	\inferrule{Nil}. $\mpP=\mpNil$ and $\stEnv=\stEnvEmpty$, so
	$\tyJudge{\tyEnv}{\mpNil}{\stEnvEmpty}$.\\
	\inferrule{If}. $\mpP=\mpIf{e'}{\mpPi}{\mpPii}$,
	$\tyJudge{\tyEnv\tyEnvComp\tyEnvMap{\mpx}{\tyGround}}{e'}{\tyBool}$,
	$\tyJudge{\tyEnv\tyEnvComp\tyEnvMap{\mpx}{\tyGround}}{\mpPi}{\stEnv}$, and
	$\tyJudge{\tyEnv\tyEnvComp\tyEnvMap{\mpx}{\tyGround}}{\mpPii}{\stEnv}$.
	By \cref{lem:sub-expr},
	$\tyJudge{\tyEnv}{e'\subst{\mpx}{e}}{\tyBool}$.
	By I.H.,
	$\tyJudge{\tyEnv}{\mpPi\subst{\mpx}{e}}{\stEnv}$ and
	$\tyJudge{\tyEnv}{\mpPii\subst{\mpx}{e}}{\stEnv}$.
	So,
	$\tyJudge{\tyEnv}{\mpP\subst{\mpx}{e}}{\stEnv}$.\\
	\inferrule{Par}.
	$\mpP=\mpPi\mpPar\mpPii$,
	$\tyJudge{\tyEnv\tyEnvComp\tyEnvMap{\mpx}{\tyGround}}{\mpPi}{\stEnvi}$,
	$\tyJudge{\tyEnv\tyEnvComp\tyEnvMap{\mpx}{\tyGround}}{\mpPii}{\stEnvii}$, and
	$\stEnv=\stEnvi\stEnvComp\stEnvii$.
	By I.H.,
	$\tyJudge{\tyEnv}{\mpPi\subst{\mpx}{e}}{\stEnvi}$ and
	$\tyJudge{\tyEnv}{\mpPii\subst{\mpx}{e}}{\stEnvii}$.
	So,
	$\tyJudge{\tyEnv}{\mpP\subst{\mpx}{e}}{\stEnv}$.\\
	\inferrule{Res}.
	$\mpP=\mpRes{\mpS}{\mpPi}$,
	$\tyJudge{\tyEnv\tyEnvComp\tyEnvMap{\mpx}{\tyGround}}{\mpPi}{\stEnvi}$,
	$\predPApp{\stEnvi[\mpS]}$, and
	$\stEnvi\setminus\mpS=\stEnv$.
	By I.H.,
	$\tyJudge{\tyEnv}{\mpPi\subst{\mpx}{e}}{\stEnvi}$.
	So, $\tyJudge{\tyEnv}{\mpRes{\mpS}{\mpPi}\subst{\mpx}{e}}{\stEnv}$.\\
	\inferrule{Sum}.
	$\mpP=\mpSum{\mpPrefix_i\mpSeq\mpP[i]}{i\in I}$,
	for all $j\in I$ $\tyJudgePrefix{\tyEnv\tyEnvComp\tyEnvMap{\mpx}{\tyGround}}{\mpPrefix_j\mpSeq\mpP[j]}{\stEnv}$;
	the other hypotheses are independent of $\tyEnv\tyEnvComp\tyEnvMap{\mpx}{\tyGround}$
	so it suffices to show that
	$\tyJudgePrefix{\tyEnv}{(\mpPrefix_j\mpSeq\mpP[j])\subst{\mpx}{e}}{\stEnv}$
	for $j\in I$.
	
	If $\mpPrefix_j=\mpSel{\mpC}{\roleQ}{\stLab}{\mpCi}{}$, then
	$\mpC\neq\mpCi$,
	$\stFmt{\roleQ{!}\stChoice{\stLab}{\stEnvApp{\stEnv}{\mpCi}}\stSeq\stTii}\preType\stEnvApp{\stEnv}{\mpC}$,
	and
	$\tyJudge{\tyEnv\tyEnvComp\tyEnvMap{\mpx}{\tyGround}}{\mpP[j]}{
				(\stEnv\setminus\mpC\setminus\mpCi)
				\stEnvComp \stEnvMap{\mpC}{\stTii}
				}$.
	By I.H.,
	$\tyJudge{\tyEnv}{\mpP[j]\subst{\mpx}{e}}{
				(\stEnv\setminus\mpC\setminus\mpCi)
				\stEnvComp \stEnvMap{\mpC}{\stTii}
				}$.\\
	So, $\tyJudgePrefix{\tyEnv}{(\mpPrefix_j\mpSeq\mpP[j])\subst{\mpx}{e}}{\stEnv}$.
	
	If $\mpPrefix_j=\mpSel{\mpC}{\roleQ}{\stLab}{e'}{}$, then
	$\stFmt{\roleQ{!}\stChoice{\stLab}{\tyGroundi}\stSeq\stTii}
				\preType \stEnvApp{\stEnv}{\mpC}$,
	$\tyJudge{\tyEnv\tyEnvComp\tyEnvMap{\mpx}{\tyGround}}{e'}{\tyGroundi}$, and
				$
				\tyJudge{\tyEnv\tyEnvComp\tyEnvMap{\mpx}{\tyGround}}{\mpP[j]}{
				(\stEnv\setminus\mpC)
				\stEnvComp \stEnvMap{\mpC}{\stTii}
				}$.
	By I.H.,
	$\tyJudge{\tyEnv}{\mpP[j]\subst{\mpx}{e}}{
				(\stEnv\setminus\mpC)
				\stEnvComp \stEnvMap{\mpC}{\stTii}
				}$.
	By \cref{lem:sub-expr},
	$\tyJudge{\tyEnv}{e'\subst{\mpx}{e}}{\tyGround}$.
	So, $\tyJudgePrefix{\tyEnv}{\mpPrefix_j\mpSeq\mpP[j]\subst{\mpx}{e}}{\stEnv}$.
	
	If $\mpPrefix_j=\mpBra{\mpC}{\roleQ}{\stLab}{\mpy}{}$, then
	$\stFmt{\roleQ{?}\stChoice{\stLab}{\stTi}\stSeq\stTii}
				\preType \stEnvApp{\stEnv}{\mpC}$ and\\
				$\tyJudge{\tyEnv\tyEnvComp\tyEnvMap{\mpx}{\tyGround}}{\mpP[j]}{
				(\stEnv\setminus\mpC)
				\stEnvComp \stEnvMap{\mpC}{\stTii}
				\stEnvComp \stEnvMap{\mpy}{\stTi}
				}$.
	By I.H., $\tyJudge{\tyEnv}{\mpP[j]\subst{\mpx}{e}}{
				(\stEnv\setminus\mpC)
				\stEnvComp \stEnvMap{\mpC}{\stTii}
				\stEnvComp \stEnvMap{\mpy}{\stTi}
				}$.
	So, $\tyJudgePrefix{\tyEnv}{\mpPrefix_j\mpSeq\mpP[j]\subst{\mpx}{e}}{\stEnv}$.
	
	If $\mpPrefix_j=\mpBra{\mpC}{\roleQ}{\stLab}{\mpx'}{}$, then
	$\stFmt{\roleQ{?}\stChoice{\stLab}{\tyGroundi}\stSeq\stTii}
				\preType \stEnvApp{\stEnv}{\mpC}$ and\\
				$\tyJudge{\tyEnv\tyEnvComp\tyEnvMap{\mpx}{\tyGround}
				\tyEnvComp\tyEnvMap{\mpx'}{\tyGroundi}
				}{\mpP[j]}{
				(\stEnv\setminus\mpC)
				\stEnvComp \stEnvMap{\mpC}{\stTii}
				}$.
	By I.H., $\tyJudge{\tyEnv
				\tyEnvComp\tyEnvMap{\mpx'}{\tyGroundi}
				}{\mpP[j]\subst{\mpx}{e}}{
				(\stEnv\setminus\mpC)
				\stEnvComp \stEnvMap{\mpC}{\stTii}
				}$.
	So, $\tyJudgePrefix{\tyEnv}{\mpPrefix_j\mpSeq\mpP[j]\subst{\mpx}{e}}{\stEnv}$.\\
	\inferrule{Sub}.
	$\stEnv=\stEnvi\stEnvComp\stEnvii$,
	$\stEnviii\tySub\stEnvi$,
	$\stEnvEndP{\stEnvii}$, and
	$\tyJudge{\tyEnv\tyEnvComp\tyEnvMap{\mpx}{\tyGround}}{\mpP}{\stEnviii}$.
	By I.H.,
	$\tyJudge{\tyEnv}{\mpP\subst{\mpx}{e}}{\stEnviii}$
	so
	$\tyJudge{\tyEnv}{\mpP\subst{\mpx}{e}}{\stEnvi\stEnvComp\stEnvii}$.
\qed\end{proof}

\begin{proof}[2]
	We proceed by induction on the typing judgement,
	$\tyJudge{\tyEnv\tyEnvComp\tyEnvMap{\mpX}{\stEnvi}}{\mpP}{\stEnv}$:\\
	\inferrule{Rec}. $\mpP=\mpRec{\mpX'}{\mpPi}$,
	$\tyJudge{\tyEnv\tyEnvComp\tyEnvMap{\mpX}{\stEnvi}\tyEnvComp\tyEnvMap{\mpX'}{\stEnv}}{\mpPi}{\stEnv}$,
	and\\$\dom{\stEnv\setminus\stEnd}\subseteq\chan{\mpPi,\tyEnv\tyEnvComp\tyEnvMap{\mpX}{\stEnvi}}$,
	so by I.H.,
	$\tyJudge{\tyEnv\tyEnvComp\tyEnvMap{\mpX'}{\stEnv}}{\mpPi\subst{\mpX}{\mpQ}}{\stEnv}$.\\
	$\chan{\mpPi,\tyEnv\tyEnvComp\tyEnvMap{\mpX}{\stEnvi}}\subseteq\chan{\mpPi\subst{\mpX}{\mpQ},\tyEnv}$
	as $\dom{\stEnvi\setminus\stEnd}\subseteq\chan{\mpQ,\tyEnv}$,
	so $\tyJudge{\tyEnv}{\mpP\subst{\mpX}{\mpQ}}{\stEnv}$.\\
	\inferrule{Var}. $\mpP=\mpX'$.
	If $\mpX=\mpX'$, then
	$\mpP\subst{\mpX}{\mpQ}=\mpQ$,
	$\stEnv=\stEnvi$, and
	$\tyJudge{\tyEnv}{\mpQ}{\stEnvi}$.
	Otherwise,
	 $\mpP\subst{\mpX}{\mpQ}=\mpX'$.
	and $\tyEnvApp{\tyEnv\tyEnvComp\tyEnvMap{\mpX}{\stEnvi}}{\mpX'}=\stEnv$,
	so $\tyEnvApp{\tyEnv}{\mpX'}=\stEnv$,
	so $\tyJudge{\tyEnv}{\mpX'}{\stEnv}$.\\
	\inferrule{Nil}. $\mpP=\mpNil$ and $\stEnv=\stEnvEmpty$, so
	$\tyJudge{\tyEnv}{\mpNil}{\stEnvEmpty}$.\\
	\inferrule{If}. $\mpP=\mpIf{e}{\mpPi}{\mpPii}$,
	$\tyJudge{\tyEnv\tyEnvComp\tyEnvMap{\mpX}{\stEnvi}}{e}{\tyBool}$,
	$\tyJudge{\tyEnv\tyEnvComp\tyEnvMap{\mpX}{\stEnvi}}{\mpPi}{\stEnv}$, and
	$\tyJudge{\tyEnv\tyEnvComp\tyEnvMap{\mpX}{\stEnvi}}{\mpPii}{\stEnv}$.
	By \cref{lem:str-expr},
	$\tyJudge{\tyEnv}{e}{\tyBool}$.
	By I.H.,
	$\tyJudge{\tyEnv}{\mpPi\subst{\mpX}{\mpQ}}{\stEnv}$ and
	$\tyJudge{\tyEnv}{\mpPii\subst{\mpX}{\mpQ}}{\stEnv}$.
	So,
	$\tyJudge{\tyEnv}{\mpP\subst{\mpX}{\mpQ}}{\stEnv}$.\\
	\inferrule{Par}.
	$\mpP=\mpPi\mpPar\mpPii$,
	$\tyJudge{\tyEnv\tyEnvComp\tyEnvMap{\mpX}{\stEnvi}}{\mpPi}{\stEnv[l]}$,
	$\tyJudge{\tyEnv\tyEnvComp\tyEnvMap{\mpX}{\stEnvi}}{\mpPii}{\stEnv[r]}$, and
	$\stEnv=\stEnv[l]\stEnvComp\stEnv[r]$.
	By I.H.,
	$\tyJudge{\tyEnv}{\mpPi\subst{\mpX}{\mpQ}}{\stEnv[l]}$ and
	$\tyJudge{\tyEnv}{\mpPii\subst{\mpX}{\mpQ}}{\stEnv[r]}$.
	So,
	$\tyJudge{\tyEnv}{\mpP\subst{\mpX}{\mpQ}}{\stEnv}$.\\
	\inferrule{Res}.
	$\mpP=\mpRes{\mpS}{\mpPi}$,
	$\tyJudge{\tyEnv\tyEnvComp\tyEnvMap{\mpX}{\stEnvi}}{\mpPi}{\stEnvii}$,
	$\predPApp{\stEnvii[\mpS]}$, and
	$\stEnvii\setminus\mpS=\stEnv$.
	By I.H.,
	$\tyJudge{\tyEnv}{\mpPi\subst{\mpX}{\mpQ}}{\stEnvii}$.
	So, $\tyJudge{\tyEnv}{\mpRes{\mpS}{\mpPi}\subst{\mpX}{\mpQ}}{\stEnv}$.\\
	\inferrule{Sum}.
	$\mpP=\mpSum{\mpPrefix_i\mpSeq\mpP[i]}{i\in I}$,
	for all $j\in I$ $\tyJudgePrefix{\tyEnv\tyEnvComp\tyEnvMap{\mpX}{\stEnvi}}{\mpPrefix_j\mpSeq\mpP[j]}{\stEnv}$;
	the other hypotheses are independent of $\tyEnv\tyEnvComp\tyEnvMap{\mpX}{\stEnvi}$
	so it suffices to show that
	$\tyJudgePrefix{\tyEnv}{(\mpPrefix_j\mpSeq\mpP[j])\subst{\mpX}{\mpQ}}{\stEnv}$
	for $j\in I$.
	
	If $\mpPrefix_j=\mpSel{\mpC}{\roleQ}{\stLab}{\mpCi}{}$, then
	$\mpC\neq\mpCi$,
	$\stFmt{\roleQ{!}\stChoice{\stLab}{\stEnvApp{\stEnv}{\mpCi}}\stSeq\stTii}\preType\stEnvApp{\stEnv}{\mpC}$,
	and
	$\tyJudge{\tyEnv\tyEnvComp\tyEnvMap{\mpX}{\stEnvi}}{\mpP[j]}{
				(\stEnv\setminus\mpC\setminus\mpCi)
				\stEnvComp \stEnvMap{\mpC}{\stTii}
				}$.
	By I.H.,
	$\tyJudge{\tyEnv}{\mpP[j]\subst{\mpX}{\mpQ}}{
				(\stEnv\setminus\mpC\setminus\mpCi)
				\stEnvComp \stEnvMap{\mpC}{\stTii}
				}$.\\
	So, $\tyJudgePrefix{\tyEnv}{(\mpPrefix_j\mpSeq\mpP[j])\subst{\mpX}{\mpQ}}{\stEnv}$.
	
	If $\mpPrefix_j=\mpSel{\mpC}{\roleQ}{\stLab}{e}{}$, then
	$\stFmt{\roleQ{!}\stChoice{\stLab}{\tyGround}\stSeq\stTii}
				\preType \stEnvApp{\stEnv}{\mpC}$,
	$\tyJudge{\tyEnv\tyEnvComp\tyEnvMap{\mpX}{\stEnvi}}{e}{\tyGround}$, and
				$
				\tyJudge{\tyEnv\tyEnvComp\tyEnvMap{\mpX}{\stEnvi}}{\mpP[j]}{
				(\stEnv\setminus\mpC)
				\stEnvComp \stEnvMap{\mpC}{\stTii}
				}$.
	By I.H.,
	$\tyJudge{\tyEnv}{\mpP[j]\subst{\mpX}{\mpQ}}{
				(\stEnv\setminus\mpC)
				\stEnvComp \stEnvMap{\mpC}{\stTii}
				}$.
	By \cref{lem:str-expr},
	$\tyJudge{\tyEnv}{e}{\tyGround}$.
	So, $\tyJudgePrefix{\tyEnv}{\mpPrefix_j\mpSeq\mpP[j]\subst{\mpX}{\mpQ}}{\stEnv}$.
	
	If $\mpPrefix_j=\mpBra{\mpC}{\roleQ}{\stLab}{\mpy}{}$, then
	$\stFmt{\roleQ{?}\stChoice{\stLab}{\stTi}\stSeq\stTii}
				\preType \stEnvApp{\stEnv}{\mpC}$ and\\
				$\tyJudge{\tyEnv\tyEnvComp\tyEnvMap{\mpX}{\stEnvi}}{\mpP[j]}{
				(\stEnv\setminus\mpC)
				\stEnvComp \stEnvMap{\mpC}{\stTii}
				\stEnvComp \stEnvMap{\mpy}{\stTi}
				}$.
	By I.H., $\tyJudge{\tyEnv}{\mpP[j]\subst{\mpX}{\mpQ}}{
				(\stEnv\setminus\mpC)
				\stEnvComp \stEnvMap{\mpC}{\stTii}
				\stEnvComp \stEnvMap{\mpy}{\stTi}
				}$.
	So, $\tyJudgePrefix{\tyEnv}{\mpPrefix_j\mpSeq\mpP[j]\subst{\mpX}{\mpQ}}{\stEnv}$.
	
	If $\mpPrefix_j=\mpBra{\mpC}{\roleQ}{\stLab}{\mpx}{}$, then
	$\stFmt{\roleQ{?}\stChoice{\stLab}{\tyGround}\stSeq\stTii}
				\preType \stEnvApp{\stEnv}{\mpC}$ and\\
				$\tyJudge{\tyEnv\tyEnvComp\tyEnvMap{\mpX}{\stEnvi}
				\tyEnvComp\tyEnvMap{\mpx}{\tyGround}
				}{\mpP[j]}{
				(\stEnv\setminus\mpC)
				\stEnvComp \stEnvMap{\mpC}{\stTii}
				}$.
	By I.H., $\tyJudge{\tyEnv
				\tyEnvComp\tyEnvMap{\mpx}{\tyGround}
				}{\mpP[j]\subst{\mpX}{\mpQ}}{
				(\stEnv\setminus\mpC)
				\stEnvComp \stEnvMap{\mpC}{\stTii}
				}$.
	So, $\tyJudgePrefix{\tyEnv}{\mpPrefix_j\mpSeq\mpP[j]\subst{\mpX}{\mpQ}}
	{\stEnv}$.\\
	\inferrule{Sub}.
	$\stEnv=\stEnv[l]\stEnvComp\stEnv[r]$,
	$\stEnvii\tySub\stEnv[l]$,
	$\stEnvEndP{\stEnv[r]}$, and
	$\tyJudge{\tyEnv\tyEnvComp\tyEnvMap{\mpX}{\stEnvi}}{\mpP}{\stEnvii}$.
	By I.H.,
	$\tyJudge{\tyEnv}{\mpP\subst{\mpX}{\mpQ}}{\stEnvii}$
	so
	$\tyJudge{\tyEnv}{\mpP\subst{\mpX}{\mpQ}}{\stEnv[l]\stEnvComp\stEnv[r]}$.
\qed\end{proof}

\begin{proof}[3]
	Every rule is invariant up to permutations of free channels
	(if the channels are free in the conclusion of the rule).
	In particular,
	if $\tyJudge{\tyEnv}{\mpP}{\stEnv\stEnvComp\stEnvMap{\mpy}{\stT}}$,
	$\fpv{\mpP}=\emptyset$, and
	$\mpC\not\in\dom{\stEnv}$, then
	$\tyJudge{\tyEnv}{\mpP\subst{\mpy}{\mpC}}{\stEnv\stEnvComp\stEnvMap{\mpC}{\stT}}$.
\qed\end{proof}

\begin{lemma}[Inversion]
\label{lem:inv}
	Suppose that $\tyJudge{\tyEnv}{\mpP}{\stEnv}$.
	\begin{enumerate}
		\item If $\mpP=\mpNil$, then $\stEnvEndP{\stEnv}$.
		\item If $\mpP=\mpPi\mpPar\mpPii$, then
		$\tyJudge{\tyEnv}{\mpPi}{\stEnvi}$ and
		$\tyJudge{\tyEnv}{\mpPii}{\stEnvii}$, where
		$\stEnvi\stEnvComp\stEnvii=\stEnv$.
		\item If $\mpP=\mpRes{\mpS}{\mpPi}$, then
		there exists $\stEnvi$ with
		$\stEnv=\stEnvi\setminus\mpS$,
		$\predPApp{\stEnvi[\mpS]}$ and
		$\tyJudge{\tyEnv}{\mpPi}{\stEnvi}$.
		\item If $\mpP=\mpIf{e}{\mpPi}{\mpPii}$, then
		$\tyJudge{\tyEnv}{e}{\tyBool}$,
		$\tyJudge{\tyEnv}{\mpPi}{\stEnv}$, and
		$\tyJudge{\tyEnv}{\mpPii}{\stEnv}$.
		\item If $\mpP=\mpRec{\mpX}{\mpPi}$, then
		$\tyJudge{\tyEnv\tyEnvComp\tyEnvMap{\mpX}{\stEnvi}}{\mpPi}{\stEnvi}$,
		$\stEnvEndP{\stEnvii}$
		with $\stEnvi\stEnvComp\stEnvii\tySub\stEnv$ and
		$\dom{\stEnv\setminus\stEnd}=\dom{\stEnvi\setminus\stEnd}\subseteq\chan{\mpP,\tyEnv}$.
		\item If $\mpP=\mpSum{\mpPrefix_i\mpSeq\mpP[i]}{i\in I}$,
		then there exists $\stEnvi$ such that:
		$\stEnvi\tySub\stEnv$;
		if $\mpPrefix_i=\mpSel{\mpC}{\roleQ}{\stLab}{\mpCi}{}$, then
		$\stFmt{\roleQ{!}\stChoice{\stLab}{\stEnvApp{\stEnvi}{\mpCi}}\stSeq\stTi}\preType\stEnvApp{\stEnvi}{\mpC}$,
		$\mpC\neq\mpCi$, and
		$\tyJudge{\tyEnv}{\mpP[i]}{(\stEnvi\setminus\mpC\setminus\mpCi)\stEnvComp\stEnvMap{\mpC}{\stTi}}$;
		if $\mpPrefix_i=\mpSel{\mpC}{\roleQ}{\stLab}{e}{}$, then
		$\stFmt{\roleQ{!}\stChoice{\stLab}{\tyGround}\stSeq\stTi}
		\preType\stEnvApp{\stEnvi}{\mpC}$,
		$\tyJudge{\tyEnv}{e}{\tyGround}$, and
		$\tyJudge{\tyEnv}{\mpP[i]}{(\stEnvi\setminus\mpC)\stEnvComp\stEnvMap{\mpC}{\stTi}}$;
		if $\mpPrefix_i=\mpBra{\mpC}{\roleQ}{\stLab}{\mpx}{}$, then
		$\stFmt{\roleQ{?}\stChoice{\stLab}{\tyGround}\stSeq\stTi}\preType\stEnvApp{\stEnvi}{\mpC}$, and
		$\tyJudge{\tyEnv\tyEnvComp\tyEnvMap{\mpx}{\tyGround}}{\mpP[i]}{(\stEnvi\setminus\mpC)\stEnvComp\stEnvMap{\mpC}{\stTi}}$;
		if $\mpPrefix_i=\mpBra{\mpC}{\roleQ}{\stLab}{\mpy}{}$, then
		$\stFmt{\roleQ{?}\stChoice{\stLab}{\stT}\stSeq\stTi}\preType\stEnvApp{\stEnvi}{\mpC}$, and
		$\tyJudge{\tyEnv}{\mpP[i]}{(\stEnvi\setminus\mpC)\stEnvComp\stEnvMap{\mpC}{\stTi}
		\stEnvComp\stEnvMap{\mpy}{\stT}}$;
		if $\mpPrefix_i$ acts along channel $\mpC$
		and $\stFmt{\roleQ{\dagger}\stChoice{\stLab}{}}\preType\stEnvApp{\stEnvi}{\mpC}$,
		then $\mpFmt{\mpChanRole{\mpC}{\roleQ}{\dagger}\stLab}\preCalc\mpP$.
	\end{enumerate}
\end{lemma}

\begin{proof}[1]
	Induction on the derivation of $\tyJudge{\tyEnv}{\mpNil}{\stEnv}$:\\
	\inferrule{Nil}. $\stEnv=\stEnvEmpty$ and $\stEnvEndP{\stEnvEmpty}$.\\
	\inferrule{Sub}. $\stEnvi\stEnvComp\stEnvii\tySub\stEnv$,
	$\tyJudge{\tyEnv}{\mpNil}{\stEnvi}$ and $\stEnvEndP{\stEnvii}$.
	By I.H., $\stEnvEndP{\stEnvi}$.
	For $\stEnvMap{\mpC}{\stT}\in\stEnv$,
	there is $\stEnvMap{\mpC}{\stTi}\in\stEnvi\stEnvComp\stEnvii$
	with $\stEnd\tySub\stTi\tySub\stT$.
	So, $\stEnvEndP{\stEnv}$.
\qed\end{proof}

\begin{proof}[2]
	Induction on the derivation of $\tyJudge{\tyEnv}{\mpPi\mpPar\mpPii}{\stEnv}$:\\
	\inferrule{Par}. $\tyJudge{\tyEnv}{\mpPi}{\stEnvi}$,
	$\tyJudge{\tyEnv}{\mpPii}{\stEnvii}$, and
	$\stEnvi\stEnvComp\stEnvii=\stEnv$.\\
	\inferrule{Sub}. $\stEnvi\stEnvComp\stEnvii\tySub\stEnv$,
	$\tyJudge{\tyEnv}{\mpPi\mpPar\mpPii}{\stEnvi}$ and $\stEnvEndP{\stEnvii}$.
	By I.H.,
	$\tyJudge{\tyEnv}{\mpPi}{\stEnv[l]}$,
	$\tyJudge{\tyEnv}{\mpPii}{\stEnv[r]}$, and
	$\stEnv[l]\stEnvComp\stEnv[r]=\stEnvi$.
	By \inferrule{Sub},
	$\tyJudge{\tyEnv}{\mpPii}{(\stEnv\restriction\dom{\stEnv[r]})\stEnvComp\stEnvii}$,
	$\tyJudge{\tyEnv}{\mpPi}{(\stEnv\restriction\dom{\stEnv[l]})}$, and
	$(\stEnv\restriction\dom{\stEnv[l]})\stEnvComp
	(\stEnv\restriction\dom{\stEnv[r]})\stEnvComp\stEnvii=\stEnv$.
\qed\end{proof}

\begin{proof}[3]
	Induction on the derivation of $\tyJudge{\tyEnv}{\mpRes{\mpS}{\mpPi}}{\stEnv}$:\\
	\inferrule{Res}.
	$\tyJudge{\tyEnv}{\mpPi}{\stEnvi}$,
	$\predPApp{\stEnvi[\mpS]}$, and
	$\stEnv=\stEnvi\setminus\mpS$.\\
	\inferrule{Sub}. $\stEnvi\stEnvComp\stEnvii\tySub\stEnv$,
	$\tyJudge{\tyEnv}{\mpRes{\mpS}{\mpPi}}{\stEnvi}$
	 and $\stEnvEndP{\stEnvii}$.
	 By I.H.,
	 $\tyJudge{\tyEnv}{\mpPi}{\stEnviii}$,
	$\predPApp{\stEnviii[\mpS]}$, and
	$\stEnvi=\stEnviii\setminus\mpS$.
	By \inferrule{Sub},
	$\tyJudge{\tyEnv}{\mpPi}{\stEnviii[\mpS]\stEnvComp\stEnv}$.
	$(\stEnviii[\mpS]\stEnvComp\stEnv)_{\mpS}=\stEnviii[\mpS]$
	and
	$(\stEnviii[\mpS]\stEnvComp\stEnv)\setminus\mpS=\stEnv$ as,
	by the Barendregt convention,
	the sessions in $\stEnviii$ are distinct from $\mpS$.
\qed\end{proof}

\begin{proof}[4]
	Induction on the derivation of $\tyJudge{\tyEnv}{\mpIf{e}{\mpPi}{\mpPii}}{\stEnv}$:\\
	\inferrule{If}. $\tyJudge{\tyEnv}{e}{\tyBool}$,
	$\tyJudge{\tyEnv}{\mpPi}{\stEnv}$, and
	$\tyJudge{\tyEnv}{\mpPii}{\stEnv}$.\\
	\inferrule{Sub}. $\stEnvi\stEnvComp\stEnvii\tySub\stEnv$,
	$\tyJudge{\tyEnv}{\mpIf{e}{\mpPi}{\mpPii}}{\stEnvi}$
	 and $\stEnvEndP{\stEnvii}$.
	 By I.H., $\tyJudge{\tyEnv}{e}{\tyBool}$,
	$\tyJudge{\tyEnv}{\mpPi}{\stEnvi}$, and
	$\tyJudge{\tyEnv}{\mpPii}{\stEnvi}$.
	By \inferrule{Sub},
	$\tyJudge{\tyEnv}{\mpPi}{\stEnv}$ and
	$\tyJudge{\tyEnv}{\mpPii}{\stEnv}$.
\qed\end{proof}

\begin{proof}[5]
	Induction on the derivation of $\tyJudge{\tyEnv}{\mpRec{\mpX}{\mpPi}}{\stEnv}$:\\
	\inferrule{Rec}. $\tyJudge{\tyEnv\tyEnvComp\tyEnvMap{\mpX}{\stEnv}}{\mpPi}{\stEnv}$ and $\chan{\mpP,\tyEnv}\supseteq\dom{\stEnv\setminus\stEnd}$.\\
	\inferrule{Sub}. $\stEnvi\stEnvComp\stEnvii\tySub\stEnv$,
	$\tyJudge{\tyEnv}{\mpRec{\mpX}{\mpPi}}{\stEnvi}$
	 and $\stEnvEndP{\stEnvii}$.
	 By I.H.,
	 $\tyJudge{\tyEnv\tyEnvComp\tyEnvMap{\mpX}{\stEnviii[0]}}{\mpPi}{\stEnviii[0]}$,
	 $\stEnvEndP{\stEnviii[1]}$,
	 $\stEnviii[0]\stEnvComp\stEnviii[1]\tySub\stEnvi$,
	 and $\chan{\mpP,\tyEnv}\supseteq\dom{\stEnvi\setminus\stEnd}$.
	 So, $\stEnvEndP{\stEnviii[1]\stEnvComp\stEnvii}$,
	 $\stEnviii[0]\stEnvComp\stEnviii[1]\stEnvComp\stEnvii\tySub\stEnv$, and
	 $\chan{\mpP,\tyEnv}\supseteq\dom{\stEnv\setminus\stEnd}$.
\qed\end{proof}

\begin{proof}[6]
	Induction on the derivation of $\tyJudge{\tyEnv}{\mpSum{\mpPrefix_i\mpSeq\mpP[i]}{i\in I}}{\stEnv}$:\\
	\inferrule{Sum}. With $\stEnv=\stEnvi$, these are precisely
	the hypotheses of $\inferrule{Sum}$,
	\inferrule{$!$Chan}, \inferrule{$?$Chan},
	\inferrule{$!$Val}, and \inferrule{$?$Val}.\\
	\inferrule{Sub}.
	$\stEnviii\stEnvComp\stEnvii\tySub\stEnv$,
	$\tyJudge{\tyEnv}{\mpSum{\mpPrefix_i\mpSeq\mpP[i]}{i\in I}}{\stEnviii}$
	 and $\stEnvEndP{\stEnvii}$.
	By I.H., there exists $\stEnvi$ such that:\\
		$\stEnvi\tySub\stEnviii$;
		if $\mpPrefix_i=\mpSel{\mpC}{\roleQ}{\stLab}{\mpCi}{}$, then
		$\stFmt{\roleQ{!}\stChoice{\stLab}{\stEnvApp{\stEnvi}{\mpCi}}\stSeq\stTi}\preType\stEnvApp{\stEnvi}{\mpC}$,
		$\mpC\neq\mpCi$, and
		$\tyJudge{\tyEnv}{\mpP[i]}{(\stEnvi\setminus\mpC\setminus\mpCi)\stEnvComp\stEnvMap{\mpC}{\stTi}}$;
		if $\mpPrefix_i=\mpSel{\mpC}{\roleQ}{\stLab}{e}{}$, then
		$\stFmt{\roleQ{!}\stChoice{\stLab}{\tyGround}\stSeq\stTi}
		\preType\stEnvApp{\stEnvi}{\mpC}$,
		$\tyJudge{\tyEnv}{e}{\tyGround}$, and
		$\tyJudge{\tyEnv}{\mpP[i]}{(\stEnvi\setminus\mpC)\stEnvComp\stEnvMap{\mpC}{\stTi}}$;
		if $\mpPrefix_i=\mpBra{\mpC}{\roleQ}{\stLab}{\mpx}{}$, then
		$\stFmt{\roleQ{?}\stChoice{\stLab}{\tyGround}\stSeq\stTi}\preType\stEnvApp{\stEnvi}{\mpC}$, and
		$\tyJudge{\tyEnv\tyEnvComp\tyEnvMap{\mpx}{\tyGround}}{\mpP[i]}{(\stEnvi\setminus\mpC)\stEnvComp\stEnvMap{\mpC}{\stTi}}$;
		if $\mpPrefix_i=\mpBra{\mpC}{\roleQ}{\stLab}{\mpy}{}$, then
		$\stFmt{\roleQ{?}\stChoice{\stLab}{\stT}\stSeq\stTi}\preType\stEnvApp{\stEnvi}{\mpC}$, and
		$\tyJudge{\tyEnv}{\mpP[i]}{(\stEnvi\setminus\mpC)\stEnvComp\stEnvMap{\mpC}{\stTi}
		\stEnvComp\stEnvMap{\mpy}{\stT}}$;
		if $\mpPrefix_i$ acts along channel $\mpC$
		and $\stFmt{\roleQ{\dagger}\stChoice{\stLab}{}}\preType\stEnvApp{\stEnvi}{\mpC}$,
		then $\mpFmt{\mpChanRole{\mpC}{\roleQ}{\dagger}\stLab}\preCalc\mpP$.\\
		$\chan{\mpP}\subseteq\dom{\stEnvi}$
		are disjoint from $\stEnvii$, so
		the above is true with $\stEnvi\stEnvComp\stEnvii$
		for $\stEnvi$, and
		$\stEnvi\stEnvComp\stEnvii\tySub\stEnv$.
\qed\end{proof}
\subjectReduction*

\begin{proof}[1]
	Induction on the derivation of $\mpP\prestruct\mpPi$
	and use \cref{lem:inv}:\\
	$\mpP\prestruct\mpP$. $\tyJudge{\tyEnv}{\mpP}{\stEnv}$.\\
	$\mpP\mpPar\mpNil\prestruct\mpP$.
	$\tyJudge{\tyEnv}{\mpP}{\stEnvi}$ and
	$\tyJudge{\tyEnv}{\mpNil}{\stEnvii}$
	with $\stEnvi\stEnvComp\stEnvii=\stEnv$.
	$\stEnvEndP{\stEnvii}$, so by \inferrule{Sub},
	$\tyJudge{\tyEnv}{\mpP}{\stEnv}$.\\
	$\mpP\prestruct\mpP\mpPar\mpNil$.
	$\tyJudge{\tyEnv}{\mpNil}{\stEnvEmpty}$ so
	$\tyJudge{\tyEnv}{\mpP\mpPar\mpNil}{\stEnv}$.\\
	$\mpP\mpPar(\mpQ\mpPar\mpR)\prestruct(\mpP\mpPar\mpQ)\mpPar\mpR$.
	$\tyJudge{\tyEnv}{\mpP}{\stEnvi}$ and
	$\tyJudge{\tyEnv}{\mpQ\mpPar\mpR}{\stEnvii}$
	with $\stEnvi\stEnvComp\stEnvii=\stEnv$.
	$\tyJudge{\tyEnv}{\mpQ}{\stEnv[l]}$ and
	$\tyJudge{\tyEnv}{\mpR}{\stEnv[r]}$
	with $\stEnv[l]\stEnvComp\stEnv[r]=\stEnvii$.
	So
	$\tyJudge{\tyEnv}{\mpP\mpPar\mpQ}{\stEnvi\stEnvComp\stEnv[l]}$,
	so
	$\tyJudge{\tyEnv}{(\mpP\mpPar\mpQ)\mpPar\mpR}{\stEnvi\stEnvComp\stEnv[l]\stEnvComp\stEnv[r]}$.\\
	$(\mpP\mpPar\mpQ)\mpPar\mpR\prestruct\mpP\mpPar(\mpQ\mpPar\mpR)$.
	$\tyJudge{\tyEnv}{\mpP\mpPar\mpQ}{\stEnvi}$ and
	$\tyJudge{\tyEnv}{\mpR}{\stEnvii}$
	with $\stEnvi\stEnvComp\stEnvii=\stEnv$.
	$\tyJudge{\tyEnv}{\mpP}{\stEnv[l]}$ and
	$\tyJudge{\tyEnv}{\mpQ}{\stEnv[r]}$
	with $\stEnv[l]\stEnvComp\stEnv[r]=\stEnvi$.
	So
	$\tyJudge{\tyEnv}{\mpQ\mpPar\mpR}{\stEnv[r]\stEnvComp\stEnvii}$,
	so
	$\tyJudge{\tyEnv}{\mpP\mpPar(\mpQ\mpPar\mpR)}{\stEnv[l]\stEnvComp\stEnv[r]\stEnvComp\stEnvii}$.\\
	$\mpP\mpPar\mpQ\prestruct\mpQ\mpPar\mpP$.
	$\tyJudge{\tyEnv}{\mpP}{\stEnv[l]}$ and
	$\tyJudge{\tyEnv}{\mpQ}{\stEnv[r]}$
	with $\stEnv[l]\stEnvComp\stEnv[r]=\stEnv=\stEnv[r]\stEnvComp\stEnv[l]$.
	So
	$\tyJudge{\tyEnv}{\mpQ\mpPar\mpP}{\stEnv}$.\\
	$\mpRes{\mpS}{\mpNil}\prestruct\mpNil$.
	$\tyJudge{\tyEnv}{\mpNil}{\stEnvi}$
	with $\stEnvi\setminus\mpS=\stEnv$.
	So, $\stEnvEndP{\stEnvi}$.
	So, $\stEnvEndP{\stEnv}$.
	$\tyJudge{\tyEnv}{\mpNil}{\stEnvEmpty}$,
	so $\tyJudge{\tyEnv}{\mpNil}{\stEnv}$
	by \inferrule{Sub}.\\
	$\mpNil\prestruct\mpRes{\mpS}{\mpNil}$.
	$\predPApp{\stEnvEmpty}$ and $\stEnv[\mpS]=\stEnvEmpty$
	by the Barendregt convention.
	So,
	$\tyJudge{\tyEnv}{\mpRes{\mpS}{\mpNil}}{\stEnv}$.\\
	$\mpS\not\in\fs{\mpP}$ and
	$\mpRes{\mpS}{(\mpP\mpPar\mpQ)}\prestruct\mpP\mpPar\mpRes{\mpS}{\mpQ}$.
	$\tyJudge{\tyEnv}{\mpP\mpPar\mpQ}{\stEnvi}$
	with $\stEnv=\stEnvi\setminus\mpS$ and $\predPApp{\stEnvi[\mpS]}$.
	$\tyJudge{\tyEnv}{\mpP}{\stEnv[l]}$,
	$\tyJudge{\tyEnv}{\mpQ}{\stEnv[r]}$, and
	$\stEnv[l]\stEnvComp\stEnv[r]=\stEnvi$.
	$\dom{\stEnv[l]\setminus\stEnd}\subseteq\chan{\mpP,\tyEnv}$,
	which does not mention $\mpS$ by the Barendregt convention in $\tyEnv$,
	so $\stEnvEndP{\stEnv[l,\mpS]}$
	and
	$\tyJudge{\tyEnv}{\mpP}{\stEnv[l]\setminus\mpS}$.
	By \inferrule{Sub}, $\tyJudge{\tyEnv}{\mpQ}{\stEnv[l,\mpS]\stEnvComp\stEnv[r]}$.
	$\stEnvi[\mpS]=(\stEnv[l,\mpS]\stEnvComp\stEnv[r])_{\mpS}$,
	so by \inferrule{Res},
	$\tyJudge{\tyEnv}{\mpRes{\mpS}{\mpQ}}{\stEnv[r]\setminus\mpS}$.
	So,
	$\tyJudge{\tyEnv}{\mpP\mpPar\mpRes{\mpS}{\mpQ}}{\stEnv}$.\\
	$\mpS\not\in\fs{\mpP}$ and
	$\mpP\mpPar\mpRes{\mpS}{\mpQ}\prestruct\mpRes{\mpS}{(\mpP\mpPar\mpQ)}$.
	$\tyJudge{\tyEnv}{\mpP}{\stEnv[l]}$,
	$\tyJudge{\tyEnv}{\mpRes{\mpS}{\mpQ}}{\stEnv[r]}$, and
	$\stEnv=\stEnv[l]\stEnvComp\stEnv[r]$.
	$\tyJudge{\tyEnv}{\mpQ}{\stEnvi}$,
	$\predPApp{\stEnvi[\mpS]}$, and
	$\stEnv[r]=\stEnvi\setminus\mpS$.
	So,
	$\tyJudge{\tyEnv}{\mpP\mpPar\mpQ}{\stEnv[l]\stEnvComp\stEnvi}$,
	so
	$\tyJudge{\tyEnv}{\mpRes{\mpS}{(\mpP\mpPar\mpQ)}}{\stEnv}$.\\
	$\mpRes{\mpS}{\mpRes{\mpSi}{\mpP}}\prestruct\mpRes{\mpS}{\mpRes{\mpSi}{\mpP}}$.
	$\tyJudge{\tyEnv}{\mpRes{\mpSi}{\mpP}}{\stEnvi}$,
	$\stEnv=\stEnvi\setminus\mpS$,
	$\predPApp{\stEnvi[\mpS]}$.
	$\tyJudge{\tyEnv}{\mpP}{\stEnvii}$,
	$\stEnvi=\stEnvii\setminus\mpSi$,
	$\predPApp{\stEnvii[\mpSi]}$.
	$\stEnvii[\mpS]=\stEnvi[\mpS]$,
	$\stEnvii[\mpSi]=(\stEnvi\setminus\mpS)_{\mpSi}$, and
	$\stEnvii\setminus\mpS\setminus\mpSi=\stEnv$, so
	$\tyJudge{\tyEnv}{\mpRes{\mpS}{\mpP}}{\stEnvii\setminus\mpS}$ and
	$\tyJudge{\tyEnv}{\mpRes{\mpSi}{\mpRes{\mpS}{\mpP}}}{\stEnv}$.\\
	$\mpP\prestruct\mpPi\prestruct\mpPii$.
	$\tyJudge{\tyEnv}{\mpP}{\stEnv}$, so by I.H.,
	$\tyJudge{\tyEnv}{\mpPi}{\stEnv}$, so by I.H.,
	$\tyJudge{\tyEnv}{\mpPii}{\stEnv}$.\\
	$\mpP\prestruct\mpQ$ and $\mpRes{\mpS[1],\dots,\mpS[n]}{(\mpP\mpPar\mpR)}\prestruct\mpRes{\mpS[1],\dots,\mpS[n]}{(\mpP\mpPar\mpR)}$.\\
	$\tyJudge{\tyEnv}{\mpRes{\mpS[1],\dots,\mpS[n]}{(\mpP\mpPar\mpR)}}{\stEnv}$,
	so
	$\tyJudge{\tyEnv}{\mpP}{\stEnvi}$,
	$\tyJudge{\tyEnv}{\mpR}{\stEnvii}$,
	$\stEnv=\stEnvi\stEnvComp\stEnvii\setminus\mpS[1]\setminus\dots\setminus\mpS[n]$, and
	$\forall\mpSi\in\{\mpS[i]\}_{1\leq i\leq n}:\predPApp{(\stEnvi\stEnvComp\stEnvii)_{\mpSi}}$.
	By I.H.,
	$\tyJudge{\tyEnv}{\mpQ}{\stEnvi}$,
	so by \inferrule{Par} and multiple applications of $\inferrule{Res}$,
	$\tyJudge{\tyEnv}{\mpRes{\mpS[1],\dots,\mpS[n]}{(\mpQ\mpPar\mpR)}}{\stEnv}$.\\
	$\mpRec{\mpX}{\mpP}\prestruct\mpP\subst{\mpX}{\mpRec{\mpX}{\mpP}}$.
	$\tyJudge{\tyEnv}{\mpRec{\mpX}{\mpP}}{\stEnvi}$,
	$\tyJudge{\tyEnv\tyEnvComp\tyEnvMap{\mpX}{\stEnvi}}{\mpP}{\stEnvi}$,
	$\stEnvEndP{\stEnvii}$, and
	$\stEnvi\stEnvComp\stEnvii\tySub\stEnv$.
	By substitution lemma,
	$\tyJudge{\tyEnv}{\mpP\subst{\mpX}{\mpRec{\mpX}{\mpP}}}{\stEnvi}$.
	By \inferrule{Sub},
	$\tyJudge{\tyEnv}{\mpP\subst{\mpX}{\mpRec{\mpX}{\mpP}}}{\stEnv}$.
\qed\end{proof}

\begin{lemma}[No Three-Participant Locks]
	\label{lem:pi-obs}
	If $\tyJudge{\tyEnv}{\mpRes{\mpS}{\mpQ}}{\stEnv}$
	and $\predP\subseteq\stEnvLivePred$,
	then $\mpRes{\mpS}{\mpQ}$ is
	not of the form in the 3-plock.
\end{lemma}

\begin{proof}
	By inversion,
	$\tyJudge{\tyEnv}{\mpQ}{\stEnvi}$
	where $\stEnvi[\mpS]$ is live and hence
	3-plock free.
	
	By induction on $\mpQ$ and inversion,
	if $\obsvNew{\mpQ}{\mpS}{\roleP[1]}{\roleP[2]}{\dagger}$
	and $\mpChanRole{\mpS}{\roleP[1]}\in\dom{\stEnvi}$, then either:
	\[
	\stSum{\roleP[2]}{i\in I}{\stChoice{\stLab[i]}{\stS[i]}\stSeq\stT[i]}{\dagger}\tySub\stEnvApp{\stEnvi}{\mpChanRole{\mpS}{\roleP[1]}}
	\;\text{or}\;
	\unfoldOne{\stEnvApp{\stEnvi}{\mpChanRole{\mpS}{\roleP[1]}}}=\stEnd
	\]
	
	Suppose that
	$\mpQ=
	\mpSum{\mpPrefix_i\mpSeq\mpQ[i]}{i\in I}$
	with $\mpChanRole{\mpChanRole{\mpS}{\roleP}}{\roleQ[i]}{\dagger_i}\preCalc{\mpPrefix_i}$ for $i\in I$.
	
	By inversion,
	\[
	\stSum{\roleQ[i]}{i\in I}{\stChoice{\stLab[i]}{\stS[i]}\stSeq\stT[i]}{\dagger_i}\tySub
	\unfoldOne{\stEnvApp{\stEnv}{\mpChanRole{\mpS}{\roleP}}}
	\]
	
	If $\mpRes{\mpS}{\mpQ}$ has a 3-plock,
	then for $i\in I$ there exists $\roleR[i]$ and $\dagger'_i$ such that:
	\begin{enumerate}
	\item $\obsvNew{\mpQ}{\mpS}{\roleQ[i]}{\roleR[i]}{\dagger'_i}$;
	\item
	if $\roleR[i]=\roleP$, then $\dagger'_i=\dagger_i$;
	\item if $\roleR[i]\neq\roleP$, then there is $\dagger''_i$ s.t.
	$\obsvNew{\mpQ}{\mpS}{\roleR[i]}{\roleP}{\dagger''_i}$.
	\end{enumerate}
	So for $i \in I$,
	\[
	\stSum{\roleR[i]}{j\in I_i}{\stChoice{\stLab[i,j]}{\stS[i,j]}\stSeq\stT[i,j]}{\dagger'_i}\tySub\stEnvApp{\stEnvi}{\mpChanRole{\mpS}{\roleQ[i]}}
	\,\text{or}\,
	\unfoldOne{\stEnvApp{\stEnvi}{\mpChanRole{\mpS}{\roleQ[i]}}}=\stEnd
	\,\text{or}\,
	\mpChanRole{\mpS}{\roleQ[i]}\not\in\dom{\stEnvi}
	\]
	If $\roleR[i]=\roleP$, then
	\[
	\stSum{\roleP}{j\in I'_i}{\stChoice{\stLabi[i,j]}{\stSi[i,j]}\stSeq\stTi[i,j]}{\dagger''_i}\tySub\stEnvApp{\stEnvi}{\mpChanRole{\mpS}{\roleR[i]}}
	\,\text{or}\,
	\unfoldOne{\stEnvApp{\stEnvi}{\mpChanRole{\mpS}{\roleR[i]}}}=\stEnd
	\,\text{or}\,
	\mpChanRole{\mpS}{\roleR[i]}\not\in\dom{\stEnvi}
	\]
	By inversion of subtyping,
	for $i \in I$
	$\stEnvApp{\stEnvi}{\mpChanRole{\mpS}{\roleQ[i]}}$
	only has the action $\roleR\stFmt{\dagger'_i}$
	and
	$\stEnvApp{\stEnvi}{\mpChanRole{\mpS}{\roleR[i]}}$
	only has the action $\roleP\stFmt{\dagger''_i}$
	and
	$\stEnvApp{\stEnv}{\mpChanRole{\mpS}{\roleP}}$
	only has the actions
	$\{\roleQ[i]\stFmt{\dagger}\}_{i\in I}$
	Therefore, $\stEnvi$ has a 3-plock,
	contradicting liveness.
\qed\end{proof}

\begin{theorem}[Subject Reduction]
\label{thm:pi-sr}
	Let $\predP$ in \inferrule{Res} \cref{fig:typing-rules}
be \RC-safe.
	Suppose that $\tyJudge{\tyEnv}{\mpP}{\stEnv}$
	and that $\stEnv$ is safe.
	\begin{enumerate}
		\item $\mpP\,\,\not\!\!\mpMoveErr$.
		\item If $\mpP\mpMoveTau\mpPi$,
		then
		$\tyJudge{\tyEnv}{\mpPi}{\stEnv}$.
		\item If $\mpP\mpMoveCommS{\mpS}{\roleP}{\roleQ}{\stLab}{}\mpPi$,
		then
		there is $\stS$ such that
		$\stEnv\gtMove[\ltsSendRecvS{\mpS}{\roleP}{\roleQ}{\stChoice{\stLab}{\stS}}]\,\stEnvi$, and
		$\tyJudge{\tyEnv}{\mpPi}{\stEnvi}$.
	\end{enumerate}
\end{theorem}

\begin{proof}
	Induction on the derivation of $\mpP\mpMoveGen\mpPi$:\\
	\inferrule{R-Cond}. $\mpP=\mpIf{e}{\mpP[\mpTrue]}{\mpP[\mpFalse]}$,
	$\mpPi=\mpP[v]$, and $v\in\{\mpTrue,\mpFalse\}$.
	
	By inversion, $\tyJudge{\tyEnv}{\mpPi}{\stEnv}$.\\
	\inferrule{Ctx-I},\inferrule{Ctx-II},\inferrule{Ctx-III}.
	$\mpP=\mpRes{\mpSi[1],\dots,\mpSi[n]}{(\mpQ\mpPar\mpR)}$,
	$\mpPi=\mpRes{\mpSi[1],\dots,\mpSi[n]}{(\mpQi\mpPar\mpR)}$, and
	$\mpQ\mpMoveGen\mpQi$.
	
	By inversion,
	\[\begin{array}{c}\tyJudge{\tyEnv}{\mpQ}{\stEnvi}\\
	\tyJudge{\tyEnv}{\mpR}{\stEnvii}\\
	\stEnv=(\stEnvi\stEnvComp\stEnvii)\setminus\mpSi[1]\setminus\dots\setminus\mpSi[n]\\
	\forall\mpSii\in\{\mpSi[i]\}_{1\leq i\leq n}.\quad\predPApp{(\stEnvi\stEnvComp\stEnvii)_{\mpSii}}\end{array}\]
	So, $\stEnvi$ is safe.
	
	If $\stEnvAnnotGenericSym=\tau$, then
	by I.H.,
	\[\tyJudge{\tyEnv}{\mpQi}{\stEnvi}\]
	So by
	\inferrule{Res} and \inferrule{Par},
	\[\tyJudge{\tyEnv}{\mpRes{\mpSi[1],\dots,\mpSi[n]}{(\mpQi\mpPar\mpR)}}{\stEnv}\]
	If $\stEnvAnnotGenericSym=\ltsSendRecvS{\mpS}{\roleP}{\roleQ}{\stLab}$,
	then, by I.H., there exists $\stS$ such that
	\[\stEnvi\gtMove[\ltsSendRecvS{\mpS}{\roleP}{\roleQ}{\stChoice{\stLab}{\stS}}]\stEnviii
	\,\text{and}\,\tyJudge{\tyEnv}{\mpQi}{\stEnviii}\]
	So, \[\tyJudge{\tyEnv}{\mpQ\mpPar\mpR}{\stEnviii\stEnvComp\stEnvii}\]
	For $\mpSii\in\{\mpSi[i]\}_{1\leq i\leq n}$,
	\[\stEnvi[\mpSii]\gtMoveStar(\stEnviii\stEnvComp\stEnvii)_{\mpSii}\]
	So $\predPApp{(\stEnviii\stEnvComp\stEnvii)_{\mpSii}}$.
	
	By \inferrule{Par} and \inferrule{Res},
	\[\tyJudge{\tyEnv}{\mpRes{\mpSi[1],\dots,\mpSi[n]}{(\mpQi\mpPar\mpR)}}{(\stEnviii\stEnvComp\stEnvii)\setminus\mpSi[1]\setminus\dots\setminus\mpSi[n]}\]
	If $\mpS\in\{\mpSi[i]\}_{1\leq i\leq n}$, then
	\[\stEnv=(\stEnviii\stEnvComp\stEnvii)\setminus\mpSi[1]\setminus\dots\setminus\mpSi[n]\]
	If $\mpS\not\in\{\mpSi[i]\}_{1\leq i\leq n}$, then
	\[\stEnv\gtMove[\ltsSendRecvS{\mpS}{\roleP}{\roleQ}{\stChoice{\stLab}{\stS}}](\stEnviii\stEnvComp\stEnvii)\setminus\mpSi[1]\setminus\dots\setminus\mpSi[n]\]
	\inferrule{Cong}. $\mpP\prestruct\mpPi\mpMoveGen\mpQi\prestruct\mpQ$.
	
	By subject precongruence,
	\[\tyJudge{\tyEnv}{\mpPi}{\stEnv}\]
	By I.H., if $\mpPi\mpMoveTau\mpQi$ then
	$\tyJudge{\tyEnv}{\mpQi}{\stEnv}$, so
	$\tyJudge{\tyEnv}{\mpQ}{\stEnv}$.
	
	By I.H., if
	\[\mpPi\mpMoveCommS{\mpS}{\roleP}{\roleQ}{\stLab}{}\mpQi\]
	Then
	there is $\stS$ such that
	\[\stEnv\gtMove[\ltsSendRecvS{\mpS}{\roleP}{\roleQ}{\stChoice{\stLab}{\stS}}]\stEnvi\,\text{and}\,
	\tyJudge{\tyEnv}{\mpQi}{\stEnvi}\]
	So
	\[\tyJudge{\tyEnv}{\mpQ}{\stEnvi}\]
	\inferrule{R-Val}. $\mpP=\mpPi\mpPar\mpPii$,
	$\mpBra{\mpChanRole{\mpS}{\roleQ}}{\roleP}{\stLab}{\mpx}{\mpQi}\preCalc\mpPi$,
	$\mpSel{\mpChanRole{\mpS}{\roleP}}{\roleQ}{\stLab}{e}{\mpQii}\preCalc\mpPii$,
	$\eval{e}{v}$, and
	$\mpQ=\mpQi\subst{\mpx}{v}\mpPar\mpQii$.
	
	By inversion,
	\[
	\tyJudge{\tyEnv}{\mpPi}{\stEnv[l]},\,
	\tyJudge{\tyEnv}{\mpPii}{\stEnv[r]},\,\text{and}\,
	\stEnv=\stEnv[l]\stEnvComp\stEnv[r]
	\]
	By inversion,
	there is
	\[\stEnvi[l]\stEnvComp
	\stEnvMap{\mpChanRole{\mpS}{\roleQ}}{\stT[\roleQ]}
	\tySub\stEnv[l]\]
	Such that
	\[\roleP\stFmt{?}\stChoice{\stLab}{\tyGround}\stSeq\stTi[\roleQ]\preType\stT[\roleQ]\]	
	\[\tyJudge{(\tyEnv\tyEnvComp\tyEnvMap{\mpx}{\tyGround})}{\mpQi}{(
	\stEnvi[l]\stEnvComp
	\stEnvMap{\mpChanRole{\mpS}{\roleQ}}{\stTi[\roleQ]}
	)}\]
	By inversion,
	there is
	\[\stEnvi[r]\stEnvComp
	\stEnvMap{\mpChanRole{\mpS}{\roleP}}{\stT[\roleP]}
	\tySub\stEnv[r]\]
	Such that
	\[\roleQ\stFmt{!}\stChoice{\stLab}{\tyGroundi}\stSeq\stTi[\roleP]\preType\stT[\roleP]\]
	\[\tyJudge{\tyEnv}{\mpQii}{(
	\stEnvi[r]\stEnvComp
	\stEnvMap{\mpChanRole{\mpS}{\roleP}}{\stTi[\roleP]})
	}\]
	\[\tyJudge{\tyEnv}{e}{\tyGroundi}\]
	By inversion of subtyping,
	\[\roleQ\stFmt{!}\stChoice{\stLab}{\tyGroundi}\stSeq\stTii[\roleP]\in
	\stEnvApp{\stEnv}{\mpChanRole{\mpS}{\roleP}}\]
	\[\stTi[\roleP]\tySub\stTii[\roleP]\]
	\[\roleP\stFmt{?}\preType\stEnvApp{\stEnv}{\mpChanRole{\mpS}{\roleQ}}\]
	$\stEnv$ is safe, so 
	$\roleP\stFmt{?}\stChoice{\stLab}{\tyGroundi}\stSeq\stTii[\roleQ]
	\preType\stEnvApp{\stEnv}{\mpChanRole{\mpS}{\roleQ}}$.
	
	So by inversion of subtyping and uniqueness of labels,
	\[\tyGroundi=\tyGround\quad\text{and}\quad\stTi[\roleQ]\tySub\stTii[\roleQ]\]
	We have
	\[\stEnv\gtMove[\ltsSendRecvS{\mpS}{\roleP}{\roleQ}{\stLab}{\tyGround}]\stEnvii\]
	\[
	\stEnvi[l]\stEnvComp\stEnvi[r]\stEnvComp
	\stEnvMap{\mpChanRole{\mpS}{\roleQ}}{\stT[\roleQ]}\stEnvComp
	\stEnvMap{\mpChanRole{\mpS}{\roleP}}{\stT[\roleP]}
	\gtMove[\ltsSendRecvS{\mpS}{\roleP}{\roleQ}{\stLab}{\tyGround}]
	\stEnvi[l]\stEnvComp\stEnvi[r]\stEnvComp
	\stEnvMap{\mpChanRole{\mpS}{\roleQ}}{\stTi[\roleQ]}\stEnvComp
	\stEnvMap{\mpChanRole{\mpS}{\roleP}}{\stTi[\roleP]}
	\]
	So
	\[
	\stEnvi[l]\stEnvComp\stEnvi[r]\stEnvComp
	\stEnvMap{\mpChanRole{\mpS}{\roleQ}}{\stTi[\roleQ]}\stEnvComp
	\stEnvMap{\mpChanRole{\mpS}{\roleP}}{\stTi[\roleP]}
	\tySub
	\stEnvii
	\]
	By \cref{lem:eval-type},
	$\tyJudge{\tyEnv}{v}{\tyGround}$.
	
	By substitution,
	\[\tyJudge{\tyEnv}{\mpQi\subst{\mpx}{v}}{
	\stEnvi[l]\stEnvComp
	\stEnvMap{\mpChanRole{\mpS}{\roleQ}}{\stTi[\roleQ]}
	}\]
	So,
	\[\tyJudge{\tyEnv}{\mpQi\subst{\mpx}{v}\mpPar\mpQii}{(
	\stEnvi[l]\stEnvComp\stEnvi[r]\stEnvComp
	\stEnvMap{\mpChanRole{\mpS}{\roleQ}}{\stTi[\roleQ]}\stEnvComp
	\stEnvMap{\mpChanRole{\mpS}{\roleP}}{\stTi[\roleP]})
	}\]
	So,
	\[\tyJudge{\tyEnv}{\mpQi\subst{\mpx}{v}\mpPar\mpQii}{
	\stEnvii
	}\]
	\inferrule{R-Chan}. $\mpP=\mpPi\mpPar\mpPii$,
	$\mpBra{\mpChanRole{\mpS}{\roleQ}}{\roleP}{\stLab}{\mpy}{\mpQi}\preCalc\mpPi$,
	$\mpSel{\mpChanRole{\mpS}{\roleP}}{\roleQ}{\stLab}{\mpC}{\mpQii}\preCalc\mpPii$, and
	$\mpQ=\mpQi\subst{\mpy}{\mpC}\mpPar\mpQii$.
	
	By inversion,
	\[\tyJudge{\tyEnv}{\mpPi}{\stEnv[l]},\,
	\tyJudge{\tyEnv}{\mpPii}{\stEnv[r]},\, \text{and}\,
	\stEnv=\stEnv[l]\stEnvComp\stEnv[r]\]
	By inversion,
	there is
	\[\stEnvi[l]\stEnvComp
	\stEnvMap{\mpChanRole{\mpS}{\roleQ}}{\stT[\roleQ]}
	\tySub\stEnv[l]\]
	Such that
	\[\roleP\stFmt{?}\stChoice{\stLab}{\stT}\stSeq\stTi[\roleQ]\preType\stT[\roleQ]\]
	\[\tyJudge{\tyEnv}{\mpQi}{(
	\stEnvi[l]\stEnvComp
	\stEnvMap{\mpChanRole{\mpS}{\roleQ}}{\stTi[\roleQ]}
	\stEnvComp{\mpy}{\stT})
	}\]
	By inversion,
	there is
	\[\stEnvi[r]\stEnvComp
	\stEnvMap{\mpChanRole{\mpS}{\roleP}}{\stT[\roleP]}
	\stEnvComp
	\stEnvMap{\mpC}{\stTi}
	\tySub\stEnv[r]\]
	Such that
	\[\roleQ\stFmt{!}\stChoice{\stLab}{\stTi}\stSeq\stTi[\roleP]\preType\stT[\roleP]\quad\text{and}\quad
	\tyJudge{\tyEnv}{\mpQii}{(
	\stEnvi[r]\stEnvComp
	\stEnvMap{\mpChanRole{\mpS}{\roleP}}{\stTi[\roleP]})}\]
	By inversion of subtyping,
	\[
	\roleQ\stFmt{!}\stChoice{\stLab}{\stTi[1]}\stSeq\stTii[\roleP]\in
	\stEnvApp{\stEnv}{\mpChanRole{\mpS}{\roleP}}
	,\,
	\stTi[\roleP]\tySub\stTii[\roleP],\,\stTi[1]\tySub\stTi,\, \text{and}\,
	\roleP\stFmt{?}\preType\stEnvApp{\stEnv}{\mpChanRole{\mpS}{\roleQ}}\]
	$\stEnv$ is safe, so 
	\[
	\roleP\stFmt{?}\stChoice{\stLab}{\stT[1]}\stSeq\stTii[\roleQ]
	\preType\stEnvApp{\stEnv}{\mpChanRole{\mpS}{\roleQ}}
	\,
	\text{with}\,\stT[1]\tySub\stTi[1]\]
	So by inversion of subtyping and uniqueness of labels,
	\[
	\stT\tySub\stT[1]\tySub\stTi[1]\tySub\stTi\quad\text{and}\quad \stTi[\roleQ]\tySub\stTii[\roleQ]
	\]
	We have $\stEnv\gtMove[\ltsSendRecvS{\mpS}{\roleP}{\roleQ}{\stChoice{\stLab}{\stTi[1]}}]\stEnvii$ and
	\[\begin{array}{cl}
	&\stEnvi[l]\stEnvComp\stEnvi[r]\stEnvComp\stEnvMap{\mpC}{\stT}\stEnvComp
	\stEnvMap{\mpChanRole{\mpS}{\roleQ}}{\stT[\roleQ]}\stEnvComp
	\stEnvMap{\mpChanRole{\mpS}{\roleP}}{\stT[\roleP]}\\
	\gtMove[\ltsSendRecvS{\mpS}{\roleP}{\roleQ}{\stChoice{\stLab}{\stTi}}]&
	\stEnvi[l]\stEnvComp\stEnvi[r]\stEnvComp\stEnvMap{\mpC}{\stT}\stEnvComp
	\stEnvMap{\mpChanRole{\mpS}{\roleQ}}{\stTi[\roleQ]}\stEnvComp
	\stEnvMap{\mpChanRole{\mpS}{\roleP}}{\stTi[\roleP]}
	\end{array}
	\]
	So
	\[
	\stEnvi[l]\stEnvComp\stEnvi[r]\stEnvComp\stEnvMap{\mpC}{\stT}\stEnvComp
	\stEnvMap{\mpChanRole{\mpS}{\roleQ}}{\stTi[\roleQ]}\stEnvComp
	\stEnvMap{\mpChanRole{\mpS}{\roleP}}{\stTi[\roleP]}
	\tySub
	\stEnvii
	\]
	By substitution,
	\[
	\tyJudge{\tyEnv}{\mpQi\subst{\mpy}{\mpC}}{(
	\stEnvi[l]\stEnvComp
	\stEnvMap{\mpChanRole{\mpS}{\roleQ}}{\stTi[\roleQ]}
	\stEnvComp\stEnvMap{\mpC}{\stT})
	}
	\]
	So,
	\[\tyJudge{\tyEnv}{\mpQi\subst{\mpy}{\mpC}\mpPar\mpQii}{(
	\stEnvi[l]\stEnvComp\stEnvi[r]\stEnvComp\stEnvMap{\mpC}{\stT}\stEnvComp
	\stEnvMap{\mpChanRole{\mpS}{\roleQ}}{\stTi[\roleQ]}\stEnvComp
	\stEnvMap{\mpChanRole{\mpS}{\roleP}}{\stTi[\roleP]})
	}
	\]
	So,
	\[
	\tyJudge{\tyEnv}{\mpQi\subst{\mpy}{\mpC}\mpPar\mpQii}{
	\stEnvii
	}
	\]
	
	Errors:\\
	\inferrule{Cong}. If $\mpP\prestruct\mpPi$,
	then $\tyJudge{\tyEnv}{\mpPi}{\stEnv}$,
	so $\mpPi\not\mpMoveErr$ by I.H.\\
	\inferrule{E-Ctx}. Similar to \inferrule{R-Ctx},
	$\mpP$ is typed by a safe context, so by I.H.,
	$\mpP\not\mpMoveErr$.\\
	\inferrule{E-Cond}. $\mpP=\mpIf{e}{\mpPi}{\mpPii}$.
	By inversion, $\tyJudge{\tyEnv}{e}{\tyBool}$.
	By \cref{lem:eval-type}, if $\eval{e}{v}$
	then $v\in\{\mpTrue,\mpFalse\}$.\\
	\inferrule{E-Eval}. $\mpSel{\mpC}{\roleP}{\stLab}{e}{}\preCalc\mpP$.
	By inversion, $\tyJudge{\tyEnv}{e}{\tyGround}$.
	By \cref{lem:eval-type}, $\neg\eval{e}{\mpErr}$.\\
	\inferrule{E-M-I}. $\mpP=\mpPi\mpPar\mpPii$,
	$\mpBra{\mpChanRole{\mpS}{\roleQ}}{\roleP}{\stLab}{\mpx}{\mpQi}\preCalc{\mpPi}$, and
	$\mpSel{\mpChanRole{\mpS}{\roleP}}{\roleQ}{\stLab}{\mpC}{\mpQii}\preCalc{\mpPii}$.
	
	By inversion,
	\[\roleP\stFmt{?}\stChoice{\stLab}{\tyGround}\preType\stT\tySub\stEnvApp{\stEnv}{\mpChanRole{\mpS}{\roleQ}}\]
	\[\roleQ\stFmt{!}\stChoice{\stLab}{\stTii}\preType\stTi\tySub\stEnvApp{\stEnv}{\mpChanRole{\mpS}{\roleQ}}\]
	By inversion of subtyping,
	\[\roleP\stFmt{?}\preType\stEnvApp{\stEnv}{\mpChanRole{\mpS}{\roleQ}}\,
	\text{and}\,
	\roleQ\stFmt{!}\stChoice{\stLab}{\stTii[1]}\preType\stEnvApp{\stEnv}{\mpChanRole{\mpS}{\roleP}}\]
	$\stEnv$ is safe, so
	\[\roleP\stFmt{?}\stChoice{\stLab}{\stTii[2]}\preType\stEnvApp{\stEnv}{\mpChanRole{\mpS}{\roleQ}}\]
	By inversion of subtyping,
	$\roleP\stFmt{?}\stChoice{\stLab}{\stTii[2]}\preType\stT$.
	This is a contradiction as labels in a sum type must be unique
	and $\stTii[2]\neq\tyGround$.
	Therefore, this case cannot occur.\\
	\inferrule{E-M-II}. $\mpP=\mpPi\mpPar\mpPii$,
	$\mpBra{\mpChanRole{\mpS}{\roleQ}}{\roleP}{\stLab}{\mpy}{\mpQi}\preCalc{\mpPi}$, and
	$\mpSel{\mpChanRole{\mpS}{\roleP}}{\roleQ}{\stLab}{e}{\mpQii}\preCalc{\mpPii}$.
	
	By inversion,
	\[\roleP\stFmt{?}\stChoice{\stLab}{\stTii}\preType\stT\tySub\stEnvApp{\stEnv}{\mpChanRole{\mpS}{\roleQ}}\quad\text{and}\quad
	\roleQ\stFmt{!}\stChoice{\stLab}{\tyGround}\preType\stTi\tySub\stEnvApp{\stEnv}{\mpChanRole{\mpS}{\roleQ}}
	\]
	By inversion of subtyping,
	\[\roleP\stFmt{?}\preType\stEnvApp{\stEnv}{\mpChanRole{\mpS}{\roleQ}}
	\quad\text{and}\quad
	\roleQ\stFmt{!}\stChoice{\stLab}{\tyGround}\preType\stEnvApp{\stEnv}{\mpChanRole{\mpS}{\roleP}}
	\]
	$\stEnv$ is safe, so
	\[
	\roleP\stFmt{?}\stChoice{\stLab}{\tyGround}\preType\stEnvApp{\stEnv}{\mpChanRole{\mpS}{\roleQ}}
	\]
	By inversion of subtyping,
	$\roleP\stFmt{?}\stChoice{\stLab}{\tyGround}\preType\stT$.
	This is a contradiction as labels in a sum type must be unique
	and $\stTii\neq\tyGround$.
	Therefore, this case cannot occur.\\
	\inferrule{E-M-III}. $\mpP=\mpPi\mpPar\mpPii$,
	$\mpBra{\mpChanRole{\mpS}{\roleQ}}{\roleP}{\stLabi}{}{}\preCalc{\mpPi}$,
	$\mpSel{\mpChanRole{\mpS}{\roleP}}{\roleQ}{\stLab}{}{}\preCalc{\mpPii}$, and
	$\mpBra{\mpChanRole{\mpS}{\roleQ}}{\roleP}{\stLab}{}{}\not\preCalc{\mpPi}$.
	
	By inversion,
	\[\roleP\stFmt{?}\stLabi\preType\stT\tySub\stEnvApp{\stEnv}{\mpChanRole{\mpS}{\roleQ}},\,
	\roleP\stFmt{?}\stLab\not\preType\stT,\,\text{and}\,
	\roleQ\stFmt{!}\stLab\preType\stTi\tySub\stEnvApp{\stEnv}{\mpChanRole{\mpS}{\roleQ}}\]
	By inversion of subtyping,
	\[\roleP\stFmt{?}\preType\stEnvApp{\stEnv}{\mpChanRole{\mpS}{\roleQ}}
	\quad\text{and}\quad
	\roleQ\stFmt{!}\stLab\preType\stEnvApp{\stEnv}{\mpChanRole{\mpS}{\roleP}}\]
	By safety of $\stEnv$,
	$\roleP\stFmt{?}\stLab\preType\stEnvApp{\stEnv}{\mpChanRole{\mpS}{\roleQ}}$.
	
	By inversion of subtyping,
	$\roleP\stFmt{?}\stLab\preType\stT$.
	This is a contradiction, so this case cannot occur.
\qed\end{proof}

\errorFreedom*

 \begin{proof}
 By \cref{thm:subj-red},
 there exists $\stEnvi$ such that
 $\stEnvSafeP{\stEnvi}$ and 
 $\tyJudge{\tyEnv}{\mpPi}{\stEnvi}$.
 Processes containing $\mpErr$ as a subterm are untypable,
 thus $\mpPi$ contains no error.
 \qed
 \end{proof}

\subjectCorrectness*

\begin{proof}
	Let $\mpP\mpMoveStar\mpPi$.
	By \cref{thm:pi-sr},
	there exists $\stEnvi$ s.t.
	$\stEnv\,\gtMoveStar\,\stEnvi$
	and
	$\tyJudge{\tyEnv}{\mpPi}{\stEnvi}$.
	$\stEnv$ is safe, so $\stEnvi$ is safe.
	So by Lem.~\ref{lem:pi-no-err},
	$\mpPi$ contains no $\mpErr$.
	If $\predP\subseteq\stEnvLivePred$,
	then, by \cref{lem:pi-obs},
	$\mpPi$ has no 3-plock.
\qed\end{proof}

\begin{proposition}[Empty Context is Necessary in Reduction Closed]
\label{prop:empty-ctx-safe}
	Without the condition $\stEnvEmpty\in\predP$ for \RC
	$\predP$,
	subject reduction fails.
\end{proposition}

\begin{proof}
	Take $\predP=\{\stEnv\suchthat\stEnv\text{ is safe and non-terminating}\}$.
	This was an example property in \cite{POPL19LessIsMore}.
	Consider the process
	\[
		\mpP
		=
		\mpRec{\mpX}{\mpSel{\mpChanRole{\mpS}{\roleP}}{\roleQ}{\stLab}{\mpNum{0}}{\mpX}}
		\mpPar
		\mpRec{\mpX}{\mpBra{\mpChanRole{\mpS}{\roleQ}}{\roleP}{\stLab}{\mpx}{\mpX}}
	\]
	We have that
	\[
		\tyJudge{\tyEnvEmpty}{\mpP}{(
		\stEnvMap{\mpChanRole{\mpS}{\roleP}}{
		\stRec{\stRecVar}{\roleQ!\stLab\stSeq\stRecVar}
		}
		\stEnvComp
		\stEnvMap{\mpChanRole{\mpS}{\roleQ}}{
		\stRec{\stRecVar}{\roleP?\stLab\stSeq\stRecVar}
		}
		)}
	\]
	\[
		\mpP\mpMove\mpP\mpPar\mpRes{\mpSi}{\mpNil}
	\]
	If
	$
		\tyJudge{\tyEnvEmpty}{\mpP\mpPar\mpRes{\mpSi}{\mpNil}}{\stEnv}
	$
	then
	there is $\stEnvi$ such that
	$\tyJudge{\tyEnvEmpty}{\mpNil}{\stEnvi}$
	and $\stEnvi[\mpSi]$ is non-terminating.
	We must have that $\stEnvEndP{\stEnvi}$
	so $\stEnvi[\mpSi]$ is not non-terminating.
	Therefore, subject reduction fails.
\qed\end{proof}

\begin{example}[Leader Election with Delegation]
	\label{ex:typing-rules-full}
	Recall
	\begingroup
	\small
	\[
		\begin{array}{lcl}
			\mpP[i]&=&
				\mpSel{\mpChanRole{\mpS}{\roleP[i]}}{\roleP[(i+4)\%5]}{\stLabFmt{elect}}{\mpChanRole{\mpS[i]}{\roleP}}{}+
				\mpBra{\mpChanRole{\mpS}{\roleP[i]}}{\roleP[(i+1)\%5]}{\stLabFmt{elect}}{\mpy}{\mpPi[i]\!\left(\mpy\right)}+
				\\&&
				\mpSel{\mpChanRole{\mpS}{\roleP[i]}}{\roleP[(i+3)\%5]}{\stLabFmt{elect}}{\mpChanRole{\mpS[i]}{\roleP}}{}
			\\
			\mpPi[i]\!\left(\mpC\right)&=&
					\mpSel{\mpChanRole{\mpS}{\roleP[i]}}{\roleP[(i+2)\%5]}{\stLabFmt{elect}}{\mpChanRole{\mpS[i]}{\roleP}}{
					\mpSel{\mpC}{\roleS}{\stLabFmt{token}}{\mpNum{i}}{}
					} +
					\\&&
					\mpBra{\mpChanRole{\mpS}{\roleP[i]}}{\roleP[(i+3)\%5]}{\stLabFmt{elect}}{\mpyi}{
						\mpBra{\mpChanRole{\mpS}{\roleP[i]}}{\roleP[(i+2)\%5]}{\stLabFmt{elect}}{\mpyii}{\mpPii[i]\!\left(\mpC,\mpyi,\mpyii\right)}}\\
			\mpPii[i]\!\left(\mpC,\mpCi,\mpCii\right)&=&\mpSel{\mpChanRole{\mpS[i]}{\roleP}}{\roleS}{\stLabFmt{token}}{\mpNum{i}}{
						\mpSel{\mpC}{\roleS}{\stLabFmt{token}}{\mpNum{i}}{
						\mpSel{\mpCi}{\roleS}{\stLabFmt{token}}{\mpNum{i}}{
						\mpSel{\mpCii}{\roleS}{\stLabFmt{token}}{\mpNum{i}}{}
						}
						}
						}\\
			&&\text{for}\;0\leq i \leq 4
			\\
			\mpQ&=&\mpBigPar{0\leq i< 5}{
			\mpBra{\mpChanRole{\mpS[i]}{\roleS}}{\roleP}{\stLabFmt{token}}{\mpx}{}
			}\\
			\mpP[\text{\tiny lead}]&=&\mpP[0]\mpPar\mpP[1]\mpPar\mpP[2]\mpPar\mpP[3]\mpPar\mpP[4]\mpPar\mpQ
		\end{array}
	\]
	\[
\stEnv[\text{\tiny lead}] =
\cup_{0\leq i\leq 4}
\set{\stEnvMap{\mpChanRole{\mpS[i]}{\roleP}}{\stT}
	\stEnvComp
	\stEnvMap{\mpChanRole{\mpS[i]}{\roleS}}{\stTi}\stEnvComp
        \stEnvMap{\mpChanRole{\mpS}{\roleP[i]}}{\stT[i]}}
\]
where for $0\leq i \leq 4$,
\[
	\begin{array}{rcl}
		\stT&=&
		\roleS\stFmt{!}\stChoice{\stLabFmt{token}}{\tyInt}
		\\
\stTi&=&
		\roleP\stFmt{?}\stChoice{\stLabFmt{token}}{\tyInt}\\
		\stT[i]&=&
		\stSum{}{}{
		\begin{array}{l}
			\roleP[(i+4)\%5]!\stChoice{\stLabFmt{elect}}{\stT}\\
			\roleP[(i+3)\%5]!\stChoice{\stLabFmt{elect}}{\stT}\\
			\roleP[(i+1)\%5]?\stChoice{\stLabFmt{elect}}{\stT}\stSeq\stTi[i]
		\end{array}
		}{}
		\\
		\stTi[i]&=&
		\stSum{}{}{
		\begin{array}{l}
			\roleP[(i+2)\%5]!\stChoice{\stLabFmt{elect}}{\stT}\\
			\roleP[(i+3)\%5]?\stChoice{\stLabFmt{elect}}{\stT}\stSeq
			\roleP[(i+2)\%5]?\stChoice{\stLabFmt{elect}}{\stT}
		\end{array}
		}{}
\end{array}        
\]
	\endgroup
We derive the typing judgement $\tyJudge{\tyEnvEmpty}{\mpP[\text{\tiny lead}]}{\stEnv[\text{\tiny lead}]}$.
\begingroup
\fontsize{6}{8}\selectfont
\[
\begin{array}{@{\tyEnvEmpty\,\stFmt{\vdash}\,}l@{\,:\,}ll}
	\mpNil&\stEnvEmpty
	&\inferrule{Nil}
	\\
	\mpNil&
	\set{
	\stEnvMap{\mpChanRole{\mpS}{\roleP[i]}}{\stEnd}
	\stEnvComp
	\stEnvMap{\mpChanRole{\mpS[i]}{\roleP}}{\stEnd}
	}
	&\inferrule{Sub}
	\\
	\mpNil&
	\set{
	\stEnvMap{\mpChanRole{\mpS}{\roleP[i]}}{\stEnd}
	\stEnvComp
	\stEnvMap{\mpChanRole{\mpS[i]}{\roleP}}{\stEnd}
	\stEnvComp
	\stEnvMap{\mpC}{\stEnd}
	}
	&\inferrule{Sub}
	\\
	\mpNil&
	\set{\begin{array}{l}
	\stEnvMap{\mpChanRole{\mpS}{\roleP[i]}}{\stEnd}
	\stEnvComp
	\stEnvMap{\mpChanRole{\mpS[i]}{\roleP}}{\stEnd}
	\stEnvComp\\
	\stEnvMap{\mpC}{\stEnd}
	\stEnvComp
	\stEnvMap{\mpCi}{\stEnd}
	\stEnvComp\\
	\stEnvMap{\mpCii}{\stEnd}
	\end{array}}
	&\inferrule{Sub}
	\\
	\mpSel{\mpCii}{\roleS}{\stLabFmt{token}}{\mpNum{i}}{}
	&
	\set{\begin{array}{l}
	\stEnvMap{\mpChanRole{\mpS}{\roleP[i]}}{\stEnd}
	\stEnvComp
	\stEnvMap{\mpChanRole{\mpS[i]}{\roleP}}{\stEnd}
	\stEnvComp\\
	\stEnvMap{\mpC}{\stEnd}
	\stEnvComp
	\stEnvMap{\mpCi}{\stEnd}
	\stEnvComp\\
	\stEnvMap{\mpCii}{\roleS!\stChoice{\stLabFmt{token}}{\tyInt}}
	\end{array}}
	&\inferrule{Sum}
	\\
	\mpSel{\mpCi}{\roleS}{\stLabFmt{token}}{\mpNum{i}}{
						\mpSel{\mpCii}{\roleS}{\stLabFmt{token}}{\mpNum{i}}{}
						}
	&
	\set{\begin{array}{l}
	\stEnvMap{\mpChanRole{\mpS}{\roleP[i]}}{\stEnd}
	\stEnvComp
	\stEnvMap{\mpChanRole{\mpS[i]}{\roleP}}{\stEnd}
	\stEnvComp\\
	\stEnvMap{\mpC}{\stEnd}
	\stEnvComp\\
	\stEnvMap{\mpCi}{\roleS!\stChoice{\stLabFmt{token}}{\tyInt}}
	\stEnvComp
	\stEnvMap{\mpCii}{\roleS!\stChoice{\stLabFmt{token}}{\tyInt}}
	\end{array}}
	&\inferrule{Sum}
	\\
	{
						\mpSel{\mpC}{\roleS}{\stLabFmt{token}}{\mpNum{i}}{
						\mpSel{\mpCi}{\roleS}{\stLabFmt{token}}{\mpNum{i}}{
						\mpSel{\mpCii}{\roleS}{\stLabFmt{token}}{\mpNum{i}}{}
						}
						}
						}
	&
	\set{\begin{array}{l}
	\stEnvMap{\mpChanRole{\mpS}{\roleP[i]}}{\stEnd}
	\stEnvComp
	\stEnvMap{\mpChanRole{\mpS[i]}{\roleP}}{\stEnd}
	\stEnvComp\\
	\stEnvMap{\mpC}{\roleS!\stChoice{\stLabFmt{token}}{\tyInt}}
	\stEnvComp
	\stEnvMap{\mpCi}{\roleS!\stChoice{\stLabFmt{token}}{\tyInt}}
	\stEnvComp\\
	\stEnvMap{\mpCii}{\roleS!\stChoice{\stLabFmt{token}}{\tyInt}}
	\end{array}}
	&\inferrule{Sum}
	\\
	\mpPii[i]\!\left(\mpC,\mpCi,\mpCii\right)
	&
	\set{\begin{array}{l}
	\stEnvMap{\mpChanRole{\mpS}{\roleP[i]}}{\stEnd}
	\stEnvComp
	\stEnvMap{\mpChanRole{\mpS[i]}{\roleP}}{\roleS!\stChoice{\stLabFmt{token}}{\tyInt}}
	\stEnvComp\\
	\stEnvMap{\mpC}{\roleS!\stChoice{\stLabFmt{token}}{\tyInt}}
	\stEnvComp
	\stEnvMap{\mpCi}{\roleS!\stChoice{\stLabFmt{token}}{\tyInt}}
	\stEnvComp\\
	\stEnvMap{\mpCii}{\roleS!\stChoice{\stLabFmt{token}}{\tyInt}}
	\end{array}}
	&\inferrule{Sum}
	\\
	\mpBra{\mpChanRole{\mpS}{\roleP[i]}}{\roleP[(i+2)\%5]}{\stLabFmt{elect}}{\mpyii}{\mpPii[i]\!\left(\mpC,\mpyi,\mpyii\right)}
	&
	\set{\begin{array}{l}
	\stEnvMap{\mpChanRole{\mpS}{\roleP[i]}}{
	\roleP[(i+2)\% 5]?\stChoice{\stLabFmt{elect}}{\stT}
	}
	\stEnvComp
	\stEnvMap{\mpChanRole{\mpS[i]}{\roleP}}{\roleS!\stChoice{\stLabFmt{token}}{\tyInt}}
	\stEnvComp\\
	\stEnvMap{\mpC}{\roleS!\stChoice{\stLabFmt{token}}{\tyInt}}
	\stEnvComp
	\stEnvMap{\mpyi}{\roleS!\stChoice{\stLabFmt{token}}{\tyInt}}
	\end{array}}
	&\inferrule{Sum}
	\\
	\mpPi[i]\!\left(\mpC\right)
	&
	\set{\begin{array}{l}
	\stEnvMap{\mpChanRole{\mpS}{\roleP[i]}}{\stTi[i]}
	\stEnvComp
	\stEnvMap{\mpChanRole{\mpS[i]}{\roleP}}{\roleS!\stChoice{\stLabFmt{token}}{\tyInt}}
	\stEnvComp\\
	\stEnvMap{\mpC}{\roleS!\stChoice{\stLabFmt{token}}{\tyInt}}\end{array}}
	&\inferrule{Sum}
	\\
	\mpP[i]
	&
	\set{
	\stEnvMap{\mpChanRole{\mpS}{\roleP[i]}}{\stT[i]}
	\stEnvComp
	\stEnvMap{\mpChanRole{\mpS[i]}{\roleP}}{\stT}}
	&\inferrule{Sum}
	\\
	\mpBigPar{0\leq i \leq 4}{\mpP[i]}
	&
	\bigcup_{0\leq i\leq 4}\set{
	\stEnvMap{\mpChanRole{\mpS}{\roleP[i]}}{\stT[i]}
	\stEnvComp
	\stEnvMap{\mpChanRole{\mpS[i]}{\roleP}}{\stT}}
	&
	\inferrule{Par}
	\\
	\mpBra{\mpChanRole{\mpS[i]}{\roleS}}{\roleP}{\stLabFmt{token}}{\mpx}{}
	&\set{\stEnvMap{\mpChanRole{\mpS[i]}{\roleS}}{\stTi}}
	&\inferrule{Sum}
	\\
	\mpQ
	&\bigcup_{0\leq i \leq 4}\set{\stEnvMap{\mpChanRole{\mpS[i]}{\roleS}}{\stTi}}
	&\inferrule{Par}
	\\
	\mpP[\text{\tiny lead}]
	&
	\stEnv[\text{\tiny lead}]
	&\inferrule{Par}
\end{array}
\]
\endgroup
\end{example}

\begin{example}[Typing Cycle]
\label{ex:cycle-typing}
Recall the types from
\cref{ex:cycle-sem-type}
and the processes from \cref{ex:cycle}.
\begingroup
\small
\[\begin{array}{lcl}
	\stT[i]&=&
	\stRec{\stRecVar}{\left(
		\stSum{\roleP[(i+j)\%5]}{j\in\{-1,1\}}{\stChoice{\stLab}{\tyInt}\stSeq\stRecVar}{?}\right.}\\&\stFmt{+}&\left.
		\stSum{\roleP[(i+j)\%5]}{j\in\{-1,1\}}{\stChoice{\stLab}{\tyInt}\stSeq
		\stSum{\roleP[(i+j)\%5]}{j\in\{-1,1\}}{\stChoice{\stLab}{\tyInt}\stSeq\stRecVar}{?}
		}{!}\right)
	\\
	\stTi[i]&=&\stSum{\roleP[(i+j)\%5]}{j\in\{-1,1\}}{\stChoice{\stLab}{\tyInt}\stSeq\stT[i]}{?}
	\end{array}
\]
\[
	\stEnv[\text{\tiny cycle}] =\set{
	\stEnvMap{\mpChanRole{\mpS}{\roleP[0]}}{\stT[0]}
	\stEnvComp
	\stEnvMap{\mpChanRole{\mpS}{\roleP[1]}}{\stTi[1]}
	\stEnvComp
	\stEnvMap{\mpChanRole{\mpS}{\roleP[2]}}{\stT[2]}
	\stEnvComp
	\stEnvMap{\mpChanRole{\mpS}{\roleP[3]}}{\stTi[3]}
	\stEnvComp
	\stEnvMap{\mpChanRole{\mpS}{\roleP[4]}}{\stTi[4]}}
\]
\[
		\begin{array}{lcl}
			\mpP[i]&=&\mpRec{\mpX}{\left(
			\mpSum{\mpBra{\mpChanRole{\mpS}{\roleP[i]}}{\roleP[(i+j)\%5]}{\stLab}{\mpx}{\mpX}}{j\in\{-1,1\}}
			+\right.}\\&&\;
			\left.\mpSum{\mpSel{\mpChanRole{\mpS}{\roleP[i]}}{\roleP[(i+j)\%5]}{\stLab}{\mpNum{0}}{
				\mpSum{\mpBra{\mpChanRole{\mpS}{\roleP[i]}}{\roleP[(i+j)\%5]}{\stLab}{\mpx}{\mpX}}{j\in\{-1,1\}}
			}}{j\in\{-1,1\}}
			\right)
			\\
			\mpQ[i]&=&\mpSum{\mpBra{\mpChanRole{\mpS}{\roleP[i]}}{\roleP[(i+j)\%5]}{\stLab}{\mpx}{\mpP[i]}}{j\in\{-1,1\}}
			\\
			\mpP[\text{\tiny cycle}]&=&\mpP[0]\mpPar\mpQ[1]\mpPar\mpP[2]\mpPar\mpQ[3]\mpPar\mpQ[4]
		\end{array}
	\]
\endgroup
We derive $\tyJudge{\tyEnvEmpty}{\mpP[\text{\tiny cycle}]}{\stEnv[\text{\tiny cycle}]}$.
\begingroup
\fontsize{8}{10}\selectfont
\[
\begin{array}{r@{\,\stFmt{\vdash}\,}l@{\,:\,}ll}
	\set{
		\tyEnvMap{\mpX}{\set{\stEnvMap{\mpChanRole{\mpS}{\roleP[i]}}{\stT[i]}}}
	}
	&
	\mpX
	&
	\set{\stEnvMap{\mpChanRole{\mpS}{\roleP[i]}}{\stT[i]}}
	&
	\inferrule{Var}
	\\
	\set{
		\tyEnvMap{\mpX}{\set{\stEnvMap{\mpChanRole{\mpS}{\roleP[i]}}{\stT[i]}}}
	}
	&
	\mpSum{\mpBra{\mpChanRole{\mpS}{\roleP[i]}}{\roleP[(i+j)\%5]}{\stLab}{\mpx}{\mpX}}{j\in\{-1,1\}}
	&
	{
		{\set{\stEnvMap{\mpChanRole{\mpS}{\roleP[i]}}{\stTi[i]}}}
	}
	&
	\inferrule{Sum}
	\\
	\set{
		\tyEnvMap{\mpX}{\set{\stEnvMap{\mpChanRole{\mpS}{\roleP[i]}}{\stT[i]}}}
	}
	&
	\left(\begin{array}{l}
			\mpSum{\mpBra{\mpChanRole{\mpS}{\roleP[i]}}{\roleP[(i+j)\%5]}{\stLab}{\mpx}{\mpX}}{j\in\{-1,1\}}
			+\\\;
			\mpSum{\mpSel{\mpChanRole{\mpS}{\roleP[i]}}{\roleP[(i+j)\%5]}{\stLab}{\mpNum{0}}{\ }}{j\in\{-1,1\}}\\\quad
			\mpSum{\mpBra{\mpChanRole{\mpS}{\roleP[i]}}{\roleP[(i+j)\%5]}{\stLab}{\mpx}{\mpX}}{j\in\{-1,1\}}
			\end{array}
			\right)
	&
	{
		{\set{\stEnvMap{\mpChanRole{\mpS}{\roleP[i]}}{\unfoldOne{\stT[i]}}}}
	}
	&
	\inferrule{Sum}
	\\
	\set{
		\tyEnvMap{\mpX}{\set{\stEnvMap{\mpChanRole{\mpS}{\roleP[i]}}{\stT[i]}}}
	}
	&
	\left(\begin{array}{l}
			\mpSum{\mpBra{\mpChanRole{\mpS}{\roleP[i]}}{\roleP[(i+j)\%5]}{\stLab}{\mpx}{\mpX}}{j\in\{-1,1\}}
			+\\\;
			\mpSum{\mpSel{\mpChanRole{\mpS}{\roleP[i]}}{\roleP[(i+j)\%5]}{\stLab}{\mpNum{0}}{\ }}{j\in\{-1,1\}}\\\quad
			\mpSum{\mpBra{\mpChanRole{\mpS}{\roleP[i]}}{\roleP[(i+j)\%5]}{\stLab}{\mpx}{\mpX}}{j\in\{-1,1\}}
			\end{array}
			\right)
	&
	{
		{\set{\stEnvMap{\mpChanRole{\mpS}{\roleP[i]}}{\stT[i]}}}
	}
	&
	\inferrule{Sub}
	\\
	\tyEnvEmpty
	&
	\mpP[i]
	&
	{
		{\set{\stEnvMap{\mpChanRole{\mpS}{\roleP[i]}}{\stT[i]}}}
	}
	&
	\inferrule{Rec}
	\\
	\tyEnvEmpty
	&
	\mpQ[i]
	&
	{
		{\set{\stEnvMap{\mpChanRole{\mpS}{\roleP[i]}}{\stTi[i]}}}
	}
	&
	\inferrule{Sum}
	\\
	\tyEnvEmpty
	&
	\mpP[\text{\tiny cycle}]
	&
	\stEnv[\text{\tiny cycle}]
	&
	\inferrule{Par}
\end{array}
\]
\endgroup
\end{example}

\section{Appendix for Preciseness (\S~4)}
\label{app:precise}
\subsection{Characteristics}

\begin{definition}[Complementary Types]\rm
\label{def:char-ctx-app}\ \\
	Fix a finite set of session types $\mathcal{T}$
	such that for all $\stT\in\mathcal{T}$,
	if $\stTi$ is a payload type in $\stT$ then $\stTi\in\mathcal{T}$.
	
	Let $L$ be the set of labels appearing within $\mathcal{T}$.
	Let $\roleSet$ be the set of participants appearing within $\mathcal{T}$.
	
	Enumerate:
	\[
	\begin{array}{rcl}
	\roleSet&=&\{\roleR[i]\}_{0\leq i\leq n}
	\\
	\{\stChoice{\stLab}{\stS}\suchthat\stLab\in L,\stS\in\mathcal{T}\cup\{\tyInt,\tyBool\}\}
	&=&\{\stChoice{\stLab[i]}{\stS[i]}\}_{0\leq i < N}
	\end{array}
	\]
	Let
	\[
	Act=\{I\subseteq\{0,\dots,N-1\}\suchthat \forall i\neq j\in I.\; \stLab[i]\neq\stLab[j]\}
	\]
	the set of all message sets, where messages have distinct labels.
	
	Let $\roleP$ and $\roleFmt{sch}$ be fresh roles,
	and $\stLabii$ be a fresh label
	\ie
	$\roleP,\roleFmt{sch}\not\in\roleSet$ and
	$\stLabii\not\in L$.
	
	We define the following functions below
	for $\roleQ\in\roleSet$, $\dagger\in\{!,?\}$,
	$I\in Act$, and all $\stT$, $\{\stT[i]\}_{i\in I}$:
	$\syncBra{\roleQ}{\stT}$;
	$\syncSel{\roleQ}{\stT}$;
	$\charTAct{\roleQ}{I}{\dagger}{\stT}$;
	$\charTActi{\dagger}{\stT}$;
	$\charTwait{\roleQ}$;
	$\syncSch{\roleQ}{\stT}$;
	$\charSchTest{\roleQ}{\dagger}{\stT}$; and
	$\charSchTrigger{\roleQ}{I}{\dagger}{\{\stT[i]\}_{i\in I}}$.
	
	\[
			\begin{array}{rcl}
				\syncBra{\roleQ}{\stT}&=&
				\roleQ\stFmt{?}\stLab\stSeq\stT
		\\
				\syncSel{\roleR[k]}{\stT}& =& 
				\roleR[0]\stFmt{!}{\stLab}\stSeq
				\dots
				\roleR[k-1]\stFmt{!}\stLab\stSeq
				\roleR[k+1]\stFmt{!}\stLab\stSeq
				\dots
				\roleR[n]\stFmt{!}\stLab\stSeq\stT
 			\\
 				\charTAct{\roleQ}{I}{\dagger}{\stT}&=&
 				\stSum{\roleP}{i\in I}{
 					\stChoice{\stLab[i]}{\stS[i]}\stSeq
 					\syncSel{\roleQ}{
 						\roleFmt{sch}\stFmt{!}\stLab[i]\stSeq\stT
 					}
 				}{\dagger}
				\\
 				\charTActi{\dagger}{\stT}& = &
 					\stFmt{
 					\roleFmt{sch}!\stLab\stSeq\stT
 					+
 						\roleP{\dagger}\stLabii
 					}
				 \\				
\charTwait{\roleQ}&=&
					\stRec{\stRecVar}{\left(
						\stSum{\roleFmt{sch}}{
							I \in Act,\dagger\in\{!,?\}
						}{
							\stLab[(I,\dagger)]\stSeq\charTAct{\roleQ}{I}{\dagger}{\stRecVar}
						}{?}
						\right.}
						\\
						&&\quad\quad\stFmt{+}\stSum{\roleFmt{sch}}{
							\dagger\in\{!,?\}
						}{
							\stLabi[\dagger]\stSeq\charTActi{\dagger}{\stRecVar}
						}{?}
						\\
						& &\quad\quad\stFmt{+}\stSum{\roleFmt{sch}}{
							\roleR\in\roleSet\setminus\{\roleQ\}
						}{
							\stLab[i]\stSeq
							\syncBra{\roleR[i]}{\stRecVar}
						}{?}
						\stFmt{+}\stFmt{
							\left.
							\roleFmt{sch}?\stLab[\text{\tiny end}]
							\right)
						}

			\end{array}
		\]
		\[
			\begin{array}{rcl}
			\syncSch{\roleR[k]}{\stT} & = & 
				\roleR[0]\stFmt{!}{\stLab[k]}
				\dots
				\roleR[k-1]\stFmt{!}\stLab[k]\stSeq
				\roleR[k+1]\stFmt{!}\stLab[k]
				\dots
				\roleR[n]\stFmt{!}\stLab[k]\stSeq\stT
\\
				\charSchTrigger{\roleQ}{I}{\dagger}{\{\stT[i]\}_{i\in
					I}} & = & 
				\stFmt{
						\roleQ!
						\stLab[(I,\dagger)]\stSeq
						\syncSch{\roleQ}{
							\stSum{\roleQ}{i\in I}{\stLab[i]\stSeq\stT[i]}{?}
						}
				}
                       \\
                       \charSchTest{\roleQ}{\dagger}{\stT}
					&=& 
				\stFmt{
					\roleQ!\stLabi[\dagger]\stSeq\roleQ?\stLab\stSeq\stT
				}
			\end{array}
		\]
	
	The closed type $\charTwait{\roleQ}$ will be the type
	of $\roleQ$ in a complementary context.
	
	We define $\charSch{\stT}$ for $\mathcal{T}$-valid types $\stT$,
	by the following recursive definition:
	\[
				\charSch{\stEnd}=
					\roleR[0]\stFmt{!}{\stLab[\text{\tiny end}]}
					\dots
					\roleR[n]\stFmt{!}\stLab[\text{\tiny end}]
					\qquad
				\charSch{\stRecVar}=\stRecVar
				\qquad
				\charSch{\stRec{\stRecVar}{\stT}}=\stRec{\stRecVar}{\charSch{\stT}}
	\]
	\[\begin{array}{l}
				\charSch{\stSum{\roleQ}{(\roleQ,\dagger)\in J,\;i\in I_{(\roleQ,\dagger)}}{
					
						\stChoice{\stLab[i]}{\stS[i]}\stSeq\stT[i,\roleQ,\dagger]
					
				}{\dagger}}=\\
				\quad\stFmt{
					\sum_{(\roleQ,\dagger)\in\roleSet\times\{!,?\}\setminus J}
					\charSchTest{\roleQ}{\overline{\dagger}}{
						\sum_{(\roleQ,\dagger)\in J}
						{\charSchTrigger{\roleQ}{I_{(\roleQ,\dagger)}}{\overline{\dagger}}{\{\charSch{\stT[i,\roleQ,\dagger]}\}_{i\in I_{(\roleQ,\dagger)}}}}
					}
				}
	\end{array}
	\]
	We write $\compEnv{\mpS}{\stT}=
	\bigcup_{\roleQ\in\roleSet}\set{\stEnvMap{\mpChanRole{\mpS}{\roleQ}}{\charTwait{\roleQ}}}
	\stEnvComp
	\stEnvMap{\mpChanRole{\mpS}{\roleFmt{sch}}}{\charSch{\stT}}$,
	the complementary context for $\stT$ and $\mpS$,
	for $\mathcal{T}$-valid $\stT$.
\end{definition}

\begin{itemize}
\item
Our live contexts for $\stT$ are
$\stEnvMap{\mpChanRole{\mpS}{\roleP}}{\stT}
\stEnvComp\compEnv{\mpS}{\stT}$.
\item $\syncSel{\roleQ}{\stT}$ sends a synchronisation message
from $\roleQ$ to the other participants, with continuation $\stT$.
\item Dually, $\syncBra{\roleQ}{\stT}$ receives a synchronisation message
from $\roleQ$, with continuation $\stT$.
\item $\charTAct{\roleQ}{I}{\dagger}{\stT}$
allows $\roleQ$ to communicate with $\roleP$
with action $\dagger$,
message set $\{\stChoice{\stLab[i]}{\stS[i]}\}_{i\in I}$, before
synchronising with the other participants and continuing with $\stT$.
\item $\charTActi{\dagger}{\stT}$ allows $\roleQ$
to test for the action of $\roleQ\stFmt{\overline{\dagger}}$
in $\roleP$, causing an error if it is present
and returning to $\roleFmt{sch}$ otherwise.
\item
for all $\roleQ\in\roleSet$,
$\charTwait{\roleQ}$ responds to commands from $\roleFmt{sch}$
to either:
terminate;
check if $\roleP$ erroneously contains the action
$\roleQ\stFmt{\overline{\dagger}}$;
communicate with $\roleP$ with $\dagger$
and the message set $\{\stChoice{\stLab[i]}{\stS[i]}\}_{i\in I}$;
or synchronise with another participant.
\item $\syncSch{\roleQ}{\stT}$ lets
$\roleFmt{sch}$ command all participants in $\roleSet\setminus\{\roleQ\}$
to receive a synchronisation from $\roleQ$,
then continues with $\stT$.
\item $\charSchTest{\roleQ}{\dagger}{\stT}$ lets $\roleFmt{sch}$ command 
$\roleQ$ to test $\roleP$ for the presence of $\roleQ\stFmt{\overline{\dagger}}$.
\item $\charSchTrigger{\roleQ}{I}{\dagger}{\{\stT[i]\}_{i\in I}}$
lets $\roleFmt{sch}$ command 
$\roleQ$ communicate with $\roleP$ with action $\stFmt{\dagger}$ and
the message set $\{\stChoice{\stLab[i]}{\stS[i]}\}_{i\in I}$.
\item $\charSch{\stT}$ is the scheduling type.
It tells the other participants how to communicate with $\roleP$
to check for erroneous actions and
to make all acceptable actions.
\end{itemize}

\begin{lemma}[Variables]
	If $\stT$ is $\mathcal{T}$-valid, then
	$\gtFv{\charSch{\stT}}=\gtFv{\stT}$.
\end{lemma}

\begin{proof}
	Induction on $\stT$:
	$\gtFv{\charSch{\stEnd}}=\emptyset$;
	$\gtFv{\charSch{\stRecVar}}=\{\stRecVar\}$;
	\[\begin{array}{cl}
	&\gtFv{\charSch{\stRec{\stRecVar}{\stT}}}\\=&\gtFv{\charSch{\stT}}\setminus\{\stRecVar\}\\=&\gtFv{\stT}\setminus\{\stRecVar\}\\=&\gtFv{\stRec{\stRecVar}{\stT}}
	\end{array}\]
	\[\begin{array}{cl}
	&\gtFv{\charSch{\stSum{\roleQ[i]}{i\in I}{\stChoice{\stLab[i]}{\stS[i]}\stSeq\stT[i]}{\dagger_i}}}\\=&
	\bigcup_{i\in I}\gtFv{\charSch{\stT[i]}}\\=&
	\bigcup_{i\in I}\gtFv{\stT[i]}\\=&
	\gtFv{\stSum{\roleQ[i]}{i\in I}{\stChoice{\stLab[i]}{\stS[i]}\stSeq\stT[i]}{\dagger_i}}\end{array}\]
\qed\end{proof}

\begin{lemma}[Scheduler Substitution]
	If $\stT$ and $\stTi$ are $\mathcal{T}$-valid, then
	$\charSch{\tySubst{\stT}{\stRecVar}{\stTi}}=
	\tySubst{\charSch{\stT}}{\stRecVar}{\charSch{\stTi}}$
\end{lemma}

\begin{proof}
	Induction on $\stT$:\\
	$\stT=\stRecVar$.
	$\charSch{\tySubst{\stRecVar}{\stRecVar}{\stTi}}=
	\charSch{\stTi}=
	\tySubst{\stRecVar}{\stRecVar}{\charSch{\stTi}}=
	\tySubst{\charSch{\stRecVar}}{\stRecVar}{\charSch{\stTi}}$.\\
	$\stT=\stRecVari$.
	$\charSch{\tySubst{\stRecVari}{\stRecVar}{\stTi}}=
	\charSch{\stRecVari}=
	\stRecVari=
	\tySubst{\stRecVari}{\stRecVar}{\charSch{\stTi}}=
	\tySubst{\charSch{\stRecVari}}{\stRecVar}{\charSch{\stTi}}$.\\
	$\stT=\stEnd$
	\[\begin{array}{cl}
	&\charSch{\tySubst{\stEnd}{\stRecVar}{\stTi}}\\=&
	\charSch{\stEnd}\\=&
	\roleR[0]\stFmt{!}{\stLab[\text{\tiny end}]}
					\dots
					\roleR[n]\stFmt{!}\stLab[\text{\tiny end}]\\=&
	\tySubst{\roleR[0]\stFmt{!}{\stLab[\text{\tiny end}]}
					\dots
					\roleR[n]\stFmt{!}\stLab[\text{\tiny end}]}{\stRecVar}{\charSch{\stTi}}\\=&
	\tySubst{\charSch{\stEnd}}{\stRecVar}{\charSch{\stTi}}
	\end{array}\]
	$\stT=\stRec{\stRecVari}{\stTii}$.
	\[\begin{array}{cl}
	&\charSch{\tySubst{\stRec{\stRecVari}{\stTii}}{\stRecVar}{\stTi}}\\=&
	\stRec{\stRecVari}{\charSch{\tySubst{\stTii}{\stRecVar}{\stTi}}}\\=&
	\stRec{\stRecVari}{(\tySubst{\charSch{\stTii}}{\stRecVar}{\stTi})}\\=&
	\tySubst{(\stRec{\stRecVari}{\charSch{\stTii}})}{\stRecVar}{\stTi}\\=&
	\tySubst{\charSch{\stRec{\stRecVari}{\stTii}}}{\stRecVar}{\stTi}
	\end{array}\]
	$\stT=
					\stSum{\roleQ}{(\roleQ,\dagger)\in J,\;i\in I_{(\roleQ,\dagger)}}{
						\stChoice{\stLab[i]}{\stS[i]}\stSeq\stT[i,\roleQ,\dagger]
					}{\dagger}
				$.
	\[\begin{array}{cl}
	&\charSch{\tySubst{\stT}{\stRecVar}{\stTi}}\\=&
	\charSch{
					\stSum{\roleQ}{(\roleQ,\dagger)\in J,\;i\in I_{(\roleQ,\dagger)}}{
						\stChoice{\stLab[i]}{\stS[i]}\stSeq\tySubst{\stT[i,\roleQ,\dagger]}{\stRecVar}{\stTi}
					}{\dagger}
			}\\=&
	\stFmt{
					\sum_{(\roleQ,\dagger)\in\roleSet\times\{!,?\}\setminus J}
					\charSchTest{\roleQ}{\overline{\dagger}}{}
				}\\&
	\qquad
	\stFmt{
						\sum_{(\roleQ,\dagger)\in J}
						{\charSchTrigger{\roleQ}{I_{(\roleQ,\dagger)}}{\overline{\dagger}}{\{\charSch{\tySubst{\stT[i,\roleQ,\dagger]}{\stRecVar}{\stTi}}\}_{i\in I_{(\roleQ,\dagger)}}}}
					)}
	\\=&
	\stFmt{
					\sum_{(\roleQ,\dagger)\in\roleSet\times\{!,?\}\setminus J}
					\charSchTest{\roleQ}{\overline{\dagger}}{}
				}\\&
				\qquad\stFmt{
						\sum_{(\roleQ,\dagger)\in J}
						{\charSchTrigger{\roleQ}{I_{(\roleQ,\dagger)}}{\overline{\dagger}}{\{\charSch{\tySubst{\charSch{\stT[i,\roleQ,\dagger]}}{\stRecVar}{\charSch{\stTi}}}\}_{i\in I_{(\roleQ,\dagger)}}}}
					)}
	\\=&
	\stFmt{
					\sum_{(\roleQ,\dagger)\in\roleSet\times\{!,?\}\setminus J}
					\charSchTest{\roleQ}{\overline{\dagger}}{}
				}\\&
				\qquad\stFmt{
						\sum_{(\roleQ,\dagger)\in J}
						{\charSchTrigger{\roleQ}{I_{(\roleQ,\dagger)}}{\overline{\dagger}}{\{\charSch{\charSch{\stT[i,\roleQ,\dagger]}}\}_{i\in I_{(\roleQ,\dagger)}}}}
					)\subst{\stRecVar}{\charSch{\stTi}}}
	\\=&\tySubst{\charSch{\stT}}{\stRecVar}{\charSch{\stTi}}
	\end{array}\]
\qed\end{proof}

\begin{lemma}
	If $\stT$ is $\mathcal{T}$-valid, then
	$\unfoldOne{\charSch{\stT}}=\charSch{\unfoldOne{\stT}}$.
\end{lemma}

\begin{proof}
	Induction on $\depth{\stT}$:
	If $\stT$ is not recursive, then neither is $\charSch{\stT}$,
	so\\
	$\unfoldOne{\charSch{\stT}}=\charSch{\stT}=\charSch{\unfoldOne{\stT}}$.\\
	If $\stT=\stRec{\stRecVar}{\stTi}$,
	then $\charSch{\stT}=\stRec{\stRecVar}{\charSch{\stTi}}$.
	$\charSch{\unfoldOne{\stT}}
	=\charSch{\unfoldOne{\tySubst{\stTi}{\stRecVar}{\stT}}}
	=\unfoldOne{\charSch{\tySubst{\stTi}{\stRecVar}{\stT}}}$ by I.H.
	and\\
	$\unfoldOne{\charSch{\tySubst{\stTi}{\stRecVar}{\stT}}}
	=\unfoldOne{\tySubst{\charSch{\stTi}}{\stRecVar}{\charSch{\stT}}}
	=\unfoldOne{\charSch{\stT}}$.
\qed\end{proof}

\propComplCtx*

\begin{proof}
	For $\stT=\stEnd$ the only reduction sequence is:\\
	\[
	\begin{array}{cl}
	&\compEnv{\mpS}{\stEnd}\stEnvComp\stEnvMap{\mpChanRole{\mpS}{\roleP}}{\stEnd}
	\\=&
	\bigcup_{0\leq i \leq n}\set{\stEnvMap{\mpChanRole{\mpS}{\roleQ[i]}}{\charTwait{\roleQ[i]}}}
	\stEnvComp
	\stEnvMap{\mpChanRole{\mpS}{\roleP}}{\stEnd}\stEnvComp\stEnvMap{\mpChanRole{\mpS}{\roleFmt{sch}}}{
	\roleR[0]\stFmt{!}{\stLab[\text{\tiny end}]}
					\dots
					\roleR[n]\stFmt{!}\stLab[\text{\tiny end}]
	}
	\\
	\gtMove&
	\bigcup_{0\leq i \leq 0}\set{\stEnvMap{\mpChanRole{\mpS}{\roleQ[i]}}{\stEnd}}
	\stEnvComp\\&
	\bigcup_{1\leq i \leq n}\set{\stEnvMap{\mpChanRole{\mpS}{\roleQ[i]}}{\charTwait{\roleQ[i]}}}
	\stEnvComp
	\stEnvMap{\mpChanRole{\mpS}{\roleP}}{\stEnd}\stEnvComp\stEnvMap{\mpChanRole{\mpS}{\roleFmt{sch}}}{
	\roleR[1]\stFmt{!}{\stLab[\text{\tiny end}]}
					\dots
					\roleR[n]\stFmt{!}\stLab[\text{\tiny end}]
	}
	\\
	\gtMove&\dots\\
	\gtMove&
	\bigcup_{0\leq i \leq j-1}\set{
	\stEnvMap{\mpChanRole{\mpS}{\roleQ[i]}}{\stEnd}}
	\cup\\&
	\bigcup_{j \leq i \leq n}\set{\stEnvMap{\mpChanRole{\mpS}{\roleQ[i]}}{\charTwait{\roleQ[i]}}}
	\stEnvComp
	\stEnvMap{\mpChanRole{\mpS}{\roleP}}{\stEnd}\stEnvComp\stEnvMap{\mpChanRole{\mpS}{\roleFmt{sch}}}{
	\roleR[j]\stFmt{!}{\stLab[\text{\tiny end}]}
					\dots
					\roleR[n]\stFmt{!}\stLab[\text{\tiny end}]
	}
	\\
	\gtMove&\dots\\
	\gtMove&
	\bigcup_{0\leq i \leq n}\stEnvMap{\mpChanRole{\mpS}{\roleQ[i]}}{\stEnd}
	\stEnvComp
	\stEnvMap{\mpChanRole{\mpS}{\roleP}}{\stEnd}\stEnvComp\stEnvMap{\mpChanRole{\mpS}{\roleFmt{sch}}}{
	\stEnd
	}
	\end{array}\]
	None of these contexts have label or payload mismatches,
	and this is the only fair path, and it is live.
	So, $\compEnv{\mpS}{\stEnd}\stEnvComp\stEnvMap{\mpChanRole{\mpS}{\roleP}}{\stEnd}$ is safe and live.
	
	$\compEnv{\mpS}{\stT}\stEnvComp\stEnvMap{\mpChanRole{\mpS}{\roleP}}{\stT}$
	is safe and live iff
	$\compEnv{\mpS}{\unfoldOne{\stT}}\stEnvComp\stEnvMap{\mpChanRole{\mpS}{\roleP}}{\unfoldOne{\stT}}$
	is safe and live.
	So, we need only consider the case where $\stT$ is a sum.
	Suppose that $\stT=
					\stSum{\roleQ}{(\roleQ,\dagger)\in J,\;i\in I_{(\roleQ,\dagger)}}{
						\stChoice{\stLab[i]}{\stS[i]}\stSeq\stT[i,\roleQ,\dagger]
					}{\dagger}
				$:
	\[
		\begin{array}{cl}
		&\left\{
			\stEnv\suchthat
			\compEnv{\mpS}{\stT}
			\stEnvComp
			\stEnvMap{\mpChanRole{\mpS}{\roleP}}{\stT}
			\gtMoveStar\stEnv\text{ in exactly }3+2\cdot|\roleSet|\text{ steps}
		\right\}
		\\=&
		\left\{
			\compEnv{\mpS}{\stT}
			\stEnvComp
			\stEnvMap{\mpChanRole{\mpS}{\roleP}}{\stT[i,\roleQ,\dagger]}
			\suchthat
			(\roleQ,\dagger)\in J\;\wedge\;
			i\in I_{(\roleQ,\dagger)}
		\right\}
		\end{array}
	\]
	Every participant is involved in at least one of the
	$3+2\cdot|\roleSet|$ transitions to any of the environments above
	(via the synchronisation communications).
	If $\stEnv$
	is reached in fewer than $3+2\cdot|\roleSet|$ transitions
	then we observe that it contains no label mismatches or deadlocks.
	
	If $\compEnv{\mpS}{\stT}
	\stEnvComp
	\stEnvMap{\mpChanRole{\mpS}{\roleP}}{\stT}
	\gtMoveStar\stEnv$,
	then we assume that the path is of length less than $3+2\cdot|\roleSet|$,
	and so $\stEnv$ has no label mismatches and every participant will act in the next
	$3+2\cdot|\roleSet|$ transitions.
	Therefore, $\stEnvMap{\mpChanRole{\mpS}{\roleP}}{\stT}\stEnvComp\compEnv{\mpS}{\stT}$ is safe and live.
\qed\end{proof}

\begin{definition}[Rank]\rm
\label{def:rank}
	We define the rank of type $\stT$
	to be $0$ if all payloads are $\stEnd$
	or basic types,
	and $1+$max rank of payload type, otherwise.
	Define the rank of a type $\stS$,
	written $\rank{\stS}$,
	recursively to be:
	$\rank{\stS}=0$ if $\stS\in\{\stEnd,\tyInt,\tyBool\}$;
	$\rank{\stS}=1+\max(\{0\}\cup\{\rank{\stSi}\suchthat \stSi\;\text{is a payload in}\;\stS\})$.
	
	Define the strict partial order
	$\stT\rankRel\stTi$
	iff $\rank{\stT} < \rank{\stTi}$
	or ($\rank{\stT}=\rank{\stTi}$ and $|\stT|<|\stTi|$).
	This relation is well-founded,
	thus we may perform induction on it.
\end{definition}
Note that the complementary types have rank $1$.

Also note that proof by induction using
well-foundedness of $\rankRel$
is equivalent to proof by induction on $\rank{\stT}$
with proof by induction on $|\stT|$ at every inductive case and base case.
We do proof by induction using
well-foundedness of $\rankRel$
to avoid unnecessary repetition.

\propCharProcWellDef*

\begin{proof}
	All recursive calls in \cref{def:char-proc}
	use parameters that are strictly smaller with respect to $\rankRel$
	\eg all types in $\compEnv{\mpS}{\stT}$ have rank at most $\rank{\stT}$.
	By well-foundedness of $\rankRel$,
	$\charP{\mpC}{\stT}$ terminates.
	\qed
\end{proof}

\begin{lemma}[Characteristic Variables]
	If $\stT$ is $\mathcal{T}$-valid,
	then\\
	\begin{enumerate*}
	\item $\mpC\in\chan{\charP{\mpC}{\stT}}$ if $\unfoldOne{\stT}$ is a sum
	and
	\item $\fpv{\charP{\mpC}{\stT}}=\{\mpX_{\stRecVar}\suchthat \stRecVar\in\gtFv{\stT}\}$
	\end{enumerate*}.
\end{lemma}

\begin{proof}
	We proceed by induction on the well-founded partial order $\rankRel$:\\
	$\fpv{\charP{\mpC}{\stEnd}}=\emptyset$.\\
	$\fpv{\charP{\mpC}{\stRecVar}}=\{\mpX_{\stRecVar}\}$.\\
	$\chan{\charP{\mpC}{\stRec{\stRecVar}{\stT}}}=\chan{\charP{\mpC}{\stT}}$ and
	$\fpv{\charP{\mpC}{\stRec{\stRecVar}{\stT}}}=\fpv{\charP{\mpC}{\stRec{\stRecVar}{\stT}}}\setminus\{\mpX_{\stRecVar}\}=
	\{\mpX_{\stRecVari}\suchthat\stRecVari\in\gtFv{\stRec{\stRecVar}{\stT}}\}$.\\
	$\mpC\in\chan{\charP{\mpC}{\stSum{\roleQ[i]}{i\in I}{\stChoice{\stLab[i]}{\stS[i]}\stSeq\stT[i]}{\dagger_i}}}$ and
	payload types are closed, so by I.H.,
	\[\begin{array}{cl}&\fpv{\charP{\mpC}{\stSum{\roleQ[i]}{i\in I}{\stChoice{\stLab[i]}{\stS[i]}\stSeq\stT[i]}{\dagger_i}}}\\=&
	\bigcup_{i\in I}\fpv{\charP{\mpC}{\stT[i]}}\\=&\bigcup_{i\in I}\{\mpX_{\stRecVar}\suchthat\stRecVar\in\gtFv{\stT[i]}\}\\=&
	\{\mpX_{\stRecVar}\suchthat\stRecVar\in{\gtFv{\charP{\mpC}{\stSum{\roleQ[i]}{i\in I}{\stChoice{\stLab[i]}{\stS[i]}\stSeq\stT[i]}{\dagger_i}}}}\}
	\end{array}\]
\qed\end{proof}

\thmCharProc*

\begin{proof}
	We proceed by induction on the well-founded partial order $\rankRel$:\\
	$\stT=\stEnd$.
	$\charP{\mpC}{\stT}=\mpNil$.
	By \inferrule{End},
	$\tyJudge{\tyEnvEmpty}{\mpNil}{\stEnvEmpty}$.
	By \inferrule{Sub},
	$\tyJudge{\tyEnvEmpty}{\mpNil}{\set{\stEnvMap{\mpC}{\stEnd}}}$.\\
	$\stT=\stRecVar$.
	$\charP{\mpC}{\stRecVar}=\mpX_{\stRecVar}$.
	By \inferrule{Var}, $
	\tyJudge{
	\tyEnvMap{\mpX_{\stRecVar}}{\set{\stEnvMap{\mpC}{\stT[\stRecVar]}}}
	}
	{\mpX_{\stRecVar}}
	{\set{\stEnvMap{\mpC}{\stT[\stRecVar]}}}
	$.\\
	$\stT=\stRec{\stRecVar}{\stTi}$.
	$\stT$ is guarded otherwise
	$\stT\stFmt{\subst{\stRecVar[1]}{\stT[1]}\dots\subst{\stRecVar[n]}{\stT[n]}}$
	wouldn't be closed,\\
	so $\unfoldOne{\stT}$ is defined.
	By I.H.,
	\[\tyJudge
	{\left(\begin{array}{l}
		\{\tyEnvMap{\mpX_{\stRecVar[i]}}{(\stEnvMap{\mpC}{\stT[i]})}\}_{i\leq n}\tyEnvComp\\
		\tyEnvMap{\mpX_{\stRecVar}}{
			\set{\stEnvMap{\mpC}{
				\stT\stFmt{\subst{\stRecVar[1]}{\stT[1]}\dots\subst{\stRecVar[n]}{\stT[n]}}
			}}
		}
	\end{array}\right)}
	{\charP{\mpC}{\stTi}}
	{
		\set{
			\stEnvMap{\mpC}{
				\tySubst{\stTi}{\stRecVar}{\stT}\stFmt{\subst{\stRecVar[1]}{\stT[1]}\dots\subst{\stRecVar[n]}{\stT[n]}}
			}
		}
	}\]
	By \inferrule{Sub},
	\[\tyJudge
	{\left(\begin{array}{l}
		\{\tyEnvMap{\mpX_{\stRecVar[i]}}{(\stEnvMap{\mpC}{\stT[i]})}\}_{i\leq n}\tyEnvComp\\
		\tyEnvMap{\mpX_{\stRecVar}}{
			\set{\stEnvMap{\mpC}{
				\stT\stFmt{\subst{\stRecVar[1]}{\stT[1]}\dots\subst{\stRecVar[n]}{\stT[n]}}
			}}
		}
	\end{array}\right)}
	{\charP{\mpC}{\stTi}}
	{
		\set{
			\stEnvMap{\mpC}{
				\stT\stFmt{\subst{\stRecVar[1]}{\stT[1]}\dots\subst{\stRecVar[n]}{\stT[n]}}
			}
		}
	}\]
	If $\unfoldOne{\stT}=\stEnd$, then
	$\set{
			\stEnvMap{\mpC}{
				\stT\stFmt{\subst{\stRecVar[1]}{\stT[1]}\dots\subst{\stRecVar[n]}{\stT[n]}}
			}
		}\setminus\stEnd=\stEnvEmpty$.\\
	If $\unfoldOne{\stT}=\stRecVari$, then $\mpC\in\dom{\stEnvMap{\mpC}{\stT[\stRecVari]}}$
	and $\mpX_{\stRecVari}\in\fpv{\charP{\mpC}{\stT}}$.\\
	If $\unfoldOne{\stT}$ is a sum type, then $\mpC\in\chan{\charP{\mpC}{\stT}}$.\\
	So, \[\dom{\set{
			\stEnvMap{\mpC}{
				\stT\stFmt{\subst{\stRecVar[1]}{\stT[1]}\dots\subst{\stRecVar[n]}{\stT[n]}}
			}
		}\setminus\stEnd}\subseteq\chan{\charP{\mpC}{\stT},\{\tyEnvMap{\mpX_{\stRecVari}}{(\stEnvMap{\mpC}{\stT[\stRecVari]})}\}_{\stRecVari\in\gtFv{\stT}}}\]
	So by
	\inferrule{Rec},
	\[\tyJudge
	{
		\{\tyEnvMap{\mpX_{\stRecVari}}{\set{\stEnvMap{\mpC}{\stT[\stRecVari]}}}\}_{\stRecVari\in\gtFv{\stT}}
	}
	{\charP{\mpC}{\stT}}
	{
		\set{
			\stEnvMap{\mpC}{
				\tySubstBig{\stT}{\stRecVari}{\stT[\stRecVari]}{\stRecVari\in\gtFv{\stT}}
			}
		}
	}\]
	$\stT=\stSum{\roleQ[i]}
			{i\in I}
			{\stChoice{\stLab[i]}{\stS[i]}}{\dagger_i}$.
			So,
	\[\charP{\mpC}{\stT}=
	\mpRes{\mpS[i]\suchthat i\in I'}{
				\left(
				\mpSum{\mpPrefix_i\mpSeq\mpP[i]}{i\in I}
				\mpPar
				\mpBigPar{i\in I''}{
					\charPM{\compEnv{\mpS[i]}{\stS[i]}}
					}\right)}
	\]
	where
	$I'=\{i\in I\suchthat \dagger_i={!},\stS[i]\not\in\{\tyInt,\tyBool\}\}$,
	$I'' = \{i\in I'\suchthat \stS[i]\neq\stEnd\}$, and
	\[\begin{array}{l}
	\mpPrefix_i\mpSeq\mpP[i]=\\\;\left\{
	\begin{array}{l}
				\mpSel{\mpC}{\roleQ[i]}{\stLab[i]}{\mpChanRole{\mpS[i]}{\roleP}}{\left(
						\charP{\mpC}{\stT[i]}
						\mpPar
						\charPM{\bigcup_{j\in I'\setminus\{i\}}\stEnvMap{\mpChanRole{\mpS[j]}{\roleP}}{\stS[j]}}
					\right)}
					\;\quad\dagger_i={!},\stS[i]\not\in\{\tyInt,\tyBool\}\\
				\mpSel{\mpC}{\roleQ[i]}{\stLab[i]}{v_{\stS[i]}}{\left(
						\charP{\mpC}{\stT[i]}
						\mpPar
						\charPM{\bigcup_{j\in I'}\stEnvMap{\mpChanRole{\mpS[j]}{\roleP}}{\stS[j]}}
					\right)}
					\;\dagger_i={!},\stS[i]\in\{\tyInt,\tyBool\}
					\\
				\mpBra{\mpC}{\roleQ[i]}{\stLab[i]}{\mpy}{\left(
						\charP{\mpC}{\stT[i]}
						\mpPar
						\charP{\mpy}{\stS[i]}
						\mpPar
						\charPM{\bigcup_{j\in I'}\stEnvMap{\mpChanRole{\mpS[j]}{\roleP}}{\stS[j]}}
					\right)}\;\dagger_i={?},\stS[i]\not\in\{\tyInt,\tyBool\}\\
				\mpBra{\mpC}{\roleQ[i]}{\stLab[i]}{\mpx}{\left(
						\charP{\mpC}{\stT[i]}
						\mpPar
						\mpP[{\stS[i]}]\!\left(\mpx\right)
						\mpPar
						\charPM{\bigcup_{j\in I'}\stEnvMap{\mpChanRole{\mpS[j]}{\roleP}}{\stS[j]}}
					\right)}\;
					\dagger_i={?},\stS[i]\in\{\tyInt,\tyBool\}
			\end{array}
			\right.\end{array}
	\]
	Let $\tyEnv=\{\tyEnvMap{\mpX_{\stRecVar_i}}{\stT[i]}\}_{i\leq n}$ and, for all $\stTi$,
	$\anonSubst{\stTi}=\stTi\stFmt{\subst{\stRecVar[1]}{\stT[1]}\dots\subst{\stRecVar[n]}{\stT[n]}}$.\\
	By I.H. and weakening,
	\[
		\tyJudge{\tyEnv}{\charP{\mpC}{\stT[i]}}{\set{\stEnvMap{\mpC}{\anonSubst{\stT[i]}}}}
	\]
	for $i\in I$.
	\[
		\tyJudge{\tyEnv}{\charPM{\compEnv{\mpS[i]}{\stS[i]}}}{\compEnv{\mpS[i]}{\stS[i]}}
	\]
	for $i\in I''$.
	\[
		\tyJudge{\tyEnv}{\charP{\mpChanRole{\mpS[j]}{\roleP}}{\stS[j]}}{\set{\stEnvMap{\mpChanRole{\mpS[j]}{\roleP}}{\stS[j]}}}
	\]
	for $j \in I'$.
	\[
		\tyJudge{\tyEnv}{\charP{\mpy}{\stS[i]}}{\set{\stEnvMap{\mpy}{\stS[i]}}}
	\]
	for $i\in I$ such that $\dagger_i={?}$ and $\stS[i]$ is a session type.
	\[
		\tyJudge{\tyEnvMap{\mpx}{\tyGround}}{\mpP[{\tyGround}]\!\left(\mpx\right)}{\stEnvEmpty}
	\]
	for $\tyGround\in\{\tyInt,\tyBool\}$.\\
	So by \inferrule{Par}, for $i$ s.t. $\dagger_i={!}$
	and $\stS[i]$ is a session type:
	\[
		\tyJudge{\tyEnv}{
		\left(
						\charP{\mpC}{\stT[i]}
						\mpPar
						\charPM{\bigcup_{j\in I'\setminus\{i\}}\stEnvMap{\mpChanRole{\mpS[j]}{\roleP}}{\stS[j]}}
					\right)
		}{\left(
			\bigcup_{j\in I'\setminus\{i\}}\set{
			\stEnvMap{\mpChanRole{\mpS[j]}{\roleP}}{\stS[j]}}
			\stEnvComp
			\stEnvMap{\mpC}{\anonSubst{\stT[i]}}
		\right)}
	\]
	By \inferrule{Par}, for $i$ s.t. $\dagger_i={!}$
	and $\stS[i]$ is a basic type:
	\[
		\tyJudge{\tyEnv}{
		\left(
						\charP{\mpC}{\stT[i]}
						\mpPar
						\charPM{\bigcup_{j\in I'}\stEnvMap{\mpChanRole{\mpS[j]}{\roleP}}{\stS[j]}}
					\right)
		}{\left(
			\bigcup_{j\in I'}
			\set{\stEnvMap{\mpChanRole{\mpS[j]}{\roleP}}{\stS[j]}}
			\stEnvComp
			\stEnvMap{\mpC}{\anonSubst{\stT[i]}}
		\right)}
	\]
	By \inferrule{Par}, for $i$ s.t. $\dagger_i={?}$
	and $\stS[i]$ is a session type:
	\[
		\tyJudge{\tyEnv}{
		\left(
						\charP{\mpC}{\stT[i]}
						\mpPar
						\charP{\mpy}{\stS[i]}
						\mpPar
						\charPM{\bigcup_{j\in I'}\stEnvMap{\mpChanRole{\mpS[j]}{\roleP}}{\stS[j]}}
					\right)
		}{\left(\begin{array}{l}
			\bigcup_{j\in I'}
			\set{\stEnvMap{\mpChanRole{\mpS[j]}{\roleP}}{\stS[j]}}
			\stEnvComp\\
			\stEnvMap{\mpC}{\anonSubst{\stT[i]}}
			\stEnvComp\\
			\stEnvMap{\mpy}{\stS[i]}
		\end{array}\right)}
	\]
	By \inferrule{Par} and weakening, for $i$ s.t. $\dagger_i={?}$
	and $\stS[i]$ is a basic type:
	\[\small
		\tyJudge{(\tyEnv\tyEnvComp\tyEnvMap{\mpx}{\stS[i]})}{
		\left(
						\charP{\mpC}{\stT[i]}
						\mpPar
						\mpP[{\stS[i]}]\!\left(\mpx\right)
						\mpPar
						\charPM{\bigcup_{j\in I'}\stEnvMap{\mpChanRole{\mpS[j]}{\roleP}}{\stS[j]}}
					\right)
		}{\left(
			\bigcup_{j\in I'}
			\set{\stEnvMap{\mpChanRole{\mpS[j]}{\roleP}}{\stS[j]}}
			\stEnvComp
			\stEnvMap{\mpC}{\anonSubst{\stT[i]}}
		\right)}
	\]
	So by \inferrule{Sum},
	\[
		\tyJudge{\tyEnv}{\mpSum{\mpPrefix_i\mpSeq\mpP[i]}{i\in I}}{
			\left(
			\bigcup_{j\in I'}
			\set{\stEnvMap{\mpChanRole{\mpS[j]}{\roleP}}{\stS[j]}}
			\stEnvComp
			\stEnvMap{\mpC}{\anonSubst{\stT}}
		\right)
		}
	\]
	By \inferrule{Par},
	\[\small
		\tyJudge{\tyEnv}{\left(
		\mpSum{\mpPrefix_i\mpSeq\mpP[i]}{i\in I}
		\mpPar
		\mpBigPar{i\in I''}{
					\charPM{\compEnv{\mpS[i]}{\stS[i]}}
					}
		\right)}{
			\left(
			\bigcup_{i\in I'}
			\set{\stEnvMap{\mpChanRole{\mpS[i]}{\roleP}}{\stS[i]}}
			\cup
			\bigcup_{i\in I''}
			\set{\compEnv{\mpS[i]}{\stS[i]}}
			\stEnvComp
			\stEnvMap{\mpC}{\anonSubst{\stT}}
		\right)
		}
	\]
	By \inferrule{Res},
	\[
		\tyJudge{\tyEnv}{\charP{\mpC}{\stT}}{\set{\stEnvMap{\mpC}{\anonSubst{\stT}}}}
	\]
\qed\end{proof}

\begin{lemma}[Substitution]
\label{lem:char-substitution}
	If $\stT$ and $\stTi$ are $\mathcal{T}$-valid, then:\\
	(1) $\charP{\mpC}{\tySubst{\stT}{\mpX_{\stRecVar}}{\stTi}}=\charP{\mpC}{\stT}\subst{\mpX_{\stRecVar}}{\charP{\mpC}{\stTi}}$;\\
	(2) $\charP{\mpC}{\stT}\prestruct\charP{\mpC}{\unfoldOne{\stT}}$.
\end{lemma}

\begin{proof}[1]
	Induct on $\stT$, since payload types have no free variables,
	so the only recursive case is on subterms of $\stT$.
\qed\end{proof}

\begin{proof}[2]
	Induction on $\unfoldOne{\stT}$:
	if $\stT$ is not a recursion, then
	$\unfoldOne{\stT}=\stT$;
	if $\stT=\stRec{\stRecVar}{\stTi}$, then
	\[
	\begin{array}{cl}
	&\charP{\mpC}{\stT}\\=&
	\mpRec{\mpX_{\stRecVar}}{\charP{\mpC}{\stTi}}\\\prestruct&
	\charP{\mpC}{\stTi}\subst{\mpX_{\stRecVar}}{\charP{\mpC}{\stT}}\\=&
	\charP{\mpC}{\tySubst{\stTi}{\stRecVar}{\stT}}
	\end{array}\]
	By I.H., $\charP{\mpC}{\stT}\prestruct\charP{\mpC}{\tySubst{\stTi}{\stRecVar}{\stT}}\prestruct\charP{\mpC}{\unfoldOne{\stT}}$.
\qed\end{proof}

\subsection{Negation of Subtyping}

\thmNegationSubtyping*

\begin{proof}[Proof for $\stTi\tyNot\stT\implies\stTi\tyNotSub\stT$]
	We induct on the derivation of $\stTi\tyNot\stT$,
	that $\stTi\tyNotSub\stT$.
	
	\begin{itemize}
		\item \inferrule{$\stEnd$-L}
			$\stEnd\tyNotSub\stSum{\roleQ[i]}{i\in I}{\stChoice{\stLab[i]}{\stS[i]}\stSeq\stT[i]}{\dagger_i}$
			by Lem.~\ref{lem:sub-inv-2}.
			
		\item \inferrule{$\stEnd$-R}
			$\stSum{\roleQ[i]}{i\in I}{\stChoice{\stLab[i]}{\stS[i]}\stSeq\stT[i]}{\dagger_i}\tyNotSub\stEnd$
			by Lem.~\ref{lem:sub-inv-2}.
			
		\item \inferrule{$\mu$L}
			By I.H., $\tySubst{\stTi}{\stRecVar}{\stRec{\stRecVar}{\stTi}}\tyNotSub\stT$.
			By Lem.~\ref{lem:sub-inv},
			$\stRec{\stRecVar}{\stTi}\tyNotSub\stT$.
			
		\item \inferrule{$\mu$R}
			By I.H., $\stT\tyNotSub\tySubst{\stT}{\stRecVar}{\stRec{\stRecVar}{\stT}}$.
			By Lem.~\ref{lem:sub-inv},
			$\stT\tyNotSub\stRec{\stRecVar}{\stT}$.
			
		\item \inferrule{${!}$L}
			Let $\stTi=\stSum{\roleQ[i]}{i\in I'}{\stChoice{\stLab[i]}{\stS[i]}\stSeq\stT[i]}{\dagger_i}$
			and
			$\stT=\stSum{\roleQ[i]}{i\in I}{\stChoice{\stLab[i]}{\stS[i]}\stSeq\stT[i]}{\dagger_i}$
			such that
			there exists $i'\in I'$ such that
			$\stFmt{\dagger_{i'}}=\stFmt{{!}}$
			and, for all $i\in I$,
			$\roleQ[i']\stFmt{{!}}\stLab[i']
			\neq
			\roleQ[i]\stFmt{\dagger_i}\stLab[i]$.
			By Lem.~\ref{lem:sub-inv-2},
			$\stTi\tyNotSub\stT$.
			
		\item \inferrule{${?}$L}
			Let $\stTi=\stSum{\roleQ[i]}{i\in I'}{\stChoice{\stLab[i]}{\stS[i]}\stSeq\stT[i]}{\dagger_i}$
			and
			$\stT=\stSum{\roleQ[i]}{i\in I}{\stChoice{\stLab[i]}{\stS[i]}\stSeq\stT[i]}{\dagger_i}$
			such that
			there exists $i'\in I'$ such that
			$\stFmt{\dagger_{i'}}=\stFmt{{?}}$
			and, for all $i\in I$,
			$\roleQ[i']\stFmt{{?}}
			\neq
			\roleQ[i]\stFmt{\dagger_i}$.
			By Lem.~\ref{lem:sub-inv-2},
			$\stTi\tyNotSub\stT$.
			
		\item \inferrule{${?}$R}
			Let $\stTi=\stSum{\roleQ[i]}{i\in I'}{\stChoice{\stLab[i]}{\stS[i]}\stSeq\stT[i]}{\dagger_i}$
			and
			$\stT=\stSum{\roleQ[i]}{i\in I}{\stChoice{\stLab[i]}{\stS[i]}\stSeq\stT[i]}{\dagger_i}$
			such that
			there exists $i\in I$ such that
			$\stFmt{\dagger_{i}}=\stFmt{{?}}$
			and, for all $i'\in I'$,
			$\roleQ[i]\stFmt{{?}}\stLab[i]
			\neq
			\roleQ[i']\stFmt{\dagger_{i'}}\stLab[i']$.
			By Lem.~\ref{lem:sub-inv-2},
			$\stTi\tyNotSub\stT$.
			
		\item \inferrule{${!}$R}
			Let $\stTi=\stSum{\roleQ[i]}{i\in I'}{\stChoice{\stLab[i]}{\stS[i]}\stSeq\stT[i]}{\dagger_i}$
			and
			$\stT=\stSum{\roleQ[i]}{i\in I}{\stChoice{\stLab[i]}{\stS[i]}\stSeq\stT[i]}{\dagger_i}$
			such that
			there exists $i\in I$ such that
			$\stFmt{\dagger_{i}}=\stFmt{{!}}$
			and, for all $i'\in I'$,
			$\roleQ[i]\stFmt{{!}}
			\neq
			\roleQ[i']\stFmt{\dagger_{i'}}$.
			By Lem.~\ref{lem:sub-inv-2},
			$\stTi\tyNotSub\stT$.
			
		\item \inferrule{LR${?}$}
			Let $\stTi=\stSum{\roleQ[i]}{i\in I'}{\stChoice{\stLab[i]}{\stS[i]}\stSeq\stT[i]}{\dagger_i}$
			and
			$\stT=\stSum{\roleQ[i]}{i\in I}{\stChoice{\stLab[i]}{\stS[i]}\stSeq\stT[i]}{\dagger_i}$
			such that
			there exist $j\in J$ and $j'\in J'$
			such that
			$\roleQ[j]=\roleQ[j']$,
			$\stLab[j]=\stLab[j']$,
			$\stFmt{\dagger_j}=\stFmt{\dagger_{j'}}=\stFmt{{?}}$,
			and $\stS[j']\tyNot\stS[j]$.
			If both of $\stS[j]$ and $\stS[j']$ are
			basic types, then $\stS[j]\neq\stS[j']$
			so $\stS[j']\tyNotSub\stS[j]$.
			If exactly one is a basic type, then $\stS[j']\tyNotSub\stS[j]$,
			by Lem.~\ref{lem:sub-inv-2}.
			If both are session types, then
			$\stS[j']\tyNotSub\stS[j]$
			by I.H.
			By Lem.~\ref{lem:sub-inv-2},
			$\stTi\tyNotSub\stT$.
			
		\item \inferrule{LR${!}$}
			Let $\stTi=\stSum{\roleQ[i]}{i\in I'}{\stChoice{\stLab[i]}{\stS[i]}\stSeq\stT[i]}{\dagger_i}$
			and
			$\stT=\stSum{\roleQ[i]}{i\in I}{\stChoice{\stLab[i]}{\stS[i]}\stSeq\stT[i]}{\dagger_i}$
			such that
			there exist $j\in J$ and $j'\in J'$
			such that
			$\roleQ[j]=\roleQ[j']$,
			$\stLab[j]=\stLab[j']$,
			$\stFmt{\dagger_j}=\stFmt{\dagger_{j'}}=\stFmt{{!}}$,
			and $\stS[j]\tyNot\stS[j']$.
			If both of $\stS[j]$ and $\stS[j']$ are
			basic types, then $\stS[j]\neq\stS[j']$
			so $\stS[j]\tyNotSub\stS[j']$.
			If exactly one is a basic type, then $\stS[j]\tyNotSub\stS[j']$,
			by Lem.~\ref{lem:sub-inv-2}.
			If both are session types, then
			$\stS[j]\tyNotSub\stS[j']$
			by I.H.
			By Lem.~\ref{lem:sub-inv-2},
			$\stTi\tyNotSub\stT$.
			
	\item \inferrule{LR}
			Let $\stTi=\stSum{\roleQ[i]}{i\in I'}{\stChoice{\stLab[i]}{\stS[i]}\stSeq\stT[i]}{\dagger_i}$
			and
			$\stT=\stSum{\roleQ[i]}{i\in I}{\stChoice{\stLab[i]}{\stS[i]}\stSeq\stT[i]}{\dagger_i}$
			such that
			there exist $j\in J$ and $j'\in J'$
			such that
			$\roleQ[j]=\roleQ[j']$,
			$\stLab[j]=\stLab[j']$,
			$\stFmt{\dagger_j}=\stFmt{\dagger_{j'}}$,
			and $\stT[j']\tyNot\stT[j]$.
			By Lem.~\ref{lem:sub-inv-2},
			$\stTi\tyNotSub\stT$.
	\end{itemize}
	Therefore, $\stTi\tyNot\stT\implies\stTi\tyNotSub\stT$.
\qed\end{proof}

\begin{proof}[Proof for $\stTi\tyNotSub\stT\implies\stTi\tyNot\stT$]
	Suppose that $\neg\stSi\tyNot\stS$,
	we construct a corecursive derivation of $\stSi\tySub\stS$.
	Consider the following cases,
	picking the first one that applies:
	\begin{itemize}
		\item If $\stS$ or $\stSi$ are basic types,
		then neither \inferrule{$\tyGround$-L}
		nor \inferrule{$\tyGround$-R} apply,
		hence $\stS=\stSi$ is a basic type.
		Therefore, $\stSi\tySub\stS$.
		\item If $\stSi=\stRec{\stRecVar}{\stTii}$,
		then, since \inferrule{$\mu$L} does not apply,
		$\neg\tySubst{\stTii}{\stRecVar}{\stSi}\tyNot\stS$.
		We corecursively construct a derivation of
		$\tySubst{\stTii}{\stRecVar}{\stSi}\tySub\stS$
		and conclude
		$\stSi\tySub\stS$ by \inferrule{S$\mu$L}.
		
		\item If $\stS=\stRec{\stRecVar}{\stTii}$,
		then, since \inferrule{$\mu$R} does not apply,
		$\neg\stSi\tyNot\tySubst{\stTii}{\stRecVar}{\stS}$.
		We corecursively construct a derivation of
		$\stSi\tySub\tySubst{\stTii}{\stRecVar}{\stS}$
		and conclude
		$\stSi\tySub\stS$ by \inferrule{S$\mu$R}.
		
		\item If $\stS=\stSi=\stEnd$,
		then $\stSi\tySub\stS$ by \inferrule{SEnd}.
		
		\item If
		$\stS$ and $\stSi$ are sum types, then write:
		\[
		\stS=\stFmt{\sum_{k\in K}}\stSum{\roleQ[k]}{i\in I_k}{\stChoice{\stLab[i]}{\stS[i]}\stSeq\stT[i]}{\dagger_k}
		\qquad
		\stSi=\stFmt{\sum_{k\in K'}}\stSum{\roleQ[k]}{i\in I'_k}{\stChoice{\stLab[i]}{\stSi[i]}\stSeq\stTi[i]}{\dagger_k}
		\]
		with $\{\roleQ[k]\stFmt{\dagger_k}\}_{k\in K}$ distinct
		and $\{\roleQ[k]\stFmt{\dagger_k}\}_{k\in K'}$ distinct.
		Since \inferrule{${?}$L}, \inferrule{${!}$L},
		\inferrule{${?}$R}, and \inferrule{${!}$R} do not apply,
		we have that
		$\{\roleQ[k]\stFmt{\dagger_k}\}_{k\in K}
		=\{\roleQ[k]\stFmt{\dagger_k}\}_{k\in K'}$,
		so without loss of generality we may assume that $K=K'$,
		and, by reindexing, we have that
		$I'_k\subset I_k$ for all $k\in K$ such that $\stFmt{\dagger_k}=\stFmt{{!}}$
		and
		$I_k\subset I'_k$ for all $k\in K$ such that $\stFmt{\dagger_k}=\stFmt{{?}}$.
		\inferrule{LR${?}$} does not apply,
		so, for all $k\in K$ and $i\in I_k$ such that $\stFmt{\dagger_k}=\stFmt{{?}}$,
		$\neg\stSi[i]\tyNot\stS[i]$ from which we find
		a corecursive derivation of $\stSi[i]\tySub\stS[i]$.
		\inferrule{LR${!}$} does not apply,
		so, for all $k\in K$ and $i\in I'_k$ such that $\stFmt{\dagger_k}=\stFmt{{!}}$,
		$\neg\stS[i]\tyNot\stSi[i]$ from which we find
		a corecursive derivation of $\stS[i]\tySub\stSi[i]$.
		\inferrule{LR} does not apply,
		so, for all $k\in K$ and $i\in I_k\cap I'_k$,
		$\neg \stTi[i]\tyNot\stT[i]$
		from which we find a corecursive derivation of
		$\stTi[i]\tySub\stT[i]$.
		We now conclude that $\stSi\tySub\stS$
		by \inferrule{S$\Sigma^\star$}.
	\end{itemize}
\qed\end{proof}

\subsection{Completeness}

\thmCharFidelity*

\begin{proof}[1]
	Assume that the communicating participants are unfolded,
	by characteristic substitution.
	
	Suppose that
	\[
	\stEnv=\stEnvii\stEnvComp
	\stEnvMap{\mpChanRole{\mpS}{\roleQ}}{\stT[\roleQ]}\stEnvComp
	\stEnvMap{\mpChanRole{\mpS}{\roleP}}{\stT[\roleP]}
	\]
	\[
	\roleP\stFmt{?}\stChoice{\stLab}{\stSi}\stSeq\stTi[\roleQ]\preType\stT[\roleQ]
	,\,
	\roleQ\stFmt{!}\stChoice{\stLab}{\stS}\stSeq\stTi[\roleP]\preType\stT[\roleP],
	\,\text{and}\,
	\stSi\tySub\stS
	\]
	So that
	\[
	\stEnv\gtMove\stEnvii\stEnvComp
	\stEnvMap{\mpChanRole{\mpS}{\roleQ}}{\stTi[\roleQ]}\stEnvComp
	\stEnvMap{\mpChanRole{\mpS}{\roleP}}{\stTi[\roleP]}
	\]
	\[
	\charP{\mpChanRole{\mpS}{\roleQ}}{\stT[\roleQ]}=
	\mpCtxApp{\mpCtx}{\mpQi}
	\]
	with
	\[
	\mpBra{\mpChanRole{\mpS}{\roleQ}}{\roleP}{\stLab}{d}{(\charP{\mpChanRole{\mpS}{\roleQ}}{\stTi[\roleQ]}\mpPar\mpRi)}\preCalc{\mpQi}
	\]
	\[
	\charP{\mpChanRole{\mpS}{\roleP}}{\stT[\roleP]}=
	\mpCtxApp{\mpCtx'}{\mpQii}
	\]
	with
	\[
	\mpSel{\mpChanRole{\mpS}{\roleP}}{\roleQ}{\stLab}{\mpz}{(\charP{\mpChanRole{\mpS}{\roleP}}{\stTi[\roleP]}\mpPar\mpRii)}\preCalc{\mpQii}
	\]
	So,
	\[\begin{array}{cl}
	&\charPM{\stEnv}\\\prestruct&
	\mpCtxApp{\mpCtx''}{\mpQi\mpPar\mpQii}\mpPar\charPM{\stEnvii}
	\\\mpMove&
	\mpCtxApp{\mpCtx'''}{\charP{\mpChanRole{\mpS}{\roleQ}}{\stTi[\roleQ]}\mpPar\charP{\mpChanRole{\mpS}{\roleP}}{\stTi[\roleP]}}\mpPar\charPM{\stEnvii}
	\\\prestruct&
	\charPM{\stEnvii\stEnvComp
	\stEnvMap{\mpChanRole{\mpS}{\roleQ}}{\stTi[\roleQ]}\stEnvComp
	\stEnvMap{\mpChanRole{\mpS}{\roleP}}{\stTi[\roleP]}}\mpPar\mpR
	\end{array}\]
	Since
	characteristic processes have no free expressions
	and no free channels except their parameter.
	
	Note that this holds even when $\stSi\tyNotSub\stS$
	and that $\chan{\mpR}=\emptyset$.
\qed\end{proof}

\begin{proof}[2]
	Suppose that $\roleQ\stFmt{!}\stChoice{\stLab}{\stT}\preType\stT[\roleR]$
	and
	$\roleR\stFmt{?}\stChoice{\stLab}{\stTi}\preType\stT[\roleQ]$.
	So,
	\[
		\charP{\mpChanRole{\mpS}{\roleQ}}{\stT[\roleQ]}
		\prestruct
		\mpRes{\mpSi[1],\dots,\mpSi[n]}{
		\left(
			\mpR\mpPar
			\mpQ
		\right)
		}
	\]
	\[
		\mpBra{\mpChanRole{\mpS}{\roleQ}}{\roleR}{\stLab}{\mpy}{
			\mpQi
		}\preCalc{\mpQ}
	\]
	\[
		\mpQi\prestruct
		\mpRi\mpPar\charP{\mpy}{\stTi}
	\]
	\[
		\chan{\charP{\mpy}{\stTi}}=\{\mpy\}
		\qquad
		\chan{\mpRi}\not\ni\mpy
	\]
	\[
		\charP{\mpChanRole{\mpS}{\roleR}}{\stT[\roleR]}
		\prestruct
		\mpRes{\mpSiii}{\mpRes{\mpSii[1],\dots,\mpSii[m]}{
		\left(
			\mpRii\mpPar
			\mpQii
		\right)
		}}
	\]
	\[
		\mpSel{\mpChanRole{\mpS}{\roleR}}{\roleQ}{\stLab}{\mpChanRole{\mpSiii}{\roleP}}{
			\mpQiii
		}\preCalc{\mpQii}
	\]
	\[
		\mpSiii\not\in\fs{\mpQiii}
	\]
	If $\stT=\stEnd$, then $\mpSiii\not\in\fs{\mpRii}$.
	If $\stT\neq\stEnd$, then $\mpRii=\charPM{\compEnv{\mpSiii}{\stT}}\mpPar\mpRiii$
	and $\mpSiii\not\in\fs{\mpRiii}$.
	So,
	\[\begin{array}{cl}
		&\charP{\mpChanRole{\mpS}{\roleQ}}{\stT[\roleQ]}
		\mpPar
		\charP{\mpChanRole{\mpS}{\roleR}}{\stT[\roleR]}\\
		\mpMove&
		\mpRes{\mpSiii,
		\mpSii[1],\dots,\mpSii[m],
		\mpSi[1],\dots,\mpSi[n]
		}{
		\left(
			\mpRii\mpPar
			\mpR\mpPar
			\mpQiii\mpPar
			\mpRi\mpPar\charP{\mpChanRole{\mpSiii}{\roleP}}{\stTi}
		\right)
		}\\
		\prestruct&
		\left\{
			\begin{array}{ll}
				\mpRes{
				\mpSii[1],\dots,\mpSii[m],
				\mpSi[1],\dots,\mpSi[n]
				}{
		\left(
			\mpRii\mpPar
			\mpR\mpPar
			\mpQiii\mpPar
			\mpRi\mpPar\charP{\mpChanRole{\mpSiii}{\roleP}}{\stTi}
		\right)
		}\\\quad\mpPar\mpRes{\mpSiii}{\charP{\mpChanRole{\mpSiii}{\roleP}}{\stTi}}
		&\stT=\stEnd\\[1ex]
		\mpRes{
		\mpSii[1],\dots,\mpSii[m],
		\mpSi[1],\dots,\mpSi[n]
		}{
		\left(
			\mpRiii\mpPar
			\mpR\mpPar
			\mpQiii\mpPar
			\mpRi\mpPar\charP{\mpChanRole{\mpSiii}{\roleP}}{\stTi}
		\right)
		}\\\quad\mpPar\mpRes{\mpSiii}{
		\charPM{\compEnv{\mpSiii}{\stT}\stEnvComp\stEnvMap{\mpChanRole{\mpSiii}{\roleP}}{\stTi}}}
		&\stT\neq\stEnd\\
			\end{array}
		\right.
	\end{array}\]
\qed\end{proof}

\begin{proof}[3]
Label mismatch:\\
Suppose that $\roleQ\stFmt{!}\stChoice{\stLab}{}\preType\stT[\roleR]$,
	$\roleR\stFmt{?}\stChoice{\stLabi}{}\preType\stT[\roleQ]$,
	and
	$\roleR\stFmt{?}\stChoice{\stLab}{}\not\preType\stT[\roleQ]$.
	So,
	\[
		\charP{\mpChanRole{\mpS}{\roleQ}}{\stT[\roleQ]}
		\prestruct
		\mpRes{\mpSi[1],\dots,\mpSi[n]}{
		\left(
			\mpR\mpPar
			\mpQ
		\right)
		}
	\]
	\[
		\mpBra{\mpChanRole{\mpS}{\roleQ}}{\roleR}{\stLabi}{}{}\preCalc{\mpQ}
		\qquad
		\mpBra{\mpChanRole{\mpS}{\roleQ}}{\roleR}{\stLab}{}{}\not\preCalc{\mpQ}
	\]
	\[
		\charP{\mpChanRole{\mpS}{\roleR}}{\stT[\roleR]}
		\prestruct
		\mpRes{\mpSii[1],\dots,\mpSii[m]}{
		\left(
			\mpRi\mpPar
			\mpQi
		\right)
		}
	\]
	\[
		\mpSel{\mpChanRole{\mpS}{\roleR}}{\roleQ}{\stLab}{}{}\preCalc{\mpQi}
	\]
	So,
	\[
		\mpQ\mpPar\mpQi\mpMove\mpErr
	\]
	\[\begin{array}{cl}
		&\charP{\mpChanRole{\mpS}{\roleR}}{\stT[\roleR]}
		\mpPar
		\charP{\mpChanRole{\mpS}{\roleQ}}{\stT[\roleQ]}\\
		\prestruct&
		\mpRes{\mpSii[1],\dots,\mpSii[m]}{\mpRes{\mpSi[1],\dots,\mpSi[n]}{
		\left(
			\mpR\mpPar
			\mpRi\mpPar
			\mpQ\mpPar
			\mpQi
		\right)
		}}\\
		\mpMove&
		\mpRes{
		\mpSii[1],\dots,\mpSii[m],
		\mpSi[1],\dots,\mpSi[n]
		}{
		\left(
			\mpR\mpPar
			\mpRi\mpPar
			\mpErr
		\right)
		}
	\end{array}\]
	Payload Mismatch (Send Basic, Receive Session):\\
	Suppose that $\roleQ\stFmt{!}\stChoice{\stLab}{\tyGround}\preType\stT[\roleR]$ and
	$\roleR\stFmt{?}\stChoice{\stLab}{\stT}\preType\stT[\roleQ]$.
	So,
	\[
		\charP{\mpChanRole{\mpS}{\roleQ}}{\stT[\roleQ]}
		\prestruct
		\mpRes{\mpSi[1],\dots,\mpSi[n]}{
		\left(
			\mpR\mpPar
			\mpQ
		\right)
		}
	\]
	\[
		\mpBra{\mpChanRole{\mpS}{\roleQ}}{\roleR}{\stLab}{\mpy}{}\preCalc{\mpQ}
	\]
	\[
		\charP{\mpChanRole{\mpS}{\roleR}}{\stT[\roleR]}
		\prestruct
		\mpRes{\mpSii[1],\dots,\mpSii[m]}{
		\left(
			\mpRi\mpPar
			\mpQi
		\right)
		}
	\]
	\[
		\mpSel{\mpChanRole{\mpS}{\roleR}}{\roleQ}{\stLab}{v_{\tyGround}}{}\preCalc{\mpQi}
	\]
	So,
	\[
		\mpQ\mpPar\mpQi\mpMove\mpErr
	\]
	\[\begin{array}{cl}
		&\charP{\mpChanRole{\mpS}{\roleR}}{\stT[\roleR]}
		\mpPar
		\charP{\mpChanRole{\mpS}{\roleQ}}{\stT[\roleQ]}\\
		\prestruct&
		\mpRes{
		\mpSii[1],\dots,\mpSii[m],
		\mpSi[1],\dots,\mpSi[n]
		}{
		\left(
			\mpR\mpPar
			\mpRi\mpPar
			\mpQ\mpPar
			\mpQi
		\right)
		}\\
		\mpMove&
		\charP{\mpChanRole{\mpS}{\roleR}}{\stT[\roleR]}
		\mpPar
		\charP{\mpChanRole{\mpS}{\roleQ}}{\stT[\roleQ]}
		\prestruct
		\mpRes{
		\mpSii[1],\dots,\mpSii[m],
		\mpSi[1],\dots,\mpSi[n]
		}{
		\left(
			\mpR\mpPar
			\mpRi\mpPar
			\mpErr
		\right)
		}
	\end{array}\]
		Payload Mismatch (Send Session, Receive Basic):\\
	Suppose that $\roleQ\stFmt{!}\stChoice{\stLab}{\stT}\preType\stT[\roleR]$ and
	$\roleR\stFmt{?}\stChoice{\stLab}{\tyGround}\preType\stT[\roleQ]$.
	So,
	\[
		\charP{\mpChanRole{\mpS}{\roleQ}}{\stT[\roleQ]}
		\prestruct
		\mpRes{\mpSi[1],\dots,\mpSi[n]}{
		\left(
			\mpR\mpPar
			\mpQ
		\right)
		}
	\]
	\[
		\mpBra{\mpChanRole{\mpS}{\roleQ}}{\roleR}{\stLab}{\mpx}{}\preCalc{\mpQ}
	\]
	\[
		\charP{\mpChanRole{\mpS}{\roleR}}{\stT[\roleR]}
		\prestruct
		\mpRes{\mpSii[1],\dots,\mpSii[m]}{
		\left(
			\mpRi\mpPar
			\mpQi
		\right)
		}
	\]
	\[
		\mpSel{\mpChanRole{\mpS}{\roleR}}{\roleQ}{\stLab}{\mpChanRole{\mpSiii}{\roleP}}{}\preCalc{\mpQi}
	\]
	So,
	\[
		\mpQ\mpPar\mpQi\mpMove\mpErr
	\]
	\[\begin{array}{cl}
		&\charP{\mpChanRole{\mpS}{\roleR}}{\stT[\roleR]}
		\mpPar
		\charP{\mpChanRole{\mpS}{\roleQ}}{\stT[\roleQ]}\\
		\prestruct&
		\mpRes{
		\mpSii[1],\dots,\mpSii[m],
		\mpSi[1],\dots,\mpSi[n]
		}{
		\left(
			\mpR\mpPar
			\mpRi\mpPar
			\mpQ\mpPar
			\mpQi
		\right)
		}\\
		\mpMove&
		\mpRes{
		\mpSii[1],\dots,\mpSii[m],
		\mpSi[1],\dots,\mpSi[n]
		}{
		\left(
			\mpR\mpPar
			\mpRi\mpPar
			\mpErr
		\right)
		}
	\end{array}\]
	Payload Mismatch (Send and Receive Basic):\\
	Suppose that $\roleQ\stFmt{!}\stChoice{\stLab}{\tyGround}\preType\stT[\roleR]$,
	$\roleR\stFmt{?}\stChoice{\stLab}{\tyGroundi}\preType\stT[\roleQ]$, and
	$\tyGround\neq\tyGroundi$.
	So,
	\[
		\charP{\mpChanRole{\mpS}{\roleQ}}{\stT[\roleQ]}
		\prestruct
		\mpRes{\mpSi[1],\dots,\mpSi[n]}{
		\left(
			\mpR\mpPar
			\mpQ
		\right)
		}
	\]
	\[
		\mpBra{\mpChanRole{\mpS}{\roleQ}}{\roleR}{\stLab}{\mpx}{\mpR}\preCalc{\mpQ}
	\]
	\[
		\mpR\prestruct\mpP[\tyGroundi]\!\left(\mpx\right)\mpPar\mpRii
	\]
	\[
		\charP{\mpChanRole{\mpS}{\roleR}}{\stT[\roleR]}
		\prestruct
		\mpRes{\mpSii[1],\dots,\mpSii[m]}{
		\left(
			\mpRi\mpPar
			\mpQi
		\right)
		}
	\]
	\[
		\mpSel{\mpChanRole{\mpS}{\roleR}}{\roleQ}{\stLab}{v_{\tyGround}}{}\preCalc{\mpQi}
	\]
	So,
	\[
		\mpQ\mpPar\mpQi\mpMove
		\mpRiii\mpPar\mpP[\tyGroundi]\!\left(v_{\tyGround}\right)
	\]
	\[
		\mpP[\tyGroundi]\!\left(v_{\tyGround}\right)\mpMove\mpErr
	\]
	\[\begin{array}{cl}
		&\charP{\mpChanRole{\mpS}{\roleR}}{\stT[\roleR]}
		\mpPar
		\charP{\mpChanRole{\mpS}{\roleQ}}{\stT[\roleQ]}\\
		\prestruct&
		\mpRes{
		\mpSii[1],\dots,\mpSii[m],
		\mpSi[1],\dots,\mpSi[n]
		}{
		\left(
			\mpR\mpPar
			\mpRi\mpPar
			\mpQ\mpPar
			\mpQi
		\right)
		}\\
		\mpMove&
		\mpRes{
		\mpSii[1],\dots,\mpSii[m],
		\mpSi[1],\dots,\mpSi[n]
		}{
		\left(
			\mpR\mpPar
			\mpRi\mpPar
			\mpRii\mpPar
			\mpErr
		\right)
		}
	\end{array}\]
\qed\end{proof}

\begin{proof}[4]
	Suppose that $\stEnv$ has a 3-plock,
	so
	\[\unfoldOne{\stEnvApp{\stEnv}{\mpChanRole{\mpS}{\roleP}}}
=\stSum{\roleQ[i]}{i\in I}{\stChoice{\stLab[i]}{\stS[i]}\stSeq\stT[i]}{\dagger_i}\]
For all $i\in I$,
if $\mpChanRole{\mpS}{\roleQ[i]}\in\dom{\stEnv}$ then either:
\begin{enumerate}
	\item $\unfoldOne{\stEnvApp{\stEnv}{\mpChanRole{\mpS}{\roleQ[i]}}}
=\stEnd
$; or
	\item $\unfoldOne{\stEnvApp{\stEnv}{\mpChanRole{\mpS}{\roleQ[i]}}}
=\stSum{\roleR[i]}{j\in J_i}{\stChoice{\stLab[i,j]}{\stS[i,j]}\stSeq\stT[i,j]}{\dagger'_i}$
and if $\mpChanRole{\mpS}{\roleR[i]}\in\dom{\stEnv}$ then either:
	\begin{enumerate}
		\item $\roleR[i]=\roleP$ and $\dagger'_i=\dagger_i$; or
		\item $\unfoldOne{\stEnvApp{\stEnv}{\mpChanRole{\mpS}{\roleR[i]}}}
=\stEnd
$; or
		\item $\unfoldOne{\stEnvApp{\stEnv}{\mpChanRole{\mpS}{\roleR[i]}}}
=\stSum{\roleP}{j\in J'_i}{\stChoice{\stLabi[i,j]}{\stSi[i,j]}\stSeq\stTi[i,j]}{\dagger''_i}$.
	\end{enumerate}
\end{enumerate}
	Assume that $\mpChanRole{\mpS}{\roleQ[i]}\in\dom{\stEnv}$
	and $\mpChanRole{\mpS}{\roleR[i]}\in\dom{\stEnv}$
	for $i\in I$ by adding $\stEnvMap{\mpChanRole{\mpS}{\roleQ[i]}}{\stEnd}$
	and $\stEnvMap{\mpChanRole{\mpS}{\roleQ[i]}}{\stEnd}$ to $\stEnv$
	if necessary, preserving $\charPM{\stEnv}$ up to $\equiv$,
	since $\mpP\equiv\mpP\mpPar\mpNil$.
	
	It is immediate that
	\[
		\obsvNew{\charP{\mpChanRole{\mpS}{\roleQ[i]}}{\stEnvApp{\stEnv}{\mpChanRole{\mpS}{\roleQ[i]}}}}{\mpS}{\mpQ[i]}{\mpR[i]}{\dagger'_i}
	\]
	and $\roleP=\roleR[i]$ or
	\[
		\obsvNew{\charP{\mpChanRole{\mpS}{\roleR[i]}}{\stEnvApp{\stEnv}{\mpChanRole{\mpS}{\roleR[i]}}}}{\mpS}{\mpR[i]}{\mpP}{\dagger'_i}
	\]
	Also,
	\[
		\charP{\mpChanRole{\mpS}{\roleP}}{\stEnvApp{\stEnv}{\mpChanRole{\mpS}{\roleP}}}
		\prestruct
		\charP{\mpChanRole{\mpS}{\roleP}}{\unfoldOne{\stEnvApp{\stEnv}{\mpChanRole{\mpS}{\roleP}}}}
		=
		\mpRes{\mpSi[1],\dots,\mpSi[m]}{\left(\mpSum{\mpPrefix_i\mpSeq\mpQ[i]}{i\in I}\mpPar\mpR\right)}
	\]
	Therefore,
	\[
	\mpRes{\mpS}{\charPM{\mpS}{\stEnv}}
	\prestruct
	\mpRes{\mpSi[1],\dots,\mpSi[m],\mpS}{\left(
		\mpSum{\mpPrefix_i\mpSeq\mpQ[i]}{i\in I}\mpPar\mpR
		\mpPar\mpBigPar{\mpC\in\dom{\stEnv}\setminus\{\mpChanRole{\mpS}{\roleP}\}}{\charP{\mpChanRole{\mpS}{\mpC}}{\stEnvApp{\stEnv}{\mpC}}}
	\right)}
	\]
	This has a 3-plock.
\qed\end{proof}

Below we write $\threeparticipantlock{\stEnv}{\mpS}$
if $\stEnv$ has a 3-plock on $\mpS$.

\begin{restatable}[Complementary Contexts]{lemma}{thmComplCtxFidelity}
\label{thm:comp-fid}
	If $\stT$ and $\stTi$ are $\mathcal{T}$-valid and closed, then:\\
	(1) $\compEnv{\mpS}{\unfoldOne{\stT}}=\unfoldOne{\compEnv{\mpS}{\stT}}$;
	(2) $\compEnv{\mpS}{\stEnd}\gtMoveStar\stEnvi$
	with $\stEnvEndP{\stEnvi}$;\\
	(3) if $\roleQ\stFmt{\dagger}\stChoice{\stLab}{\stS}\stSeq\stTii\preType\stT$ then
	$\compEnv{\mpS}{\stT}\stEnvComp\stEnvMap{\mpChanRole{\mpS}{\roleP}}{\stT}
	\gtMoveStar
	\compEnv{\mpS}{\stTii}\stEnvComp\stEnvMap{\mpChanRole{\mpS}{\roleP}}{\stTii}$;\\
	(4) if $\roleQ\stFmt{\dagger}\preType\stT$ and $\roleQ\stFmt{\dagger}\not\preType\stTi$ or $\stTi=\stEnd$,
	then $\compEnv{\mpS}{\stT}\stEnvComp\stEnvMap{\mpChanRole{\mpS}{\roleP}}{\stTi}
	\gtMoveStar\stEnvi$ and
	$\threeparticipantlock{\stEnvi}{\mpS}$.\\
	(5) if $\roleQ\stFmt{!}\stLab\preType\stTi$ and $\roleQ\stFmt{!}\stLab\not\preType\stT$ or $\stT=\stEnd$,
	then $\compEnv{\mpS}{\stT}\stEnvComp\stEnvMap{\mpChanRole{\mpS}{\roleP}}{\stTi}
	\gtMoveStar\stEnvi$ with a label mismatch;\\
	(6) if $\roleQ\stFmt{?}\stLab\preType\stTi$ and $\roleQ\stFmt{?}\not\preType\stT$ or $\stT=\stEnd$,
	then $\compEnv{\mpS}{\stT}\stEnvComp\stEnvMap{\mpChanRole{\mpS}{\roleP}}{\stTi}
	\gtMoveStar\stEnvi$ with a label mismatch;\\
	(7) if $\roleQ\stFmt{!}\stChoice{\stLab}{\stSi}\preType\stTi$,
	$\roleQ\stFmt{!}\stChoice{\stLab}{\stS}\preType\stT$,
	and $\stS\tyNotSub\stSi$, then
	$\compEnv{\mpS}{\stT}\stEnvComp\stEnvMap{\mpChanRole{\mpS}{\roleP}}{\stTi}
	\gtMoveStar\stEnvi$ with a payload mismatch
	of sending $\stSi$ when $\stS$ is received;\\
	(8) if $\roleQ\stFmt{?}\stChoice{\stLab}{\stSi}\preType\stTi$,
	$\roleQ\stFmt{?}\stChoice{\stLab}{\stS}\preType\stT$,
	and $\stSi\tyNotSub\stS$, then
	$\compEnv{\mpS}{\stT}\stEnvComp\stEnvMap{\mpChanRole{\mpS}{\roleP}}{\stTi}
	\gtMoveStar\stEnvi$ with a payload mismatch
	of receiving $\stSi$ when $\stS$ is sent;\\
	(9) if $\stT=\stEnd$ and $\stTi$ is a sum, or
	$\stTi=\stEnd$ and $\stT$ is a sum,
	then\\
	$\compEnv{\mpS}{\stT}\stEnvComp\stEnvMap{\mpChanRole{\mpS}{\roleP}}{\stTi}
	\gtMoveStar\stEnvi$ and
	$\threeparticipantlock{\stEnvi}{\mpS}$.
\end{restatable}

\begin{proof}[1]
	This is immediate from \cref{lem:char-substitution}.
\qed\end{proof}

\begin{proof}[2]
	\[\begin{array}{cl}
	&\compEnv{\mpS}{\stEnd}\\
	=&\bigcup_{0\leq i\leq n}\set{\stEnvMap{\mpChanRole{\mpS}{\roleR[i]}}{\charTwait{\roleR[i]}}}
	\stEnvComp
	\stEnvMap{\mpChanRole{\mpS}{\roleFmt{sch}}}{
	\roleR[0]!\stLab[\text{\tiny end}]\stSeq\dots
	\roleR[n]!\stLab[\text{\tiny end}]
	}
	\\
	\gtMove&
	\bigcup_{0\leq i\leq 0}\set{\stEnvMap{\mpChanRole{\mpS}{\roleR[i]}}{\stEnd}}
	\cup
	\bigcup_{1\leq i\leq n}\set{\stEnvMap{\mpChanRole{\mpS}{\roleR[i]}}{\charTwait{\roleR[i]}}}
	\stEnvComp
	\stEnvMap{\mpChanRole{\mpS}{\roleFmt{sch}}}{
	\roleR[1]!\stLab[\text{\tiny end}]\stSeq\dots
	\roleR[n]!\stLab[\text{\tiny end}]
	}
	\\
	\gtMove&
	\bigcup_{0\leq i\leq 1}\set{\stEnvMap{\mpChanRole{\mpS}{\roleR[i]}}{\stEnd}}
	\cup
	\bigcup_{2\leq i\leq n}\set{\stEnvMap{\mpChanRole{\mpS}{\roleR[i]}}{\charTwait{\roleR[i]}}}
	\stEnvComp
	\stEnvMap{\mpChanRole{\mpS}{\roleFmt{sch}}}{
	\roleR[2]!\stLab[\text{\tiny end}]\stSeq\dots
	\roleR[n]!\stLab[\text{\tiny end}]
	}
	\\
	\gtMove&\dots\\
	\gtMove&
	\bigcup_{0\leq i\leq j-1}\set{\stEnvMap{\mpChanRole{\mpS}{\roleR[i]}}{\stEnd}}
	\cup
	\bigcup_{j\leq i\leq n}\set{\stEnvMap{\mpChanRole{\mpS}{\roleR[i]}}{\charTwait{\roleR[i]}}}
	\stEnvComp
	\stEnvMap{\mpChanRole{\mpS}{\roleFmt{sch}}}{
	\roleR[j]!\stLab[\text{\tiny end}]\stSeq\dots
	\roleR[n]!\stLab[\text{\tiny end}]
	}
	\\
	\gtMove&\dots\\
	\gtMove&
	\bigcup_{0\leq i\leq n-1}\set{\stEnvMap{\mpChanRole{\mpS}{\roleR[i]}}{\stEnd}}
	\cup
	\bigcup_{n\leq i\leq n}\set{\stEnvMap{\mpChanRole{\mpS}{\roleR[i]}}{\charTwait{\roleR[i]}}}
	\stEnvComp
	\stEnvMap{\mpChanRole{\mpS}{\roleFmt{sch}}}{
	\roleR[n]!\stLab[\text{\tiny end}]
	}
	\\
	\gtMove&
	\bigcup_{0\leq i\leq n}\set{\stEnvMap{\mpChanRole{\mpS}{\roleR[i]}}{\stEnd}}
	\stEnvComp
	\stEnvMap{\mpChanRole{\mpS}{\roleFmt{sch}}}{
	\stEnd
	}
	\\
	\end{array}
	\]
\qed\end{proof}

\begin{proof}[3]
	Suppose that
	\[\stT=\stSum{\roleQ}{(\roleQ,\dagger)\in J,\;i\in I_{(\roleQ,\dagger)}}{
						\stChoice{\stLab[i]}{\stS[i]}\stSeq\stT[i,\roleQ,\dagger]
				}{\dagger}
\]
	Pick $(\roleQ,\dagger)\in J$.
	Pick $(\roleQi,\dagger')\not\in J$.
	Let
	\[
		\stTi = \stFmt{
						\sum_{(\roleQ,\dagger)\in J}
						{\charSchTrigger{\roleQ}{I_{(\roleQ,\dagger)}}{\overline{\dagger}}{\{\charSch{\stT[i,\roleQ,\dagger]}\}_{i\in I_{(\roleQ,\dagger)}}}}
					}
	\]
	\[\begin{array}{cl}
	&\compEnv{\mpS}{\stT}\\
	=&
	\bigcup_{0\leq j \leq n}
	\set{\stEnvMap{\mpChanRole{\mpS}{\roleR[j]}}{\charTwait{\roleR[j]}}}
	\stEnvComp
	\stEnvMap{\mpChanRole{\mpS}{\roleFmt{sch}}}{\charSch{\stT}}
	\\
	\gtMove&
	\bigcup_{0\leq j \leq n,\,\roleR[j]\neq\roleQi}
	\set{\stEnvMap{\mpChanRole{\mpS}{\roleR[j]}}{\charTwait{\roleR[j]}}}
	\stEnvComp
	\stEnvMap{\mpChanRole{\mpS}{\roleFmt{sch}}}{
		\roleQi?\stLab\stSeq
						\sum_{(\roleQ,\dagger)\in J}
						{\stTi}
	}
	\stEnvComp
	\stEnvMap{\mpChanRole{\mpS}{\roleQi}}{\charTActi{\overline{\dagger}}{
		\charTwait{\roleQi}
	}}
	\\
	\gtMove&
	\bigcup_{0\leq j \leq n}
	\set{\stEnvMap{\mpChanRole{\mpS}{\roleR[j]}}{\charTwait{\roleR[j]}}}
	\stEnvComp
	\stEnvMap{\mpChanRole{\mpS}{\roleFmt{sch}}}{\stTi}
	\\[2ex]
	\gtMove&
	\bigcup_{0\leq j \leq n,\,\roleR[j]\neq\roleQ}
	\set{\stEnvMap{\mpChanRole{\mpS}{\roleR[j]}}{\charTwait{\roleR[j]}}}
	\stEnvComp\\&
	\stEnvMap{\mpChanRole{\mpS}{\roleFmt{sch}}}{\syncSch{\roleQ}{
		\stSum{\roleQ}{i\in I_{(\roleQ,\dagger)}}{
			\stLab[i]\stSeq\charSch{\stT[i,\roleQ,\dagger]}
		}{?}
	}}
	\stEnvComp
	\stEnvMap{\mpChanRole{\mpS}{\roleQ}}{
		\charTAct{\roleQ}{I_{(\roleQ,\dagger)}}{\overline{\dagger}}{\charTwait{\roleQ}}
	}
	\\[2ex]\gtMoveStar&
	\bigcup_{0\leq j \leq n,\,\roleR[j]\neq\roleQ}
	\set{\stEnvMap{\mpChanRole{\mpS}{\roleR[j]}}{\syncBra{\roleQ}{\charTwait{\roleR[j]}}}}
	\stEnvComp\\&
	\stEnvMap{\mpChanRole{\mpS}{\roleFmt{sch}}}{
		\stSum{\roleQ}{i\in I_{(\roleQ,\dagger)}}{
			\stLab[i]\stSeq\charSch{\stT[i,\roleQ,\dagger]}
		}{?}
	}
	\stEnvComp
	\stEnvMap{\mpChanRole{\mpS}{\roleQ}}{
		\charTAct{\roleQ}{I_{(\roleQ,\dagger)}}{\overline{\dagger}}{\charTwait{\roleQ}}
	}
	\end{array}
	\]
	So for $k\in I_{(\roleQ,\dagger)}$,
	\[
		\begin{array}{cl}
		&
		\compEnv{\mpS}{\stT}\stEnvComp
		\stEnvMap{\mpChanRole{\mpS}{\roleP}}{\stT}
		\\[2ex]
		\gtMoveStar&
	\bigcup_{0\leq j \leq n,\,\roleR[j]\neq\roleQ}\set{
	\stEnvMap{\mpChanRole{\mpS}{\roleR[j]}}{\syncBra{\roleQ}{\charTwait{\roleR[j]}}}}
	\stEnvComp\\&
	\stEnvMap{\mpChanRole{\mpS}{\roleP}}{\stT[k,\roleQ,\dagger]}
		\stEnvComp
		\stEnvMap{\mpChanRole{\mpS}{\roleFmt{sch}}}{
		\stSum{\roleQ}{i\in I_{(\roleQ,\dagger)}}{
			\stLab[i]\stSeq\charSch{\stT[i,\roleQ,\dagger]}
		}{?}
	}
	\stEnvComp
	\stEnvMap{\mpChanRole{\mpS}{\roleQ}}{
		\syncSel{\roleQ}{
						\roleFmt{sch}\stFmt{!}\stLab[k]\stSeq\charTwait{\roleQ}
					}
	}
	\\[2ex]
	\gtMoveStar&
	\bigcup_{0\leq j \leq n,\,\roleR[j]\neq\roleQ}
	\set{\stEnvMap{\mpChanRole{\mpS}{\roleR[j]}}{\charTwait{\roleR[j]}}}
	\stEnvComp\\&
	\stEnvMap{\mpChanRole{\mpS}{\roleP}}{\stT[k,\roleQ,\dagger]}
		\stEnvComp
		\stEnvMap{\mpChanRole{\mpS}{\roleFmt{sch}}}{
		\stSum{\roleQ}{i\in I_{(\roleQ,\dagger)}}{
			\stLab[i]\stSeq\charSch{\stT[i,\roleQ,\dagger]}
		}{?}
	}
	\stEnvComp
	\stEnvMap{\mpChanRole{\mpS}{\roleQ}}{
						\roleFmt{sch}\stFmt{!}\stLab[k]\stSeq\charTwait{\roleQ}
	}
	\\[2ex]\gtMove&
	\bigcup_{0\leq j \leq n}\set{
	\stEnvMap{\mpChanRole{\mpS}{\roleR[j]}}{\charTwait{\roleR[j]}}}
	\stEnvComp\\&
	\stEnvMap{\mpChanRole{\mpS}{\roleP}}{\stT[k,\roleQ,\dagger]}
		\stEnvComp
		\stEnvMap{\mpChanRole{\mpS}{\roleFmt{sch}}}{
			\charSch{\stT[k,\roleQ,\dagger]}
	}
	\\[2ex]=&\compEnv{\mpS}{\stT[k,\roleQ,\dagger]}
	\stEnvComp
	\stEnvMap{\mpChanRole{\mpS}{\roleP}}{\stT[k,\roleQ,\dagger]}
		\end{array}
	\]
\qed\end{proof}

\begin{proof}[4]
	Write $\stT$ as
	\[\stT=\stSum{\roleQ}{(\roleQ,\dagger)\in J,\;i\in I_{(\roleQ,\dagger)}}{
						\stChoice{\stLab[i]}{\stS[i]}\stSeq\stT[i,\roleQ,\dagger]
				}{\dagger}
\]
	Let $(\roleQ,\dagger)\in J$
	such that $\roleQ\stFmt{\dagger}\not\preType\stTi$.
	\[\begin{array}{cl}
	&\compEnv{\mpS}{\stT}
	\stEnvComp
	\stEnvMap{\mpChanRole{\mpS}{\roleP}}{\stTi}
	\\[2ex]\gtMoveStar&
	\bigcup_{0\leq j \leq n,\,\roleR[j]\neq\roleQ}
	\set{\stEnvMap{\mpChanRole{\mpS}{\roleR[j]}}{\syncBra{\roleQ}{\charTwait{\roleR[j]}}}}
	\stEnvComp
	\stEnvMap{\mpChanRole{\mpS}{\roleQ}}{
		\charTAct{\roleQ}{I_{(\roleQ,\dagger)}}{\overline{\dagger}}{\charTwait{\roleQ}}
	}
	\stEnvComp
	\\&
	\stEnvMap{\mpChanRole{\mpS}{\roleP}}{\stTi}
	\stEnvComp
	\stEnvMap{\mpChanRole{\mpS}{\roleFmt{sch}}}{
		\stSum{\roleQ}{i\in I_{(\roleQ,\dagger)}}{
			\stLab[i]\stSeq\charSch{\stT[i,\roleQ,\dagger]}
		}{?}
	}
	\end{array}
	\]
	Write $\stTi=\stSum{\roleQi[i]}{i\in I}{\stChoice{\stLabi[i]}{\stSi[i]}\stSeq\stTi[i]}{\dagger'_i}$.
	For $i\in I$,
	if $\roleQi[i]=\roleQ$ then $\dagger'_i\neq\dagger$
	and the only action of $\charTAct{\roleQ}{I_{(\roleQ,\dagger)}}{\overline{\dagger}}{\charTwait{\roleQ}}$ is $\roleP\stFmt{\overline{\dagger}}=\roleP\stFmt{\dagger'_i}$.
	For $i\in I$,
	if $\roleQi[i]\neq\roleQ$ then
	the only action of 
	$\syncBra{\roleQ}{\charTwait{\roleR[j]}}$ is
	$\roleQ\stFmt{?}$
	and
	the only action of $\charTAct{\roleQ}{I_{(\roleQ,\dagger)}}{\overline{\dagger}}{\charTwait{\roleQ}}$ is $\roleP\stFmt{\overline{\dagger}}$.
	
	So, this context has a 3-plock on $\mpS$.
\qed\end{proof}

\begin{proof}[5 and 6]
	Write $\stT$ as
	\[\stT=\stSum{\roleQ}{(\roleQ,\dagger)\in J,\;i\in I_{(\roleQ,\dagger)}}{
						\stChoice{\stLab[i]}{\stS[i]}\stSeq\stT[i,\roleQ,\dagger]
				}{\dagger}
\]
	Let $(\roleQ,\dagger)\not\in J$
	such that $\roleQ\stFmt{\dagger}\preType\stTi$.
	\[\begin{array}{cl}
	&\compEnv{\mpS}{\stT}
	\stEnvComp
	\stEnvMap{\mpChanRole{\mpS}{\roleP}}{\stTi}
	\\\gtMoveStar&
	\bigcup_{0\leq j \leq n,\,\roleR[j]\neq\roleQ}
	\set{\stEnvMap{\mpChanRole{\mpS}{\roleR[j]}}{\charTwait{\roleR[j]}}}
	\stEnvComp\\&
	\stEnvMap{\mpChanRole{\mpS}{\roleP}}{\stTi}
	\stEnvComp
	\stEnvMap{\mpChanRole{\mpS}{\roleFmt{sch}}}{
		\roleQ?\stLab\stSeq
						\sum_{(\roleQi,\dagger')\in J}
						{\stTi}
	}
	\stEnvComp\\[2ex]&
	\stEnvMap{\mpChanRole{\mpS}{\roleQ}}{
	\roleFmt{sch}!\stLab\stSeq\charTwait{\roleQ}
					+
	\roleP{\overline{\dagger}}\stLabii
	}
	\end{array}
	\]
	This is a label mismatch, as:\\
	if $\stFmt{\dagger}=\stFmt{!}$
	and $\roleQ\stFmt{!}\stLabiii\preType\stTi$, then 
	$\stLabii\neq\stLabiii$ as $\stLabii$ does not appear in $\mathcal{T}$-valid types;\\
	if $\stFmt{\dagger}=\stFmt{?}$ then
	$\roleQ\stFmt{?}\preType\stTi$
	and
	$\roleQ\stFmt{?}\stLabii\not\preType\stTi$
	as $\stLabii$ does not appear in $\mathcal{T}$-valid types.
	
	Let $(\roleQ,\dagger)\in J$
	such that $\roleQ\stFmt{\dagger}\preType\stTi$.
	\[\begin{array}{cl}
	&\stEnvMap{\mpChanRole{\mpS}{\roleP}}{\stTi}\stEnvComp\compEnv{\mpS}{\stT}
	\\[2ex]\gtMoveStar&
	\stEnvMap{\mpChanRole{\mpS}{\roleP}}{\stTi}
	\stEnvComp
	\stEnvMap{\mpChanRole{\mpS}{\roleFmt{sch}}}{
		\stSum{\roleQ}{i\in I_{(\roleQ,\dagger)}}{
			\stLab[i]\stSeq\charSch{\stT[i,\roleQ,\dagger]}
		}{?}
	}
	\stEnvComp\\&
	\stEnvMap{\mpChanRole{\mpS}{\roleQ}}{
		\charTAct{\roleQ}{I_{(\roleQ,\dagger)}}{\overline{\dagger}}{\charTwait{\roleQ}}
	}
	\stEnvComp
	\bigcup_{0\leq j \leq n,\,\roleR[j]\neq\roleQ}
	\stEnvMap{\mpChanRole{\mpS}{\roleR[j]}}{\syncBra{\roleQ}{\charTwait{\roleR[j]}}}
	\end{array}
	\]
	
	If $\stFmt{\dagger}=\stFmt{!}$, $\roleQ\stFmt{!}\stLabii\preType\stTi$,
	and $\stLabii\not\in\{\stLab[i]\}_{i\in I_{(\roleQ,!)}}$,
	then there is a label mismatch between $\mpChanRole{\mpS}{\roleP}$
	and $\mpChanRole{\mpS}{\roleQ}$,
	since $\charTAct{\roleQ}{I_{(\roleQ,!)}}{?}{\charTwait{\roleQ}}$
	can receive from $\roleP$ but cannot receive $\stLabii$.
	
	If $\stFmt{\dagger}=\stFmt{?}$, $k\in I_{(\roleQ,?)}$, and
	$\roleQ\stFmt{?}\stLab[k]\not\preType\stTi$,
	then there is a label mismatch between $\mpChanRole{\mpS}{\roleP}$
	and $\mpChanRole{\mpS}{\roleQ}$,
	since $\charTAct{\roleQ}{I_{(\roleQ,?)}}{!}{\charTwait{\roleQ}}$
	can send $\stLab[k]$ to $\roleP$
	and $\stTii$ can receive from $\roleQ$ but cannot receive $\stLab[k]$.
\qed\end{proof}

\begin{proof}[7 and 8]
	Write $\stT$ as
	\[\stT=\stSum{\roleQ}{(\roleQ,\dagger)\in J,\;i\in I_{(\roleQ,\dagger)}}{
						\stChoice{\stLab[i]}{\stS[i]}\stSeq\stT[i,\roleQ,\dagger]
				}{\dagger}
\]
	Let $(\roleQ,\dagger)\in J$ and $k\in I_{(\roleQ,\dagger)}$
	such that $\roleQ\stFmt{\dagger}\stLab[k]\preType\stTi$.
	\[\begin{array}{cl}
	&\compEnv{\mpS}{\stT}
	\stEnvComp
	\stEnvMap{\mpChanRole{\mpS}{\roleP}}{\stTi}
	\\[2ex]\gtMoveStar&
	\bigcup_{0\leq j \leq n,\,\roleR[j]\neq\roleQ}
	\set{\stEnvMap{\mpChanRole{\mpS}{\roleR[j]}}{\syncBra{\roleQ}{\charTwait{\roleR[j]}}}}
	\stEnvComp
	\stEnvMap{\mpChanRole{\mpS}{\roleQ}}{
		\charTAct{\roleQ}{I_{(\roleQ,\dagger)}}{\overline{\dagger}}{\charTwait{\roleQ}}
	}
	\stEnvComp\\&
	\stEnvMap{\mpChanRole{\mpS}{\roleP}}{\stTi}
	\stEnvComp
	\stEnvMap{\mpChanRole{\mpS}{\roleFmt{sch}}}{
		\stSum{\roleQ}{i\in I_{(\roleQ,\dagger)}}{
			\stLab[i]\stSeq\charSch{\stT[i,\roleQ,\dagger]}
		}{?}
	}
	\end{array}
	\]
	
	If $\stFmt{\dagger}=\stFmt{!}$, $\roleQ\stFmt{!}\stChoice{\stLab[k]}{\stSi}\preType\stTi$,
	and $\stS[k]\tyNotSub\stSi$,
	then there is a payload mismatch between $\mpChanRole{\mpS}{\roleP}$
	and $\mpChanRole{\mpS}{\roleQ}$
	of sending $\stSi$ and receiving $\stS[k]$,
	since $\charTAct{\roleQ}{I_{(\roleQ,!)}}{?}{\charTwait{\roleQ}}$
	can receive
	$\stChoice{\stLab[k]}{\stS[k]}$
	from $\roleP$
	and
	$\stTi$ can send $\stChoice{\stLab[k]}{\stSi}$
	to $\roleQ$
	but $\stS[k]\tyNotSub\stSi$.
	
	If $\stFmt{\dagger}=\stFmt{?}$, $\roleQ\stFmt{?}\stChoice{\stLab[k]}{\stSi}\preType\stTi$,
	and $\stSi\tyNotSub\stS[k]$,
	then there is a payload mismatch between $\mpChanRole{\mpS}{\roleP}$
	and $\mpChanRole{\mpS}{\roleQ}$
	of sending $\stS[k]$ and receiving $\stSi$,
	since $\charTAct{\roleQ}{I_{(\roleQ,?)}}{!}{\charTwait{\roleQ}}$
	can send
	$\stChoice{\stLab[k]}{\stS[k]}$
	to $\roleP$
	and
	$\stTi$ can receive $\stChoice{\stLab[k]}{\stSi}$
	from $\roleQ$
	but $\stSi\tyNotSub\stS[k]$.
\qed\end{proof}

\begin{proof}[9]
	Suppose that $\stTi=\stEnd$ and $\stT$ is a sum.
	Write $\stT$ as
	\[\stT=\stSum{\roleQ}{(\roleQ,\dagger)\in J,\;i\in I_{(\roleQ,\dagger)}}{
						\stChoice{\stLab[i]}{\stS[i]}\stSeq\stT[i,\roleQ,\dagger]
				}{\dagger}
\]
	Let $(\roleQ,\dagger)\in J$.
	\[\begin{array}{cl}
	&
	\compEnv{\mpS}{\stT}
	\stEnvComp
	\stEnvMap{\mpChanRole{\mpS}{\roleP}}{\stTi}
	\\[2ex]\gtMoveStar&
	\bigcup_{0\leq j \leq n,\,\roleR[j]\neq\roleQ}
	\set{\stEnvMap{\mpChanRole{\mpS}{\roleR[j]}}{\syncBra{\roleQ}{\charTwait{\roleR[j]}}}}
	\stEnvComp
	\stEnvMap{\mpChanRole{\mpS}{\roleQ}}{
		\charTAct{\roleQ}{I_{(\roleQ,\dagger)}}{\overline{\dagger}}{\charTwait{\roleQ}}
	}
	\stEnvComp\\&
	\stEnvMap{\mpChanRole{\mpS}{\roleP}}{\stTi}
	\stEnvComp
	\stEnvMap{\mpChanRole{\mpS}{\roleFmt{sch}}}{
		\stSum{\roleQ}{i\in I_{(\roleQ,\dagger)}}{
			\stLab[i]\stSeq\charSch{\stT[i,\roleQ,\dagger]}
		}{?}
	}
	\end{array}
	\]
	The only action of $\charTAct{\roleQ}{I_{(\roleQ,\dagger)}}{\overline{\dagger}}{\charTwait{\roleQ}}$ is $\roleP\stFmt{\overline{\dagger}}$.
	So, this is a three-party lock.
	
	Suppose that $\stT=\stEnd$ and $\stTi$ is a sum.
	\[
	\begin{array}{cl}
	&\compEnv{\mpS}{\stT}
	\stEnvComp
	\stEnvMap{\mpChanRole{\mpS}{\roleP}}{\stTi}
	\\[2ex]\gtMoveStar&
	\bigcup_{0\leq j \leq n}\set{
	\stEnvMap{\mpChanRole{\mpS}{\roleR[j]}}{\stEnd}}
	\stEnvComp
	\stEnvMap{\mpChanRole{\mpS}{\roleP}}{\stTi}
	\stEnvComp
	\stEnvMap{\mpChanRole{\mpS}{\roleFmt{sch}}}{
		\stEnd
	}
	\end{array}
	\]
	This has a 3-plock on $\mpS$.
\qed\end{proof}

\thmCharStrict*

\begin{proof}
	Suppose that $\stTi\tyNotSub\stT$ are $\mathcal{T}$-valid.
	So, $\stTi\tyNot\stT$.
	We proceed by induction on the derivation of $\stTi\tyNot\stT$,
	using \cref{thm:char-fid} and \cref{thm:comp-fid}
	at the inductive and base cases.
	
	Consider the last rule in the derivation:\\
	\inferrule{$\mu$L}.
	$\stTi=\stRec{\stRecVar}{\stTii}$
	and $\tySubst{\stTii}{\stRecVar}{\stTi}\tyNot\stT$.
	
	By I.H.,
	\[
		\begin{array}{cl}
			&\mpRes{\mpS}{\left(
			\charPM{\compEnv{\mpS}{\stT}}
			\mpPar
			\charP{\mpChanRole{\mpS}{\roleP}}{\stTi}
			\right)}
			\\
			\prestruct&
			\mpRes{\mpS}{\left(
			\charPM{\compEnv{\mpS}{\stT}}
			\mpPar
			\charP{\mpChanRole{\mpS}{\roleP}}{\tySubst{\stTii}{\stRecVar}{\stTi}}
			\right)}
		\end{array}
	\]
	is not safe or is not 3-plock free.\\
	\inferrule{$\mu$R}.
	$\stT=\stRec{\stRecVar}{\stTii}$
	and $\stTi\tyNot\tySubst{\stTii}{\stRecVar}{\stT}$.
	
	By I.H.,
	\[
		\begin{array}{cl}
			&\mpRes{\mpS}{\left(
			\charPM{\compEnv{\mpS}{\stT}}
			\mpPar
			\charP{\mpChanRole{\mpS}{\roleP}}{\stTi}
			\right)}
			\\
			\prestruct&
			\mpRes{\mpS}{\left(
			\charPM{\compEnv{\mpS}{\tySubst{\stTii}{\stRecVar}{\stT}}}
			\mpPar
			\charP{\mpChanRole{\mpS}{\roleP}}{\stTi}
			\right)}
		\end{array}
	\]
	is not safe or is not 3-plock free.\\
	\inferrule{$!$L}.
	$\roleQ\stFmt{!}\stLab\preType\stTi$
	and
	$\roleQ\stFmt{!}\stLab\not\preType\stT$.
	
	By \cref{thm:comp-fid},
	$\compEnv{\mpS}{\stT}\stEnvComp\stEnvMap{\mpChanRole{\mpS}{\roleP}}{\stTi}
	\gtMoveStar\stEnv$
	with a label mismatch.
	
	By \cref{thm:char-fid},
	$\charPM{\compEnv{\mpS}{\stT}\stEnvComp\stEnvMap{\mpChanRole{\mpS}{\roleP}}{\stTi}}
	\mpMoveStar\mpQ\mpPar\charPM{\stEnv}
	\mpMoveStar\mpQ\mpPar\mpCtxApp{\mpCtx}{\mpErr}$.
	
	So $\charPM{\compEnv{\mpS}{\stT}\stEnvComp\stEnvMap{\mpChanRole{\mpS}{\roleP}}{\stTi}}$ is not safe.\\
	\inferrule{$?$L}.
	$\roleQ\stFmt{?}\preType\stTi$
	and
	$\roleQ\stFmt{?}\not\preType\stT$.
	
	By \cref{thm:comp-fid},
	$\compEnv{\mpS}{\stT}\stEnvComp\stEnvMap{\mpChanRole{\mpS}{\roleP}}{\stTi}
	\gtMoveStar\stEnv$
	with a label mismatch.
	
	By \cref{thm:char-fid},
	$\charPM{\compEnv{\mpS}{\stT}\stEnvComp\stEnvMap{\mpChanRole{\mpS}{\roleP}}{\stTi}}
	\mpMoveStar\mpQ\mpPar\charPM{\stEnv}
	\mpMoveStar\mpQ\mpPar\mpCtxApp{\mpCtx}{\mpErr}$.
	
	So $\charPM{\compEnv{\mpS}{\stT}\stEnvComp\stEnvMap{\mpChanRole{\mpS}{\roleP}}{\stTi}}$ is not safe.\\
	\inferrule{$!$R}.
	$\roleQ\stFmt{!}\preType\stT$
	and
	$\roleQ\stFmt{!}\not\preType\stTi$.
	
	By \cref{thm:comp-fid},
	$\compEnv{\mpS}{\stT}\stEnvComp\stEnvMap{\mpChanRole{\mpS}{\roleP}}{\stTi}
	\gtMoveStar\stEnv$
	which has a 3-plock.
	
	By \cref{thm:char-fid},
	$\mpRes{\mpS}{\charPM{\compEnv{\mpS}{\stT}\stEnvComp\stEnvMap{\mpChanRole{\mpS}{\roleP}}{\stTi}}}
	\mpMoveStar\mpRes{\mpS}{\left(\mpQ\mpPar\charPM{\stEnv}\right)}$
	which has a 3-plock.\\
	\inferrule{$?$R}.
	$\roleQ\stFmt{?}\stLab\preType\stT$
	and
	$\roleQ\stFmt{?}\stLab\not\preType\stTi$.
	
	If $\roleQ\stFmt{?}\not\preType\stTi$ then
	by \cref{thm:comp-fid},
	$\compEnv{\mpS}{\stT}\stEnvComp\stEnvMap{\mpChanRole{\mpS}{\roleP}}{\stTi}
	\gtMoveStar\stEnv$
	which has a 3-plock.
	
	By \cref{thm:char-fid},
	$\mpRes{\mpS}{\charPM{\compEnv{\mpS}{\stT}\stEnvComp\stEnvMap{\mpChanRole{\mpS}{\roleP}}{\stTi}}}
	\mpMoveStar\mpRes{\mpS}{\left(\mpQ\mpPar\charPM{\stEnv}\right)}$
	which has a 3-plock.
	
	If $\roleQ\stFmt{?}\preType\stTi$
	then by \cref{thm:comp-fid},
	$\compEnv{\mpS}{\stT}\stEnvComp\stEnvMap{\mpChanRole{\mpS}{\roleP}}{\stTi}
	\gtMoveStar\stEnv$
	with a label mismatch.
	
	By \cref{thm:char-fid},
	$\charPM{\compEnv{\mpS}{\stT}\stEnvComp\stEnvMap{\mpChanRole{\mpS}{\roleP}}{\stTi}}
	\mpMoveStar\mpQ\mpPar\charPM{\stEnv}
	\mpMoveStar\mpQ\mpPar\mpCtxApp{\mpCtx}{\mpErr}$.
	
	So $\charPM{\compEnv{\mpS}{\stT}\stEnvComp\stEnvMap{\mpChanRole{\mpS}{\roleP}}{\stTi}}$ is not safe.\\
	\inferrule{$\stEnd$-L}.
	$\stTi=\stEnd$ and $\stT=\stSum{\roleQ[i]}{i\in I}{\stChoice{\stLab[i]}{\stS[i]}\stSeq\stT[i]}{\dagger_i}$ then
	by \cref{thm:comp-fid},\\
	$\compEnv{\mpS}{\stT}\stEnvComp\stEnvMap{\mpChanRole{\mpS}{\roleP}}{\stTi}
	\gtMoveStar\stEnv$
	which has a 3-plock.
	
	By \cref{thm:char-fid},
	$\mpRes{\mpS}{\charPM{\compEnv{\mpS}{\stT}\stEnvComp\stEnvMap{\mpChanRole{\mpS}{\roleP}}{\stTi}}}
	\mpMoveStar\mpRes{\mpS}{\left(\mpQ\mpPar\charPM{\stEnv}\right)}$
	which has a 3-plock.
	\\
	\inferrule{$\stEnd$-R}.
	$\stT=\stEnd$ and $\stTi=\stSum{\roleQ[i]}{i\in I}{\stChoice{\stLabiii[i]}{\stSiii[i]}\stSeq\stT[i]}{\dagger_i}$ then
	by \cref{thm:comp-fid},\\
	$\compEnv{\mpS}{\stT}\stEnvComp\stEnvMap{\mpChanRole{\mpS}{\roleP}}{\stTi}
	\gtMoveStar\stEnv$
	which has a 3-plock.
	
	By \cref{thm:char-fid},
	$\mpRes{\mpS}{\charPM{\compEnv{\mpS}{\stT}\stEnvComp\stEnvMap{\mpChanRole{\mpS}{\roleP}}{\stTi}}}
	\mpMoveStar\mpRes{\mpS}{\left(\mpQ\mpPar\charPM{\stEnv}\right)}$
	which has a 3-plock.
	\\
	\inferrule{LR?}.
	$\roleQ\stFmt{?}\stChoice{\stLab}{\stSi}\preType\stTi$,
	$\roleQ\stFmt{?}\stChoice{\stLab}{\stS}\preType\stT$, and
	$\stSi\tyNot\stS$ then
	by \cref{thm:comp-fid},\\
	$\compEnv{\mpS}{\stT}\stEnvComp\stEnvMap{\mpChanRole{\mpS}{\roleP}}{\stTi}
	\gtMoveStar\stEnv$
	with a payload mismatch, sending $\stS$ and receiving $\stSi$.
	
	If either $\stS$ or $\stSi$ are basic types then
	by \cref{thm:char-fid},
	$\charPM{\compEnv{\mpS}{\stT}\stEnvComp\stEnvMap{\mpChanRole{\mpS}{\roleP}}{\stTi}}
	\mpMoveStar\mpQ\mpPar\charPM{\stEnv}
	\mpMoveStar\mpQ\mpPar\mpCtxApp{\mpCtx}{\mpErr}$.
	
	So $\charPM{\compEnv{\mpS}{\stT}\stEnvComp\stEnvMap{\mpChanRole{\mpS}{\roleP}}{\stTi}}$ is not safe.
	
	If both $\stS$ and $\stSi$ are session types and $\stS\neq\stEnd$ then
	by \cref{thm:char-fid},
	\[\charPM{\compEnv{\mpS}{\stT}\stEnvComp\stEnvMap{\mpChanRole{\mpS}{\roleP}}{\stTi}}
	\mpMoveStar\mpQ\mpPar\mpRes{\mpSi}{\charPM{\compEnv{\mpSi}{\stS}\stEnvComp\stEnvMap{\mpChanRole{\mpSi}{\roleP}}{\stSi}}}\]
	By I.H.,
	$\mpRes{\mpSi}{\charPM{\compEnv{\mpSi}{\stS}\stEnvComp\stEnvMap{\mpChanRole{\mpSi}{\roleP}}{\stSi}}}$
	is not safe or is not 3-plock free.
	
	If $\stSi$ is a session type and $\stS=\stEnd$ then
	\[
	\begin{array}{cl}
		&\charPM{\compEnv{\mpS}{\stT}\stEnvComp\stEnvMap{\mpChanRole{\mpS}{\roleP}}{\stTi}}\\
	\mpMoveStar&\mpQ\mpPar\mpRes{\mpSi}{\charP{\mpChanRole{\mpSi}{\roleP}}{\stSi}}\\
	\prestruct&
	\mpQ\mpPar\mpRes{\mpSii[1],\dots,\mpSii[n],\mpSi}{\left(\mpSum{\mpPrefix_i\mpSeq\mpQ[i]}{i\in I}\mpPar\mpNil\right)}\\
	&\forall i\in I,\;\ltsSubject{\mpPrefix_i}=\mpChanRole{\mpSi}{\roleP}
	\;\text{and}\;
	\forall\roleR\in\roleSet,\;\mpChanRole{\mpSi}{\roleR}\not\in\chan{\mpSum{\mpPrefix_i\mpSeq\mpQ[i]}{i\in I}}
	\end{array}
	\]
	Since $\stSi\tyNot\stEnd$, this has a 3-plock.
	
	Therefore,
	$\charPM{\compEnv{\mpS}{\stT}\stEnvComp\stEnvMap{\mpChanRole{\mpS}{\roleP}}{\stTi}}$ is not safe or is not 3-plock free.
	\\
	\inferrule{LR!}.
	$\roleQ\stFmt{!}\stChoice{\stLab}{\stSi}\preType\stTi$,
	$\roleQ\stFmt{!}\stChoice{\stLab}{\stS}\preType\stT$, and
	$\stS\tyNot\stSi$ then
	by \cref{thm:comp-fid},\\
	$\compEnv{\mpS}{\stT}\stEnvComp\stEnvMap{\mpChanRole{\mpS}{\roleP}}{\stTi}
	\gtMoveStar\stEnv$
	with a payload mismatch, sending $\stSi$ and receiving $\stS$.
	
	If either $\stS$ or $\stSi$ are basic types then
	by \cref{thm:char-fid},
	\[\charPM{\compEnv{\mpS}{\stT}\stEnvComp\stEnvMap{\mpChanRole{\mpS}{\roleP}}{\stTi}}
	\mpMoveStar\mpQ\mpPar\charPM{\stEnv}
	\mpMoveStar\mpQ\mpPar\mpCtxApp{\mpCtx}{\mpErr}\]
	So $\charPM{\compEnv{\mpS}{\stT}\stEnvComp\stEnvMap{\mpChanRole{\mpS}{\roleP}}{\stTi}}$ is not safe.
	
	If both $\stS$ and $\stSi$ are session types and $\stSi\neq\stEnd$ then
	by \cref{thm:char-fid},
	\[\charPM{\compEnv{\mpS}{\stT}\stEnvComp\stEnvMap{\mpChanRole{\mpS}{\roleP}}{\stTi}}
	\mpMoveStar\mpQ\mpPar\mpRes{\mpSi}{\charPM{\compEnv{\mpSi}{\stSi}\stEnvComp\stEnvMap{\mpChanRole{\mpSi}{\roleP}}{\stS}}}\]
	
	By I.H.,
	$\mpRes{\mpSi}{\charPM{\compEnv{\mpSi}{\stSi}\stEnvComp\stEnvMap{\mpChanRole{\mpSi}{\roleP}}{\stS}}}$
	is not safe or is not 3-plock free.
	
	If $\stS$ is a session type and $\stSi=\stEnd$ then
	\[
	\begin{array}{cl}
		&\charPM{\compEnv{\mpS}{\stT}\stEnvComp\stEnvMap{\mpChanRole{\mpS}{\roleP}}{\stTi}}\\
	\mpMoveStar&\mpQ\mpPar\mpRes{\mpSi}{\charP{\mpChanRole{\mpSi}{\roleP}}{\stS}}\\
	\prestruct&
	\mpQ\mpPar\mpRes{\mpSii[1],\dots,\mpSii[n],\mpSi}{\left(\mpSum{\mpPrefix_i\mpSeq\mpQ[i]}{i\in I}\mpPar\mpNil\right)}\\
	&\forall i\in I,\;\ltsSubject{\mpPrefix_i}=\mpChanRole{\mpSi}{\roleP}
	\;\text{and}\;
	\forall\roleR\in\roleSet,\;\mpChanRole{\mpSi}{\roleR}\not\in\chan{\mpSum{\mpPrefix_i\mpSeq\mpQ[i]}{i\in I}}
	\end{array}
	\]
	Since $\stS\tyNot\stEnd$, this has a 3-plock.
	
	Therefore,
	$\charPM{\compEnv{\mpS}{\stT}\stEnvComp\stEnvMap{\mpChanRole{\mpS}{\roleP}}{\stTi}}$ is not safe or is not 3-plock free.\\
	\inferrule{LR}.
	$\roleQ\stFmt{\dagger}\stChoice{\stLab}{\stSi}\stSeq\stTiii\preType\stTi$,
	$\roleQ\stFmt{\dagger}\stChoice{\stLab}{\stS}\stSeq\stTii\preType\stT$, and
	$\stTiii\tyNot\stTii$.
	
	If one of $\stS$ and $\stSi$ are basic types and
	\inferrule{LR?} or \inferrule{LR!} apply,
	then we derive an error as above without appealing to the I.H.
	
	Otherwise,
	\[\begin{array}{cl}
	&\mpRes{\mpS}{\charPM{\compEnv{\mpS}{\stT}\stEnvComp\stEnvMap{\mpChanRole{\mpS}{\roleP}}{\stTi}}}\\
	\mpMoveStar&
	\mpQ\mpPar\mpRes{\mpS}{\charPM{\compEnv{\mpS}{\stTii}\stEnvComp\stEnvMap{\mpChanRole{\mpS}{\roleP}}{\stTiii}}}
	\end{array}\]
	By I.H., this is not safe or is not 3-plock free.
\qed\end{proof}

\section{Appendix for the Extensions (\S~5)}

\label{app:res-type}
\subsection{Restricted Typing System}

\begin{lemma}[Inversion]
	Suppose that $\tyJudge{\tyEnv}{\mpP}{\stEnv}$.
	\begin{enumerate}
		\item If $\mpP=\mpNil$, then $\stEnvEndP{\stEnv}$.
		\item If $\mpP=\mpPi\mpPar\mpPii$, then
		$\tyJudgeRes{\tyEnv}{\mpPi}{\stEnvi}$ and
		$\tyJudgeRes{\tyEnv}{\mpPii}{\stEnvii}$, where
		$\stEnvi=\stEnv\restriction\chan{\mpPi}$,
		$\stEnvii=\stEnv\restriction\chan{\mpPii}$ and
		$\dom{\stEnvi}\cap\dom{\stEnvii}=\emptyset$.
		\item If $\mpP=\mpRes{\mpS}{\mpPi}$, then
		there exists $\stEnvi$ with
		$\stEnv=\stEnvi\setminus\mpS$,
		$\predPApp{\stEnvi[\mpS]}$ and
		$\tyJudgeRes{\tyEnv}{\mpPi}{\stEnvi}$.
		\item If $\mpP=\mpIf{e}{\mpPi}{\mpPii}$, then
		$\tyJudgeRes{\tyEnv}{e}{\tyBool}$,
		$\tyJudgeRes{\tyEnv}{\mpPi}{\stEnv}$, and
		$\tyJudgeRes{\tyEnv}{\mpPii}{\stEnv}$.
		\item If $\mpP=\mpSum{\mpPrefix_i\mpSeq\mpP[i]}{i\in I}$,
		then there exists $\mpC$ and
		$\stEnvi$ such that:
		$\{\ltsSubject{\mpPrefix_i}\suchthat i\in I\}=\{\mpC\}$;
		$\stEnvi\tySub\stEnv$;
		$\stEnvEndP{\stEnvi\setminus\mpC}$;
		if $\mpPrefix_i=\mpSel{\mpC}{\roleQ}{\stLab}{\mpCi}{}$, then
		$\stFmt{\roleQ{!}\stChoice{\stLab}{\stEnvApp{\stEnvi}{\mpCi}}\stSeq\stTi}\preType\stEnvApp{\stEnvi}{\mpC}$,
		$\mpC\neq\mpCi$, and
		$\tyJudge{\tyEnv}{\mpP[i]}{(\stEnvi\setminus\mpC\setminus\mpCi)\stEnvComp\stEnvMap{\mpC}{\stTi}}$;
		if $\mpPrefix_i=\mpSel{\mpC}{\roleQ}{\stLab}{e}{}$, then
		$\stFmt{\roleQ{!}\stChoice{\stLab}{\tyGround}\stSeq\stTi}
		\preType\stEnvApp{\stEnvi}{\mpC}$,
		$\tyJudge{\tyEnv}{e}{\tyGround}$, and
		$\tyJudge{\tyEnv}{\mpP[i]}{(\stEnvi\setminus\mpC)\stEnvComp\stEnvMap{\mpC}{\stTi}}$;
		if $\mpPrefix_i=\mpBra{\mpC}{\roleQ}{\stLab}{\mpx}{}$, then
		$\stFmt{\roleQ{?}\stChoice{\stLab}{\tyGround}\stSeq\stTi}\preType\stEnvApp{\stEnvi}{\mpC}$, and
		$\tyJudge{\tyEnv\tyEnvComp\tyEnvMap{\mpx}{\tyGround}}{\mpP[i]}{(\stEnvi\setminus\mpC)\stEnvComp\stEnvMap{\mpC}{\stTi}}$;
		if $\mpPrefix_i=\mpBra{\mpC}{\roleQ}{\stLab}{\mpy}{}$, then
		$\stFmt{\roleQ{?}\stChoice{\stLab}{\stT}\stSeq\stTi}\preType\stEnvApp{\stEnvi}{\mpC}$, and
		$\tyJudge{\tyEnv}{\mpP[i]}{(\stEnvi\setminus\mpC)\stEnvComp\stEnvMap{\mpC}{\stTi}
		\stEnvComp\stEnvMap{\mpy}{\stT}}$;
		if $\stFmt{\roleQ{\dagger}\stChoice{\stLab}{}}\preType\stEnvApp{\stEnvi}{\mpC}$,
		then $\mpFmt{\mpChanRole{\mpC}{\roleQ}{\dagger}\stLab}\preCalc\mpP$.
	\end{enumerate}
\end{lemma}

\begin{proof}
	Similar to the proof of subject inversion, differing only at the choice constructor:
	
	We proceed by induction on the derivation of
	$\tyJudgeRes{\tyEnv}{
		\mpSum{\mpPrefix_i\mpSeq\mpP[i]}{i\in I}
	}{\stEnv}$.
	
	\inferrule{Sub}.
	$\tyJudgeRes{\tyEnv}{
		\mpSum{\mpPrefix_i\mpSeq\mpP[i]}{i\in I}
	}{\stEnviii}$, $\stEnvEndP{\stEnvii}$, and
	$\stEnviii\stEnvComp\stEnvii=\stEnv$.
	By I.H.,
	there exists $\mpC$ and
		$\stEnvi$ such that:
		$\{\ltsSubject{\mpPrefix_i}\suchthat i\in I\}=\{\mpC\}$;
		$\stEnvi\tySub\stEnviii$;
		$\stEnvEndP{\stEnvi\setminus\mpC}$;
		if $\mpPrefix_i=\mpSel{\mpC}{\roleQ}{\stLab}{\mpCi}{}$, then
		$\stFmt{\roleQ{!}\stChoice{\stLab}{\stEnvApp{\stEnvi}{\mpCi}}\stSeq\stTi}\preType\stEnvApp{\stEnvi}{\mpC}$,
		$\mpC\neq\mpCi$, and
		$\tyJudgeRes{\tyEnv}{\mpP[i]}{(\stEnvi\setminus\mpC\setminus\mpCi)\stEnvComp\stEnvMap{\mpC}{\stTi}}$;
		if $\mpPrefix_i=\mpSel{\mpC}{\roleQ}{\stLab}{e}{}$, then
		$\stFmt{\roleQ{!}\stChoice{\stLab}{\tyGround}\stSeq\stTi}
		\preType\stEnvApp{\stEnvi}{\mpC}$,
		$\tyJudgeRes{\tyEnv}{e}{\tyGround}$, and
		$\tyJudgeRes{\tyEnv}{\mpP[i]}{(\stEnvi\setminus\mpC)\stEnvComp\stEnvMap{\mpC}{\stTi}}$;
		if $\mpPrefix_i=\mpBra{\mpC}{\roleQ}{\stLab}{\mpx}{}$, then
		$\stFmt{\roleQ{?}\stChoice{\stLab}{\tyGround}\stSeq\stTi}\preType\stEnvApp{\stEnvi}{\mpC}$, and
		$\tyJudgeRes{\tyEnv\tyEnvComp\tyEnvMap{\mpx}{\tyGround}}{\mpP[i]}{(\stEnvi\setminus\mpC)\stEnvComp\stEnvMap{\mpC}{\stTi}}$;
		if $\mpPrefix_i=\mpBra{\mpC}{\roleQ}{\stLab}{\mpy}{}$, then
		$\stFmt{\roleQ{?}\stChoice{\stLab}{\stT}\stSeq\stTi}\preType\stEnvApp{\stEnvi}{\mpC}$, and
		$\tyJudgeRes{\tyEnv}{\mpP[i]}{(\stEnvi\setminus\mpC)\stEnvComp\stEnvMap{\mpC}{\stTi}
		\stEnvComp\stEnvMap{\mpy}{\stT}}$;
		if $\stFmt{\roleQ{\dagger}\stChoice{\stLab}{}}\preType\stEnvApp{\stEnvi}{\mpC}$,
		then $\mpFmt{\mpChanRole{\mpC}{\roleQ}{\dagger}\stLab}\preCalc\mpP$.
		Let $\stEnvi[1]=\stEnvi\stEnvComp\stEnvii\tySub\stEnviii\stEnvComp\stEnvii\tySub\stEnv$.
		So by \inferrule{Sub}:
		$\stEnvi[1]\tySub\stEnv$;
		$\stEnvEndP{\stEnvi[1]\setminus\mpC}$;
		if $\mpPrefix_i=\mpSel{\mpC}{\roleQ}{\stLab}{\mpCi}{}$, then
		$\stFmt{\roleQ{!}\stChoice{\stLab}{\stEnvApp{\stEnvi}{\mpCi}}\stSeq\stTi}\preType\stEnvApp{\stEnvi[1]}{\mpC}$,
		$\mpC\neq\mpCi$, and
		$\tyJudgeRes{\tyEnv}{\mpP[i]}{(\stEnvi[1]\setminus\mpC\setminus\mpCi)\stEnvComp\stEnvMap{\mpC}{\stTi}}$;
		if $\mpPrefix_i=\mpSel{\mpC}{\roleQ}{\stLab}{e}{}$, then
		$\stFmt{\roleQ{!}\stChoice{\stLab}{\tyGround}\stSeq\stTi}
		\preType\stEnvApp{\stEnvi[1]}{\mpC}$,
		$\tyJudgeRes{\tyEnv}{e}{\tyGround}$, and
		$\tyJudgeRes{\tyEnv}{\mpP[i]}{(\stEnvi[1]\setminus\mpC)\stEnvComp\stEnvMap{\mpC}{\stTi}}$;
		if $\mpPrefix_i=\mpBra{\mpC}{\roleQ}{\stLab}{\mpx}{}$, then
		$\stFmt{\roleQ{?}\stChoice{\stLab}{\tyGround}\stSeq\stTi}\preType\stEnvApp{\stEnvi[1]}{\mpC}$, and
		$\tyJudgeRes{\tyEnv\tyEnvComp\tyEnvMap{\mpx}{\tyGround}}{\mpP[i]}{(\stEnvi\setminus\mpC)\stEnvComp\stEnvMap{\mpC}{\stTi}}$;
		if $\mpPrefix_i=\mpBra{\mpC}{\roleQ}{\stLab}{\mpy}{}$, then
		$\stFmt{\roleQ{?}\stChoice{\stLab}{\stT}\stSeq\stTi}\preType\stEnvApp{\stEnvi[1]}{\mpC}$, and
		$\tyJudgeRes{\tyEnv}{\mpP[i]}{(\stEnvi[1]\setminus\mpC)\stEnvComp\stEnvMap{\mpC}{\stTi}
		\stEnvComp\stEnvMap{\mpy}{\stT}}$;
		if $\stFmt{\roleQ{\dagger}\stChoice{\stLab}{}}\preType\stEnvApp{\stEnvi[1]}{\mpC}$,
		then $\mpFmt{\mpChanRole{\mpC}{\roleQ}{\dagger}\stLab}\preCalc\mpP$.
		
		\inferrule{$\text{Sum}^*$}.
		We have that $\stEnv=\stEnvi\stEnvComp\stEnvMap{\mpC}{\stT}$
		and $\stEnvEndP{\stEnvi}$.
		For $i\in I$,
		$\stEnvApp{\stEnv}{\ltsSubject{\mpPrefix_i}}$
		is a sum type by
		\inferrule{$!$Chan},
		\inferrule{$?$Chan},
		\inferrule{$!$Val}, and
		\inferrule{$?$Val},
		so $\ltsSubject{\mpPrefix_i}=\mpC$.
		The rest are precisely the hypotheses of \inferrule{$\text{Sum}^*$},
		\inferrule{$!$Chan},
		\inferrule{$?$Chan},
		\inferrule{$!$Val}, and
		\inferrule{$?$Val}.
\qed\end{proof}

\propNoSessionDelegation*

\begin{proof}
	Suppose that
	$\tyJudgeRes{\tyEnvi}{\mpPi}{\stEnvi}$ and $\mpSel{\mpCi}{\roleP}{\stLab}{\mpC}{\mpPii}\preCalc{\mpPi}$.
	Induct on the derivation of
	$\tyJudgeRes{\tyEnvi}{\mpPi}{\stEnvi}$.
	$\mpPi$ is a sum process, so the last step was either
	\inferrule{Sub} or \inferrule{$\text{Sum}^*$}.
	If the last rule was \inferrule{Sub}, then
	$\tyJudgeRes{\tyEnvi}{\mpPi}{\stEnvii}$
	for some $\stEnvii\tySub\stEnvi$, so
	$\stEnd\tySub\stEnvApp{\stEnvii}{\mpC}\tySub\stEnvApp{\stEnvi}{\mpC}$
	by the I.H.
	If the last rule was
	\inferrule{$\text{Sum}^*$}, then
	$\stEnvEndP{\stEnvi\setminus\mpCi}$
	and
	$\tyJudgePrefix{\tyEnvi}{
	\mpSel{\mpCi}{\roleP}{\stLab}{\mpC}{\mpPii}}{
	\stEnvi}$.
	By \inferrule{$!$Chan},
	$\mpC\neq\mpCi$ so $\mpC\in\dom{\stEnvi\setminus\mpCi}$
	so $\stEnd\tySub\stEnvApp{\stEnvi}{\mpC}$.
\qed	
\end{proof}

\subjectReductionRes*

\begin{proof}[1]
	Same as proof of (1) \cref{thm:subj-red},
	since precongruence cannot change
	choice processes or act under them.
\qed\end{proof}

As with \cref{thm:pi-sr}, we strengthen (2) to the following:

\begin{theorem}[Subject Reduction]
\label{thm:pi-res-sr}
Let $\predP$ in \inferrule{Res} \cref{fig:typing-rules}
be \RC-safe.
	Suppose that $\tyJudgeRes{\tyEnv}{\mpP}{\stEnv}$
	and that $\stEnv$ is safe.
	\begin{enumerate}
		\item $\mpP\,\,\not\!\!\mpMoveErr$.
		\item If $\mpP\mpMoveTau\mpPi$,
		then
		$\tyJudgeRes{\tyEnv}{\mpPi}{\stEnv}$.
		\item If $\mpP\mpMoveCommS{\mpS}{\roleP}{\roleQ}{\stLab}{}\mpPi$,
		then
		there is $\stS$ such that
		$\stEnv\gtMove[\ltsSendRecvS{\mpS}{\roleP}{\roleQ}{\stChoice{\stLab}{\stS}}]\,\stEnvi$, and
		$\tyJudgeRes{\tyEnv}{\mpPi}{\stEnvi}$.
	\end{enumerate}
\end{theorem}

\begin{proof}
	Same as the proof of \cref{thm:pi-sr}.
\qed\end{proof}

The advantage of using \inferrule{$\text{Sum}^*$}
is \cref{prop:reducts},
which is the key for proving \cref{thm:sess-fid} and
\cref{thm:typing-prop-df-live}.

\begin{restatable}[Reducts]{proposition}{subjectReducts}
\label{prop:reducts}
Let $\predP$ in \inferrule{Res} in \cref{fig:typing-rules}
be \RC-safe.
Suppose that $\mpP$ is guarded,
$\tyJudgeRes{\tyEnvEmpty}{\mpP}{
	\left(\stEnv\stEnvComp\stEnvMap{\mpC}{\stT}\right)
	}$, and $\roleP\stFmt{\dagger}\preType\stT$,
	then $\mpP\mpMoveTauStar\mpCtxApp{\mpCtx}{\mpQ}$
	with $\mpChanRole{\mpC}{\roleP}{\dagger}\preCalc{\mpQ}$
	and
	$\tyJudgeRes{\tyEnvEmpty}{\mpCtxApp{\mpCtx}{\mpQ}}{
		\left(\stEnv\stEnvComp\stEnvMap{\mpC}{\stT}\right)
	}$.
\end{restatable}

\begin{proof}
	Induction on $\depth{\mpP}$
	and inversion.
	We use $\tau$-transitions so that
	$\tyJudgeRes{\tyEnvEmpty}{\mpCtxApp{\mpCtx}{\mpQ}}{(\stEnv\stEnvComp\stEnvMap{\mpC}{\stT})}$
	follows from subject reduction.
	\\
	$\mpP=\mpNil$. $\stEnd\tySub\stT$ so $\roleP\stFmt{\dagger}\not\preType\stT$.\\
	$\mpP=\mpX$. $\mpX\in\dom{\stEnvEmpty}$, which cannot occur.\\
	$\mpP=\mpRec{\mpX}{\mpPi}$.
	$\mpP\prestruct\mpPi\subst{\mpX}{\mpPi}$.
	
	By I.H., $\mpP\prestruct\mpPi\subst{\mpX}{\mpPi}
	\mpMoveTauStar\mpCtxApp{\mpCtx}{\mpQ}$
	with $\mpChanRole{\mpC}{\roleP}{\dagger}\preCalc{\mpQ}$,
	so $\mpP\mpMoveTauStar\mpCtxApp{\mpCtx}{\mpQ}$.\\
	$\mpP=\mpPi\mpPar\mpPii$.
	
	By inversion,
	there are $\stEnvi$ and $\stEnvii$ such that $\stEnv\stEnvComp\stEnvMap{\mpC}{\stT}=\stEnvi\stEnvComp\stEnvii$,
	$\tyJudgeRes{\tyEnvEmpty}{\mpPi}{\stEnvi}$, and
	$\tyJudgeRes{\tyEnvEmpty}{\mpPii}{\stEnvii}$.
	
	If $\mpC\in\dom{\stEnvi}$ then by I.H.,
	$\mpPi\mpMoveTauStar\mpRes{\mpS[1],\dots,\mpS[n]}{\left(\mpQ\mpPar\mpPiii\right)}$
	with $\mpChanRole{\mpC}{\roleP}{\dagger}\preCalc{\mpQ}$, so
	$\mpP\mpMoveTauStar\mpRes{\mpS[1],\dots,\mpS[n]}{\left(\mpQ\mpPar\mpPii\mpPar\mpPiii\right)}$.
	
	Otherwise $\mpC\in\dom{\stEnvii}$ so by I.H.,
	$\mpPii\mpMoveTauStar\mpRes{\mpS[1],\dots,\mpS[n]}{\left(\mpQ\mpPar\mpPiii\right)}$
	with $\mpChanRole{\mpC}{\roleP}{\dagger}\preCalc{\mpQ}$, so
	$\mpP\mpMoveTauStar\mpRes{\mpS[1],\dots,\mpS[n]}{\left(\mpQ\mpPar\mpPi\mpPar\mpPiii\right)}$.\\
	$\mpP=\mpIf{e}{\mpP[\mpTrue]}{\mpP[\mpFalse]}$.
	
	By inversion, $\tyJudge{\tyEnvEmpty}{e}{\tyBool}$ and
	$\tyJudgeRes{\tyEnvEmpty}{\mpP[v]}{(\stEnv\stEnvComp\stEnvMap{\mpC}{\stT})}$
	for $v\in\{\mpTrue,\mpFalse\}$.
	
	So, $\eval{e}{v}$ for some $v\in\{\mpTrue,\mpFalse\}$.
	
	By I.H., $\mpP[v]\mpMoveTauStar\mpCtxApp{\mpCtx}{\mpQ}$
	with $\mpChanRole{\mpC}{\roleP}{\dagger}\preCalc{\mpQ}$.
	
	So, $\mpP\mpMoveTau\mpP[v]\mpMoveTauStar\mpCtxApp{\mpCtx}{\mpQ}$
	with $\mpChanRole{\mpC}{\roleP}{\dagger}\preCalc{\mpQ}$.\\
	$\mpP=\mpRes{\mpS}{\mpPi}$.
	
	By inversion,
	there is $\stEnvi$ such that $\stEnv\stEnvComp\stEnvMap{\mpC}{\stT}=\stEnvi\setminus\mpS$,
	$\predPApp{\stEnvi[\mpS]}$, and
	$\tyJudgeRes{\tyEnvEmpty}{\mpPi}{\stEnvi}$.
	
	By I.H.,
	$\mpPi\mpMoveTauStar\mpRes{\mpS[1],\dots,\mpS[n]}{\left(\mpQ\mpPar\mpPii\right)}$
	with $\mpChanRole{\mpC}{\roleP}{\dagger}\preCalc{\mpQ}$.
	
	So,
	$\mpP\mpMoveTauStar\mpRes{\mpS,\mpS[1],\dots,\mpS[n]}{\left(\mpQ\mpPar\mpPii\right)}$.
\qed\end{proof}

\sessionFidelity*

We strengthen session fidelity:

\begin{theorem}[Session Fidelity]
\label{thm:strong-sess-fid}
	Let $\predP$ in \inferrule{Res} \cref{fig:typing-rules}
be \RC-safe.
	Suppose that $\mpP$ is guarded, $\stEnv$ is safe, and
	$\tyJudgeRes{\tyEnvEmpty}{\mpP}{\stEnv}$.
	If $\stEnv\gtMove[\ltsSendRecvS{\mpS}{\roleP}{\roleQ}{\stChoice{\stLab}{\stS}}]$, then
	there exists $\mpPi,\stEnvi,\stLabi,\stSi$
	such that:\\
	$\stEnv\gtMove[\ltsSendRecvS{\mpS}{\roleP}{\roleQ}{\stChoice{\stLabi}{\stSi}}]\stEnvi$,
	$\mpP\mpMoveTauStar\cdot\mpMoveCommS{\mpS}{\roleP}{\roleQ}{\stLabi}{}\mpPi$, and
	$\tyJudgeRes{\tyEnvEmpty}{\mpPi}{\stEnvi}$.
\end{theorem}

\begin{proof}
	This is an application of \cref{prop:reducts}
	and inversion.
	Suppose that $\mpP$ is guarded, $\stEnv$ is safe,
	$\tyJudgeRes{\tyEnvEmpty}{\mpP}{\stEnv}$, and
	$\stEnv\gtMove[\ltsSendRecvS{\mpS}{\roleP}{\roleQ}{\stChoice{\stLab}{\stS}}]$.
	So, $\roleQ\stFmt{!}\in\unfoldOne{\stEnvApp{\stEnv}{\mpChanRole{\mpS}{\roleP}}}$.
	By \cref{prop:reducts},
	\[
	\mpP\mpMoveTauStar\mpRes{\mpSi[1],\dots\mpSi[n]}{
	\left(\mpQ\mpPar\mpPi\right)
	}
	\;\text{and}\;
	\mpChanRole{\mpChanRole{\mpS}{\roleP}}{\roleQ}{!}\preCalc{\mpQ}
	\]
	By inversion there are $\stEnvi$ and $\stEnvii$
	such that
	$\tyJudgeRes{\tyEnvEmpty}{\mpQ}{\stEnvi}$,
	$\tyJudgeRes{\tyEnvEmpty}{\mpPi}{\stEnvii}$,
	$\predPApp{(\stEnvi\stEnvComp\stEnvii)_{\mpSi[i]}}$ for $1\leq i\leq n$, and
	$\stEnv = (\stEnvi\stEnvComp\stEnvii)\setminus\mpSi[1]\setminus\dots\setminus\mpSi[n]$.
	
	By inversion and inversion of subtyping,
	\[\dom{\stEnvi\setminus\stEnd}=\{\mpChanRole{\mpS}{\roleP}\}\]
	$\roleP\stFmt{?}\in\unfoldOne{\stEnvApp{\stEnv}{\mpChanRole{\mpS}{\roleQ}}}$,
	so $\mpChanRole{\mpS}{\roleQ}\in\dom{\stEnvii}$.
	
	By \cref{prop:reducts},
	\[
	\mpPi\mpMoveTauStar\mpRes{\mpSii[1],\dots\mpSii[m]}{
	\left(\mpQi\mpPar\mpPii\right)}
	\;\text{and}\;
	\mpChanRole{\mpChanRole{\mpS}{\roleQ}}{\roleP}{?}\preCalc{\mpQi}
	\]
	So,
	\[
		\mpP\mpMoveTauStar
		\mpRes{\mpSi[1],\dots,\mpSi[n],\mpSii[1],\dots\mpSii[m]}{
	\left(\mpQ\mpPar\mpQi\mpPar\mpPii\right)}
	\]
	Write
	\[
		\mpQ=\mpSum{\mpPrefix_i\mpSeq\mpQ[i]}{i\in I}
		\quad
		\mpQi=\mpSum{\mpPrefix'_i\mpSeq\mpQi[i]}{i\in J}
	\]
	$\mpP$ is safe, so
	\[
	\begin{array}{rcl}
	&
	\mpRes{\mpSi[1],\dots,\mpSi[n],\mpSii[1],\dots\mpSii[m]}{
	\left(\mpQ\mpPar\mpQi\mpPar\mpPii\right)}
	\\
	\not\mpMoveErr&\mpRes{\mpSi[1],\dots,\mpSi[n],\mpSii[1],\dots\mpSii[m]}{
	\left(\mpErr\mpPar\mpPii\right)}
	\end{array}
	\]
	
	If there is $k\in I$ such that
	$\mpPrefix_k=\mpSel{\mpChanRole{\mpS}{\roleP}}{\roleQ}{\stLabi}{e}{}$
	then, since $\mpP$ is safe,
	there is $k'\in J$ such that
	$\mpPrefix'_{k'}=\mpBra{\mpChanRole{\mpS}{\roleQ}}{\roleP}{\stLabi}{\mpx}{}$.
	By inversion,
	$\tyJudge{\tyEnvEmpty}{e}{\tyGround}$ for some $\tyGround$,
	so $\eval{e}{v}$ for some $v$.
	So,
	\[\begin{array}{cl}
		&\mpRes{\mpSi[1],\dots,\mpSi[n],\mpSii[1],\dots\mpSii[m]}{
	\left(\mpQ\mpPar\mpQi\mpPar\mpPii\right)}\\
	\mpMoveCommS{\mpS}{\roleP}{\roleQ}{\stLabi}{}&
		\mpRes{\mpSi[1],\dots,\mpSi[n],\mpSii[1],\dots\mpSii[m]}{
	\left(\mpQ[k]\mpPar\mpQi[k]\subst{\mpx}{v}\mpPar\mpPii\right)}
	\end{array}\]
	By subject reduction, there exist $\stEnv[1]$ and $\stSi$ such that
	$\stEnv\gtMove[\ltsSendRecvS{\mpS}{\roleP}{\roleQ}{\stLabi}{\stSi}]\stEnv[1]$
	and
	\[\tyJudgeRes{\tyEnvEmpty}{\mpRes{\mpSi[1],\dots,\mpSi[n],\mpSii[1],\dots\mpSii[m]}{
	\left(\mpQ[k]\mpPar\mpQi[k]\subst{\mpx}{v}\mpPar\mpPii\right)}}{\stEnv[1]}\]
	
	Otherwise, there is $k\in I$ such that
	$\mpPrefix_k=\mpSel{\mpChanRole{\mpS}{\roleP}}{\roleQ}{\stLabi}{\mpChanRole{\mpSii}{\roleR}}{}$
	then, since $\mpP$ is safe,
	there is $k'\in J$ such that
	$\mpPrefix'_{k'}=\mpBra{\mpChanRole{\mpS}{\roleQ}}{\roleP}{\stLabi}{\mpy}{}$.
	So,
	\[\begin{array}{cl}
		&\mpRes{\mpSi[1],\dots,\mpSi[n],\mpSii[1],\dots\mpSii[m]}{
	\left(\mpQ\mpPar\mpQi\mpPar\mpPii\right)}\\
	\mpMoveCommS{\mpS}{\roleP}{\roleQ}{\stLabi}{}&
		\mpRes{\mpSi[1],\dots,\mpSi[n],\mpSii[1],\dots\mpSii[m]}{
	\left(\mpQ[k]\mpPar\mpQi[k]\subst{\mpy}{\mpChanRole{\mpSii}{\roleR}}\mpPar\mpPii\right)}
	\end{array}\]
	By subject reduction, there exist $\stEnv[1]$ and $\stSi$ such that
	$\stEnv\gtMove[\ltsSendRecvS{\mpS}{\roleP}{\roleQ}{\stLabi}{\stSi}]\stEnv[1]$
	and
	\[\tyJudgeRes{\tyEnvEmpty}{\mpRes{\mpSi[1],\dots,\mpSi[n],\mpSii[1],\dots\mpSii[m]}{
	\left(\mpQ[k]\mpPar\mpQi[k]\subst{\mpy}{\mpChanRole{\mpSii}{\roleR}}\mpPar\mpPii\right)}}{\stEnv[1]}\]
\qed\end{proof}

\begin{lemma}[End Typing]
\label{lem:end-type-df}
	If $\tyJudgeRes{\tyEnvEmpty}{\mpP}{\stEnv}$,
	$\mpP\mpMoveNot$,
	$\stEnv$ is safe,
	$\mpP$ guarded,
	and $\predP\subseteq\stEnvDFPred$,
	then
	$\mpP\prestruct\mpNil$,
	$\chan{\mpP}=\emptyset$,
	$\stEnvEndP{\stEnv}$.
\end{lemma}

\begin{proof}
	Induction on $\depth{\mpP}$,
	using inversion
	to note that sum processes cannot appear in
	$\mpP$.
	By session fidelity, $\stEnv\mpMoveNot$
	so by deadlock-freedom
	$\stEnvEndP{\stEnv}$.
	\\
	$\mpP=\mpNil$. $\mpNil\prestruct\mpNil$
	and $\chan{\mpNil}=\emptyset$.\\
	$\mpP=\mpIf{e}{\mpPi}{\mpPii}$.
	$\tyJudge{\tyEnvEmpty}{e}{\tyGround}$ for some $\tyGround$,
	so $\eval{e}{v}$ for some $v$,
	so $\mpP\mpMove$.\\
	$\mpP=\mpRec{\mpX}{\mpPi}$.
	$\mpP\prestruct\mpPi\subst{\mpX}{\mpP}$
	so apply I.H.\\
	$\mpP=\mpPi\mpPar\mpPii$.
	By inversion,
	$\tyJudge{\tyEnvEmpty}{\mpPi}{\stEnvi}$,
	$\tyJudge{\tyEnvEmpty}{\mpPii}{\stEnvii}$, and
	$\stEnv=\stEnvi\stEnvComp\stEnvii$.
	By I.H., $\chan{\mpP}=\chan{\mpPi}\cup\chan{\mpPii}$,
	$\mpPi\prestruct\mpNil$,
	$\mpPii\prestruct\mpNil$, so
	$\mpP\prestruct\mpNil$.\\
	$\mpP=\mpSum{\mpPrefix_i\mpSeq\mpP[i]}{i\in I}$.
	By inversion, $\neg\stEnvEndP{\stEnv}$.\\
	$\mpP=\mpRes{\mpS}{\mpPi}$.
	By inversion,
	there is $\stEnvi$ such that
	$\tyJudgeRes{\tyEnvEmpty}{\mpPi}{\stEnvi}$,
	$\predPApp{\stEnvi[\mpS]}$,
	$\stEnv=\stEnvi\setminus\mpS$, and
	$\mpPi\mpMoveNot$.
	By I.H., $\mpP\prestruct\mpRes{\mpS}{\mpNil}\prestruct\mpNil$
	and $\chan{\mpP}\subseteq\chan{\mpPi}=\emptyset$.
\qed\end{proof}

\thmTypingProperties*

\begin{proof}[1]
	If $\mpP\mpMoveNot$, then $\mpP\prestruct\mpNil$
	by \cref{lem:end-type-df}.
	So $\mpP$ is deadlock-free.
\qed\end{proof}

\begin{proof}[2]
	If $\mpP=\mpRes{\mpSi[1],\dots,\mpSi[n]}{\left(\mpQ\mpPar\mpPi\right)}$
	with $\mpQ$ a sum process over channel $\mpC$, then
	invert typing to get
	$\tyJudgeRes{\tyEnvEmpty}{\mpPi}{\stEnvi}$,
	$\tyJudgeRes{\tyEnvEmpty}{\mpQ}{(\stEnvii\stEnvComp\stEnvMap{\mpC}{\stT})}$,
	with $\stEnvi\stEnvComp\stEnvMap{\mpC}{\stT}$ live,
	$\stEnvEndP{\stEnvii}$, and $\stT\tyNotSub\stEnd$.
	
	Use \cref{prop:fair-path}
	and
	\cref{thm:strong-sess-fid}
	to obtain a fair path
	$\{\stEnv[i]\}_{i\in N}$
	and a path $\{\mpP[i]\}_{i\in N}$
	such that
	$\stEnvi=\stEnv[0]$,
	$\mpPi=\mpP[0]$, and
	for $i\in N$
	$\tyJudgeRes{\tyEnvEmpty}{\mpPi[i]}{\stEnv[i]}$.
	
	If
	$\{(\stEnv[i]\stEnvComp\stEnvMap{\mpC}{\stT})\}_{i\in N}$
	is live, then there is $j\in N$
	such that
	$(\stEnv[j]\stEnvComp\stEnvMap{\mpC}{\stT})\gtMove[\stEnvAnnotGenericSym]$
	with $\mpC\in\ltsSubject{\stEnvAnnotGenericSym}$.
	
	If $\{(\stEnv[i]\stEnvComp\stEnvMap{\mpC}{\stT})\}_{i\in N}$
	is not live, then it is not fair.
	Since $\{\stEnv[i]\}_{i\in N}$ is fair, there is $j\in N$
	such that
	$(\stEnv[j]\stEnvComp\stEnvMap{\mpC}{\stT})\gtMove[\stEnvAnnotGenericSym]$
	with $\mpC\in\ltsSubject{\stEnvAnnotGenericSym}$.
	
	Write $\stEnvAnnotGenericSym=\mpChanRole{\mpChanRole{\mpS}{\roleP}}{\roleQ}\stChoice{\stLab}{\stS}$ so
	$\mpC=\mpChanRole{\mpS}{\roleP}$
	or $\mpC=\mpChanRole{\mpS}{\roleQ}$.
	So
	$\roleP\stFmt{?}\in\unfoldOne{\stEnvApp{\stEnv}{\mpChanRole{\mpS}{\roleQ}}}$
	and $\mpChanRole{\mpChanRole{\mpS}{\roleP}}{\roleQ}{!}\preCalc{\mpQ}$.
	By \cref{prop:reducts} and inversion,
	\[
	\mpP[j]\mpMoveTauStar\mpRes{\mpSii[1],\dots\mpSii[m]}{
	\left(\mpQi\mpPar\mpPii\right)}
	\]
	If $\mpC=\mpChanRole{\mpS}{\roleP}$,
	then there is $\stLabi$ such that
	$\mpChanRole{\mpChanRole{\mpS}{\roleQ}}{\roleP}{?}\preCalc{\mpQi}$ and
	$\mpChanRole{\mpChanRole{\mpS}{\roleP}}{\roleQ}{!}\stLabi\preCalc{\mpQ}$.
	\[
		\mpRes{\mpSi[1],\dots,\mpSi[n]}{\left(\mpCtxHole\mpPar\mpPi\right)}
		\mpMoveStar
		\mpRes{\mpSi[1],\dots,\mpSi[n],\mpSii[1],\dots,\mpSii[m]}{\left(\mpCtxHole\mpPar\mpQi\mpPar\mpPii\right)}
	\]
	By safety,
	$\mpChanRole{\mpChanRole{\mpS}{\roleQ}}{\roleP}{?}\stLabi\preCalc{\mpQi}$.
	
	If $\mpC=\mpChanRole{\mpS}{\roleQ}$,
	then there is $\stLabi$ such that
	$\mpChanRole{\mpChanRole{\mpS}{\roleQ}}{\roleP}{?}\preCalc{\mpQ}$ and
	$\mpChanRole{\mpChanRole{\mpS}{\roleP}}{\roleQ}{!}\stLabi\preCalc{\mpQi}$.
	\[
		\mpRes{\mpSi[1],\dots,\mpSi[n]}{\left(\mpCtxHole\mpPar\mpPi\right)}
		\mpMoveStar
		\mpRes{\mpSi[1],\dots,\mpSi[n],\mpSii[1],\dots,\mpSii[m]}{\left(\mpCtxHole\mpPar\mpQi\mpPar\mpPii\right)}
	\]
	By safety,
	$\mpChanRole{\mpChanRole{\mpS}{\roleQ}}{\roleP}{?}\stLabi\preCalc{\mpQ}$.
	
	Therefore, $\mpP$ is live.
\qed\end{proof}

\begin{example}[Leader Election without Delegation]
\label{ex:leader-election-res-detail}
Recall \cref{ex:leader-election-res}.

Let $\predP=\stEnvSafePred\cap\stEnvLivePred$ in \inferrule{Res} in \cref{fig:typing-rules}.

\[
	\mpR[\text{\tiny lead}]=\mpR[0]\mpPar\mpR[1]\mpPar\mpR[2]\mpPar\mpR[3]\mpPar\mpR[4]\quad\text{where, for}\,0\leq i\leq 4
\]
\begingroup
	\small
	\[
		\begin{array}{rcl}
			\mpR[i]&=&
				\mpSel{\mpChanRole{\mpS}{\roleP[i]}}{\roleP[(i+4)\%5]}{\stLabFmt{elect}}{\mpNum{i}}{}+
				\mpBra{\mpChanRole{\mpS}{\roleP[i]}}{\roleP[(i+1)\%5]}{\stLabFmt{elect}}{\mpx}{\mpRi[i]}+
				\mpSel{\mpChanRole{\mpS}{\roleP[i]}}{\roleP[(i+3)\%5]}{\stLabFmt{elect}}{\mpNum{i}}{}
			\\
			\mpRi[i]&=&
					\mpSel{\mpChanRole{\mpS}{\roleP[i]}}{\roleP[(i+2)\%5]}{\stLabFmt{elect}}{\mpNum{i}}{} +
					\mpBra{\mpChanRole{\mpS}{\roleP[i]}}{\roleP[(i+3)\%5]}{\stLabFmt{elect}}{\mpxi}{
						\mpBra{\mpChanRole{\mpS}{\roleP[i]}}{\roleP[(i+2)\%5]}{\stLabFmt{elect}}{\mpxii}{}}
		\end{array}
	\]
	
	\[
\stEnv[\text{\tiny lead},\mpS] =
\cup_{0\leq i\leq 4}
\set{
        \stEnvMap{\mpChanRole{\mpS}{\roleP[i]}}{\stT[i]}}
\]
where for $0\leq i \leq 4$,
\[
	\begin{array}{rcl}
		\stT[i]&=&
		\stSum{}{}{
		\begin{array}{l}
			\roleP[(i+4)\%5]!\stChoice{\stLabFmt{elect}}{\tyInt}\\
			\roleP[(i+3)\%5]!\stChoice{\stLabFmt{elect}}{\tyInt}\\
			\roleP[(i+1)\%5]?\stChoice{\stLabFmt{elect}}{\tyInt}\stSeq\stTi[i]
		\end{array}
		}{}
		\\
		\stTi[i]&=&
		\stSum{}{}{
		\begin{array}{l}
			\roleP[(i+2)\%5]!\stChoice{\stLabFmt{elect}}{\tyInt}\\
			\roleP[(i+3)\%5]?\stChoice{\stLabFmt{elect}}{\tyInt}\stSeq
			\roleP[(i+2)\%5]?\stChoice{\stLabFmt{elect}}{\tyInt}
		\end{array}
		}{}
\end{array}        
\]
	\endgroup
	We derive the typing judgement $\tyJudgeRes{\tyEnvEmpty}{\mpR[\text{\tiny lead}]}{\stEnv[\text{\tiny lead},\mpS]}$.
\begingroup
\small
\[
\begin{array}{@{\tyEnvEmpty\,\stFmt{{\vdash}^{\star}}\,}l@{\,:\,}ll}
	\mpNil&\stEnvEmpty
	&\inferrule{Nil}
	\\
	\mpNil&
	\set{
	\stEnvMap{\mpChanRole{\mpS}{\roleP[i]}}{\stEnd}
	}
	&\inferrule{Sub}
	\\
	\mpBra{\mpChanRole{\mpS}{\roleP[i]}}{\roleP[(i+2)\%5]}{\stLabFmt{elect}}{\mpxii}{}
	&
	\set{
	\stEnvMap{\mpChanRole{\mpS}{\roleP[i]}}{
	\roleP[(i+2)\% 5]?\stChoice{\stLabFmt{elect}}{\stT}
	}
	}
	&\inferrule{$\text{Sum}^*$}
	\\
	\mpRi[i]
	&
	\set{
	\stEnvMap{\mpChanRole{\mpS}{\roleP[i]}}{\stTi[i]}
	}
	&\inferrule{$\text{Sum}^*$}
	\\
	\mpR[i]
	&
	\set{
	\stEnvMap{\mpChanRole{\mpS}{\roleP[i]}}{\stT[i]}
	}
	&\inferrule{$\text{Sum}^*$}
	\\
	\mpR[\text{\tiny lead}]
	&
	\stEnv[\text{\tiny lead},\mpS]
	&
	\inferrule{Par}
\end{array}
\]
\endgroup

\[
	\mpQ[\text{\tiny lead}]=\mpRec{\mpX}{
	\mpRes{\mpS}{\left(
	\mpQ[0]\mpPar\mpQ[1]\mpPar\mpQ[2]\mpPar\mpQ[3]\mpPar\mpQ[4]\right)}}\quad\text{where, for}\,0\leq i\leq 4
\]
\begingroup
	\small
	\[
		\begin{array}{rcl}
			\mpQ[i]&=&
				\mpSel{\mpChanRole{\mpS}{\roleP[i]}}{\roleP[(i+4)\%5]}{\stLabFmt{elect}}{\mpNum{i}}{}+
				\mpBra{\mpChanRole{\mpS}{\roleP[i]}}{\roleP[(i+1)\%5]}{\stLabFmt{elect}}{\mpx}{\mpQi[i]}+
				\mpSel{\mpChanRole{\mpS}{\roleP[i]}}{\roleP[(i+3)\%5]}{\stLabFmt{elect}}{\mpNum{i}}{}
			\\
			\mpQi[i]&=&
					\mpSel{\mpChanRole{\mpS}{\roleP[i]}}{\roleP[(i+2)\%5]}{\stLabFmt{elect}}{\mpNum{i}}{} +
					\mpBra{\mpChanRole{\mpS}{\roleP[i]}}{\roleP[(i+3)\%5]}{\stLabFmt{elect}}{\mpxi}{
						\mpBra{\mpChanRole{\mpS}{\roleP[i]}}{\roleP[(i+2)\%5]}{\stLabFmt{elect}}{\mpxii}{
						\mpX
						}}
		\end{array}
	\]
	\endgroup
	We derive $\tyJudgeRes{\tyEnvEmpty}{\mpQ[\text{\tiny lead}]}{\stEnvEmpty}$.
	
	\begingroup
\small
\[
\begin{array}{@{\set{\tyEnvMap{\mpX}{\tyEnvEmpty}}\,\stFmt{{\vdash}^{\star}}\,}l@{\,:\,}ll}
	\mpX&\stEnvEmpty
	&\inferrule{Var}
	\\
	\mpX&
	\set{
	\stEnvMap{\mpChanRole{\mpS}{\roleP[i]}}{\stEnd}
	}
	&\inferrule{Sub}
	\\
	\mpBra{\mpChanRole{\mpS}{\roleP[i]}}{\roleP[(i+2)\%5]}{\stLabFmt{elect}}{\mpxii}{\mpX}
	&
	\set{
	\stEnvMap{\mpChanRole{\mpS}{\roleP[i]}}{
	\roleP[(i+2)\% 5]?\stChoice{\stLabFmt{elect}}{\stT}
	}
	}
	&\inferrule{$\text{Sum}^*$}
	\\
	\mpQi[i]
	&
	\set{
	\stEnvMap{\mpChanRole{\mpS}{\roleP[i]}}{\stTi[i]}
	}
	&\inferrule{$\text{Sum}^*$}
	\\
	\mpQ[i]
	&
	\set{
	\stEnvMap{\mpChanRole{\mpS}{\roleP[i]}}{\stT[i]}
	}
	&\inferrule{$\text{Sum}^*$}
	\\
	\mpBigPar{0\leq i \leq 4}{\mpQ[i]}
	&
	\stEnv[\text{\tiny lead},\mpS]
	&
	\inferrule{Par}
	\\
	\mpRes{\mpS}{\left(\mpBigPar{0\leq i \leq 4}{\mpQ[i]}\right)}
	&
	\stEnvEmpty
	&
	\inferrule{Res}
\end{array}
\]
\[
	\tyJudgeRes{\tyEnvEmpty}{\mpQ[\text{\tiny lead}]}{\stEnvEmpty}
	\qquad
	\inferrule{Rec}
\]
\endgroup
\end{example}

\begin{example}[Calculating Exponentials]
	Our typing system is capable of typing
	processes that can compute non-trivial functions.
	The processes below are typable by
	safe contexts, and
	(in some sense)
	compute addition,
	multiplication, and exponentiation,
	respectively.
	\[\begin{array}{l}
			\mpFmt{Add}\!\left(\mpS,\roleP,\roleQ,\roleR\right)
			=\\\quad\left(
			\begin{array}{l}\left.\mpRec{\mpXi}{
			\mpSum{
			\left(\begin{array}{l}
				\mpBra{\mpChanRole{\mpS}{\roleP}}{\roleR}
				{\stLabFmt{inp}}{\mpx}
				{
				\mpBra{\mpChanRole{\mpS}{\roleP}}{\roleR}
				{\stLabFmt{inp}}{\mpxi}{\ }
				}\\
				\;\mpSel{\mpChanRole{\mpS}{\roleP}}{\roleQ}
				{\stLabFmt{start}}{\mpNum{0}}
				{
				\mpSel{\mpChanRole{\mpS}{\roleP}}{\roleQ}
				{\stLabFmt{inp}}{\mpx}
				{\ }
				}
				\\
				\;\;
				\mpRec{\mpX}{
				\mpBra{\mpChanRole{\mpS}{\roleP}}{\roleQ}
				{\stLabFmt{count}}{\mpxii}
				{
				\mpBra{\mpChanRole{\mpS}{\roleP}}{\roleQ}
				{\stLabFmt{acc}}{\mpxiii}
				{\ }
				}
				}
				\\
				\;\;\;
				\mathsf{if}\,{\mpxii < \mpxi}\,\mathsf{then}
				\\
				\;\;\;\;
				{
				\mpSel{\mpChanRole{\mpS}{\roleP}}{\roleQ}
				{\stLabFmt{count}}{\mpxii}
				{
				\mpSel{\mpChanRole{\mpS}{\roleP}}{\roleQ}
				{\stLabFmt{inp}}{\mpxiii}
				{
				\mpX
				}
				}
				}
				\\
				\;\;\;
				\mathsf{else}
				\\
				\;\;\;\;
				{
				{
				\mpSel{\mpChanRole{\mpS}{\roleP}}{\roleR}{\stLabFmt{res}}{\mpxiii}{\mpXi}
				}
				}
				\\
				\mpBra{\mpChanRole{\mpS}{\roleP}}{\roleR}
				{\stLabFmt{halt}}{\mpx}{
				\mpSel{\mpChanRole{\mpS}{\roleP}}{\roleQ}
				{\stLabFmt{halt}}{\mpNum{0}}{}
				}
			\end{array}\right)
			}{}}\right|
			\\
			\\
			\mpRec{\mpX}{
			\mpSum{
			\left(
			\begin{array}{l}
				\mpBra{\mpChanRole{\mpS}{\roleQ}}{\roleP}{\stLabFmt{start}}{\mpxii}{
				\mpBra{\mpChanRole{\mpS}{\roleQ}}{\roleP}{\stLabFmt{inp}}{\mpx}{\ }
				}
				\\
				\;
				\mpSel{\mpChanRole{\mpS}{\roleQ}}{\roleP}{\stLabFmt{count}}{\mpNum{0}}{
				\mpSel{\mpChanRole{\mpS}{\roleQ}}{\roleP}{\stLabFmt{acc}}{\mpx}{
				\mpX
				}
				}
				\\
				\mpBra{\mpChanRole{\mpS}{\roleQ}}{\roleP}{\stLabFmt{count}}{\mpxii}{
				\mpBra{\mpChanRole{\mpS}{\roleQ}}{\roleP}{\stLabFmt{inp}}{\mpx}{\ }
				}
				\\
				\;
				\mpSel{\mpChanRole{\mpS}{\roleQ}}{\roleP}{\stLabFmt{count}}{\mpSucc{\mpxii}}{\ }
				\\
				\;\;
				\mpSel{\mpChanRole{\mpS}{\roleQ}}{\roleP}{\stLabFmt{acc}}{\mpSucc{\mpx}}{
				\mpX
				}
				\\
				\mpBra{\mpChanRole{\mpS}{\roleQ}}{\roleP}{\stLabFmt{halt}}{\mpxii}{\mpNil}
			\end{array}
			\right)
			}{}
			}
			\end{array}\right)\end{array}
			\]
			\[\begin{array}{l}
			\mpFmt{Mul}\!\left(\mpS,\roleP,\roleQ,\rolePi,\roleQi,\roleR\right)
			=\\\quad\left(
			\begin{array}{l}\left.\mpRec{\mpXi}{
			\mpSum{
			\left(\begin{array}{l}
				\mpBra{\mpChanRole{\mpS}{\roleP}}{\roleR}
				{\stLabFmt{inp}}{\mpx}
				{
				\mpBra{\mpChanRole{\mpS}{\roleP}}{\roleR}
				{\stLabFmt{inp}}{\mpxi}{\ }
				}\\
				\;\mpSel{\mpChanRole{\mpS}{\roleP}}{\roleQ}
				{\stLabFmt{start}}{\mpNum{0}}
				{
				\mpSel{\mpChanRole{\mpS}{\roleP}}{\roleQ}
				{\stLabFmt{inp}}{\mpx}
				{\ }
				}
				\\
				\;\;
				\mpRec{\mpX}{
				\mpBra{\mpChanRole{\mpS}{\roleP}}{\roleQ}
				{\stLabFmt{count}}{\mpxii}
				{
				\mpBra{\mpChanRole{\mpS}{\roleP}}{\roleQ}
				{\stLabFmt{acc}}{\mpxiii}
				{\ }
				}
				}
				\\
				\;\;\;
				\mathsf{if}\,{\mpxii < \mpxi}\,\mathsf{then}
				\\
				\;\;\;\;
				{
				\mpSel{\mpChanRole{\mpS}{\roleP}}{\roleQ}
				{\stLabFmt{count}}{\mpxii}
				{
				\mpSel{\mpChanRole{\mpS}{\roleP}}{\roleQ}
				{\stLabFmt{inp}}{\mpxiii}
				{
				\mpX
				}
				}
				}
				\\
				\;\;\;
				\mathsf{else}
				\\
				\;\;\;\;
				{
				{
				\mpSel{\mpChanRole{\mpS}{\roleP}}{\roleQ}{\stLabFmt{stop}}{\mpNil}{
				\mpSel{\mpChanRole{\mpS}{\roleP}}{\roleR}{\stLabFmt{res}}{\mpxiii}{\mpXi}}
				}
				}
				\\
				\mpBra{\mpChanRole{\mpS}{\roleP}}{\roleR}
				{\stLabFmt{halt}}{\mpx}{
				\mpSel{\mpChanRole{\mpS}{\roleP}}{\roleQ}
				{\stLabFmt{halt}}{\mpNum{0}}{}
				}
			\end{array}\right)
			}{}}\right|
			\\
			\\
			\left.
			\mpRec{\mpX}{
			\mpSum{
			\left(
			\begin{array}{l}
				\mpBra{\mpChanRole{\mpS}{\roleQ}}{\roleP}{\stLabFmt{start}}{\mpxii}{
				\mpBra{\mpChanRole{\mpS}{\roleQ}}{\roleP}{\stLabFmt{inp}}{\mpx}{\ }
				}
				\\
				\;
				\mpSel{\mpChanRole{\mpS}{\roleQ}}{\roleP}{\stLabFmt{count}}{\mpNum{0}}{
				\mpSel{\mpChanRole{\mpS}{\roleQ}}{\roleP}{\stLabFmt{acc}}{\mpNum{0}}{\ }
				}
				\\
				\;\;
				\mpRec{\mpXi}{
				\mpSum{\left(
				\begin{array}{l}
					\mpBra{\mpChanRole{\mpS}{\roleQ}}{\roleP}{\stLabFmt{count}}{\mpxii}
					{\ }
					\\\;
					\mpBra{\mpChanRole{\mpS}{\roleQ}}{\roleP}{\stLabFmt{inp}}{\mpxi}
					{\ }
					\\\;\;
					\mpSel{\mpChanRole{\mpS}{\roleQ}}{\roleP}{\stLabFmt{count}}{\mpSucc{\mpxii}}
					{\ }
					\\\;\;\;
					\mpSel{\mpChanRole{\mpS}{\roleQ}}{\rolePi}{\stLabFmt{inp}}{\mpx}
						{\ }
					\\\;\;\;\;
					\mpSel{\mpChanRole{\mpS}{\roleQ}}{\rolePi}{\stLabFmt{inp}}{\mpxi}
						{\ }
					\\\;\;\;\;\;
					\mpBra{\mpChanRole{\mpS}{\roleQ}}{\rolePi}{\stLabFmt{res}}{\mpxiii}
						{
							\mpSel{\mpChanRole{\mpS}{\roleQ}}{\roleP}{\stLabFmt{acc}}{\mpxiii}
						{
						\mpXi				
						}
						}
					\\
					\mpBra{\mpChanRole{\mpS}{\roleQ}}{\roleP}{\stLabFmt{stop}}{\mpx}{\mpX}
				\end{array}\right)
				}{}}
				\\
				\mpBra{\mpChanRole{\mpS}{\roleQ}}{\roleP}{\stLabFmt{halt}}{\mpx}{
				\mpSel{\mpChanRole{\mpS}{\roleQ}}{\rolePi}{\stLabFmt{halt}}{\mpx}{}
				}
			\end{array}
			\right)
			}{}
			}\right|
			\\\\
			\mpFmt{Add\!\left(\mpS,\rolePi,\roleQi,\roleQ\right)}
			\end{array}\right)\end{array}
			\]
			\[\begin{array}{l}
			\mpFmt{Exp}\!\left(\mpS,\roleP,\roleQ,\rolePi,\roleQi,\rolePii,\roleQii,\roleR\right)
			=\\\quad\left(
			\begin{array}{l}\left.\mpRec{\mpXi}{
			\mpSum{
			\left(\begin{array}{l}
				\mpBra{\mpChanRole{\mpS}{\roleP}}{\roleR}
				{\stLabFmt{inp}}{\mpx}
				{
				\mpBra{\mpChanRole{\mpS}{\roleP}}{\roleR}
				{\stLabFmt{inp}}{\mpxi}{\ }
				}\\
				\;\mpSel{\mpChanRole{\mpS}{\roleP}}{\roleQ}
				{\stLabFmt{start}}{\mpNum{0}}
				{
				\mpSel{\mpChanRole{\mpS}{\roleP}}{\roleQ}
				{\stLabFmt{inp}}{\mpx}
				{\ }
				}
				\\
				\;\;
				\mpRec{\mpX}{
				\mpBra{\mpChanRole{\mpS}{\roleP}}{\roleQ}
				{\stLabFmt{count}}{\mpxii}
				{
				\mpBra{\mpChanRole{\mpS}{\roleP}}{\roleQ}
				{\stLabFmt{acc}}{\mpxiii}
				{\ }
				}
				}
				\\
				\;\;\;
				\mathsf{if}\,{\mpxii < \mpxi}\,\mathsf{then}
				\\
				\;\;\;\;
				{
				\mpSel{\mpChanRole{\mpS}{\roleP}}{\roleQ}
				{\stLabFmt{count}}{\mpxii}
				{
				\mpSel{\mpChanRole{\mpS}{\roleP}}{\roleQ}
				{\stLabFmt{inp}}{\mpxiii}
				{
				\mpX
				}
				}
				}
				\\
				\;\;\;
				\mathsf{else}
				\\
				\;\;\;\;
				{
				{
				\mpSel{\mpChanRole{\mpS}{\roleP}}{\roleQ}{\stLabFmt{stop}}{\mpNil}{
				\mpSel{\mpChanRole{\mpS}{\roleP}}{\roleR}{\stLabFmt{res}}{\mpxiii}{\mpXi}}
				}
				}
				\\
				\mpBra{\mpChanRole{\mpS}{\roleP}}{\roleR}
				{\stLabFmt{halt}}{\mpx}{
				\mpSel{\mpChanRole{\mpS}{\roleP}}{\roleQ}
				{\stLabFmt{halt}}{\mpNum{0}}{}
				}
			\end{array}\right)
			}{}}\right|
			\\
			\\
			\left.
			\mpRec{\mpX}{
			\mpSum{
			\left(
			\begin{array}{l}
				\mpBra{\mpChanRole{\mpS}{\roleQ}}{\roleP}{\stLabFmt{start}}{\mpxii}{
				\mpBra{\mpChanRole{\mpS}{\roleQ}}{\roleP}{\stLabFmt{inp}}{\mpx}{\ }
				}
				\\
				\;
				\mpSel{\mpChanRole{\mpS}{\roleQ}}{\roleP}{\stLabFmt{count}}{\mpNum{0}}{
				\mpSel{\mpChanRole{\mpS}{\roleQ}}{\roleP}{\stLabFmt{acc}}{\mpNum{1}}{\ }
				}
				\\
				\;\;
				\mpRec{\mpXi}{
				\mpSum{\left(
				\begin{array}{l}
					\mpBra{\mpChanRole{\mpS}{\roleQ}}{\roleP}{\stLabFmt{count}}{\mpxii}
					{\ }
					\\\;
					\mpBra{\mpChanRole{\mpS}{\roleQ}}{\roleP}{\stLabFmt{inp}}{\mpxi}
					{\ }
					\\\;\;
					\mpSel{\mpChanRole{\mpS}{\roleQ}}{\roleP}{\stLabFmt{count}}{\mpSucc{\mpxii}}
					{\ }
					\\\;\;\;
					\mpSel{\mpChanRole{\mpS}{\roleQ}}{\rolePi}{\stLabFmt{inp}}{\mpx}
						{\ }
					\\\;\;\;\;
					\mpSel{\mpChanRole{\mpS}{\roleQ}}{\rolePi}{\stLabFmt{inp}}{\mpxi}
						{\ }
					\\\;\;\;\;\;
					\mpBra{\mpChanRole{\mpS}{\roleQ}}{\rolePi}{\stLabFmt{res}}{\mpxiii}
						{
							\mpSel{\mpChanRole{\mpS}{\roleQ}}{\roleP}{\stLabFmt{acc}}{\mpxiii}
						{
						\mpXi				
						}
						}
					\\
					\mpBra{\mpChanRole{\mpS}{\roleQ}}{\roleP}{\stLabFmt{stop}}{\mpx}{\mpX}
				\end{array}\right)
				}{}}
				\\
				\mpBra{\mpChanRole{\mpS}{\roleQ}}{\roleP}{\stLabFmt{halt}}{\mpx}{
				\mpSel{\mpChanRole{\mpS}{\roleQ}}{\rolePi}{\stLabFmt{halt}}{\mpx}{}
				}
			\end{array}
			\right)
			}{}
			}\right|
			\\\\
			\mpFmt{Mul\!\left(\mpS,\rolePi,\roleQi,\rolePii,\roleQii,\roleQ\right)}
			\end{array}\right)\end{array}\]
\end{example}

\subsection{Extension Characteristics}
\label{app:extension}
\begin{definition}[Complementary Types for Separate Choice]\rm
\label{def:char-ctx-sc}\ \\
	Fix a finite set of session types $\mathcal{T}$
	such that for all $\stT\in\mathcal{T}$,
	if $\stTi$ is a payload type in $\stT$ then $\stTi\in\mathcal{T}$.
	
	Let $L$ be the set of labels appearing within $\mathcal{T}$.
	Let $\roleSet$ be the set of participants appearing within $\mathcal{T}$.
	
	Enumerate:
	\[
	\begin{array}{rcl}
	\roleSet&=&\{\roleR[i]\}_{0\leq i\leq n}
	\\
	\{\stChoice{\stLab}{\stS}\suchthat\stLab\in L,\stS\in\mathcal{T}\cup\{\tyInt,\tyBool\}\}
	&=&\{\stChoice{\stLab[i]}{\stS[i]}\}_{0\leq i < N}
	\end{array}
	\]
	Let
	\[
	Act=\{I\subseteq\{0,\dots,N-1\}\suchthat \forall i\neq j\in I.\; \stLab[i]\neq\stLab[j]\}
	\]
	the set of all message sets, where messages have distinct labels.
	
	Let $\roleP$ and $\roleFmt{sch}$ be fresh roles,
	and $\stLabii$ be a fresh label
	\ie
	$\roleP,\roleFmt{sch}\not\in\roleSet$ and
	$\stLabii\not\in L$.
	
	We define the following functions
	for $\roleQ\in\roleSet$, $\dagger\in\{!,?\}$,
	$I\in Act$, and all $\stT$, $\{\stT[i]\}_{i\in I}$:
	$\syncBra{\roleQ}{\stT}$;
	$\syncSel{\roleQ}{\stT}$;
	$\charTAct{\roleQ}{I}{\dagger}{\stT}$;
	$\charTActi{\dagger}{\stT}$;
	$\charTwait{\roleQ}$;
	$\syncSch{\roleQ}{\stT}$;
	$\charSchTest{\roleQ}{\dagger}{\stT}$; and
	$\charSchTrigger{\roleQ}{I}{\dagger}{\{\stT[i]\}_{i\in I}}$.
	
	\[
			\begin{array}{rcl}
				\syncBra{\roleQ}{\stT}&=&
				\roleQ\stFmt{?}\stLab\stSeq\stT
				\\
				\syncSel{\roleR[k]}{\stT}&=&
				\roleR[0]\stFmt{!}{\stLab}\stSeq
				\dots
				\roleR[k-1]\stFmt{!}\stLab\stSeq
				\roleR[k+1]\stFmt{!}\stLab\stSeq
				\dots
				\roleR[n]\stFmt{!}\stLab\stSeq\stT
			\\
				\charTAct{\roleQ}{I}{\dagger}{\stT}&=&
				\stSum{\roleP}{i\in I}{
					\stChoice{\stLab[i]}{\stS[i]}\stSeq
					\syncSel{\roleQ}{
						\roleFmt{sch}\stFmt{!}\stLab[i]\stSeq\stT
					}
				}{\dagger}
				\\
				\charTActi{\dagger}{\stT}&=&
					\stFmt{
					\roleFmt{sch}{\colorbox{yellow}{$\stFmt{\dagger}$}}\stLab\stSeq\stT
					+
						\roleP{\dagger}\stLabii
					}
				\\
				\charTwait{\roleQ}&=&
					\stRec{\stRecVar}{\left(
						\stSum{\roleFmt{sch}}{
							I \in Act,\dagger\in\{!,?\}
						}{
							\stLab[(I,\dagger)]\stSeq\charTAct{\roleQ}{I}{\dagger}{\stRecVar}
						}{?}
						\right.}\\
						&&\qquad\stFmt{+}\stSum{\roleFmt{sch}}{
							\dagger\in\{!,?\}
						}{
							\stLabi[\dagger]\stSeq\charTActi{\dagger}{\stRecVar}
						}{?}
						\\
						&&\qquad\stFmt{+}\stSum{\roleFmt{sch}}{
							\roleR\in\roleSet\setminus\{\roleQ\}
						}{
							\stLab[i]\stSeq
							\syncBra{\roleR[i]}{\stRecVar}
						}{?}\\
						&&\qquad\stFmt{+}\stFmt{
							\left.
							\roleFmt{sch}?\stLab[\text{\tiny end}]
							\right)
						}
			\end{array}
		\]
		\[
			\begin{array}{rcl}
			\syncSch{\roleR[k]}{\stT}&=&
				\roleR[0]\stFmt{!}{\stLab[k]}
				\dots
				\roleR[k-1]\stFmt{!}\stLab[k]\stSeq
				\roleR[k+1]\stFmt{!}\stLab[k]
				\dots
				\roleR[n]\stFmt{!}\stLab[k]\stSeq\stT\\
				\charSchTrigger{\roleQ}{I}{\dagger}{\{\stT[i]\}_{i\in I}}&=&
				\stFmt{
						\roleQ!
						\stLab[(I,\dagger)]\stSeq
						\syncSch{\roleQ}{
							\stSum{\roleQ}{i\in I}{\stLab[i]\stSeq\stT[i]}{?}
						}
				}\\
				\charSchTest{\roleQ}{\dagger}{\stT}&=&
				\stFmt{
					\roleQ!\stLabi[\dagger]\stSeq\roleQ\colorbox{yellow}{$\stFmt{\overline{\dagger}}$}\stLab\stSeq\stT
				}
			\end{array}
		\]
	
	The type $\charTwait{\roleQ}$ will be the type
	of $\roleQ$ in a complementary context.
	
	We define $\charSch{\stT}$ for $\mathcal{T}$-valid types $\stT$,
	by the following recursive definition:
	\[
				\charSch{\stEnd}=
					\roleR[0]\stFmt{!}{\stLab[\text{\tiny end}]}
					\dots
					\roleR[n]\stFmt{!}\stLab[\text{\tiny end}]
					\qquad
				\charSch{\stRecVar}=\stRecVar
				\qquad
				\charSch{\stRec{\stRecVar}{\stT}}=\stRec{\stRecVar}{\charSch{\stT}}
	\]
	\centerline{\colorbox{yellow}{$
	\begin{array}{l}
				\charSch{\stSum{\roleQ}{\roleQ\in J,\;i\in I_{\roleQ}}{
					
						\stChoice{\stLab[i]}{\stS[i]}\stSeq\stT[i,\roleQ]
					
				}{\dagger}}=\\
				\quad\stFmt{
					\sum_{\roleQ\in\roleSet\setminus J}
					\charSchTest{\roleQ}{\overline{\dagger}}{
						\stFmt{\sum_{\roleQ\in J}
						{\charSchTrigger{\roleQ}{I_{\roleQ}}{\overline{\dagger}}{\{\charSch{\stT[i,\roleQ]}\}_{i\in I_{\roleQ}}}}
					}}
				}
	\end{array}$
	}
	}
	
	We write $\compEnv{\mpS}{\stT}=
	\bigcup_{\roleQ\in\roleSet}\set{\stEnvMap{\mpChanRole{\mpS}{\roleQ}}{\charTwait{\roleQ}}}
	\stEnvComp
	\stEnvMap{\mpChanRole{\mpS}{\roleFmt{sch}}}{\charSch{\stT}}
	$ for valid $\stT$.
\end{definition}

The only place where full mixed choice was used over separate choice
is when a participant returns to the scheduler
after attempting to erroneously receive from $\roleP$.
We change this case to separate choice by having
the participant return to the scheduler by receiving instead of sending.

\begin{definition}[Complementary Types for Directed Choice]\rm
\label{def:char-ctx-dc}\ \\
	Fix a finite set of session types $\mathcal{T}$
	such that for all $\stT\in\mathcal{T}$,
	if $\stTi$ is a payload type in $\stT$ then $\stTi\in\mathcal{T}$.
	
	Let $L$ be the set of labels appearing within $\mathcal{T}$.
	Let $\roleSet$ be the set of participants appearing within $\mathcal{T}$.
	
	Enumerate:
	\[
	\begin{array}{rcl}
	\roleSet&=&\{\roleR[i]\}_{0\leq i\leq n}
	\\
	\{\stChoice{\stLab}{\stS}\suchthat\stLab\in L,\stS\in\mathcal{T}\cup\{\tyInt,\tyBool\}\}
	&=&\{\stChoice{\stLab[i]}{\stS[i]}\}_{0\leq i < N}
	\end{array}
	\]
	Let
	\[
	Act=\{I\subseteq\{0,\dots,N-1\}\suchthat \forall i\neq j\in I.\; \stLab[i]\neq\stLab[j]\}
	\]
	the set of all message sets, where messages have distinct labels.
	
	Let $\roleP$ and $\roleFmt{sch}$ be fresh roles,
	and $\stLabii$ be a fresh label
	\ie
	$\roleP,\roleFmt{sch}\not\in\roleSet$ and
	$\stLabii\not\in L$.
	
	We define the following functions
	for $\roleQ\in\roleSet$, $\dagger\in\{!,?\}$,
	$I\in Act$, and all $\stT$, $\{\stT[i]\}_{i\in I}$:
	$\syncBra{\roleQ}{\stT}$;
	$\syncSel{\roleQ}{\stT}$;
	$\charTAct{\roleQ}{I}{\dagger}{\stT}$;
	$\charTActii{\roleQ}{I}{\dagger}{\stT}$;
	$\charTwait{\roleQ}$;
	$\syncSch{\roleQ}{\stT}$;
	$\charSchTrigger{\roleQ}{I}{\dagger}{\{\stT[i]\}_{i\in I}}$; and
	$\charSchTriggeri{\roleQ}{I}{\dagger}{\{\stT[i]\}_{i\in I}}$.
	
	\[
			\begin{array}{rcl}
				\syncBra{\roleQ}{\stT}&=&
				\roleQ\stFmt{?}\stLab\stSeq\stT
				\\
				\syncSel{\roleR[k]}{\stT}&=&
				\roleR[0]\stFmt{!}{\stLab}\stSeq
				\dots
				\roleR[k-1]\stFmt{!}\stLab\stSeq
				\roleR[k+1]\stFmt{!}\stLab\stSeq
				\dots
				\roleR[n]\stFmt{!}\stLab\stSeq\stT
			\\
				\charTAct{\roleQ}{I}{\dagger}{\stT}&=&
				\stSum{\roleP}{i\in I}{
					\stChoice{\stLab[i]}{\stS[i]}\stSeq
					\syncSel{\roleQ}{
						\roleFmt{sch}\stFmt{!}\stLab[i]\stSeq\stT
					}
				}{\dagger}
				\\
				\charTActii{\roleQ}{I}{\dagger}{\stT}&=&
				\colorbox{yellow}{$
				\stSum{\roleP}{i\in I}{
					\stChoice{\stLab[i]}{\stS[i]}\stSeq
					\syncSel{\roleQ}{
						\roleFmt{sch}\stFmt{!}\stLab[i]\stSeq\stT
					}
				}{\dagger}
				+
				\roleP\stFmt{\overline{\dagger}}\stLabii$}
				\\
				\charTwait{\roleQ}&=&
					\stRec{\stRecVar}{\left(
						\stSum{\roleFmt{sch}}{
							I \in Act,\dagger\in\{!,?\}
						}{
							\stLab[(I,\dagger)]\stSeq\charTAct{\roleQ}{I}{\dagger}{\stRecVar}
						}{?}
						\right.}\\
						&&\qquad\stFmt{+}\colorbox{yellow}{$\stSum{\roleFmt{sch}}{
							I\in Act,\dagger\in\{!,?\}
						}{
							\stLabi[(I,\dagger)]\stSeq\charTActii{\roleQ}{I}{\dagger}{\stRecVar}
						}{?}$}
						\\
						&&\qquad\stFmt{+}\stSum{\roleFmt{sch}}{
							\roleR\in\roleSet\setminus\{\roleQ\}
						}{
							\stLab[i]\stSeq
							\syncBra{\roleR[i]}{\stRecVar}
						}{?}\\
						&&\qquad\stFmt{+}\stFmt{
							\left.
							\roleFmt{sch}?\stLab[\text{\tiny end}]
							\right)
						}
			\end{array}
		\]
		\[
			\begin{array}{rcl}
			\syncSch{\roleR[k]}{\stT}&=&
				\roleR[0]\stFmt{!}{\stLab[k]}
				\dots
				\roleR[k-1]\stFmt{!}\stLab[k]\stSeq
				\roleR[k+1]\stFmt{!}\stLab[k]
				\dots
				\roleR[n]\stFmt{!}\stLab[k]\stSeq\stT\\
				\charSchTrigger{\roleQ}{I}{\dagger}{\{\stT[i]\}_{i\in I}}&=&
				\stFmt{
						\roleQ!
						\stLab[(I,\dagger)]\stSeq
						\syncSch{\roleQ}{
							\stSum{\roleQ}{i\in I}{\stLab[i]\stSeq\stT[i]}{?}
						}
				}\\
				\charSchTriggeri{\roleQ}{I}{\dagger}{\{\stT[i]\}_{i\in I}}&=&
				\colorbox{yellow}{$
				\stFmt{
						\roleQ!
						\stLabi[(I,\dagger)]\stSeq
						\syncSch{\roleQ}{
							\stSum{\roleQ}{i\in I}{\stLab[i]\stSeq\stT[i]}{?}
						}
				}$}
			\end{array}
		\]
	
	The type $\charTwait{\roleQ}$ will be the type
	of $\roleQ$ in a complementary context.
	
	We define $\charSch{\stT}$ for $\mathcal{T}$-valid types $\stT$,
	by the following recursive definition:
	\[
				\charSch{\stEnd}=
					\roleR[0]\stFmt{!}{\stLab[\text{\tiny end}]}
					\dots
					\roleR[n]\stFmt{!}\stLab[\text{\tiny end}]
					\qquad
				\charSch{\stRecVar}=\stRecVar
				\qquad
				\charSch{\stRec{\stRecVar}{\stT}}=\stRec{\stRecVar}{\charSch{\stT}}
	\]
	\centerline{\colorbox{yellow}{$
	\begin{array}{l}
		\charSch{\stSum{\roleQ}{i\in I}{\stChoice{\stLab[i]}{\stS[i]}\stSeq\stT[i]}{\dagger}}
		=\stFmt{
			\sum_{i\in I}\charSchTrigger{\roleQ}{I}{\overline{\dagger}}{\{\charSch{\stT[i]}\}_{i\in I}}
		}\\[1ex]
		\charSch{\stSum{\roleQ}{i\in I}{\stChoice{\stLab[i]}{\stS[i]}\stSeq\stT[i]}{!}
		+
		\stSum{\roleQ}{i\in I'}{\stChoice{\stLab[i]}{\stS[i]}\stSeq\stT[i]}{?}}
		=\\
		\quad\stFmt{
			\sum_{i\in I}\charSchTriggeri{\roleQ}{I}{?}{\{\charSch{\stT[i]}\}_{i\in I}}
			+
			\sum_{i\in I'}\charSchTriggeri{\roleQ}{I'}{!}{\{\charSch{\stT[i]}\}_{i\in I'}}
		}\\
	\end{array}$}}
	We write $\compEnv{\mpS}{\stT}=
	\bigcup_{\roleQ\in\roleSet}\set{\stEnvMap{\mpChanRole{\mpS}{\roleQ}}{\charTwait{\roleQ}}}
	\stEnvComp
	\stEnvMap{\mpChanRole{\mpS}{\roleFmt{sch}}}{\charSch{\stT}}
	$ for valid $\stT$.
\end{definition}

Mixed choice was used over directed choice
when testing for erroneous actions.
Since types may only communicate with one participant at once,
we only have to test to see if $\roleP$ can receive from/send to $\roleQ$
when it should only be able to send to/receive from $\roleQ$.

\begin{definition}[Complementary Types for Separated Choice]\rm
\label{def:char-ctx-s}\ \\
	Fix a finite set of session types $\mathcal{T}$
	such that for all $\stT\in\mathcal{T}$,
	if $\stTi$ is a payload type in $\stT$ then $\stTi\in\mathcal{T}$.
	
	Let $L$ be the set of labels appearing within $\mathcal{T}$.
	Let $\roleSet$ be the set of participants appearing within $\mathcal{T}$.
	
	Enumerate:
	\[
	\begin{array}{rcl}
	\roleSet&=&\{\roleR[i]\}_{0\leq i\leq n}
	\\
	\{\stChoice{\stLab}{\stS}\suchthat\stLab\in L,\stS\in\mathcal{T}\cup\{\tyInt,\tyBool\}\}
	&=&\{\stChoice{\stLab[i]}{\stS[i]}\}_{0\leq i < N}
	\end{array}
	\]
	Let
	\[
	Act=\{I\subseteq\{0,\dots,N-1\}\suchthat \forall i\neq j\in I.\; \stLab[i]\neq\stLab[j]\}
	\]
	the set of all message sets, where messages have distinct labels.
	
	Let $\roleP$ and $\roleFmt{sch}$ be fresh roles,
	and $\stLabii$ be a fresh label
	\ie
	$\roleP,\roleFmt{sch}\not\in\roleSet$ and
	$\stLabii\not\in L$.
	
	We define the following functions
	for $\roleQ\in\roleSet$, $\dagger\in\{!,?\}$,
	$I\in Act$, and all $\stT$, $\{\stT[i]\}_{i\in I}$:
	$\syncBra{\roleQ}{\stT}$;
	$\syncSel{\roleQ}{\stT}$;
	$\charTAct{\roleQ}{I}{\dagger}{\stT}$;
	$\charTwait{\roleQ}$;
	$\syncSch{\roleQ}{\stT}$; and
	$\charSchTrigger{\roleQ}{I}{\dagger}{\{\stT[i]\}_{i\in I}}$.
	
	\[
			\begin{array}{rcl}
				\syncBra{\roleQ}{\stT}&=&
				\roleQ\stFmt{?}\stLab\stSeq\stT
				\\
				\syncSel{\roleR[k]}{\stT}&=&
				\roleR[0]\stFmt{!}{\stLab}\stSeq
				\dots
				\roleR[k-1]\stFmt{!}\stLab\stSeq
				\roleR[k+1]\stFmt{!}\stLab\stSeq
				\dots
				\roleR[n]\stFmt{!}\stLab\stSeq\stT
			\\
				\charTAct{\roleQ}{I}{\dagger}{\stT}&=&
				\stSum{\roleP}{i\in I}{
					\stChoice{\stLab[i]}{\stS[i]}\stSeq
					\syncSel{\roleQ}{
						\roleFmt{sch}\stFmt{!}\stLab[i]\stSeq\stT
					}
				}{\dagger}
				\\
				\charTwait{\roleQ}&=&
					\stRec{\stRecVar}{\left(
						\stSum{\roleFmt{sch}}{
							I \in Act,\dagger\in\{!,?\}
						}{
							\stLab[(I,\dagger)]\stSeq\charTAct{\roleQ}{I}{\dagger}{\stRecVar}
						}{?}
						\right.}
						\\
						&&\qquad\stFmt{+}\stSum{\roleFmt{sch}}{
							\roleR\in\roleSet\setminus\{\roleQ\}
						}{
							\stLab[i]\stSeq
							\syncBra{\roleR[i]}{\stRecVar}
						}{?}\\
						&&\qquad\stFmt{+}\stFmt{
							\left.
							\roleFmt{sch}?\stLab[\text{\tiny end}]
							\right)
						}
			\end{array}
		\]
		\[
			\begin{array}{rcl}
			\syncSch{\roleR[k]}{\stT}&=&
				\roleR[0]\stFmt{!}{\stLab[k]}
				\dots
				\roleR[k-1]\stFmt{!}\stLab[k]\stSeq
				\roleR[k+1]\stFmt{!}\stLab[k]
				\dots
				\roleR[n]\stFmt{!}\stLab[k]\stSeq\stT\\
				\charSchTrigger{\roleQ}{I}{\dagger}{\{\stT[i]\}_{i\in I}}&=&
				\stFmt{
						\roleQ!
						\stLab[(I,\dagger)]\stSeq
						\syncSch{\roleQ}{
							\stSum{\roleQ}{i\in I}{\stLab[i]\stSeq\stT[i]}{?}
						}
				}
			\end{array}
		\]
	
	The type $\charTwait{\roleQ}$ will be the type
	of $\roleQ$ in a complementary context.
	
	We define $\charSch{\stT}$ for $\mathcal{T}$-valid types $\stT$,
	by the following recursive definition:
	\[
				\charSch{\stEnd}=
					\roleR[0]\stFmt{!}{\stLab[\text{\tiny end}]}
					\dots
					\roleR[n]\stFmt{!}\stLab[\text{\tiny end}]
					\qquad
				\charSch{\stRecVar}=\stRecVar
				\qquad
				\charSch{\stRec{\stRecVar}{\stT}}=\stRec{\stRecVar}{\charSch{\stT}}
	\]
	\centerline{\colorbox{yellow}{$
		\charSch{\stSum{\roleQ}{i\in I}{\stChoice{\stLab[i]}{\stS[i]}\stSeq\stT[i]}{\dagger}}
		=
		\stFmt{
			\sum_{i\in I}\charSchTrigger{\roleQ}{I}{\overline{\dagger}}{\{\charSch{\stT[i]}\}_{i\in I}}
		}$}}
	We write $\compEnv{\mpS}{\stT}=
	\bigcup_{\roleQ\in\roleSet}\set{\stEnvMap{\mpChanRole{\mpS}{\roleQ}}{\charTwait{\roleQ}}}
	\stEnvComp
	\stEnvMap{\mpChanRole{\mpS}{\roleFmt{sch}}}{\charSch{\stT}}
	$ for valid $\stT$.
\end{definition}

We simplify further from $\DM$ as
we no longer need to test for $\roleP$ erroneously
allowing both sending and receiving.

\section{Appendix for the Typing Context Property Algorithm (\S~6)}

\label{app:impl}

\subsection{Deciding Liveness}

\propLiveness*

We begin by tightening the analysis
in \cite{UY2025}.

\begin{definition}[Action Set]\rm
	The action set of $\stT$ is the set of $\roleQ\stFmt{\dagger}\stLab$
	appearing within $\stT$, denoted $\act{\stT}$.
\end{definition}

\begin{lemma}
	$|\act{\stT}|+1\leq |\stT|$
\end{lemma}

\begin{proof}
	Induction on $\stT$:
	\begin{itemize}
		\item If $\stT=\stEnd$,
		then $|\act{\stT}|+1=1=|\stEnd|$.
		\item If $\stT=\stRecVar$,
		then $|\act{\stT}|+1=1=|\stRecVar|$.
		\item If $\stT=\stRec{\stRecVar}{\stTi}$, then
		$|\act{\stT}|+1=|\act{\stTi}|+1\leq |\stTi|=|\stT|$.
		\item If $\stT=\stSum{\roleQ[i]}{i\in I}{\stChoice{\stLab[i]}{\stS[i]}\stSeq\stT[i]}{\dagger_i}$, then
		\[
		\begin{array}{lcl}
		|\act{\stT}|+1&=&
		|\bigcup_{i\in I}\left(
		\{\roleQ[i]\stFmt{\dagger_i}\stLab[i]\}\cup\act{\stT[i]}
		\right)|+1\\
		&\leq&
		\sum_{i\in I}\left(1+|\act{\stT[i]}|\right)+1\\
		&\leq&
		\sum_{i\in I}|\stT[i]|+1\\
		&\leq&
		|\stT|
		\end{array}
		\]
	\end{itemize}
\qed\end{proof}

\begin{lemma}
	$\act{\tySubst{\stT}{\stRecVar}{\stTi}}\subseteq\act{\stT}\cup\act{\stTi}$.
\end{lemma}

\begin{proof}
	Induction on $\stT$.
\qed\end{proof}

\begin{lemma}
	If $\stT\,\gtMove\,\stTi$, then $\act{\stTi}\subseteq\act{\stT}$.
\end{lemma}

\begin{proof}
	Induction on $\stT\,\gtMove\,\stTi$.
	If $\stT=\stRec{\stRecVar}{\stTii}$, then
	$\tySubst{\stTii}{\stRecVar}{\stT}\,\gtMove\,\stTi$
	and $\act{\stTi}\subseteq\act{\tySubst{\stTii}{\stRecVar}{\stT}}$,
	by I.H.,
	and
	$\act{\tySubst{\stTii}{\stRecVar}{\stT}}\subseteq \act{\stTii}\cup\act{\stT}=\act{\stT}$.
	If $\stT=\stSum{\roleQ[i]}{i\in I}{\stChoice{\stLab[i]}{\stS[i]}\stSeq\stT[i]}{\dagger_i}$, then
	$\stTi=\stT[j]$ for some $j\in I$,
	so $\act{\stTi}\subseteq\act{\stT}$.
\qed\end{proof}

\begin{lemma}
	If $\stT\,\gtMove[\roleP\stEnvAnnotOutSym\stChoice{\stLab}{\stS}]$,
	then $\roleP\stFmt{!}\stLab\in\act{\stT}$.
	If $\stT\,\gtMove[\roleP\stEnvAnnotInSym\stChoice{\stLab}{\stS}]$,
	then $\roleP\stFmt{?}\stLab\in\act{\stT}$.
\end{lemma}

\begin{proof}
	Induction on $\stT\,\gtMove[\roleP\stEnvAnnotOutSym\stChoice{\stLab}{\stS}]$.
	If $\stT=\stRec{\stRecVar}{\stTi}$, then
	$\tySubst{\stTi}{\stRecVar}{\stT}\,\gtMove[\roleP\stEnvAnnotOutSym\stChoice{\stLab}{\stS}]$,
	so, by I.H.,
	$\roleP\stFmt{!}\stLab\in\act{\tySubst{\stTi}{\stRecVar}{\stT}}\subseteq\act{\stT}$.
	If $\stT=\stSum{\roleQ[i]}{i\in I}{\stChoice{\stLab[i]}{\stS[i]}\stSeq\stT[i]}{\dagger_i}$, then
	there exists $j\in I$ s.t.
	$\roleQ[j]\stFmt{\dagger_j}\stLab[j]=\roleP\stFmt{!}\stLab\in\act{\stT}$.
	
	The case of $\stT\,\gtMove[\roleP\stEnvAnnotInSym\stChoice{\stLab}{\stS}]$
	is similar. 
\qed\end{proof}

Say that $\stEnv\,\gtMove[\ltsSendRecvS{\mpS}{\roleP}{\roleQ}{\stLab}]$
iff there exists $\stS$ such that
$\stEnv\,\gtMove[\ltsSendRecvS{\mpS}{\roleP}{\roleQ}{\stChoice{\stLab}{\stS}}]$.
Note that, using these labels,
transitions are deterministic as types
have at most one choice per label.
(Recall well-formedness assumption from \cref{def:types})

\begin{lemma}
	If $\stEnv\,\gtMove[\ltsSendRecvS{\mpS}{\roleP}{\roleQ}{\stLab}]$,
	then $\act{\stEnvApp{\stEnv}{\mpChanRole{\mpS}{\mpP}}}\ni\roleQ\stFmt{!}\stLab$,
	and $\act{\stEnvApp{\stEnv}{\mpChanRole{\mpS}{\mpQ}}}\ni\roleP\stFmt{?}\stLab$.
\end{lemma}

\begin{proof}
	Invert context transitions to local types and apply the prior lemma.
\qed\end{proof}

\begin{lemma}
	Let $n=\sum_{\mpC\in\dom{\stEnv}}|\stEnvApp{\stEnv}{\mpC}|$.
	There are at most $\frac{n}{2}$ labels,
	of the form $\stEnvAnnotGenericSym=\stEnv\,\gtMove[\ltsSendRecvS{\mpS}{\roleP}{\roleQ}{\stLab}]$,
	such that $\stEnv\,\gtMoveStar\,\cdot\,\gtMove[\stEnvAnnotGenericSym]$.
\end{lemma}

\begin{proof}
	Let $\stEnvAnnotGenericSym$ be s.t.
	$\stEnv\,\gtMoveStar\,\cdot\,\gtMove[\stEnvAnnotGenericSym]$.
	
	Consider the map:
	\[
		F :
		\ltsSendRecvS{\mpS}{\roleP}{\roleQ}{\stLab}
		\mapsto
		\{
			(\mpChanRole{\mpS}{\mpP}, \roleQ\stFmt{!}\stLab),
			(\mpChanRole{\mpS}{\mpQ}, \roleP\stFmt{?}\stLab)
		\}
		\subseteq
		 \bigcup_{\mpC\in \dom{\stEnv}} \{\mpC\}\times\act{\stEnvApp{\stEnv}{\mpC}}
	\]
	If $\stEnvAnnotGenericSym\neq \stEnvAnnotGenericSymi$,
	then $F(\stEnvAnnotGenericSym)\cap F(\stEnvAnnotGenericSymi)=\emptyset$.
	
	Observe that
	$\bigcup_{\mpC\in \dom{\stEnv}} \{\mpC\}\times\act{\stEnvApp{\stEnv}{\mpC}}
	\leq n$.
	
	Thus, there are at most $\frac{n}{2}$ such $\stEnvAnnotGenericSym$.
\qed\end{proof}

A \emph{counterwitness} for $\stEnv$
is a fair and non-live path.

\begin{lemma}[Refining Counterwitnesses]
\label{lem:ref-counter}
	Let $m=\prod_{\mpC\in\dom{\stEnv}}|\stEnvApp{\stEnv}{\mpC}|$.
	If $\stEnv$ is not live, then
	it has a counterwitness of the form $\mathcal{P}_1\mathcal{P}^{\omega}_2$
	with $|\mathcal{P}_1|\leq m$
	and
	$0\leq |\mathcal{P}_2|\leq m\lfloor\frac{|\stEnv|}{2}\rfloor$.
\end{lemma}

\begin{proof}
	Let $\{\stEnv[i]\}_{i\in N}$
	be a fair and non-live path for $\stEnv$.
	
	By definition, find $k\in N$ and $\mpC\in\barb{\stEnv[k]}$
	such that $\mpC\not\in\bigcup_{i\geq k}\bigcup \obs{\stEnv[i]}$.
	So $\mpC\in\barb{\stEnv[k']}$
	and $\mpC\not\in\bigcup_{i\geq k'}\bigcup \obs{\stEnv[i]}$
	for all $k'\geq k$.
	
	\begin{enumerate}
		\item[] Case $N$ is finite.
		Say $N=\{0,\dots,n\}$.
		As $\{\stEnv[i]\}_{i\in N}$ is fair,
		$\obs{\stEnv[n]}=\emptyset$
		and,
		by hypothesis,
		$\mpC\in\barb{\stEnv[n]}$.
		
		$\{\stEnv[i]\}_{i\in N}$ is a path from
		$\stEnv$ to $\stEnv[n]$ in the LTS,
		which is of size at most m.
		Hence,
		there is a path from $\stEnv$
		to $\stEnv[n]$ in the LTS
		of size at most m.
		Denote this path by $\{\stEnvi[i]\}_{0\leq i\leq n'}$,
		where $\stEnvi[0]=\stEnv$,
		$\stEnvi[n']=\stEnv[n]$, and
		$n'+1 \leq m$.
		
		This path is fair,
		as $\obs{\stEnvi[n']}=\emptyset$
		so all observations must have one of their participant acting
		after being observed,
		and is not live,
		as $\mpC\in\barb{\stEnvi[n']}$.
		\\[1ex]
		
		\item[] Case N is infinite.
		Find $k\leq v<u$
		s.t.
		$\stEnv[v]=\stEnv[u]$
		and
		\[
		\begin{array}{lcl}
			\{\ltsSubject{\stEnvAnnotGenericSym}
			\suchthat
			\exists j \geq v.
			\stEnv[j]\,\gtMove[\stEnvAnnotGenericSym]\,\stEnv[j+1]\}
			&=&
			\{\ltsSubject{\stEnvAnnotGenericSym}
			\suchthat
			\exists j .
			v\leq j < u\text{ and }
			\stEnv[j]\,\gtMove[\stEnvAnnotGenericSym]\,\stEnv[j+1]\}\\
			&=&
			\bigcap_{i\in N}
			\{\ltsSubject{\stEnvAnnotGenericSym}
			\suchthat
			\exists j \geq i.
			\stEnv[j]\,\gtMove[\stEnvAnnotGenericSym]\,\stEnv[j+1]\}
		\end{array}
		\]
		This is possible as some state
		must occur infinitely many times after $k$,
		and we can pick two occurrences
		far enough apart to contain all
		of the (finitely many) labels
		that occur infinitely many times.
		
		Now, $\stEnv[0],\dots,\stEnv[v],(\stEnv[v+1],\dots,\stEnv[u])^{\omega}$
		is a fair and non-live path,
		so assume that we are using it.
		
		As in the finite case,
		we replace
		$\stEnv[0],\dots,\stEnv[v-1],\stEnv[v]$
		by a path from
		$\stEnv[0]$ to $\stEnv[v]$
		of length at most $m$.
		So, without loss of generality, $v+1\leq m$.
		
		If there are $i,i'\in N$ s.t.
		$v\leq i < i'\leq u$,
		$\stEnv[i]=\stEnv[i']$, and
		\[
			\{\stEnvAnnotGenericSym
			\suchthat
			\exists j .
			i\leq j < u
			\text{ and }
			\stEnv[j]\,\gtMove[\stEnvAnnotGenericSym]\,\stEnv[j+1]\}
			=
			\{\stEnvAnnotGenericSym
			\suchthat
			\exists j .
			i'\leq j < u
			\text{ and }
			\stEnv[j]\,\gtMove[\stEnvAnnotGenericSym]\,\stEnv[j+1]\}
		\]
		Then we may remove $\{i,i+1,\dots,i'-1\}$
		from $N$ and maintain fairness and
		inaction of $\mpC$.
		
		Each of these removals decreases the size
		of $N\cap [v,u]$,
		which is finite,
		so do this until we cannot,
		and relabel the path to have no gaps.
		
		So, for $i,i'\in N$ s.t.
		$v\leq i < i'\leq u$ and
		$\stEnv[i]=\stEnv[i']$,
		\[
			|\{\stEnvAnnotGenericSym
			\suchthat
			\exists j .
			i\leq j < u
			\text{ and }
			\stEnv[j]\,\gtMove[\stEnvAnnotGenericSym]\,\stEnv[j+1]\}|
			>
			|\{\stEnvAnnotGenericSym
			\suchthat
			\exists j .
			i'\leq j < u
			\text{ and }
			\stEnv[j]\,\gtMove[\stEnvAnnotGenericSym]\,\stEnv[j+1]\}|
		\]
		These sets may only take values from $0$ to $|\stEnv|$,
		inclusive,
		and may only take on $0$ for $i=u$,
		so each state may only occur in $\{v,v+1,\dots,u\}$
		at most $n$ times,
		except $\stEnv[u]$, which may occur at most $n+1$ times.
		Hence,
		\[
			v-u+1
			\leq
			|\stEnv|\prod_{\mpC\in\dom{\stEnv}}|\stEnvApp{\stEnv}{\mpC}|
			+1
		\]
		Let
		\[
			\begin{array}{lcl}
				\mathcal{P}_1&=&\stEnv[0],\dots,\stEnv[v]\\
				\mathcal{P}_2&=&\stEnv[v+1],\dots,\stEnv[u]\\[2ex]
				|\mathcal{P}_1|&\leq&\prod_{\mpC\in\dom{\stEnv}}|\stEnvApp{\stEnv}{\mpC}|\\
				|\mathcal{P}_2|&\leq&
				|\stEnv|
				\prod_{\mpC\in\dom{\stEnv}}|\stEnvApp{\stEnv}{\mpC}|
			\end{array}
		\]
	\end{enumerate}
\qed\end{proof}

\begin{definition}[Reachable without Observations]\rm
	We define the predicate $\reach{\stEnv}{\stEnvi}{\mpC}$
	to be true
	iff there is a path from $\stEnv$ to $\stEnvi$
	where no state contains $\mpC$ in an observation.
	
	Formally, $\reach{\stEnv}{\stEnvi}{\mpC}$
	iff there exists a sequence $\{\stEnv[i]\}_{0\leq i \leq n}$
	such that $\stEnv=\stEnv[0]$,
	$\stEnvi=\stEnv[n]$,
	for all $i<n$, $\stEnv[i]\,\gtMove\,\stEnv[i+1]$, and,
	for all $i\leq n$, $\mpC\not\in\bigcup\obs{\stEnv[i]}$.
\end{definition}

\propLiveness*

\begin{proof}
	Suppose that $\stEnv$ is not live,
	so there is a fair and non-live path
	of the form
	\[
	\stEnv[0],\dots,\stEnv[n],(\stEnv[n+1]),\dots,\stEnv[n+m])^{\omega}
	\]
	with $\stEnv[n]=\stEnv[n+m]$ or $m=0$.
	If $m=0$, then $\obs{\stEnv[n]}=\emptyset$ by fairness
	and $\barb{\stEnv[n]}\neq\emptyset$ by liveness,
	so the formula holds.
	
	Assume now that $m>0$ and $\stEnv[n]=\stEnv[n+m]$.
	
	This path is not live, so there is a state with barb $\mpC$
	s.t. $\mpC$ appears in no observation following the state;
	this is true for all following states also.
	Therefore, it is true for $\stEnv[n+1]),\dots,\stEnv[n+m]$;
	in particular, it is true for $\stEnv[n]$.
	
	$\stEnv=\stEnv[0]\,\gtMoveStar\,\stEnv[n]$
	and
	$\mpC\in \barb{\stEnv[n]}$
	and
	$\mpC\not\in \bigcup_{n\leq i \leq n+m}\bigcup\obs{\stEnv[i]}$.
	
	Therefore, $\reach{\stEnv[i]}{\stEnv[j]}{\mpC}$
	for $n\leq i \leq j \leq n+m$.
	
	The path is fair, therefore, for $X\in\obs{\stEnv[n]}$,
	there exists $i \geq n$ s.t.
	$\stEnv[i]\,\gtMove[\stEnvAnnotGenericSym]\,\stEnv[i+1]$
	and $X\cap\ltsSubject{\stEnvAnnotGenericSym}\neq\emptyset$.
	By repeating states, we assume that $n\leq i < n+m$
	and so,
	$\reach{\stEnv[n]}{\stEnv[i]}{\mpC}$
	and
	$\reach{\stEnv[i+1]}{\stEnv[n]}{\mpC}$.
	
	So, the formula holds.
	\\[2ex]
	Suppose that the formula holds.
	Find $\stEnvi$, $\mpC$, $\{\stEnv[X,l]\}_{X\in\obs{\stEnvi}}$,
	$\{\stEnv[X,r]\}_{X\in\obs{\stEnvi}}$, and
	$\{\stEnvAnnotGenericSym[X]\}_{X\in\obs{\stEnvi}}$
	s.t.
	$\stEnv\,\gtMoveStar\,\stEnvi$,
	$\mpC\in\barb{\stEnvi}$, and
	for all $X\in\obs{\stEnvi}$:\\
	$\reach{\stEnvi}{\stEnv[X,l]}{\mpC}$,
	$\reach{\stEnv[X,r]}{\stEnvi}{\mpC}$,
	$\stEnv[X,l]\,\gtMove[{\stEnvAnnotGenericSym[X]}]\,\stEnv[X,r]$,
	and $X\cap\ltsSubject{\stEnvAnnotGenericSym[X]}\neq \emptyset$.
	
	Let $\mathcal{P}$ be a path from $\stEnv$ to $\stEnvi$
	and, for $X\in\obs{\stEnvi}$,
	let $\stEnvi,\mathcal{P}_{X,l}$ be a path from $\stEnvi$ to $\stEnv[X,l]$
	never observing $\mpC$
	and $\mathcal{P}_{X,r}$ be a path from $\stEnv[X,r]$ to $\stEnvi$
	never observing $\mpC$.
	
	Enumerate $\obs{\stEnvi}=\{X_1,\dots,X_n\}$.
	Consider the path:
	\[\mathcal{P}
	\left(
	\mathcal{P}_{X_1,l}\mathcal{P}_{X_1,r}\dots
	\mathcal{P}_{X_n,l}\mathcal{P}_{X_n,r}
	\right)^\omega
	\]
	This path is not live, since $\mpC$ is a barb that
	is never observed in the loop.
	This path is fair,
	since any observation in the loop
	will either be fulfilled within the loop,
	or be part of the next occurrence of $\stEnvi$'s observations,
	which are fulfilled in every loop.
\qed\end{proof}

\thmDecide*

See artifact.
Below is a sketch of the liveness algorithm.
\begingroup
\small
\begin{lstlisting}[caption=Liveness Algorithm,xleftmargin=5.0ex,escapechar=\%]
Func LIVE?(%\stEnv%)
  VAL counter: Bool = False
  VAL stateSpace: Set[Ctx] = {%$\stEnvi$%:%$\stEnv\gtMoveStar\stEnvi$%}	
  FOR %$\mpC$% IN BARB(%\stEnv%) DO
    VAL R: Map[Ctx,Set[Ctx]] =
      {
        %$\stEnvi\mapsto\{\stEnvii\suchthat\stEnvi=\stEnv[0]\gtMove\dots\gtMove\stEnv[n]=\stEnvii,\forall i \leq n.\mpC\not\in\bigcup\obs{\stEnv[i]}\}$%:
        %\stEnvi% IN stateSpace
      }
    FOR %\stEnvi% IN stateSpace DO
      VAR acc: Bool = True
      FOR %$X$% IN OBS(%\stEnvi%) DO
        VAL acc': Bool = False
        FOR %\stEnvii% IN stateSpace DO
          FOR %$\stEnvii\gtMove[\stEnvAnnotGenericSym]\stEnviii$% DO
            IF R(%\stEnvi%,%\stEnvii%) AND R(%\stEnvii%,%\stEnviii%) AND SUBJ(%$\stEnvAnnotGenericSym$%)%$\cap X\neq\emptyset$% THEN
              acc' = True
        acc = acc AND acc'
      counter = counter OR acc
  RETURN NOT counter
\end{lstlisting}
\endgroup
\begin{theorem}[Liveness is PSPACE-Hard]
	The property $\stEnvLivePred$ is PSPACE-hard.
\end{theorem}

\begin{proof}
	$\stEnvLivePred$, restricted to types without mixed choice,
	is the standard liveness property \cite[Definition 3.11]{YHK2026}
	without mixed choice
	(\cref{prop:live-ctx}), which is PSPACE-hard \cite{UY2025}.
\qed\end{proof}

\begin{theorem}[Liveness is in PSPACE]
	Liveness can be decided in space polynomial in $|\stEnv|$.
\end{theorem}

\begin{proof}
	$|\mathcal{T}_{\stT}|=|\{\stTi\suchthat \stT\,\gtMoveStar\,\stTi\}|\leq|\stT|$,
	so can be stored in space $\mathcal{O}\left(|\stT|^2\right)$.
	
	\[
	\left\{\stEnvi\suchthat \stEnv\,\gtMoveStar\,\stEnvi\right\}
	\subseteq
	\left\{
	\left(\bigcup_{\mpC\in\dom{\stEnv}}\set{\stEnvMap{\mpC}{\stT[\mpC]}}\right)
	\suchthat
	\stT[\mpC] \in \mathcal{T}_{\stEnvApp{\stEnv}{\mpC}}
	\right\}
	= \mathcal{D}
	\]
	
	We need $\mathcal{O}\left(\sum_{\mpC\in\dom{\stEnv}}|\stEnvApp{\stEnv}{\mpC}|^2\right)=
	\mathcal{O}\left(|\stEnv|^2\right)$ space
	to iterate over $\mathcal{D}$.
	
	Since the state space is of size at most $m=|\stEnv|^{|\dom{\stEnv}|}$,
	\[\stEnv\,\gtMoveStar\,\stEnvi
	\,
	\text{iff}\,(\stEnvi\in\mathcal{D}\,
	\text{and}
	\,
	\stEnv\,\gtMoveStar\,\stEnvi\,\text{in a path of length at most $m$})
	\]
	
	Define $\reachi{\stEnv[l]}{\stEnv[r]}{t}$
	to be true
	iff $\stEnv[l]\,\gtMoveStar\,\stEnv[r]$
	in a path with at most $2^t$ edges.
	
	Define $\reachii{\stEnv[l]}{\stEnv[r]}{\mpC}{t}$
	to be true
	iff $\stEnv[l]\,\gtMoveStar\,\stEnv[r]$
	in a path with at most $2^t$ edges
	where $\mpC$ is never observed.
	
	We have that
	$\reachi{\stEnv[l]}{\stEnv[r]}{0}$ iff
	$\stEnv[l]=\stEnv[r]$
	or
	$\stEnv[l]\,\gtMove\,\stEnv[r]$.
	This can be decided in time $\mathcal{O}(n^2)$
	and space $\mathcal{O}(n)$.
	
	We have that
	$\reachi{\stEnv[l]}{\stEnv[r]}{t+1}$
	iff
	there exists $\stEnv[m]\in\mathcal{D}$ s.t.
	$\reachi{\stEnv[l]}{\stEnv[m]}{t}$
	and
	$\reachi{\stEnv[m]}{\stEnv[r]}{t}$.
	This step requires at most $|\stEnv|^2$
	space, $m$ loops, two recursive calls and a depth of $t+2$.
	
	Therefore, computing $\reachi{\stEnv[l]}{\stEnv[r]}{t}$
	takes time $\mathcal{O}(\left(2m\right)^t n^2)$
	and space $\mathcal{O}(tn^2)$.
	
	We compute
	$\reachii{\stEnv[l]}{\stEnv[r]}{t}{\mpC}$
	similarly, but we also need to ensure the $\mpC$
	is not an observation in the base case, which adds no further complexity.
	
	Now by our upper bound on the size of the state space:
	\[\begin{array}{rcl}
	\stEnv[l]\,\gtMoveStar\,\stEnv[r]
	&\iff&\reachi{\stEnv[l]}{\stEnv[r]}{|\dom{\stEnv}|\lceil \log{n}\rceil}\\[1ex]
	\reach{\stEnv[l]}{\stEnv[r]}{\mpC}&
	\iff&
	\reachii{\stEnv[l]}{\stEnv[r]}{|\dom{\stEnv}|\lceil \log{n}\rceil}{\mpC}
	\end{array}\]
	
	So, these are both computed in time
	\[\mathcal{O}((2m)^{r(\log{n}+1)}n^2)=
	\mathcal{O}(2^r n^{r^2 \log{n}+r+3})\]
	and space
	\[\mathcal{O}(rn^2\log{n})\]
	
	Now, we use the above formula to decide liveness,
	by looping over
	$\stEnvi\in\mathcal{D}$
	such that
	$\stEnv\,\gtMoveStar\,\stEnvi$
	whenever we need to introduce a state in the LTS.
	
	This takes time
	$\mathcal{O}(
	r^2 4^r n^{2 r^2 \log{n}+5r+10} 
	)
	$.
	This takes space $\mathcal{O}(rn^2\log{n})=\mathcal{O}(n^3\log{n})$.
\qed\end{proof}
Thus, $\stEnvLivePred$ is PSPACE-complete.

As stated in \cref{sec:liveness-algo},
the $\mu$-calculus model checking approach
to deciding liveness is compatible with our definition.
However, our formula differs from the existing formula
for liveness without mixed choice.
This is because we need to allow fair paths where
one participant in an enabled pair may instead
communicate with a different participant, not in the pair,
and allow live paths where a participant
have multiple distinct head actions but only takes one of them.

\[\muFmt{
	\begin{array}{lcl}
	\muPred[\text{live}]&=&
	\muGFP{\muVar}{
		\forall \mpS,\roleP,\roleQ,\stLab,\stS.
		\left(
			\left[
				\ltsSendRecvS{\mpS}{\roleP}{\roleQ}{\stLab}
			\right]
			\muVar\right.}\\
			&&\quad\muAnd
			\left(\left(
							\left\langle\ltsSend{\mpS}{\roleP}{\roleQ}{\stLab}{\stS}\right\rangle\muTrue
							\muOr
							\left\langle\ltsRecv{\mpS}{\roleP}{\roleQ}{\stLab}{\stS}\right\rangle\muTrue
			\right)\right.\\
			&&\quad\quad\muImplies
			\muLFP{\muVari}{
				\exists \rolePi,\roleQi,\stLabi.
				\left(
									\left\langle
										\ltsSendRecvS{\mpS}{\rolePi}{\roleQi}{\stLabi}
									\right\rangle
									\muTrue
				\right.}\\
					&&\quad\quad\quad\muAnd
					\muGFP{\muVarii}{
					\left(
						\exists \roleQ,\stLab.
						\left(
							\left\langle
														\ltsSendRecvS{\mpS}{\roleP}{\roleQ}{\stLab}
													\right\rangle\muTrue
													\muOr
													\left\langle
													\ltsSendRecvS{\mpS}{\roleQ}{\roleP}{\stLab}
													\right\rangle\muTrue
						\right)\right.}\\
						&&\quad\quad\quad\quad\muOr\forall \roleR,\stLab.\left(\right.\\
						&&\quad\quad\quad\quad\quad\;\;\;\left[
														\ltsSendRecvS{\mpS}{\rolePi}{\roleR}{\stLab}
													\right]\muVari\\
													&&\quad\quad\quad\quad\quad\muAnd
													\left[
														\ltsSendRecvS{\mpS}{\roleQi}{\roleR}{\stLab}
													\right]\muVari\\
													&&\quad\quad\quad\quad\quad\muAnd
													\left[
														\ltsSendRecvS{\mpS}{\roleR}{\rolePi}{\stLab}
													\right]\muVari\\
													&&\quad\quad\quad\quad\quad\muAnd
													\left[
														\ltsSendRecvS{\mpS}{\roleR}{\roleQi}{\stLab}
													\right]\muVari\\
													&&\quad\quad\quad\quad\quad\muAnd\;
													\forall \roleRi,\roleRii.
														\left[
															\ltsSendRecvS{\mpS}{\roleRi}{\roleRii}{\stLab}
													\right]\muVarii\\
&&\quad\quad\quad\quad)\\
&&\quad\quad\quad)\\
&&\quad\quad)\\
&&\quad)\\
&&)
	\end{array}}
\]

\begin{theorem}
	$\stEnv$ is live iff $\muJudge{\stEnv}{\muPred[\text{live}]}$
\end{theorem}

\begin{proof}
	Fix $\mpS$ and $\stEnv$.
	For $\roleP$ and $n$, define $\predP[\roleP,n]$ by:\\
	$\predPApp[\roleP,0]{\stEnvi}$ is false; and
	$\predPApp[\roleP,n+1]{\stEnvi}$ iff
	there exists $\stEnvAnnotGenericSym$ such that
	$\stEnvi\,\gtMove[\stEnvAnnotGenericSym]$
	and
	for all $\stEnvii$, $\stEnviii$, and $\stEnvAnnotGenericSymi$
	such that
	$\stEnvi\,\gtMoveStar\,\stEnvii\,\gtMove[\stEnvAnnotGenericSymi]\,\stEnviii$
	and
	$\ltsSubject{\stEnvAnnotGenericSym}\cap\ltsSubject{\stEnvAnnotGenericSymi}\neq\emptyset$,
	($\predPApp[\roleP,n]{\stEnviii}$ or $\mpChanRole{\mpS}{\roleP}$ occurs in the path
	$\stEnvi\,\gtMoveStar\,\stEnvii\,\gtMove[\stEnvAnnotGenericSymi]\,\stEnviii$).
	
	For $\roleP$ and $S\subseteq\{\stEnvi\suchthat\stEnv[\mpS]\,\gtMoveStar\,\stEnvi\}$, define $\predP[\roleP,S]$ by:
	$\predPApp[\roleP,S]{\stEnvi}$ iff:\\
	(1) $\stEnvi\not\in S$; and\\
	(2) there exists $\stEnvAnnotGenericSym$ such that\\
	$\stEnvi\,\gtMove[\stEnvAnnotGenericSym]$
	and
	for all $\stEnvii$, $\stEnviii$, and $\stEnvAnnotGenericSymi$
	such that
	$\stEnvi\,\gtMoveStar\,\stEnvii\,\gtMove[\stEnvAnnotGenericSymi]\,\stEnviii$
	and
	$\ltsSubject{\stEnvAnnotGenericSym}\cap\ltsSubject{\stEnvAnnotGenericSymi}\neq\emptyset$,
	($\predPApp[\roleP,S\cup\{\stEnv\}]{\stEnviii}$ or $\mpChanRole{\mpS}{\roleP}$ occurs in the path
	$\stEnvi\,\gtMoveStar\,\stEnvii\,\gtMove[\stEnvAnnotGenericSymi]\,\stEnviii$).
	
	Define $\predP^{\mpS}$ by:
	$\predPPApp[\mpS]{\stEnvi}$ iff,
	for all $\roleP$,
	if $\stEnvApp{\stEnvi}{\mpChanRole{\mpS}{\roleP}}\,\gtMove[]$, then
	there exists $n$ such that
	$\predPApp[\roleP,n]{\stEnvi}$.
	
	Define $\predPi^{\mpS}$ by:
	$\predPPiApp[\mpS,\stEnv]{\stEnvi}$ iff,
	for all $\roleP$,
	if $\stEnvApp{\stEnvi}{\mpChanRole{\mpS}{\roleP}}\,\gtMove[]$, then
	$\predPApp[\roleP,\emptyset]{\stEnvi}$.
	
	\begin{claim}
		If $\predPPApp[\mpS]{\stEnvi[\mpS]}$, then $\stEnvi[\mpS]$ is live.
	\end{claim}
	
	\begin{claim}
		If $\stEnv[\mpS]\gtMoveStar\stEnvi$ 
		and $\stEnvi$ is live, then
		$\predPPiApp[\mpS,\stEnv]{\stEnvi}$.
	\end{claim}
	
	\begin{claim}
		If $\predPPiApp[\mpS,\stEnv]{\stEnvi}$ and $\stEnv\gtMoveStar\stEnvi$, then
		$\predPPApp[\mpS]{\stEnvi}$.
	\end{claim}
	
	\begin{claim}
		$\muJudge{\stEnv}{\muPred[\text{live}]}$
		iff
		$\predPPApp[\mpS]{\stEnv[\mpS]}$
		for all $\mpS$.
	\end{claim}
	
	Therefore,
	$\stEnv$ is live iff
	$\muJudge{\stEnv}{\muPred[\text{live}]}$.
\qed\end{proof}

\subsection{Examples for Benchmarking}

{\bf 1: OAuth2 Fragment}\cite{POPL19LessIsMore}.
Safe, deadlock-free, and live.
\[
	\begin{array}{rcl}
		\mpChanRole{\mpS}{\roleFmt{s}}&\stFmt{:}&
		\stSum{}{}{
				\roleFmt{c}!\stLabFmt{login}\stSeq
				\roleFmt{a}?\stChoice{\stLabFmt{auth}}{\tyBool},
				\roleFmt{c}!\stLabFmt{cancel}
		}{}\\
		\mpChanRole{\mpS}{\roleFmt{c}}&\stFmt{:}&
		\stSum{}{}{
				\roleFmt{s}?\stLabFmt{login}\stSeq
				\roleFmt{a}!\stChoice{\stLabFmt{passwd}}{\tyString},
				\roleFmt{s}?\stLabFmt{cancel}\stSeq
				\roleFmt{a}!\stLabFmt{quit}
		}{}\\
		\mpChanRole{\mpS}{\roleFmt{a}}&\stFmt{:}&
		\stSum{}{}{
				\roleFmt{c}?\stChoice{\stLabFmt{login}}{\tyString}\stSeq
				\roleFmt{s}!\stChoice{\stLabFmt{auth}}{\tyBool},
				\roleFmt{c}?\stLabFmt{quit}
		}{}
	\end{array}
\]
{\bf 2: Recursive Two-Buyers}\cite{POPL19LessIsMore}.
Safe, deadlock-free, and not live.
\[
	\begin{array}{rcl}
		\mpChanRole{\mpS}{\roleFmt{a}}&\stFmt{:}&
				\roleFmt{s}!\stChoice{\stLabFmt{query}}{\tyString}\stSeq
				\roleFmt{s}?\stChoice{\stLabFmt{price}}{\tyInt}\stSeq
				\stRec{\stRecVar}{
				\stSum{}{}{
					\begin{array}{l}
						\roleFmt{b}!\stChoice{\stLabFmt{split}}{\tyInt}\stSeq
						\stSum{}{}{\begin{array}{l}
							\roleFmt{b}?\stChoice{\stLabFmt{yes}}{}\stSeq
							\roleFmt{s}!\stChoice{\stLabFmt{buy}}{}\\
							\roleFmt{b}?\stChoice{\stLabFmt{no}}{}\stSeq\stRecVar
							\end{array}
						}{}
						\\
						\roleFmt{b}!\stChoice{\stLabFmt{cancel}}{}\stSeq
						\roleFmt{s}!\stChoice{\stLabFmt{no}}{}
					\end{array}
				}{}
				}
				\\
		\mpChanRole{\mpS}{\roleFmt{s}}&\stFmt{:}&
				\roleFmt{a}?\stChoice{\stLabFmt{query}}{\tyString}\stSeq
				\roleFmt{a}!\stChoice{\stLabFmt{price}}{\tyInt}\stSeq
				\stSum{}{}{\begin{array}{l}
						\roleFmt{a}?\stChoice{\stLabFmt{buy}}{}\\
						\roleFmt{a}?\stChoice{\stLabFmt{no}}{}
					\end{array}
				}{}\\
			\mpChanRole{\mpS}{\roleFmt{b}}&\stFmt{:}&
			\stRec{\stRecVar}{
			\stSum{}{}{
			\begin{array}{l}
			\roleFmt{a}?\stChoice{\stLabFmt{split}}{\tyInt}\stSeq
			\stSum{}{}{
			\begin{array}{l}
			\roleFmt{a}!\stChoice{\stLabFmt{yes}}{}
			\\
			\roleFmt{a}!\stChoice{\stLabFmt{no}}{}\stSeq\stRecVar
			\end{array}
			}{}
			\\
			\roleFmt{a}?\stChoice{\stLabFmt{cancel}}{}
			\end{array}
			}{}
			}
	\end{array}
\]
{\bf 3: Recursive Map-Reduce}\cite{POPL19LessIsMore}\\
Let $n>0$.
Safe, deadlock-free, and live.
\[
\begin{array}{rcl}
\mpChanRole{\mpS}{\roleFmt{m}}&\stFmt{:}&
\stRec{\stRecVar}{
\roleFmt{w_1}!\stChoice{\stLabFmt{datum}}{\tyInt}\stSeq
\dots
\roleFmt{w_n}!\stChoice{\stLabFmt{datum}}{\tyInt}\stSeq
\stSum{}{}{
\begin{array}{l}
\roleFmt{r}?\stChoice{\stLabFmt{continue}}{\tyInt}\stSeq\stRecVar\\
\roleFmt{r}?\stChoice{\stLabFmt{stop}}{}\stSeq
\roleFmt{w_1}!\stChoice{\stLabFmt{stop}}{}\stSeq\dots
\roleFmt{w_n}!\stChoice{\stLabFmt{stop}}{}
\end{array}
}{}
}\\
\mpChanRole{\mpS}{\roleFmt{w_i}}&\stFmt{:}&
\stRec{\stRecVar}{
\stSum{}{}{\begin{array}{l}
\roleFmt{m}?\stChoice{\stLabFmt{datum}}{\tyInt}\stSeq
\roleFmt{r}!\stChoice{\stLabFmt{result}}{\tyInt}\stSeq\stRecVar
\\
\roleFmt{m}?\stChoice{\stLabFmt{stop}}{}
\end{array}}{}
}\\
&&\text{for}\;1\leq i\leq n\\
\mpChanRole{\mpS}{\roleFmt{r}}&\stFmt{:}&
\stRec{\stRecVar}{
\roleFmt{w_1}?\stChoice{\stLabFmt{result}}{\tyInt}\stSeq
\dots
\roleFmt{w_n}?\stChoice{\stLabFmt{result}}{\tyInt}\stSeq
\stSum{}{}{
\begin{array}{l}
\roleFmt{m}!\stChoice{\stLabFmt{continue}}{\tyInt}\stSeq\stRecVar\\
\roleFmt{m}!\stChoice{\stLabFmt{stop}}{}
\end{array}
}{}
}\\
\end{array}
\]
{\bf 4: Multiparty Workers}\cite{POPL19LessIsMore}
Let $n>0$.
Safe, deadlock-free, and live.
\[
\begin{array}{rcl}
\mpChanRole{\mpS}{\roleFmt{s}}&\stFmt{:}&
\roleFmt{wa_1}\stFmt{!}\stChoice{\stLabFmt{datum}}{\tyInt}\stSeq\dots
\roleFmt{wa_n}\stFmt{!}\stChoice{\stLabFmt{datum}}{\tyInt}
\\
\mpChanRole{\mpS}{\roleFmt{wa_i}}&\stFmt{:}&
\roleFmt{s}\stFmt{?}\stChoice{\stLabFmt{datum}}{\tyInt}\stSeq
\stRec{\stRecVar}{
\stSum{}{}{
\begin{array}{l}
\roleFmt{wb_i}\stFmt{!}\stChoice{\stLabFmt{datum}}{\tyInt}\stSeq
\roleFmt{wc_i}\stFmt{?}\stChoice{\stLabFmt{result}}{\tyInt}\stSeq
\stRecVar
\\
\roleFmt{wb_i}\stFmt{!}\stChoice{\stLabFmt{stop}}{}
\end{array}
}{}
}
\\
\mpChanRole{\mpS}{\roleFmt{wb_i}}&\stFmt{:}&
\stRec{\stRecVar}{
\stSum{}{}{
\begin{array}{l}
\roleFmt{wa_i}\stFmt{?}\stChoice{\stLabFmt{datum}}{\tyInt}\stSeq
\roleFmt{wc_i}\stFmt{!}\stChoice{\stLabFmt{datum}}{\tyInt}\stSeq
\stRecVar
\\
\roleFmt{wa_i}\stFmt{?}\stChoice{\stLabFmt{stop}}{}\stSeq
\roleFmt{wc_i}\stFmt{!}\stChoice{\stLabFmt{stop}}{}
\end{array}
}{}
}
\\
\mpChanRole{\mpS}{\roleFmt{wc_i}}&\stFmt{:}&
\stRec{\stRecVar}{
\stSum{}{}{
\begin{array}{l}
\roleFmt{wb_i}\stFmt{?}\stChoice{\stLabFmt{datum}}{\tyInt}\stSeq
\roleFmt{wa_i}\stFmt{!}\stChoice{\stLabFmt{result}}{\tyInt}\stSeq
\stRecVar
\\
\roleFmt{wb_i}\stFmt{?}\stChoice{\stLabFmt{stop}}{}
\end{array}
}{}
}\\
&&\text{for}\;1\leq i\leq n
\end{array}
\]
{\bf 5: Binary Counter}
Let $n>0$.
Safe, deadlock-free, and live.
\[
\begin{array}{rcl}
\mpChanRole{\mpS}{\roleFmt{bit_{-1}}}&\stFmt{:}&
\stRec{\stRecVar}{
\stSum{\roleFmt{bit_0}}{i\in\{0,1\}}{\stLab[i]\stSeq\stRecVar}{!}
}\\
\mpChanRole{\mpS}{\roleFmt{bit_n}}&\stFmt{:}&
\stRec{\stRecVar}{
\stSum{\roleFmt{bit_{n-1}}}{i\in\{0,1\}}{\stLab[i]\stSeq\stRecVar}{?}
}\\
\mpChanRole{\mpS}{\roleFmt{bit_i}}&\stFmt{:}&
\stRec{\stRecVar}{
\stSum{}{}{
\begin{array}{l}
\roleFmt{bit_{i-1}}?\stLab[0]\stSeq\roleFmt{bit_{i+1}}!\stLab[0]\stSeq\stRecVar
\\
\roleFmt{bit_{i-1}}?\stLab[1]\stSeq\roleFmt{bit_{i+1}}!\stLab[0]\stSeq
\stRec{\stRecVari}{
\stSum{}{}{
\begin{array}{l}
\roleFmt{bit_{i-1}}?\stLab[0]\stSeq\roleFmt{bit_{i+1}}!\stLab[0]\stSeq\stRecVari
\\
\roleFmt{bit_{i-1}}?\stLab[1]\stSeq\roleFmt{bit_{i+1}}!\stLab[1]\stSeq\stRecVar
\end{array}
}{}
}
\end{array}
}{}
}\\
&&\text{for}\;0\leq i<n
\end{array}
\]
{\bf 6: Bad Binary Counter}
Let $n>0$.
Safe, deadlock-free, and not live.
\[
\begin{array}{rcl}
\mpChanRole{\mpS}{\roleFmt{bit_{-1}}}&\stFmt{:}&
\stRec{\stRecVar}{
\stSum{\roleFmt{bit_0}}{i\in\{0,1\}}{\stLab[i]\stSeq\stRecVar}{!}
}\\
\mpChanRole{\mpS}{\roleFmt{bit_n}}&\stFmt{:}&
\stRec{\stRecVar}{
\stSum{}{}{\begin{array}{l}
	\roleFmt{bit_{n-1}}?\stLab[0]\stSeq\stRecVar\\
	\roleFmt{bit_{n-1}}?\stLab[1]\stSeq\stRecVar\\
	\roleFmt{bit_{0}}!\stLab[2]\stSeq\stRecVar
	\stRec{\stRecVari}{
	\roleFmt{bit_{1}}!\stLab[2]\stSeq\dots
	\roleFmt{bit_{n-1}}!\stLab[2]\stSeq
	\roleFmt{bit_{0}}!\stLab[2]\stSeq\stRecVari
	}
\end{array}}{}
}\\
\mpChanRole{\mpS}{\roleFmt{bit_i}}&\stFmt{:}&
\stRec{\stRecVar}{
\stSum{}{}{
\begin{array}{l}
\roleFmt{bit_{i-1}}?\stLab[0]\stSeq\\\quad
\stSum{}{}{
\begin{array}{l}
\roleFmt{bit_{i+1}}!\stLab[0]\stSeq\stRecVar
\\
\roleFmt{bit_n}?\stLab[2]\stSeq\stRec{\stRecVarii}{
\roleFmt{bit_n}?\stLab[2]\stSeq\stRecVarii
}
\end{array}}{}
\\
\roleFmt{bit_{i-1}}?\stLab[1]\stSeq\\\quad
\stSum{}{}{\begin{array}{l}
\roleFmt{bit_{i+1}}!\stLab[0]\stSeq\\\quad
\stRec{\stRecVari}{
\stSum{}{}{
\begin{array}{l}
\roleFmt{bit_{i-1}}?\stLab[0]\stSeq\\\quad
\stSum{}{}{
\begin{array}{l}
\roleFmt{bit_{i+1}}!\stLab[0]\stSeq\stRecVari
\\
\roleFmt{bit_n}?\stLab[2]\stSeq\stRec{\stRecVarii}{
\roleFmt{bit_n}?\stLab[2]\stSeq\stRecVarii
}
\end{array}}{}
\\
\roleFmt{bit_{i-1}}?\stLab[1]\stSeq\\\quad
\stSum{}{}{
\begin{array}{l}
\roleFmt{bit_{i+1}}!\stLab[1]\stSeq\stRecVar
\\
\roleFmt{bit_n}?\stLab[2]\stSeq\stRec{\stRecVarii}{
\roleFmt{bit_n}?\stLab[2]\stSeq\stRecVarii
}
\end{array}}{}
\\
\roleFmt{bit_n}?\stLab[2]\stSeq\stRec{\stRecVarii}{
\roleFmt{bit_n}?\stLab[2]\stSeq\stRecVarii
}
\end{array}
}{}
}\\
\roleFmt{bit_n}?\stLab[2]\stSeq\stRec{\stRecVarii}{
\roleFmt{bit_n}?\stLab[2]\stSeq\stRecVarii
}
\end{array}
}{}\\
\roleFmt{bit_n}?\stLab[2]\stSeq\stRec{\stRecVarii}{
\roleFmt{bit_n}?\stLab[2]\stSeq\stRecVarii
}
\end{array}
}{}
}\\
&&\text{for}\;0\leq i<n
\end{array}
\]
{\bf 7: Leader Election Bracket}
Let $n>0$. We define the $n$ round leader election,
tournament.
Safe, deadlock-free, and live.
\[\begin{array}{rcl}
	modRole(i,j,k,l)&=&
	(((i\,mod\,5^{k+1})\,div\,5^k+l)\cdot5^k+j)\,mod\,5^{k+1}\\&&+(i\,div\,5^{k+1})\cdot5^{k+1}\\
	candidate\!\left(i,k\right)&=&
	\stSum{\roleP[modRole(i,j,k,4)]}{0\leq j < 5^k}{\stLabFmt{elect}}{!}\\
	&&+\\
	&&\roleS[(i\,div\,5^{k+1},k)]?\stLabFmt{del}
	\\
	&&+\\
	&&\stSum{}{0\leq j < 5^k}{\begin{array}{l}
		\roleP[modRole(i,j,k,1)]?\stLabFmt{elect}\stSeq\\\quad
		\stSum{\roleP[modRole(i,j,k,2)]}{0\leq j < 5^k}
		{\stLabFmt{elect}}{!}
		\\\quad+
		\\\quad
		\stSum{}{0\leq j < 5^k}
		{
		\begin{array}{l}
		\roleP[modRole(i,j,k,3)]?
		\stLabFmt{elect}\stSeq\\\quad
		\roleS[(i\,div\,5^{k+1},k)]!\stLabFmt{elect}\stSeq\\\qquad
		candidate(i,k+1)
		\end{array}
		}{}
		\end{array}}{}
	\\
	&&\text{for}\;0\leq i < 5^n\;\text{and}\;0\leq k < n\\
	candidate\!\left(i,n\right)&=&\stEnd
	\end{array}
\]
\[
	\begin{array}{rcl}
		\mpChanRole{\mpS}{\roleP[i]}&\stFmt{:}&
		candidate\!\left(i,0\right)
		\\
		&&\text{for}\;0\leq i < 5^n
		\\
		\mpChanRole{\mpS}{\roleS[(j,k)]}&\stFmt{:}&
		\stSum{\roleP[i]}{0\leq i < 5^n}{\stLabFmt{elect}\stSeq
		\stSum{\roleP[i]}{0\leq i < 5^n}{\stLabFmt{del}}{!}
		}{?}\\
		&&\text{for}\;0\leq k < n\;\text{and}\;0\leq j < 5^{n-k-1}
	\end{array}
\]
{\bf 8: Leader Election Knockout}
We define the knockout leader election
with $n$ candidates.
Always safe.
deadlock-free and live iff $n\,mod\,4=1$.
\[
\begin{array}{rcl}
	scheduler\!\left(i\right)&=&
	\stFmt{
		\sum_{1\leq a \leq n}\{\roleP[a]?\stLabFmt{stand}\stSeq
	}\\
	&&\;
	\stFmt{
		\sum_{1\leq b \leq n}\{\roleP[b]?\stLabFmt{stand}\stSeq
	}\\
	&&\;\;
	\stFmt{
		\sum_{1\leq c \leq n}\{\roleP[c]?\stLabFmt{stand}\stSeq
	}\\
	&&\;\;\;
	\stFmt{
		\sum_{1\leq d \leq n}\{\roleP[d]?\stLabFmt{stand}\stSeq
	}\\
	&&\;\;\;\;
	\stFmt{
		\sum_{1\leq e \leq n}\{\roleP[e]?\stLabFmt{stand}\stSeq
	}\\
	&&\;\;\;\;\;
		\roleP[a]!\stLab[(b,c,d,e)]\stSeq
		\roleP[b]!\stLab[(c,d,e,a)]\stSeq
		\roleP[c]!\stLab[(d,e,a,b)]\stSeq\\
		&&\;\;\;\;\;\roleP[d]!\stLab[(e,a,b,c)]\stSeq
		\roleP[e]!\stLab[(a,b,c,d)]\stSeq\\
		&&\;\;\;\;\;
		\stFmt{
		\sum_{f\in\{a,b,c,d,e\}}\{\mpP[f]?\stLabFmt{elect}\stSeq
		}\\
		&&\;\;\;\;\;\;
		\stSum{\mpP[f]}{f\in\{a,b,c,d,e\}}{\stLabFmt{del}\stSeq
		scheduler\!\left(i-1\right)
		}{!}\\
		&&\;\;\;\;\;\stFmt{\}}\\
		&&\;\;\;\;\stFmt{\}}\\
		&&\;\;\;\stFmt{\}}\\
		&&\;\;\stFmt{\}}\\
		&&\;\stFmt{\}}\\
		&&\stFmt{\}}\\
	&&\text{for}\;i>1\\
	scheduler\!\left(1\right)&=&\stSum{\roleP[a]}{1\leq a \leq n}{
		\stLabFmt{stand}\stSeq\roleP[a]!\stLabFmt{finish}
	}{?}
\end{array}
\]
\[\small
\begin{array}{rcl}
	\mpChanRole{\mpS}{\roleP[i]}&\stFmt{:}&
	\stRec{\stRecVar}{
	\roleS\stFmt{!}\stLabFmt{stand}\stSeq}\\
	&&\left(\begin{array}{cl}
	&\roleS?\stLabFmt{finish}\\
	+&
		\stSum{\roleS}{(a,b,c,d)\in(\{1,\dots,n\}\setminus\{i\})^4}{
			\stLab[(a,b,c,d)]\stSeq\stSum{}{}{\begin{array}{l}
		\roleP[d]!\stLabFmt{elect}
		\\
		\roleP[a]?\stLabFmt{elect}\stSeq\\\quad
		\stSum{}{}{
		\begin{array}{l}
		\roleP[b]!\stLabFmt{elect}
		\\
		\roleP[c]?\stLabFmt{elect}\stSeq\\\quad
		\roleS!\stLabFmt{elect}\stSeq
		\stRecVar
		\end{array}
		}{}
		\\
		\roleS?\stLabFmt{del}
	\end{array}}{}
		}{?}
	\end{array}\right)\\
	\mpChanRole{\mpS}{\roleS}&\stFmt{:}&
	scheduler\!\left(n\right)
\end{array}
\]
{\bf 9: Server Requests}\cite{JY2020}.
Safe, deadlock-free, and live.
\[
\begin{array}{rcl}
\mpChanRole{\mpS}{\roleFmt{s}}&\stFmt{:}&
\stRec{\stRecVar}{
\stSum{}{1\leq i\leq n}{\begin{array}{l}
	\roleFmt{c_i}?
	\stLabFmt{Add}\stSeq
	\roleFmt{c_i}!\stLabFmt{Sum}
	\stSeq\stRecVar
	\\
	\roleFmt{c_i}?
	\stLabFmt{Mul}\stSeq
	\roleFmt{c_i}!\stLabFmt{Prod}
	\stSeq\stRecVar
\end{array}}{}
}
\\
\mpChanRole{\mpS}{\roleFmt{c_i}}&\stFmt{:}&
\stRec{\stRecVar}{
\stSum{}{}{
	\begin{array}{l}
	\roleFmt{r}!
	\stLabFmt{Add}\stSeq
	\roleFmt{r}?\stLabFmt{Sum}
	\stSeq\stRecVar
	\\
	\roleFmt{r}!
	\stLabFmt{Mul}\stSeq
	\roleFmt{r}?\stLabFmt{Prod}
	\stSeq\stRecVar
	\end{array}
}{}
}\\
&&\text{for}\;1\leq i\leq n
\end{array}
\]
{\bf 10: Load Balancing}\cite{majumdar_et_al:LIPIcs.CONCUR.2021.35}
Let $n>0$.
Safe, deadlock-free, and live.
\[
\begin{array}{rcl}
\mpChanRole{\mpS}{\roleFmt{s}}&\stFmt{:}&
\stRec{\stRecVar}{
\roleFmt{c}?\stLabFmt{req}\stSeq
\stSum{\roleFmt{w_i}}{1\leq i\leq n}{
\stLabFmt{req}\stSeq\stRecVar
}{!}
}\\
\mpChanRole{\mpS}{\roleFmt{c}}&\stFmt{:}&
\stRec{\stRecVar}{
\roleFmt{s}!\stLabFmt{req}\stSeq
\stSum{\roleFmt{w_i}}{1\leq i\leq n}{
\stLabFmt{reply}\stSeq\stRecVar
}{?}
}\\
\mpChanRole{\mpS}{\roleFmt{w_i}}&\stFmt{:}&
\stRec{\stRecVar}{
\roleFmt{s}?\stLabFmt{req}\stSeq
\roleFmt{c}!\stLabFmt{reply}\stSeq\stRecVar
}\\
&&\text{for}\;1\leq i\leq n
\end{array}
\]
{\bf 11: Calculator}\cite{FASE16EndpointAPI}.
Safe, deadlock-free, and live.
\[
\begin{array}{rcl}
\mpChanRole{\mpS}{\roleFmt{c}}&\stFmt{:}&
\stRec{\stRecVar}{
\stSum{}{}{
\begin{array}{l}
\roleFmt{s}!\stChoice{\stLabFmt{Add}}{\tyInt}\stSeq
\roleFmt{s}!\stChoice{\stLabFmt{Add}}{\tyInt}\stSeq
\roleFmt{s}?\stChoice{\stLabFmt{Res}}{\tyInt}\stSeq
\stRecVar\\
\roleFmt{s}!\stChoice{\stLabFmt{Bye}}{}\stSeq
\roleFmt{s}?\stChoice{\stLabFmt{Bye}}{}
\end{array}
}{}}\\
\mpChanRole{\mpS}{\roleFmt{s}}&\stFmt{:}&
\stRec{\stRecVar}{
\stSum{}{}{
\begin{array}{l}
\roleFmt{c}?\stChoice{\stLabFmt{Add}}{\tyInt}\stSeq
\roleFmt{c}?\stChoice{\stLabFmt{Add}}{\tyInt}\stSeq
\roleFmt{c}!\stChoice{\stLabFmt{Res}}{\tyInt}\stSeq
\stRecVar\\
\roleFmt{c}?\stChoice{\stLabFmt{Bye}}{}\stSeq
\roleFmt{c}!\stChoice{\stLabFmt{Bye}}{}
\end{array}
}{}}
\end{array}
\]
{\bf 12: SMTP}\cite{FASE16EndpointAPI}.
Safe, deadlock-free, and live.
\[
\begin{array}{rcl}
\mpChanRole{\mpS}{\roleFmt{c}}&\stFmt{:}&
	\roleFmt{s}\stFmt{?}\stLabFmt{220}\stSeq
	\roleFmt{s}\stFmt{!}\stLabFmt{Ehlo}\stSeq\\&&\;
	\stRec{\stRecVar}{
	\stSum{}{}{
	\begin{array}{l}
	\roleFmt{s}\stFmt{?}\stLabFmt{250d}\stSeq\stRecVar\\
	\roleFmt{s}\stFmt{?}\stLabFmt{250}\stSeq
	\roleFmt{s}!\stLabFmt{StartTls}\stSeq
	\roleFmt{s}?\stLabFmt{220}\stSeq
	\roleFmt{s}\stFmt{!}\stLabFmt{Ehlo}\stSeq
	\stRec{\stRecVari}{
	\stSum{}{}{
	\begin{array}{l}
	\roleFmt{s}\stFmt{?}\stLabFmt{250d}\stSeq\stRecVari
	\roleFmt{s}\stFmt{?}\stLabFmt{250}
	\end{array}
	}{}
	}
	\end{array}
	}{}
	}\\
\mpChanRole{\mpS}{\roleFmt{s}}&\stFmt{:}&
	\roleFmt{c}\stFmt{!}\stLabFmt{220}\stSeq
	\roleFmt{c}\stFmt{?}\stLabFmt{Ehlo}\stSeq\\&&\;
	\stRec{\stRecVar}{
	\stSum{}{}{
	\begin{array}{l}
	\roleFmt{c}\stFmt{!}\stLabFmt{250d}\stSeq\stRecVar\\
	\roleFmt{c}\stFmt{!}\stLabFmt{250}\stSeq
	\roleFmt{c}?\stLabFmt{StartTls}\stSeq
	\roleFmt{c}!\stLabFmt{220}\stSeq
	\roleFmt{c}\stFmt{?}\stLabFmt{Ehlo}\stSeq
	\stRec{\stRecVari}{
	\stSum{}{}{
	\begin{array}{l}
	\roleFmt{c}\stFmt{!}\stLabFmt{250d}\stSeq\stRecVari
	\roleFmt{c}\stFmt{!}\stLabFmt{250}
	\end{array}
	}{}
	}
	\end{array}
	}{}
	}
\end{array}
\]
{\bf 13: Circuit Breaker}\cite{DBLP:conf/ecoop/LagaillardieNY22}.
Safe, deadlock-free, and live.
\[\small
\begin{array}{rcl}
	\mpChanRole{\mpS}{\roleFmt{api}}&\stFmt{:}&
	\roleFmt{contr}\stFmt{?}\stChoice{\stLabFmt{start}}{\tyInt}\stSeq
	\stRec{\stRecVar}{
	\roleFmt{usr}\stFmt{?}\stChoice{\stLabFmt{request}}{}\stSeq
	\roleFmt{contr}\stFmt{!}\stChoice{\stLabFmt{getMode}}{}\stSeq}
	\\&&\quad
	\stSum{}{}{
	\begin{array}{l}
	\roleFmt{contr}\stFmt{?}\stChoice{\stLabFmt{up}}{}\stSeq
	\roleFmt{st}\stFmt{!}\stChoice{\stLabFmt{request}}{\tyInt}\stSeq
	\roleFmt{st}\stFmt{?}\stChoice{\stLabFmt{reply}}{\tyInt}\stSeq
	\roleFmt{usr}\stFmt{!}\stChoice{\stLabFmt{reply}}{\tyInt}\stSeq
	\stRecVar
	\\
	\roleFmt{contr}\stFmt{?}\stChoice{\stLabFmt{fail}}{\tyInt}\stSeq
	\roleFmt{usr}\stFmt{!}\stChoice{\stLabFmt{fail}}{\tyInt}\stSeq
	\stRecVar
	\end{array}
	}{}
	\\
	\mpChanRole{\mpS}{\roleFmt{contr}}&\stFmt{:}&
	\roleFmt{st}\stFmt{!}\stChoice{\stLabFmt{start}}{\tyInt}\stSeq
	\roleFmt{api}\stFmt{!}\stChoice{\stLabFmt{start}}{\tyInt}\stSeq
	\roleFmt{st}\stFmt{?}\stChoice{\stLabFmt{ping}}{\tyInt}\stSeq
	\stRec{\stRecVar}{
	\roleFmt{api}\stFmt{?}\stChoice{\stLabFmt{getMode}}{}\stSeq}
	\\&&\quad
	\stSum{}{}{
	\begin{array}{l}
	\roleFmt{api}\stFmt{!}\stChoice{\stLabFmt{up}}{}\stSeq
	\stRecVar
	\\
	\roleFmt{api}\stFmt{!}\stChoice{\stLabFmt{fail}}{\tyInt}\stSeq
	\roleFmt{st}\stFmt{!}\stChoice{\stLabFmt{restart}}{\tyInt}\stSeq
	\stRecVar
	\end{array}
	}{}
	\\
	\mpChanRole{\mpS}{\roleFmt{st}}&\stFmt{:}&
	\roleFmt{contr}\stFmt{?}\stChoice{\stLabFmt{start}}{\tyInt}\stSeq
	\roleFmt{contr}\stFmt{!}\stChoice{\stLabFmt{ping}}{\tyInt}\stSeq
	\stRec{\stRecVar}{}\\&&\quad
	\stSum{}{}{
	\begin{array}{l}
	\roleFmt{api}\stFmt{?}\stChoice{\stLabFmt{request}}{\tyInt}\stSeq
	\roleFmt{api}\stFmt{!}\stChoice{\stLabFmt{reply}}{\tyInt}\stSeq
	\stRecVar
	\\
	\roleFmt{contr}\stFmt{?}\stChoice{\stLabFmt{restart}}{\tyInt}\stSeq
	\stRecVar
	\end{array}
	}{}
	\\
		\mpChanRole{\mpS}{\roleFmt{usr}}&\stFmt{:}&
	\stRec{\stRecVar}{
	\roleFmt{api}\stFmt{!}\stChoice{\stLabFmt{request}}{}\stSeq
	\stSum{}{}{
	\begin{array}{l}
	\roleFmt{api}\stFmt{?}\stChoice{\stLabFmt{reply}}{\tyInt}\stSeq
	\stRecVar
	\\
	\roleFmt{api}\stFmt{?}\stChoice{\stLabFmt{fail}}{\tyInt}\stSeq
	\stRecVar
	\end{array}
	}{}
	}
\end{array}
\]
{\bf 14: Distributed Logging}\cite{DBLP:conf/ecoop/LagaillardieNY22}.
Safe, deadlock-free, and live.
\[
	\begin{array}{rcl}
	\mpChanRole{\mpS}{\roleFmt{contr}}&\stFmt{:}&
	\roleFmt{super}\stFmt{!}\stChoice{\stLabFmt{start}}{\tyInt}
	\stSeq\\&&\quad
	\stRec{\stRecVar}{
		\stSum{}{}{
			\begin{array}{l}
			\roleFmt{super}\stFmt{?}\stChoice{\stLabFmt{success}}{\tyInt}\stSeq\stRecVar
			\\
			\roleFmt{super}\stFmt{?}\stChoice{\stLabFmt{fail}}{\tyInt}\stSeq
			\stSum{}{}{
			\begin{array}{l}
			\roleFmt{super}\stFmt{!}\stChoice{\stLabFmt{restart}}{\tyInt}\stSeq\stRecVar
			\\
			\roleFmt{super}\stFmt{!}\stChoice{\stLabFmt{stop}}{\tyInt}
			\end{array}
			}{}
			\end{array}
		}{}
	}
	\\
	\mpChanRole{\mpS}{\roleFmt{super}}&\stFmt{:}&
	\roleFmt{contr}\stFmt{?}\stChoice{\stLabFmt{start}}{\tyInt}
	\stSeq\\&&\quad
	\stRec{\stRecVar}{
		\stSum{}{}{
			\begin{array}{l}
			\roleFmt{contr}\stFmt{!}\stChoice{\stLabFmt{success}}{\tyInt}\stSeq\stRecVar
			\\
			\roleFmt{contr}\stFmt{!}\stChoice{\stLabFmt{fail}}{\tyInt}\stSeq
			\stSum{}{}{
			\begin{array}{l}
			\roleFmt{contr}\stFmt{?}\stChoice{\stLabFmt{restart}}{\tyInt}\stSeq\stRecVar
			\\
			\roleFmt{contr}\stFmt{?}\stChoice{\stLabFmt{stop}}{\tyInt}
			\end{array}
			}{}
			\end{array}
		}{}
	}
	\end{array}
\]
{\bf 15: Travel Agency}\cite{HYH2008}.
Safe, deadlock-free, and live.
\[
\begin{array}{rcl}
\mpChanRole{\mpS}{\roleFmt{c}}&\stFmt{:}&
\roleFmt{a}\stFmt{!}\stChoice{\stLabFmt{method}}{\tyString}
\stSeq
\roleFmt{a}\stFmt{?}\stChoice{\stLabFmt{price}}{\tyInt}
\stSeq
\stRec{\stRecVar}{
\stSum{}{}{
\begin{array}{l}
\roleFmt{a}\stFmt{!}\stChoice{\stLabFmt{method}}{\tyString}
\stSeq
\roleFmt{a}\stFmt{?}\stChoice{\stLabFmt{price}}{\tyInt}\stSeq\stRecVar
\\
\roleFmt{a}\stFmt{!}\stChoice{\stLabFmt{accept}}{\tyString}
\stSeq
\roleFmt{a}\stFmt{?}\stChoice{\stLabFmt{date}}{\tyString}
\\
\roleFmt{a}\stFmt{!}\stChoice{\stLabFmt{reject}}{}
\end{array}
}{}
}
\\
\mpChanRole{\mpS}{\roleFmt{a}}&\stFmt{:}&
\roleFmt{s}\stFmt{?}\stChoice{\stLabFmt{method}}{\tyString}
\stSeq
\roleFmt{s}\stFmt{!}\stChoice{\stLabFmt{price}}{\tyInt}
\stSeq
\stRec{\stRecVar}{
\stSum{}{}{
\begin{array}{l}
\roleFmt{s}\stFmt{?}\stChoice{\stLabFmt{method}}{\tyString}
\stSeq
\roleFmt{s}\stFmt{!}\stChoice{\stLabFmt{price}}{\tyInt}\stSeq\stRecVar
\\
\roleFmt{s}\stFmt{?}\stChoice{\stLabFmt{accept}}{\tyString}
\stSeq
\roleFmt{s}\stFmt{!}\stChoice{\stLabFmt{date}}{\tyString}
\\
\roleFmt{s}\stFmt{?}\stChoice{\stLabFmt{reject}}{}
\end{array}
}{}
}
\end{array}
\]
{\bf 16: Online Wallet}\cite{NYH2013}.
Safe, deadlock-free, and live.
\[
\begin{array}{rcl}
\mpChanRole{\mpS}{\roleFmt{s}}&\stFmt{:}&
\stSum{}{}{
\begin{array}{l}
\roleFmt{a}\stFmt{?}\stLabFmt{login\_ok}\stSeq
\stRec{\stRecVar}{
\roleFmt{c}\stFmt{!}\stChoice{\stLabFmt{balance}}{\tyInt}\stSeq
\roleFmt{c}\stFmt{!}\stChoice{\stLabFmt{overdraft}}{\tyInt}\stSeq}
\\\quad
\stSum{}{}{
\begin{array}{l}
\roleFmt{c}\stFmt{!}\stChoice{\stLabFmt{payee}}{\tyString}\stSeq
\roleFmt{c}\stFmt{!}\stChoice{\stLabFmt{amount}}{\tyInt}\stSeq
\stRecVar
\\
\roleFmt{c}\stFmt{!}\stChoice{\stLabFmt{quit}}{}\stSeq
\end{array}
}{}
\\
\roleFmt{a}\stFmt{?}\stChoice{\stLabFmt{login\_fail}}{\tyString}
\end{array}
}{}
\\
\mpChanRole{\mpS}{\roleFmt{c}}&\stFmt{:}&
\roleFmt{a}\stFmt{!}\stChoice{\stLabFmt{id}}{\tyString}\stSeq
\roleFmt{a}\stFmt{!}\stChoice{\stLabFmt{pw}}{\tyString}\stSeq\\&&\quad
\stSum{}{}{
\begin{array}{l}
\roleFmt{a}\stFmt{?}\stLabFmt{login\_ok}\stSeq
\stRec{\stRecVar}{
\roleFmt{s}\stFmt{?}\stChoice{\stLabFmt{balance}}{\tyInt}\stSeq
\roleFmt{s}\stFmt{?}\stChoice{\stLabFmt{overdraft}}{\tyInt}\stSeq}
\\\quad
\stSum{}{}{
\begin{array}{l}
\roleFmt{s}\stFmt{?}\stChoice{\stLabFmt{payee}}{\tyString}\stSeq
\roleFmt{s}\stFmt{?}\stChoice{\stLabFmt{amount}}{\tyInt}\stSeq
\stRecVar
\\
\roleFmt{s}\stFmt{?}\stChoice{\stLabFmt{quit}}{}\stSeq
\end{array}
}{}
\\
\roleFmt{a}\stFmt{?}\stChoice{\stLabFmt{login\_fail}}{\tyString}
\end{array}
}{}
\\
\mpChanRole{\mpS}{\roleFmt{a}}&\stFmt{:}&
\roleFmt{c}\stFmt{?}\stChoice{\stLabFmt{id}}{\tyString}\stSeq
\roleFmt{c}\stFmt{?}\stChoice{\stLabFmt{pw}}{\tyString}\stSeq
\stSum{}{}{
\begin{array}{l}
\roleFmt{c}\stFmt{!}\stLabFmt{login\_ok}\stSeq
\roleFmt{s}\stFmt{!}\stLabFmt{login\_ok}
\\
\roleFmt{c}\stFmt{!}\stChoice{\stLabFmt{login\_fail}}{\tyString}\stSeq
\roleFmt{s}\stFmt{!}\stChoice{\stLabFmt{login\_fail}}{\tyString}
\end{array}
}{}
\end{array}
\]
{\bf 17: Chang and Robert's Leader Election with Mixed Choice}\cite{10.1145/3798224}
Let $n>1$ and let $\sigma$ be a permutation of $\{0\dots n-1\}$.
Safe, deadlock-free, and live.
\[\small
	\begin{array}{rcl}
	\stFmt{part\!\left(i\right)}&=&\stRec{\stRecVar}{
	\stSum{\roleP[(i-1)\,mod\,n]}{0\leq j<\sigma i}{
		\stLabFmt{election_j}\stSeq
		\stRecVar
	}{?}}\\
	&&\stFmt{+}\\
	&&\stSum{\roleP[(i-1)\,mod\,n]}{\sigma i< j<n}{
		\stLabFmt{election_j}\stSeq
		\roleP[(i+1)\,mod\,n]\stFmt{!}\stLabFmt{election_j}
		\stSeq \stRecVar
	}{?}\\
	&&\stFmt{+}\\
	&&\roleP[(i-1)\,mod\,n]\stFmt{?}\stLabFmt{election_{\sigma i}}\stSeq
	\roleP[(i+1)\,mod\,n]\stFmt{!}\stLabFmt{elected_{\sigma i}}
	\stSeq \stRecVar\\
	&&\stFmt{+}\\
	&&\stSum{\roleP[(i-1)\,mod\,n]}{0\leq j<n, j\neq\sigma i}{
		\stLabFmt{elected_j}\stSeq
		\roleP[(i+1)\,mod\,n]\stFmt{!}\stLabFmt{elected_j}
		\stSeq \roleS\stFmt{!}\stLabFmt{elected_j}
	}{?}\\
	&&\stFmt{+}\\
	&&\roleP[(i-1)\,mod\,n]\stFmt{?}\stLabFmt{elected_{\sigma i}}\stSeq
	\roleS!\stLabFmt{elected_j}
	\\
	\stFmt{notPart\!\left(i\right)}&=&
	\roleP[(i+1)\,mod\,n]\stFmt{!}\stLabFmt{election_{\sigma i}}\stSeq \stFmt{part\!\left(i\right)}\\
	&&\stFmt{+}\\
	&&\stSum{\roleP[(i-1)\,mod\,n]}{0\leq j<\sigma i}{
		\stLabFmt{election_j}\stSeq
		\roleP[(i+1)\,mod\,n]\stFmt{!}\stLabFmt{election_{\sigma i}}
		\stSeq \stFmt{part\!\left(i\right)}
	}{?}\\
	&&\stFmt{+}\\
	&&\stSum{\roleP[(i-1)\,mod\,n]}{\sigma i< j<n}{
		\stLabFmt{election_j}\stSeq
		\roleP[(i+1)\,mod\,n]\stFmt{!}\stLabFmt{election_j}
		\stSeq \stFmt{part\!\left(i\right)}
	}{?}\\
	&&\stFmt{+}\\
	&&\roleP[(i-1)\,mod\,n]\stFmt{?}\stLabFmt{election_{\sigma i}}\stSeq
	\roleP[(i+1)\,mod\,n]!\stLabFmt{elected_{\sigma i}}
	\stSeq \stFmt{part\!\left(i\right)}\\
	&&\stFmt{+}\\
	&&\stSum{\roleP[(i-1)\,mod\,n]}{0\leq j<n, j\neq\sigma i}{
		\stLabFmt{elected_j}\stSeq
		\roleP[(i+1)\,mod\,n]\stFmt{!}\stLabFmt{elected_j}
		\stSeq \roleS\stFmt{!}\stLabFmt{elected_j}
	}{?}\\
	&&\stFmt{+}\\
	&&\roleP[(i-1)\,mod\,n]\stFmt{?}\stLabFmt{elected_{\sigma i}}\stSeq
	\roleS!\stLabFmt{elected_j}
	\\
	\mpChanRole{\mpS}{\roleP[i]}&\stFmt{:}&\stFmt{notPart\!\left(i\right)}
	\\
	&&\text{for}\;0\leq i<n\\
	\mpChanRole{\mpS}{\roleS}&\stFmt{:}&
	\stSum{\roleP[0]}{0\leq i < n}{\stLabFmt{elected_{\sigma i}}
	\stSeq\stFmt{\roleP[1]?\stLabFmt{elected_{\sigma i}}\stSeq\dots\roleP[n-1]?\stLabFmt{elected_{\sigma i}}}}{?}
	\end{array}
\]
{\bf 18: Running example from \cite{DBLP:conf/ictac/BlechschmidtPN25}}.
Safe, deadlock-free, and live.
\[
	\begin{array}{rcl}
	\mpChanRole{\mpS}{\roleP}&\stFmt{:}&
	\roleFmt{c}\stFmt{!}\stLabFmt{lws}\stSeq
	\roleFmt{c}\stFmt{?}\stChoice{\stLabFmt{glt}}{\tyBool}
	\\
	\mpChanRole{\mpS}{\roleFmt{c}}&\stFmt{:}&
	\roleP\stFmt{?}\stLabFmt{lws}\stSeq
	\stSum{}{}{
	\begin{array}{l}
	\roleP!\stChoice{\stLabFmt{glt}}{\tyBool}\stSeq
	\roleFmt{w}!\stChoice{\stLabFmt{rls}}{}
	\\
	\roleFmt{w}!\stChoice{\stLabFmt{rqs}}{}\stSeq
	\roleFmt{w}?\stChoice{\stLabFmt{st}}{}\stSeq
	\roleP!\stChoice{\stLabFmt{glt}}{\tyBool}
	\end{array}
	}{}
	\\
	\mpChanRole{\mpS}{\roleFmt{w}}&\stFmt{:}&
	\stSum{}{}{
	\begin{array}{l}
	\roleFmt{c}\stFmt{?}\stChoice{\stLabFmt{rqs}}{}\stSeq
	\roleFmt{c}\stFmt{!}\stChoice{\stLabFmt{st}}{}
	\\
	\roleFmt{c}\stFmt{?}\stChoice{\stLabFmt{rls}}{}
	\end{array}
	}{}
	\end{array}
\]
{\bf 19: Client Server Workers}\cite{DBLP:journals/corr/abs-2604-06872}.
Safe, deadlock-free, and live.
\[\begin{array}{rcl}
	\mpChanRole{\mpS}{\roleFmt{c}}&\stFmt{:}&
	\stRec{\stRecVar}{
	\roleFmt{s}\stFmt{!}\stLabFmt{req}\stSeq
	\stSum{}{}{
	\begin{array}{l}
		\roleFmt{w_1}?\stLabFmt{res}\stSeq\stRecVar\\
		\roleFmt{w_2}?\stLabFmt{res}\stSeq\stRecVar\\
		\roleFmt{w_1}?\stLabFmt{resL}\\
	\end{array}
	}{}
	}\\
	\mpChanRole{\mpS}{\roleFmt{s}}&\stFmt{:}&
	\stRec{\stRecVar}{
	\stSum{}{}{
	\begin{array}{l}
		\roleFmt{c}?\stLabFmt{req}\stSeq
		\stSum{}{}{
		\roleFmt{w_1}!\stLabFmt{req}\stSeq\stRecVar,
		\roleFmt{w_2}!\stLabFmt{req}\stSeq\stRecVar
		}{}
		\\	
		\roleFmt{w_1}!\stLabFmt{last}\stSeq
		\roleFmt{c}?\stLabFmt{req}\stSeq
		\roleFmt{w_1}!\stLabFmt{req}\stSeq
		\roleFmt{w_2}!\stLabFmt{halt}
	\end{array}
	}{}
	}\\
	\mpChanRole{\mpS}{\roleFmt{w_1}}&\stFmt{:}&
	\stRec{\stRecVar}{
		\stSum{}{}{
		\begin{array}{l}
			\roleFmt{s}?\stLabFmt{req}\stSeq
			\roleFmt{c}!\stLabFmt{res}\stSeq
			\stRecVar
			\\
			\roleFmt{s}?\stLabFmt{last}\stSeq
			\roleFmt{s}?\stLabFmt{req}\stSeq
			\roleFmt{c}!\stLabFmt{resL}
		\end{array}
		}{}
	}\\
	\mpChanRole{\mpS}{\roleFmt{w_2}}&\stFmt{:}&
	\stRec{\stRecVar}{
		\stSum{}{}{
		\begin{array}{l}
			\roleFmt{s}?\stLabFmt{req}\stSeq
			\roleFmt{c}!\stLabFmt{res}\stSeq
			\stRecVar
			\\
			\roleFmt{s}?\stLabFmt{halt}
		\end{array}
		}{}
	}
\end{array}
\]
{\bf 20: Time Out}\cite{DBLP:journals/corr/abs-2604-06872}.
Safe, deadlock-free, and live.
\[\begin{array}{rcl}
	\mpChanRole{\mpS}{\roleP}&\stFmt{:}&
	\stSum{}{}{
	\begin{array}{l}
		\roleR?\stLabFmt{v}\stSeq
		\roleS!\stLabFmt{t}
		\\
		\roleS!\stLabFmt{t}\stSeq
		\roleR?\stLabFmt{v}
	\end{array}
	}{}
	\\
	\mpChanRole{\mpS}{\roleS}&\stFmt{:}&
	\stFmt{\roleP?\stLabFmt{t}}
	\\
	\mpChanRole{\mpS}{\roleR}&\stFmt{:}&
	\stFmt{\roleQ?\stLabFmt{v}\stSeq\roleP!\stLabFmt{v}}
	\\
	\mpChanRole{\mpS}{\roleQ}&\stFmt{:}&
	\stFmt{\roleR!\stLabFmt{v}}
\end{array}
\]
{\bf 21: Fire}\cite{DBLP:journals/corr/abs-2604-06872}.
Safe but not deadlock-free and not live due to
being in synchronous semantics.
\[\begin{array}{rcl}
	\mpChanRole{\mpS}{\roleP}&\stFmt{:}&
	\stSum{}{}{
	\begin{array}{l}
		\roleQ?\stLabFmt{fire}\stSeq
		\roleQ!\stLabFmt{fire}\stSeq
		\roleR!\stLabFmt{fire}\stSeq
		\roleR?\stLabFmt{fire}
		\\
		\roleQ!\stLabFmt{fire}\stSeq
		\roleR!\stLabFmt{fire}\stSeq
		\roleQ?\stLabFmt{fire}\stSeq
		\roleR?\stLabFmt{fire}
		\\
		\roleR?\stLabFmt{fire}\stSeq
		\roleR!\stLabFmt{fire}\stSeq
		\roleQ!\stLabFmt{fire}\stSeq
		\roleQ?\stLabFmt{fire}
	\end{array}
	}{}
	\\
	\mpChanRole{\mpS}{\roleQ}&\stFmt{:}&
	\stSum{}{}{
	\begin{array}{l}
		\roleR?\stLabFmt{fire}\stSeq
		\roleR!\stLabFmt{fire}\stSeq
		\roleP!\stLabFmt{fire}\stSeq
		\roleP?\stLabFmt{fire}
		\\
		\roleR!\stLabFmt{fire}\stSeq
		\roleP!\stLabFmt{fire}\stSeq
		\roleR?\stLabFmt{fire}\stSeq
		\roleP?\stLabFmt{fire}
		\\
		\roleP?\stLabFmt{fire}\stSeq
		\roleP!\stLabFmt{fire}\stSeq
		\roleR!\stLabFmt{fire}\stSeq
		\roleR?\stLabFmt{fire}
	\end{array}
	}{}
	\\
	\mpChanRole{\mpS}{\roleR}&\stFmt{:}&
	\stSum{}{}{
	\begin{array}{l}
		\roleP?\stLabFmt{fire}\stSeq
		\roleP!\stLabFmt{fire}\stSeq
		\roleQ!\stLabFmt{fire}\stSeq
		\roleQ?\stLabFmt{fire}
		\\
		\roleP!\stLabFmt{fire}\stSeq
		\roleQ!\stLabFmt{fire}\stSeq
		\roleP?\stLabFmt{fire}\stSeq
		\roleQ?\stLabFmt{fire}
		\\
		\roleQ?\stLabFmt{fire}\stSeq
		\roleQ!\stLabFmt{fire}\stSeq
		\roleP!\stLabFmt{fire}\stSeq
		\roleP?\stLabFmt{fire}
	\end{array}
	}{}
\end{array}
\]

\section{Appendix for Related Work}

\subsection{1-Level Projection is Flawed}

\label{app:1-level-fails}

Consider the following global types and
the projection with 1-level merge from \cite[Definition 3.6]{GJPSY2018}.

\[\begin{array}{clcl}
	&\gtG&=&\gtFmt{
		\roleP[0]\!\to\!\roleQ[0]{:}\!\left\{
			\begin{array}{l}
				\gtLab[0]\!\gtSeq\!
				\roleP[0]\!\to\!\roleQ[1]{:}\gtLab\!\gtSeq\!
				\roleP[2]\!\to\!\roleQ[2]{:}\!\left\{
					\begin{array}{l}
						\gtLab[0]\!\gtSeq\!\roleP[2]\!\to\!\roleP[0]{:}\gtLab[0]\!\gtSeq\!
						\roleP[0]\!\to\!\roleQ[1]{:}\gtLab[0]
						\\
						\gtLab[1]\!\gtSeq\!\roleP[2]\!\to\!\roleP[0]{:}\gtLab[1]\!\gtSeq\!
						\roleP[0]\!\to\!\roleQ[1]{:}\gtLab[1]
					\end{array}
				\right\}
				\\
				\gtLab[1]\!\gtSeq\!
				\roleP[0]\!\to\!\roleQ[1]{:}\gtLab\!\gtSeq\!
				\roleP[2]\!\to\!\roleQ[2]{:}\!\left\{
					\begin{array}{l}
						\gtLab[0]\!\gtSeq\!\roleP[2]\!\to\!\roleP[0]{:}\gtLab[0]\!\gtSeq\!
						\roleP[0]\!\to\!\roleQ[1]{:}\gtLab[1]
						\\
						\gtLab[1]\!\gtSeq\!\roleP[2]\!\to\!\roleP[0]{:}\gtLab[1]\!\gtSeq\!
						\roleP[0]\!\to\!\roleQ[1]{:}\gtLab[0]
					\end{array}
				\right\}
			\end{array}
		\right\}
	}
	\\
	\gtMove&\gtGi&=&\gtFmt{
		\roleP[0]\!\to\!\roleQ[0]{:}\!\left\{
			\begin{array}{l}
				\gtLab[0]\!\gtSeq\!
				\roleP[0]\!\to\!\roleQ[1]{:}\gtLab\!\gtSeq\!
				\roleP[2]\!\to\!\roleP[0]{:}\gtLab[0]\!\gtSeq\!
				\roleP[0]\!\to\!\roleQ[1]{:}\gtLab[0]
				\\
				\gtLab[1]\!\gtSeq\!
				\roleP[0]\!\to\!\roleQ[1]{:}\gtLab\!\gtSeq\!
				\roleP[2]\!\to\!\roleP[0]{:}\gtLab[0]\!\gtSeq\!
				\roleP[0]\!\to\!\roleQ[1]{:}\gtLab[1]
			\end{array}
		\right\}
	}
	\end{array}
\]
$\gtG$ is projectable and $\gtGi$ does not project onto $\roleQ[1]$.
Therefore, subject reduction fails for the typing system in
\cite[Table 5]{GJPSY2018}.

Consider the types and contexts:
\[
\begin{array}{rcl}
	\stT[{\roleP[0]}]&=&\stSum{\roleQ[0]}{i\in \{0,1\}}{
		\stLab[i]\stSeq\roleQ[1]!\stLab\stSeq\stSum{\roleP[2]}{j\in\{0,1\}}{
		\stLab[j]\stSeq\roleQ[1]!\stLab[i\,xor\,j]
		}{?}
	}{!}
	\\
	\stT[{\roleQ[0]}]&=&\stSum{\roleP[0]}{i\in\{0,1\}}{\stLab[i]}{?}
	\\
	\stT[{\roleQ[1]}]&=&\stSum{\roleP[0]}{i\in\{0,1\}}{\stLab[i]\stSeq\stSum{\roleP[0]}{j\in\{0,1\}}{\stLab[j]}{?}}{?}
	\\
	\stT[{\roleP[2]}]&=&\stSum{\roleQ[2]}{i\in\{0,1\}}{
	\stLab[i]\stSeq\roleP[0]!\stLab[i]
	}{!}
	\\
	\stT[{\roleQ[2]}]&=&\stSum{\roleP[2]}{i\in\{0,1\}}{
	\stLab[i]
	}{?}
\end{array}
\]
\[
	\begin{array}{rcl}
		\stEnv&=&
		\set{
		\stEnvMap{\mpChanRole{\mpS}{\roleP[0]}}{\stT[{\roleP[0]}]}
		\stEnvComp
		\stEnvMap{\mpChanRole{\mpS}{\roleQ[0]}}{\stT[{\roleQ[0]}]}
		\stEnvComp
		\stEnvMap{\mpChanRole{\mpS}{\roleQ[1]}}{\stT[{\roleQ[1]}]}
		\stEnvComp
		\stEnvMap{\mpChanRole{\mpS}{\roleP[2]}}{\stT[{\roleP[2]}]}
		\stEnvComp
		\stEnvMap{\mpChanRole{\mpS}{\roleQ[2]}}{\stT[{\roleQ[2]}]}
		}
		\\
		\stEnvi&=&
		\set{
		\stEnvMap{\mpChanRole{\mpS}{\roleP[0]}}{\stT[{\roleP[0]}]}
		\stEnvComp
		\stEnvMap{\mpChanRole{\mpS}{\roleQ[0]}}{\stT[{\roleQ[0]}]}
		\stEnvComp
		\stEnvMap{\mpChanRole{\mpS}{\roleQ[1]}}{\stT[{\roleQ[1]}]}
		\stEnvComp
		\stEnvMap{\mpChanRole{\mpS}{\roleP[2]}}{\roleP[0]!\stLab[i]}
		\stEnvComp
		\stEnvMap{\mpChanRole{\mpS}{\roleQ[2]}}{\stEnd}
		}
	\end{array}
\]

We have that $\gtG\upharpoonright_{\roleR}\stT[\roleR]$
for $\roleR\in\{\roleP[0],\roleP[2],\roleQ[0],\roleQ[1],\roleQ[2]\}$.

For simplicity, consider the types to be processes from \cite{GJPSY2018} by replacing 
the type syntax with the equivalent process syntax,
and the contexts to be sessions.

We have that $\stEnv\gtMove\stEnvi$.
Subject reduction would require that:
\begin{enumerate*}
\item $\gtG\gtMove\gtGii$ or $\gtG=\gtGii$,
\item $\gtGii\upharpoonright_{\roleP[2]}\stTi$,
\item $\gtGii\upharpoonright_{\roleQ[1]}\stTii$,
\item \label{cond:know-red} $\roleP[0]!\stLab[i]\tySub\stTi$, and
\item $\stT[{\roleQ[1]}]\tySub\stTii$.
\end{enumerate*}
By \ref{cond:know-red},
$\gtGii=\gtGi$.
$\gtGi$ does not project onto $\roleQ[1]$
with the projection in \cite{GJPSY2018}
since merges can only occur at the head of the types being merged.
Therefore, subject reduction \cite[Theorem 3.21]{GJPSY2018}
fails.

\subsection{Send Rule}

\label{app:send-fails}

We will use the type syntax from this paper for ease of writing,
since \cite{10.1007/978-3-031-91121-7_13}
claims that it embeds into the automata framework.

Consider the process:
\[
	\begin{array}{lcl}
	\mpP &=&
	\mpSum{
	\mpSel
		{\mpChanRole{\mpS}{\roleP}}
		{\roleQ}
		{\stLab[i]}
		{\mpChanRole{\mpS}{\roleR}}
		{
			\mpSel
			{\mpChanRole{\mpS}{\roleP}}
			{\roleQ}
			{\stLab}
			{\mpChanRole{\mpS}{\roleR}}
			{
				\mpNil
			}
		}}
	{i\in \{0,1\}}\\&\mpPar&
	\mpSum{\mpBra
		{\mpChanRole{\mpS}{\roleQ}}
		{\roleP}
		{\stLab[i]}
		{\mpy}
		{
			\mpBra
			{\mpChanRole{\mpS}{\roleQ}}
			{\roleP}
			{\stLab}
			{\mpyi}
			{
				\left(
				\begin{array}{l}
				\mpSel
				{\mpy}
				{\roleRi}
				{\stLab}
				{\mpChanRole{\mpS}{\roleR[0]}}
				{
					\mpSel
					{\mpy}
					{\roleRi}
					{\stLabi}
					{\mpChanRole{\mpS}{\roleR[1]}}
					{\mpNil}
				}
				\mpPar\\
				\mpSel
				{\mpyi}
				{\roleRi}
				{\stLab}
				{\mpChanRole{\mpS}{\roleR[2]}}
				{
					\mpSel
					{\mpyi}
					{\roleRi}
					{\stLabi}
					{\mpChanRole{\mpS}{\roleR[3]}}
					{\mpNil}
				}
				\end{array}
				\right)
			}
		}}{i\in \{0,1\}}\\&\mpPar&
	\mpBra
		{\mpChanRole{\mpS}{\roleRi}}
		{\roleR}
		{\stLab}
		{\mpy}
		{
			\mpBra
				{\mpChanRole{\mpS}{\roleRi}}
				{\roleR}
				{\stLabi}
				{\mpyi}
				{
					\mpNil
				}
		}
	\end{array}
\]
Let:
\[
	\stT[\roleR]=
	\stFmt{
			\roleRi{!}\stChoice{\stLab}{\stEnd}\stSeq\roleRi{!}\stChoice{\stLabi}{\stEnd}\stSeq\stEnd
		}
\]
Consider the typing context:
\[
\stEnv=\left\{
\begin{array}{l}
	\stEnvMap{\mpChanRole{\mpS}{\roleR[0]}}{\stEnd}
	\stEnvComp
	\stEnvMap{\mpChanRole{\mpS}{\roleR[1]}}{\stEnd}
	\stEnvComp
	\stEnvMap{\mpChanRole{\mpS}{\roleR[2]}}{\stEnd}
	\stEnvComp
	\stEnvMap{\mpChanRole{\mpS}{\roleR[3]}}{\stEnd}
	\stEnvComp\\
	\stEnvMap{\mpChanRole{\mpS}{\roleR}}{\stT[\roleR]}
	\stEnvComp
	\stEnvMap{\mpChanRole{\mpS}{\roleRi}}{
		\stFmt{
			\roleR{?}\stChoice{\stLab}{\stEnd}\stSeq\roleR{?}\stChoice{\stLabi}{\stEnd}\stSeq\stEnd
		}
	}
	\stEnvComp\\
	\stEnvMap{\mpChanRole{\mpS}{\roleP}}{
			\stIntSum{\roleQ}{i\in\{0,1\}}{\stChoice{\stLab[i]}{\stT[\roleR]}
			\stSeq
			\roleQ{!}\stChoice{\stLab}{\stT[\roleR]}
			\stSeq
			\stEnd
			}
	}
	\stEnvComp\\
	\stEnvMap{\mpChanRole{\mpS}{\roleQ}}{
			\stExtSum{\roleP}{i\in\{0,1\}}{\stChoice{\stLab[i]}{\stT[\roleR]}
			\stSeq
			\roleP{?}\stChoice{\stLab}{\stT[\roleR]}
			\stSeq
			\stEnd
			}
	}
\end{array}\right\}
\]

We have that $\tyJudge{}{\mpP}{\stEnv}$
by the typing rules in
\cite[Fig. 4]{Li2026-pl}.

\[\fontsize{6}{8}\selectfont
\begin{array}{ll}
	\tyJudge{}{\mpNil}{\set{\stEnvMap{\mpChanRole{\mpS}{\roleP}}{\stEnd}}}
	&\inferrule{PT-end}
	\\\\
	\tyJudge{}{
		\mpSel
			{\mpChanRole{\mpS}{\roleP}}
			{\roleQ}
			{\stLab}
			{\mpChanRole{\mpS}{\roleR}}
			{
				\mpNil
			}
	}{
		\set{\stEnvMap{\mpChanRole{\mpS}{\roleP}}{
			\roleQ{!}\stChoice{\stLab}{\stT[\roleR]}\stSeq\stEnd
		}
		\stEnvComp
		\stEnvMap{\mpChanRole{\mpS}{\roleR}}{\stT[\roleR]}}
	}
	&\inferrule{PT-$\Sigma$}
	\\\\
	\tyJudge{}{
		\mpSum{\mpSel
			{\mpChanRole{\mpS}{\roleP}}
			{\roleQ}
			{\stLab[i]}
			{\mpChanRole{\mpS}{\roleR}}
			{
				\mpSel
				{\mpChanRole{\mpS}{\roleP}}
				{\roleQ}
				{\stLab}
				{\mpChanRole{\mpS}{\roleR}}
				{
					\mpNil
				}
			}}{i\in \{0,1\}}
	}{\set{
		\stEnvMap{\mpChanRole{\mpS}{\roleP}}{
			\stIntSum{\roleQ}{i\in\{0,1\}}{\stChoice{\stLab[i]}{\stT[\roleR]}
			\stSeq
			\roleQ{!}\stChoice{\stLab}{\stT[\roleR]}
			\stSeq
			\stEnd
			}
	}
		\stEnvComp
		\stEnvMap{\mpChanRole{\mpS}{\roleR}}{\stT[\roleR]}}
	}
	&\inferrule{PT-$\Sigma$}\\
	\text{as we take }\mpC[0]=\mpC[1]=\mpChanRole{\mpS}{\roleR}
	\text{ so }\{\mpC[i]\}_{i\in \{0,1\}\setminus \{j\}}=\{\mpChanRole{\mpS}{\roleR}\}
	\\\\
	\tyJudge{}{\mpNil}{\set{
		\stEnvMap{\mpChanRole{\mpS}{\roleR[0]}}{\stEnd}
		\stEnvComp
		\stEnvMap{\mpChanRole{\mpS}{\roleR[1]}}{\stEnd}
		\stEnvComp
		\stEnvMap{\mpChanRole{\mpS}{\roleQ}}{\stEnd}
		\stEnvComp
		\stEnvMap{\mpy}{\stEnd}}
	}
	&\inferrule{PT-end}\\\\
	\tyJudge{}{\mpSel
					{\mpy}
					{\roleRi}
					{\stLabi}
					{\mpChanRole{\mpS}{\roleR[1]}}
					{\mpNil}}{\set{
		\stEnvMap{\mpChanRole{\mpS}{\roleR[0]}}{\stEnd}
		\stEnvComp
		\stEnvMap{\mpChanRole{\mpS}{\roleR[1]}}{\stEnd}
		\stEnvComp
		\stEnvMap{\mpChanRole{\mpS}{\roleQ}}{\stEnd}
		\stEnvComp
		\stEnvMap{\mpy}{\stFmt{\roleR{!}\stChoice{\stLabi}{\stEnd}\stSeq\stEnd}}}
	}
	&\inferrule{PT-$\Sigma$}
	\\\\
	\tyJudge{}{\mpSel
				{\mpy}
				{\roleRi}
				{\stLab}
				{\mpChanRole{\mpS}{\roleR[0]}}
				{
					\mpSel
					{\mpy}
					{\roleRi}
					{\stLabi}
					{\mpChanRole{\mpS}{\roleR[1]}}
					{\mpNil}
				}}{\set{\begin{array}{l}
		\stEnvMap{\mpChanRole{\mpS}{\roleR[0]}}{\stEnd}
		\stEnvComp
		\stEnvMap{\mpChanRole{\mpS}{\roleR[1]}}{\stEnd}
		\stEnvComp
		\stEnvMap{\mpChanRole{\mpS}{\roleQ}}{\stEnd}
		\stEnvComp\\
		\stEnvMap{\mpy}{\stFmt{\roleR{!}\stChoice{\stLab}{\stEnd}\stSeq\stFmt{\roleR{!}\stChoice{\stLabi}{\stEnd}\stSeq\stEnd}}
	}\end{array}}}&\inferrule{PT-$\Sigma$}\\\\
	\tyJudge{}{\mpNil}{\set{
		\stEnvMap{\mpChanRole{\mpS}{\roleR[2]}}{\stEnd}
		\stEnvComp
		\stEnvMap{\mpChanRole{\mpS}{\roleR[3]}}{\stEnd}
		\stEnvComp
		\stEnvMap{\mpChanRole{\mpS}{\roleQ}}{\stEnd}
		\stEnvComp
		\stEnvMap{\mpyi}{\stEnd}}
	}&\inferrule{PT-end}\\\\
	\tyJudge{}{\mpSel
					{\mpyi}
					{\roleRi}
					{\stLabi}
					{\mpChanRole{\mpS}{\roleR[3]}}
					{\mpNil}}{\set{
		\stEnvMap{\mpChanRole{\mpS}{\roleR[2]}}{\stEnd}
		\stEnvComp
		\stEnvMap{\mpChanRole{\mpS}{\roleR[3]}}{\stEnd}
		\stEnvComp
		\stEnvMap{\mpChanRole{\mpS}{\roleQ}}{\stEnd}
		\stEnvComp
		\stEnvMap{\mpyi}{\stFmt{\roleR{!}\stChoice{\stLabi}{\stEnd}\stSeq\stEnd}}}
	}&\inferrule{PT-$\Sigma$}\\\\
	\tyJudge{}{\mpSel
				{\mpyi}
				{\roleRi}
				{\stLab}
				{\mpChanRole{\mpS}{\roleR[2]}}
				{
					\mpSel
					{\mpyi}
					{\roleRi}
					{\stLabi}
					{\mpChanRole{\mpS}{\roleR[3]}}
					{\mpNil}
				}}{\set{\begin{array}{l}
		\stEnvMap{\mpChanRole{\mpS}{\roleR[2]}}{\stEnd}
		\stEnvComp
		\stEnvMap{\mpChanRole{\mpS}{\roleR[3]}}{\stEnd}
		\stEnvComp
		\stEnvMap{\mpChanRole{\mpS}{\roleQ}}{\stEnd}
		\stEnvComp\\
		\stEnvMap{\mpyi}{\stFmt{\roleR{!}\stChoice{\stLab}{\stEnd}\stSeq\stFmt{\roleR{!}\stChoice{\stLabi}{\stEnd}\stSeq\stEnd}}
	}\end{array}}}&\inferrule{PT-$\Sigma$}\\\\
	\tyJudge{}{
				\begin{array}{l}
				\mpSel
				{\mpy}
				{\roleRi}
				{\stLab}
				{\mpChanRole{\mpS}{\roleR[0]}}
				{
					\mpSel
					{\mpy}
					{\roleRi}
					{\stLabi}
					{\mpChanRole{\mpS}{\roleR[1]}}
					{\mpNil}
				}
				\mpPar\\
				\mpSel
				{\mpyi}
				{\roleRi}
				{\stLab}
				{\mpChanRole{\mpS}{\roleR[2]}}
				{
					\mpSel
					{\mpyi}
					{\roleRi}
					{\stLabi}
					{\mpChanRole{\mpS}{\roleR[3]}}
					{\mpNil}
				}
				\end{array}
	}{\set{
	\begin{array}{l}
		\stEnvMap{\mpChanRole{\mpS}{\roleR[0]}}{\stEnd}
		\stEnvComp
		\stEnvMap{\mpChanRole{\mpS}{\roleR[1]}}{\stEnd}
		\stEnvComp\\
		\stEnvMap{\mpChanRole{\mpS}{\roleR[2]}}{\stEnd}
		\stEnvComp
		\stEnvMap{\mpChanRole{\mpS}{\roleR[3]}}{\stEnd}
		\stEnvComp\\
		\stEnvMap{\mpChanRole{\mpS}{\roleQ}}{\stEnd}
		\stEnvComp\\
		\stEnvMap{\mpy}{\stT[\roleR]}
		\stEnvComp
		\stEnvMap{\mpyi}{\stT[\roleR]}
	\end{array}}
	}&\inferrule{PT-$\mpPar\mpPar$}\\\\
	\tyJudge{}{
		\mpBra
			{\mpChanRole{\mpS}{\roleQ}}
			{\roleP}
			{\stLab}
			{\mpyi}{\left(
				\begin{array}{l}
				\mpSel
				{\mpy}
				{\roleRi}
				{\stLab}
				{\mpChanRole{\mpS}{\roleR[0]}}
				{
					\mpSel
					{\mpy}
					{\roleRi}
					{\stLabi}
					{\mpChanRole{\mpS}{\roleR[1]}}
					{\mpNil}
				}
				\mpPar\\
				\mpSel
				{\mpyi}
				{\roleRi}
				{\stLab}
				{\mpChanRole{\mpS}{\roleR[2]}}
				{
					\mpSel
					{\mpyi}
					{\roleRi}
					{\stLabi}
					{\mpChanRole{\mpS}{\roleR[3]}}
					{\mpNil}
				}
				\end{array}\right)}
	}{\set{
	\begin{array}{l}
		\stEnvMap{\mpChanRole{\mpS}{\roleR[0]}}{\stEnd}
		\stEnvComp
		\stEnvMap{\mpChanRole{\mpS}{\roleR[1]}}{\stEnd}
		\stEnvComp\\
		\stEnvMap{\mpChanRole{\mpS}{\roleR[2]}}{\stEnd}
		\stEnvComp
		\stEnvMap{\mpChanRole{\mpS}{\roleR[3]}}{\stEnd}
		\stEnvComp\\
		\stEnvMap{\mpChanRole{\mpS}{\roleQ}}{\stFmt{\roleP{?\stChoice{\stLab}{\stT[\roleR]}\stSeq\stEnd}}}
		\stEnvComp\\
		\stEnvMap{\mpy}{\stT[\roleR]}
	\end{array}}
	}&\inferrule{PT-$\Sigma$}\\\\
	\tyJudge{}{
		\begin{array}{l}
		\mpSum{\mpBra
			{\mpChanRole{\mpS}{\roleQ}}
			{\roleP}
			{\stLab[i]}
			{\mpy}{
		\mpBra
			{\mpChanRole{\mpS}{\roleQ}}
			{\roleP}
			{\stLab}
			{\mpyi}{\ }}}{i\in\{0,1\}}
			\\\quad
			\left(
				\begin{array}{l}
				\mpSel
				{\mpy}
				{\roleRi}
				{\stLab}
				{\mpChanRole{\mpS}{\roleR[0]}}
				{
					\mpSel
					{\mpy}
					{\roleRi}
					{\stLabi}
					{\mpChanRole{\mpS}{\roleR[1]}}
					{\mpNil}
				}
				\mpPar\\
				\mpSel
				{\mpyi}
				{\roleRi}
				{\stLab}
				{\mpChanRole{\mpS}{\roleR[2]}}
				{
					\mpSel
					{\mpyi}
					{\roleRi}
					{\stLabi}
					{\mpChanRole{\mpS}{\roleR[3]}}
					{\mpNil}
				}
				\end{array}\right)\end{array}
	}{\set{
	\begin{array}{l}
		\stEnvMap{\mpChanRole{\mpS}{\roleR[0]}}{\stEnd}
		\stEnvComp
		\stEnvMap{\mpChanRole{\mpS}{\roleR[1]}}{\stEnd}
		\stEnvComp\\
		\stEnvMap{\mpChanRole{\mpS}{\roleR[2]}}{\stEnd}
		\stEnvComp
		\stEnvMap{\mpChanRole{\mpS}{\roleR[3]}}{\stEnd}
		\stEnvComp\\
		\stEnvMap{\mpChanRole{\mpS}{\roleQ}}{
			\stExtSum{\roleP}{i\in\{0,1\}}{
			\stChoice{\stLab[i]}{\stT[\roleR]}\stSeq
			\stFmt{\roleP{?\stChoice{\stLab}{\stT[\roleR]}\stSeq\stEnd}}}
		}
	\end{array}}
	}&\inferrule{PT-$\Sigma$}\\\\
	\tyJudge{}{
					\mpNil
	}{\set{
		\stEnvMap{\mpChanRole{\mpS}{\roleRi}}{\stEnd}
		\stEnvComp
		\stEnvMap{\mpy}{\stEnd}
		\stEnvComp
		\stEnvMap{\mpyi}{\stEnd}}
	}&\inferrule{PT-end}\\\\
	\tyJudge{}{
			\mpBra
				{\mpChanRole{\mpS}{\roleRi}}
				{\roleR}
				{\stLabi}
				{\mpyi}
				{
					\mpNil
				}
	}{\set{
		\stEnvMap{\mpChanRole{\mpS}{\roleRi}}{
			\roleR{?}\stChoice{\stLabi}{\stT[\roleR]}\stSeq\stEnd}
		\stEnvComp
		\stEnvMap{\mpy}{\stEnd}}}&\inferrule{PT-$\Sigma$}\\\\
	\tyJudge{}{
		\mpBra
		{\mpChanRole{\mpS}{\roleRi}}
		{\roleR}
		{\stLab}
		{\mpy}
		{
			\mpBra
				{\mpChanRole{\mpS}{\roleRi}}
				{\roleR}
				{\stLabi}
				{\mpyi}
				{
					\mpNil
				}
		}
	}{\set{
		\stEnvMap{\mpChanRole{\mpS}{\roleRi}}{
			\stFmt{\roleR{?}\stChoice{\stLab}{\stT[\roleR]}\stSeq
			\roleR{?}\stChoice{\stLabi}{\stT[\roleR]}\stSeq\stEnd}
		}
	}}
	&\inferrule{PT-$\Sigma$}
	\\\\
	\tyJudge{}{\mpP}{\stEnv}
\end{array}
\]
However, $\mpP$ reduces to an untypable process.
\[
\mpP\mpMove
	\begin{array}{l}
			\mpSel
			{\mpChanRole{\mpS}{\roleP}}
			{\roleQ}
			{\stLab}
			{\mpChanRole{\mpS}{\roleR}}
			{
				\mpNil
			}\mpPar\\
		{
			\mpBra
			{\mpChanRole{\mpS}{\roleQ}}
			{\roleP}
			{\stLab}
			{\mpyi}
			{
				\left(
				\begin{array}{l}
				\mpSel
				{\mpChanRole{\mpS}{\roleR}}
				{\roleRi}
				{\stLab}
				{\mpChanRole{\mpS}{\roleR[0]}}
				{
					\mpSel
					{\mpChanRole{\mpS}{\roleR}}
					{\roleRi}
					{\stLabi}
					{\mpChanRole{\mpS}{\roleR[1]}}
					{\mpNil}
				}
				\mpPar\\
				\mpSel
				{\mpyi}
				{\roleRi}
				{\stLab}
				{\mpChanRole{\mpS}{\roleR[2]}}
				{
					\mpSel
					{\mpyi}
					{\roleRi}
					{\stLabi}
					{\mpChanRole{\mpS}{\roleR[3]}}
					{\mpNil}
				}
				\end{array}
				\right)
			}
		}\mpPar\\
	\mpBra
		{\mpChanRole{\mpS}{\roleRi}}
		{\roleR}
		{\stLab}
		{\mpy}
		{
			\mpBra
				{\mpChanRole{\mpS}{\roleRi}}
				{\roleR}
				{\stLabi}
				{\mpyi}
				{
					\mpNil
				}
		}
	\end{array}
\]
This is untypable, since $\mpChanRole{\mpS}{\roleR}$
appears in two distinct parallel components.
Therefore, subject reduction fails.

\newpage

\subsubsection*{Open Access.}
This chapter is licensed under the terms of the Creative Commons Attribution-NonCommercial-NoDerivatives 4.0 International License
(\url{http://creativecommons.org/licenses/by-nc-nd/4.0/}), which permits any noncommercial use,
sharing, distribution and reproduction in any medium or format, as long as you give
appropriate credit to the original author(s) and the source, provide a link to the
Creative Commons license and indicate if you modified the licensed material. You do
not have permission under this license to share adapted material derived from this
chapter or parts of it.

The images or other third party material in this chapter are included in the
chapter’s Creative Commons license, unless indicated otherwise in a credit line to the
material. If material is not included in the chapter’s Creative Commons license and
your intended use is not permitted by statutory regulation or exceeds the permitted
use, you will need to obtain permission directly from the copyright holder.
\\[2ex]
\doclicenseImage

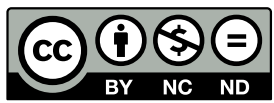
\end{document}

\endinput